\documentclass[runningheads,openany]{svmult}

\usepackage{palatino}
\usepackage{euler}
\usepackage{helvet}

\usepackage{graphicx}  
\usepackage{physprbb}  

\usepackage{proc}  
\usepackage{epsfig}

\usepackage{amssymb}
\usepackage{psfrag}
\usepackage{curves}
\usepackage{epic}
\usepackage{eepic}
\usepackage{rotating}

\makeindex             

\usepackage{amsmath}   
\DeclareMathOperator{\hc}{h.c.}
\DeclareMathOperator{\tr}{Tr}

\let\tocauthor\authorrunning

\begin{document}

\frontmatter

\thispagestyle{empty}
\parindent=0pt

{\Large\sc Blejske delavnice iz fizike \hfill Letnik~4, \v{s}t. 2--3}

\smallskip

{\large\sc Bled Workshops in Physics \hfill Vol.~4, No.~2--3}

\smallskip

\hrule

\hrule

\hrule

\vspace{0.5mm}

\hrule

\medskip
{\sc ISSN 1580--4992}

\vfill

\bigskip\bigskip
\begin{center}

{\bfseries 
{\Huge  Proceedings to the 
\smallskip
Euroconference on Symmetries 
\smallskip
Beyond the Standard Model}

\vspace{5mm}
{\Large  Portoro\v z, July 12 -- 17, 2003} 

\medskip
{\Huge (Part 2 of 2)}\vspace{10mm}}

\vfill

{\bfseries\large
Edited by

\vspace{5mm}
Norma Manko\v c Bor\v stnik\rlap{$^{1,2}$}

\smallskip

Holger Bech Nielsen\rlap{$^{3}$}

\smallskip

Colin D. Froggatt\rlap{$^{4}$}

\smallskip

Dragan Lukman\rlap{$^2$}

\bigskip

{\em\normalsize $^1$University of Ljubljana, $^2$PINT, %
$^3$ Niels Bohr Institute, $^4$ Glasgow University}

\vspace{12pt}

\vspace{3mm}

\vrule height 1pt depth 0pt width 54 mm}

\vspace*{3cm}

{\large {\sc  DMFA -- zalo\v{z}ni\v{s}tvo} \\[6pt]
{\sc Ljubljana, december 2003}}
\end{center}
\newpage

\thispagestyle{empty}
\parindent=0pt
\begin{flushright}
{\parskip 6pt
{\bfseries\large
                  The \textit{Euroconference on Symmetries Beyond  
                  the Standard Model},\\
                  12.--17. July 2003, Portoro\v z, Slovenia}

\bigskip\bigskip

{\bf was organized by}

{\parindent8pt
\textit{European Science Foundation, EURESCO office, Strassbourg}
}

\bigskip

{\bf and sponsored by}

{\parindent8pt
\textit{Ministry of Education, Science and Sport of Slovenia}

\textit{Luka Koper d.d., Koper, Slovenia}

\textit{Department of Physics, Faculty of Mathematics and Physics,
University of Ljubljana}

\textit{Primorska Institute of Natural Sciences and Technology, Koper}}

\bigskip
\medskip

{\bf Scientific Organizing Committee}

\medskip

{\parindent9pt
\textit{Norma Manko\v c Bor\v stnik (Chairperson), Ljubljana and Koper, Slovenia}

\textit{Holger Bech Nielsen (Vice Chairperson), Copenhagen, Denmark}

\textit{Colin D. Froggatt, Glasgow, United Kingdom}

\textit{Loriano Bonora, Trieste, Italy}

\textit{Roman Jackiw, Cambridge, Massachussetts, USA}

\textit{Kumar S. Narain, Trieste, Italy}

\textit{Julius Wess, Munich, Germany}

\textit{Donald L. Bennett, Copenhagen, Denmark}

\textit{Pavle Saksida, Ljubljana, Slovenia}

\textit{Astri Kleppe, Oslo, Norway}

\textit{Dragan Lukman, Koper, Slovenia} }}

\end{flushright}

\setcounter{tocdepth}{0}
\tableofcontents
\mainmatter
\parindent=20pt
\ \ \ \ \  \ 

\clearpage

\setcounter{page}{138}

\title*{Supersymmetric Grandunification and Fermion Masses}
\author{Borut Bajc}
\institute{%
Jo\v zef Stefan Institute, Jamova 39, 1001 Ljubljana, Slovenia\\
borut.bajc@ijs.si}

\titlerunning{Supersymmetric Grandunification and Fermion Masses}
\authorrunning{Borut Bajc}
\maketitle

\begin{abstract}
A short review of the status of supersymmetric grand unified 
theories and their relation to the issue of fermion masses and 
mixings is given. 
\end{abstract}

\section{Why Grandunification?}
\label{whygut}

There are essentially three reasons for trying to build grand unified 
theories (GUTs) beyond the standard model (SM).

\begin{itemize}
\item why should strong, weak and electromagnetic couplings 
in the SM be so different despite all corresponding to 
gauge symmetries?
\item there are many disconnected matter representations in the SM 
(3 families of $L$, $e^c$, $Q$, $u^c$, $d^c$)
\item quantization of electric charge (in the SM model there are two 
possible explanations - anomaly cancellation and existence of magnetic 
monopoles - both are naturally embodied in GUTs)
\end{itemize}

\section{How to check a GUT?}
\label{checkgut}

I will present here a very short review of some generic features, 
predictions and drawbacks of GUTs. Details of some topics will be 
given in the next section.

\subsection{Gauge coupling unification}

This is of course a necessary condition for any GUT to work. 
As is well known, the SM field content plus the desert assumption 
do not lead to the unification of the three gauge couplings. 
However, the idea of low-energy supersymmetry (susy), 
i.e. the minimal supersymmetric standard model (MSSM) instead 
of the SM at around TeV and again the assumption of the desert gives 
a quite precise unification of gauge couplings at 
$M_{GUT}\approx 10^{16}$ GeV \cite{Dimopoulos:1981yj}. 
Clearly there is no a-priori reason for three functions to cross 
in one point, so this fact is a strong argument for supersymmetry. 
One gets two bonuses for free in this case. First, the hierarchy 
problem gets stabilized, although not really solved, since 
the famous doublet-triplet (DT) problem still remains. 
Secondly, at least in principle one can get an insight into the 
reasons for the electroweak symmetry breaking: why the Higgs 
(other bosons in MSSM) mass squared is negative (positive) at 
low energy \cite{Inoue:1982pi}.

\subsection{Fermion masses and mixings}

Although GUTs are not theories of flavour, they bring 
constraints on the possible Yukawas. In the MSSM the 
Yukawa sector is given by 

\begin{equation}
W_Y=HQ^TY_Uu^c+\overline{ H}Q^TY_Dd^c+\overline{ H}L^TY_Ee^c\;,
\end{equation}

\noindent
and the complex $3\times 3$ generation matrices 
$Y_{U,D,E}$ are arbitrary. 
However, in a GUT the matter fields $Q$, $L$, $u^c$, $d^c$, $e^c$ 
fields live together in bigger representations, so one expects 
relations between quark and lepton Yukawa matrices. 

Take for example the SO(10) GUT. All the MSSM matter fields of each 
generation live in the same representation, the $16$ 
dimensional spinor representation, which contains and thus 
predicts also the right-handed neutrino. At the same time the minimal 
Higgs representation, the $10$ dimensional representation contains 
both doublets $H$ and $\overline{ H}$ of the MSSM (plus one color triplet 
and one antitriplet). The only renormalizable SO(10) invariant one can 
write down is thus 

\begin{equation}
W_Y=10_H16Y_{10}16\;,
\end{equation}

\noindent
which is however too restrictive, since it gives on top of the well 
satisfied (for large $\tan{\beta}$) relation $y_b=y_\tau=y_t$ for the 
third generation, the much worse predictions for the first two 
generations ($y_s=y_\mu=y_c$ and $y_d=y_e=y_u$) and no mixing 
($\theta_c=0$) at all. 

How to improve the fit? Let us mention two possibilities:

(1) Introduce new Higgs representations: although another $10_H$ can 
help with the mixing, the experimentally wrong relations $m_d=m_e$ 
and $m_s=m_\mu$ still occur, because the two bi-doublets in the two 
$10_H$ leave invariant the Pati-Salam SO(6)=SU(4)$_C$, so the leptons 
and quarks are still treated on the same footing. 
So the idea to pursue is to 
introduce bidoublets which transform nontrivially under the Pati-Salam 
SU(4) color. This can be done for example by 
introducing a $\overline{ 126}_H$, 
which couples to matter as $\Delta W_Y=\overline{ 126}_H16Y_{126}16$ and 
which gets a nontrivial vev in the $(2,2,15)_H$ SU(3) color singlet 
direction \cite{Lazarides:1980nt,Babu:1992ia}. 

(2) Another possibility is to include the effects of nonrenormalizable 
operators. These operators can cure the problem and at 
the same time ease the proton decay 
constraints. The drawback is the loss of predictivity. 

\subsection{Proton decay}

This issue is connected to 

(1) R-parity. It is needed to avoid fast proton decay. At the 
nonrenormalizable level one could for example have terms leading to 
R-parity violation of the type $16^316_H/M_{Pl}$. For this reason it is 
preferable to use the $126_H$ representation instead of the $16_H$. 
It is possible to show that such a SO(10) with $126_H$ has an exact 
R-parity \cite{Aulakh:1997ba} at all energies 
without the introduction of further symmetries. 

(2) DT splitting problem: Higgs SU(2)$_L$ doublets and 
SU(3)$_C$ triplets live usually in the same GUT multiplet; but 
while the SU(2)$_L$ doublets are light ($\approx M_W\ll M_{GUT}$), 
the SU(3)$_C$ triplets should be very heavy ($\ge M_{GUT}$) to 
avoid a too fast proton decay. For example, the proton lifetime in 
susy is proportional to $M_T^2$, which can give a lower limit 
to the triplet mass \cite{Hisano:1992jj}, although this limit depends 
on the yet unknown supersymmetry breaking sector \cite{Bajc:2002bv}. 

The solutions to the DT problem depend on the gauge group considered, 
but in general models that solve it are not minimal and necessitate 
of additional Higgs sectors. For example the missing partner mechanism 
\cite{Masiero:1982fe} in SU(5) needs at least additional 
$75_H$, $50_H$ and $\overline{ 50}_H$ representations. The same is 
true for the missing vev mechanism in SO(10) 
\cite{Dimopoulos:xm}, where the $45_H$ 
and extra $10_H$ Higgses must be introduced. Also the nice idea of 
GIFT (Goldstones Instead of Fine Tuning) \cite{Berezhiani:1995sb} 
can be implemented only by complicated models, while discrete 
symmetries for this purpose can be used with success only in 
connection with non-simple gauge groups \cite{Barbieri:1994jq}. 
Of course, although not very natural, any GUT can "solve" 
the problem phenomenologically, i.e. simply fine-tuning 
the model parameters. 

Clearly, whatever is the solution to the DT problem, the proton 
lifetime depends in a crucial (powerlike) way on the triplet mass. 
And this mass can be difficult to determine from the gauge coupling 
unification condition even in specific models 
because of the unknown model parameters 
\cite{Bachas:1995yt} or use of high representations 
\cite{Dixit:1989ff}. On top of this there can be large uncertainties 
in the triplet Yukawa couplings \cite{Dvali:1992hc}. 
All this, together with the phenomenologically completely unknown soft 
susy breaking sector, makes unfortunately proton decay not a very neat 
probe of supersymmetric grandunification \cite{Bajc:2002bv}.
Of course, if for some reason the DT mechanism is so efficient to make 
the $d=6$ operators dominant (for a recent analysis in some 
string-inspired models see \cite{Friedmann:2002ty}), 
then the situation could be simpler to analyse \cite{Mohapatra:yj}, 
although many uncertainties due to fermion mixing matrices still exist in 
realistic nonminimal models \cite{Nandi:1982ew}. 
Unfortunately there is little hope to detect proton decay in this case, 
unless $M_{GUT}$ is lower than usual \cite{Aulakh:2000sn}. 

\subsection{Magnetic monopoles}

Since magnetic monopoles 
are too heavy to be produced in colliders, the 
only hope is to find them as relics from the cosmological GUT phase 
transitions. Their density however strongly depends on the cosmological 
model considered. Unfortunately, the Rubakov-Callan effect 
\cite{Rubakov:fp} leads to the non observability 
of GUT monopoles, at least in any foreseeable future. Namely, these 
monopoles are captured by neutron stars and the resulting astrophysical 
analyses \cite{Freese:1983hz} limits the monopole flux at earth twelve 
orders of magnitude below the MACRO limit \cite{Giacomelli:2003yu}. 

This is very different from the situation in the Pati-Salam (PS) 
theory. Even in the minimal version the PS scale can be much lower 
than the GUT scale \cite{Melfo:2003xi}, as low as $10^{10}$ GeV. 
the resulting monopoles are then too light to be captured by 
neutron stars and their flux is not limited due to the Rubakov-Callan 
effect. Furthermore, MACRO results are not applicable for such light 
monopoles \cite{Giacomelli:2003yu}. 

\subsection{Low energy tests}

There are many different possible tests at low-energy, like for example 
the flavour changing neutral currents (see for example 
\cite{Barbieri:1994pv}) 
or the electric dipole moments \cite{Masina:2003iz}. In the latter 
case the exact value of the triplet mass is much less important than in 
proton decay, but the uncertainties due to the susy breaking sector 
are still present. In some of these tests like neutron-antineutron 
oscillation we can get positive signatures only for specific 
models due to very high dimensional 
operators involved \cite{Babu:2001qr}. 

\section{Fermion masses and mixings}

The regular pattern of 3 generations suggests some sort of flavour 
symmetry. 

One way, and the most ambitious one, is to consider the flavour 
symmetry group as part (subgroup) of the grand unified gauge 
symmetry (described by a simple group). In such an approach all 
three generations come from the same GUT multiplet. For example, 
in SU(8) the 216 dimensional representation gets 
decomposed under its SU(5) subgroup into three copies (generations) 
of $\overline{ 5}$ and $10$ with additional SU(5) multiplets. 
Similarly, in the SO(18) GUT, the 256 dimensional spinorial 
representation is nothing else than 8 generations of 
$(16+\overline{ 16})$ in the SO(10) language. The problem in all 
these theories is what to do with all the extra light particles 
\cite{Wilczek:1981iz}. 

Another possibility is to consider the product of the flavour 
(or, in general, extra) symmetry with the GUT symmetry (simple) 
group. In the context of SO(10) GUTs most of them use small 
representations for the Higgses, like $16_H$, $\overline{ 16}_H$ 
and $45_H$. The philosphy is to consider all terms allowed by 
symmetry, also nonrenormalizable. The DT problem can be naturally 
solved by some version of the missing vev mechanism, which however 
means that many multiplets are usually needed. Such models are quite 
successfull \cite{Albright:2000sz}, although the 
assumed symmetries are a little bit ad-hoc. There is also a huge 
number of different models with almost arbitrary flavour symmetry 
group, but unfortunately there is no room to describe them here 
(see for example the recent review \cite{Chen:2003zv}).

What we will consider in the following is instead a SO(10) GUT 
with no extra symmetry at all. We want to see how far we can go 
with just the grand unified gauge symmetry alone. To ensure automatic 
R-parity, we are forced not to use the $16_H$ and $\overline{ 16}_H$ 
Higgses, but instead a pair of $126_H$ and $\overline{ 126}_H$ 
(5 index antisymmetric representations, one self-dual, 
the other anti-self-dual; both of them are needed in order not to break 
susy at a large scale). In fact under R-parity the bosons of 
$16$ are odd, while those of $126$ are even, since 

\begin{equation}
R=(-1)^{3(B-L)+2S}
\end{equation}

\noindent
\cite{Mohapatra:su}, and the relevant vev in the SU(5) singlet 
directions have $B-L=1$ for $16_H$ ($\nu^c$), while it has 
$B-L=2$ for $126$ (the mass of $\nu^c$). 

So the rules of the game are: stick to renormalizable operators only, 
consider SO(10) as the only symmetry of the model, take the minimal 
number of multiplets (it does not mean the minimal number of fields!) 
that is able to give the correct symmetry breaking pattern
SO(10)$\to$SU(3)$\times$SU(2)$\times$U(1). Such a theory is 
given by \cite{Clark:ai} (see however \cite{Aulakh:1982sw} 
for a similar approach): on top of the usual three 
generations of $16$ dimensional matter fields, it contains four 
Higgs representations: $10_H$, $126_H$, $\overline{ 126}_H$ and
$210_H$ (4 index antisymmetric). It has been shown recently 
\cite{Aulakh:2003kg} that this theory is also the minimal GUT, 
i.e. it has the minimal number of model parameters, being still 
perfectly realistic (not in contradiction with any experiment). 

As we have seen, the $\overline{ 126}_H$ multiplet is needed both 
to help the $10_H$ multiplet in fitting the fermion masses and mixings, 
and for giving the mass to the right-handed neutrino without explicitly 
breaking R-parity. Let us now see, why the $210_H$ representation 
is needed.

The Yukawa sector is given by 

\begin{equation}
\label{wy}
W_Y=10_H16Y_{10}16+\overline{ 126}_H16Y_{126}16\;.
\end{equation}

The fields decompose under the 
SU(2)$_L\times$SU(2)$_R\times$SU(4)$_C$ subgroup as 

\begin{eqnarray}
10_H&=&(2,2,1)+(1,1,6)\;,\\
16&=&(2,1,4)+(1,2,\overline{ 4})\;,\\
\overline{ 126}_H&=&(1,3,10)+(3,1,\overline{ 10})+(2,2,15)+(1,1,6)\;.
\end{eqnarray}

The right-handed neutrino $\nu^c$ lives in $(1,2,\overline{ 4})$ of $16$, 
so it can get a large mass only through the second term in (\ref{wy}):

\begin{equation}
\label{mnr}
M_{\nu_R}=\langle (1,3,10)_{\overline{126}}\rangle Y_{126}\;,
\end{equation}

\noindent
where $\langle (1,3,10)\rangle$ is the scale of the SU(2)$_R$ 
symmetry breaking $M_R$, which we assume to be large, 
${\cal O}(M_{GUT})$. 

In order to get realistic masses we need 

\begin{eqnarray}
\langle (2,2,1)_{10}\rangle&=&
\begin{pmatrix}
   v_{10}^d
&  0
\cr   
   0
&  v_{10}^u
\cr   \end{pmatrix}
\ne 0\;,\\
\langle (2,2,15)_{\overline{126}}\rangle&=&
\begin{pmatrix}
   v_{126}^d
&  0
\cr   
   0
&  v_{126}^u
\cr   \end{pmatrix}\ne 0\;,
\end{eqnarray}

\noindent
which contribute to the light fermion masses as

\begin{eqnarray}
\label{mu}
M_U&=&v_{10}^uY_{10}+v_{126}^uY_{126}\;,\\
\label{md}
M_D&=&v_{10}^dY_{10}+v_{126}^dY_{126}\;,\\
\label{mnd}
M_{\nu_D}&=&v_{10}^uY_{10}-3v_{126}^uY_{126}\;,\\
\label{me}
M_E&=&v_{10}^dY_{10}-3v_{126}^dY_{126}\;.
\end{eqnarray}

The factor of $-3$ for leptons in the contribution from 
$\overline{126}_H$ comes automatically from the fact that 
the SU(3)$_C$ singlet in the adjoint $15$ of SU(4)$_C$ is in the 
$B-L$ direction diag$(1,1,1,-3)$. This is clearly absent in 
the contribution from $10_H$, which is a singlet under the full 
SU(4)$_C$. 

The light neutrino mass comes through the famous see-saw 
mechanism \cite{Mohapatra:1979ia}. From 

\begin{equation}
W={1\over 2}\nu^{cT}M_{\nu_R}\nu^c+\nu^{cT}M_{\nu_D}\nu_L+...
\end{equation}

\noindent
one can integrate out the heavy right-handed neutrino $\nu^c$ 
obtaining the effective mass term for the light neutrino states 
$M_N=-M_{\nu_D}^TM_{\nu_R}^{-1}M_{\nu_D}$. As we will now see, there 
is another contribution in our minimal model. 

(1) We saw that both $\langle (2,2,1)_{10}\rangle$ and 
$\langle (2,2,15)_{126}\rangle$ need to be nonzero and obviously 
${\cal O}(M_W)$. With $10_H$, $126_H$ and $\overline{126}_H$ 
Higgses one can write only two renormalizable invariants:

\begin{equation}
\label{wh}
W_H={1\over 2}M_{10}10_H^2+M_{126}126_H\overline{126}_H\;,
\end{equation}

\noindent
where $M_{10},M_{126}\approx {\cal O}(M_{GUT})$ or larger due 
to proton decay constraints. So the mass term looks like 

\begin{equation}
{1\over 2}(10_H,126_H,\overline{126}_H)
\begin{pmatrix}
   M_{10}
&  0
&  0
\cr 
   0  
&  0
&  M_{126}
\cr
   0
&  M_{126}
&  0
\cr   \end{pmatrix}
\begin{pmatrix}
   10_H
\cr
   126_H
\cr
   \overline{126}_H
\cr   \end{pmatrix}\;.
\end{equation}

Clearly all the doublets have a large positive mass, so their 
vev must be zero. Even fine-tuning cannot solve the DT problem in 
this case. So the idea to overcome this obstacle is to mix in some way 
$10_H$ with $\overline{126}_H$ ($126_H$), and after that fine-tune 
to zero one combination of doublet masses. So the new mass matrix 
should look something like 

\begin{equation}
\label{mxy}
\begin{pmatrix}
   M_{10}
&  x
&  y
\cr 
   x  
&  0
&  M_{126}
\cr
   y
&  M_{126}
&  0
\cr   \end{pmatrix}
\end{equation}

\noindent
with $x,y$ denoting such mixing. The light Higgs doublets will thus be 
linear combinations of the fields in $(2,2,1)_{10}$ and $(2,2,15)_{126_H,
\overline{126}_H}$ and this will get a nonzero vev after including the soft 
susy breaking masses. 

(2) The minimal representation that can mix $10$ and $\overline{126}$ 
is $210$, as can be seen from $10\times\overline{126}=210+1050$. 
$210$ is a 4 index antisymmetric SO(10) representation, which 
decomposes under the Pati-Salam subgroup as

\begin{equation}
210=(1,1,1)+(1,1,15)+(1,3,15)+(3,1,15)+
(2,2,6)+(2,2,10)+(2,2,\overline{10})\;.
\end{equation}

Of course one can now add other renormalizable terms to (\ref{wh}), 
and all such new terms are (in a symbolic notation)

\begin{equation}
\label{dwh}
\Delta W_H=210_H^3+210_H^2+210_H126_H\overline{126}_H+
210_H10_H126_H+210_H10_H\overline{126}_H\;.
\end{equation}

The last two terms are exactly the ones needed for the mixings 
between $10_H$ and $126_H$ ($\overline{126}_H$), i.e. contributions 
to $x,y$ in (\ref{mxy}). It is possible to show that $W_H+\Delta W_H$ 
are just enough for SO(10)$\to$SM. In the case of single-step breaking 
one thus has

\begin{eqnarray}
\langle (1,1,1)_{210}\rangle&\approx&
\langle (1,1,15)_{210}\rangle\approx
\langle (1,3,15)_{210}\rangle\approx\nonumber\\
\langle (1,3,\overline{10})_{126}\rangle&\approx&
\langle (1,3,10)_{\overline{126}}\rangle\approx
M_{GUT}\;.
\end{eqnarray}

(3) Now however there are five bidoublets that mix, 
since $(2,2,10)$ and $(2,2,\overline{10})$ from $210_H$ 
also contribute. To be honest, there is only one neutral 
component in each of these last two bidoublets, since 
their $B-L$ equals $\pm 2$, so the final mass matrix for 
the Higgs doublets is $4\times 4$. Only one eigenvalue of 
this matrix needs to be zero, and this can be achieved by 
fine-tuning. Each of the two Higgs doublets of the MSSM is 
thus a linear combination of 4 doublets, each of which has 
in general a vev of order ${\cal O}(M_W)$:

\begin{eqnarray}
\langle (2,2,1)_{10}\rangle&\approx&
\langle (2,2,15)_{\overline{126}}\rangle\approx
M_W\approx\\
\langle (2,2,15)_{126}\rangle&\approx&
\langle (2,2,10)_{210}\rangle\approx
\langle (2,2,\overline{10})_{210}\rangle\;.\nonumber
\end{eqnarray}

This mixing is nothing else than the susy version of 
\cite{Lazarides:1980nt,Babu:1992ia}.

(4) Due to all these bidoublet vevs, a SU(2)$_L$ triplet will 
also get a tiny but nonzero vev. Applying the susy constraint 
$F_{(3,1,10)_{126}}=0$ to 

\begin{equation}
W=M_{126}(3,1,10)_{126}(3,1,\overline{10})_{\overline{126}}+
(2,2,1)_{10}(2,2,\overline{10})_{210}(3,1,10)_{126}+...
\end{equation}

\noindent
one immediately gets

\begin{equation}
\langle (3,1,\overline{10})_{\overline{126}}\approx 
{\langle (2,2,1)_{10}\rangle 
\langle (2,2,\overline{10})_{210}\rangle\over
M_{126}}\approx {M_W^2\over M_{GUT}}\ne 0\;.
\end{equation}

This effect is just the susy version of 
\cite{Magg:1980ut}.

(5) Since $\nu$ lives in $(2,1,4)_{16}$, the second term in 
(\ref{wy}) gives among others also a term 
$(3,1,\overline{10})_{\overline{126}}(2,1,4)_{16}Y_{126}(2,1,4)_{16}$, 
which contributes to the light neutrino mass. So all together one gets 
for the light neutrino mass ($c$ is a 
model dependent dimensionless parameter)

\begin{equation}
\label{mn}
M_N=-M_{\nu_D}^TM_{\nu_R}^{-1}M_{\nu_D}+c{M_W^2\over M_{GUT}}Y_{126}\;.
\end{equation}

The first term is called the type I (or canonical) see-saw and is 
mediated by the SU(2)$_L$ singlet $\nu^c$, while the second is the 
type II (or non-canonical) see-saw, and is mediated by the SU(2)$_L$ 
triplet.

Equations (\ref{mu}), (\ref{md}), (\ref{mnd}), (\ref{me}), (\ref{mnr}) 
and (\ref{mn}) are all we need in the fit of known fermion masses and 
mixings and predictions of the unknown ones. A possible procedure is first 
to trade the matrices $Y_{10}$ and $Y_{126}$ for $M_U$ and $M_D$. The 
remaining freedom in $M_U$ and $M_D$ is still enough to fit $M_E$. 
But then some predictions in the neutrino sector are possible. For 
this sector we need to reproduce the experimental results 
$(\theta_l)_{12,23}\gg(\theta_q)_{12,23}$ and $(\theta_l)_{13}$ small. 
The degree of predictivity of the model however depends on the assumptions  
regarding the see-saw and on the CP phases.  

The first approach was to consider models in which type I dominates. 
It was shown that such models predict a small atmospheric neutrino 
mixing angle $\theta_{atm}=(\theta_l)_{23}$ if the CP phases are 
assumend to be small \cite{Babu:1992ia,Lee:1994je}. On the other hand, 
a large atmospheric neutrino mixing angle can be also large, if one 
allows for arbitrary CP phases and fine-tune them appropriately 
\cite{Matsuda:2001bg}.

A completely different picture emerges if one assumes that type II 
see-saw dominates. In this case even without CP violation one 
can naturally have a large atmospheric neutrino mixing angle, as 
has been first emphasized for the approximate case of second and 
third generations only in \cite{Bajc:2001fe,Bajc:2002iw}. In the three 
generation case the same result has been confirmed \cite{Goh:2003sy}. 
On top of this, a large solar neutrino mixing angle and a 
prediction of $U_{e3}\approx 0.15\pm 0.01$ (close to the 
upper experimental limit) have been obtained \cite{Goh:2003sy}.
Even allowing for general CP violation does not invalidate the 
above results: although the error bars are larger, the general 
picture of large atmospheric and solar neutrino mixing angles 
and small $U_{e3}$ still remains valid \cite{Goh:2003hf}.

It is possible to understand why type II see-saw gives so 
naturally a large atmospheric mixing angle. In type II the 
light neutrino mass matrix (\ref{mn}) is proportional to $Y_{126}$. 
From (\ref{md}) and (\ref{me}) one can easily find out, that 
$Y_{126}\propto M_D-M_E$, from which one gets \cite{Brahmachari:1997cq}

\begin{equation}
M_N\propto M_D-M_E\;.
\end{equation}

As a warm-up let us take the approximations of just 
(a) two generations, the second and the third, 
(b) neglect the masses of the second generation with 
respect of the third ($m_{s,\mu}\ll m_{b,\tau}$) and 
(c) assume that $M_D$ and $M_E$ has small mixings (this amounts 
to say, that in the basis of diagonal charged lepton mass, the smallness 
of the $({\theta_q})_{23}=\theta_{cb}$ is not caused by accidental 
cancellation of two large numbers). In this approximate set-up one gets 

\begin{equation}
M_N\propto
\begin{pmatrix}
   0
&  0
\cr   
   0
&  m_b-m_\tau
\cr   \end{pmatrix}\;.
\end{equation}

This is, in type II see-saw there is a correspondence between 
the large atmospheric angle and $b-\tau$ unification \cite{Bajc:2002iw}.

Remember here that $b-\tau$ Yukawa unification is no more automatic, 
since we have also $\overline{126}_H$ Higgs on top of the usual $10_H$. 
It is however quite well satisfied phenomenologically.

One can do better: still take $m_{s,\mu}\approx 0$, but allow a 
nonzero quark mixing. In this case the atmospheric mixing angle is

\begin{equation}
\tan{2\theta_{atm}}={\sin{2\theta_{cb}}\over 
2\sin^2{\theta_{cb}}-{m_b-m_\tau\over m_b}}\;.
\end{equation}

Since $\theta_{cb}\approx {\cal O}(10^{-2})$, one again finds out 
the correlation between the large atmospheric mixing angle and 
$b-\tau$ unification at the GUT scale. 

The result can be confirmed of course also for finite $m_{s,\mu,c}$, 
although not in a so simple and elegant way.

Of course there are many other models that predict and/or explain a 
large atmospheric mixing angle (for a recent review see for example 
\cite{King:2003jb}). What is surprising here is, however, that no 
other symmetry except the gauge SO(10) is needed whatsoever.

\section{The minimal model}

As we saw in the previos section, one can correctly fit the known 
masses and mixings, get some understanding of the light neutrino 
mass matrix, and obtain some new predictions. What we would like 
to show here is that the model considered above has less number 
of model parameters than any other GUT, and can be then called 
the minimal realistic supersymmetric grand unified theory 
(even more minimal than SU(5)!) \cite{Aulakh:2003kg}. 

The Higgs sector described by (\ref{wh}) and (\ref{dwh}) contains 
$10$ real parameters ($7$ complex parameters minus four phase 
redefinitions due to the four complex Higgs multiplets involved). 
The Yukawa sector (\ref{wy}) has two complex symmetric matrices, 
one of which can be always made diagonal and real by a unitary 
transformation of $16$ in generation space. So what remains are 
$15$ real parameters. There is on top of this also the gauge 
coupling, so all together 26 real parameters in the supersymmetric 
sector of renormalizable SO(10) GUT with three copies of matter 16 and 
Higgses in the representations $10_H$, $126_H$, $\overline{126}_H$ and 
$210_H$. We will not consider the susy breaking sector, since this is 
present in all supersymmetric theories, GUTs or not.

Before comparing with other GUTs, for example SU(5), let us count 
the number of model parameters in MSSM. There are 6 quark masses, 
3 quark mixing angles, 1 quark CP phase, 6 lepton masses, 3 lepton 
mixing angles and 3 lepton CP phases (assuming Majorana neutrinos). 
On top of this, there are 3 gauge couplings and the real $\mu$ parameter. 
Thus, all together, again 26 real parameters. They are however distributed 
differently, so that in the Yukawa sector there are only 15 parameters 
in our minimal SO(10) GUT, which has to fit 22 MSSM (at least in principle) 
measurable low-energy parameters. Although in this fitting also few 
vevs that contain parameters from the Higgs and susy breaking sector 
play a role, the minimal SO(10) is nevertheless predictive.

One can play with other SO(10) models: the renormalizable ones need 
more representations and thus have more invariants, while the 
nonrenormalizable ones (those that use $16_H$ instead of $126_H$) have 
a huge number of invariants, some of which must be very small due to 
R-symmetry constraints. Of course, with some extra discrete, global or 
local symmetry, one can forbid these unpleasant and dangerous terms, 
remaining even with a small number of parameters, but as we said, this 
is not allowed in our scheme, in which we want to obtain as much 
information as possible just from GUT gauge symmetry (and 
renormalizability).

The simplicity of the minimal renormalizable supersymmetric SU(5) 
looks as if the number of parameters here could be smaller than in 
our previous example. What however gives a large number of parameters is 
the fact, that SU(5) is not particularly suitable for the neutrino sector. 
In fact, one can play and find out, that the minimal SU(5) with nonzero 
neutrino masses is obtained adding the two index symmetric $15_H$ and 
$\overline{15}_H$, and the number of model parameters comes out to be 
39, i.e. much more than in the minimal SO(10). 

\section{Conclusion}

Before talking about flavour symmetries it is important first to know, 
what we can learn from just pure GUTs. The minimal GUT is a SO(10) 
gauge theory with representations $10_H$, $126_H$, $\overline{126}_H$, 
$210_H$ and three generations of $16$. Such a realistic theory is 
renormalizable and no extra symmetries are needed. It can fit the 
fermion masses and mixings, and can give an interesting relation 
between $b-\tau$ Yukawa unification and large atmospheric mixing angle. 
It has a testable prediction for $U_{e3}$. Due to the large 
representations involved, it is not asymptotically free, which means 
that it predicts some new physics below $M_{Pl}$. 

There are many virtues of this minimal GUT. As in any SO(10) all 
fermions of one generation are in the same representation and 
the right-handed neutrino is included automatically, thus 
explaining the tiny neutrino masses by the see-saw mechanism. 
Employing $126_H$ instead of $16_H$ to break $B-L$ maintains 
R-symmetry exact at all energies. It is economical, it employs 
the minimal number of multiplets and parameters, and thus it is 
maximally predictive. It gives a good fit to available data and gives 
a framework to better understand the differences between the mixings in 
the quark and lepton sectors.

There are of course also some drawbacks. First, in order to maintain 
predictivity, one must believe in the principle of renormalizability, 
although the suppressing parameter in the expansion $M_{GUT}/M_{Planck}$ 
is not that small. Of course, in supersymmetry these terms can be small 
and stable, but this choice is not natural in the 't Hooft sense. 
Second, the DT splitting problem is here, and 
attempts to solve it require more fields \cite{Lee:1993jw}. 
Finally, usually it is said that 
$126$ dimensional representations are not easy to get from superstring 
theories, although we are probably far from a no-go theorem. 

There are many open questions to study in the context of the minimal 
SO(10), let me mention just few of them. 
First, proton decay: although it is generically dangerous, it is 
probably still possible to fit the data with some fine-tuning of 
the model parameters as well as of soft susy breaking terms. An 
interesting question is whether the model is capable of telling us 
which type of see-saw dominates. If it is type I or mixed, can it 
still give some testable prediction for $U_{e3}$? Also, gauge 
coupling unification should be tested in some way, although large 
threshold corrections could be nasty \cite{Dixit:1989ff}. And finally, 
is there some hope to solve in this context or minimal (but still 
predictive) extensions the doublet-triplet splitting problem? 

\section*{Acknowledgements}
It is a pleasure to thank the organizers for the well organized and 
stimulating conference. I am grateful to Charan Aulakh, 
Pavel Fileviez Perez, Alejandra Melfo, Goran Senjanovi\' c and Francesco 
Vissani for a fruitful collaboration. 
I thank Goran Senjanovi\' c also for carefully reading the 
manuscript and giving several useful advises and suggestions. 
This work has been supported by the Ministry of 
Education, Science, and Sport of the Republic of Slovenia.

\newcommand{\MPdd}{\mbox{\rm d}}
\newcommand{\MPDD}{\mbox{\rm D}}

\title*{General Principles of Brane Kinematics and Dynamics}
\author{Matej Pav\v si\v c}
\institute{%
Jo\v zef Stefan Institute, Jamova 39,
1000 Ljubljana, Slovenia\\e-mail: matej.pavsic@ijs.si}

\titlerunning{General Principles of Brane Kinematics and Dynamics}
\authorrunning{Matej Pav\v si\v c}
\maketitle

\begin{abstract}

We consider branes as ``points" in an infinite dimensional brane
space ${\cal M}$ with a prescribed metric. Branes move along
the geodesics of ${\cal M}$. For a particular choice of metric
the equations of motion are equivalent to the well known
equations of the Dirac-Nambu-Goto branes (including strings).
Such theory describes ``free fall" in ${\cal M}$-space.
In the next step the metric of ${\cal M}$-space is given the
dynamical role and a corresponding kinetic term is added to the
action. So we obtain a background independent brane theory:
a space in which branes live is ${\cal M}$-space and it is not
given in advance, but comes out as a solution to the equations of
motion. The embedding space (``target space") is not separately
postulated. It is identified with the brane configuration.

\end{abstract}

\section{Introduction}

Theories of strings and higher dimensional extended objects, branes,
are very promising in explaining the origin and interrelationship of
the fundamental interactions, including gravity. But there is a cloud.
It is not clear what is a geometric principle behind string and brane
theories and how to formulate them in a background independent way.
An example of a background independent theory is general relativity
where there is no preexisting space in which the theory is formulated.
The dynamics of the 4-dimensional space (spacetime) itself results as
a solution to the equations of motion. The situation is sketched in Fig.1. 
A point particle traces
a world line in spacetime whose dynamics is governed by the Einstein-Hilbert
action. A closed string traces a world tube, but so far its has not been
clear what is the appropriate space and action for a background independent
formulation of string theory.

\begin{figure}[ht]
\setlength{\unitlength}{.4mm}
\begin{picture}(120,110)(-110,15)

\put(40,105){\line(0,-1){88}}
\put(53,78){$I[g_{\mu \nu}] = \int  \MPdd^4 x \, \sqrt{-g}\, R $}
\put(53,36){?}

\thicklines

\spline(14,67)(12,73)(15,82)(17,93)(15,101)

\spline(9,22)(7,28)(10,37)(12,48)(10,56)
\spline(19,22)(17,28)(19,37)(22,48)(20,56)
\closecurve(10,56, 12,54.5, 15,54.5, 17,55, 20,56, 17,57.5, 
        15,57.9, 12,57.5, 10,56)
\spline(9,22)(12,20.5)(14,20.3)(16,20.5)(19,22)

\end{picture}

\caption{%
To point particle there corresponds the 
Einstein-Hilbert action
in spacetime. What is a corresponding space and action for a closed string?}
\end{figure}
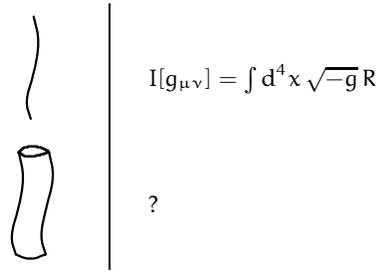

Here I will report about a formulation of string and brane theory
(see also ref. \cite{book}) which is based on the infinite dimensional brane 
space ${\cal M}$. The ``points" of this space are branes and their
coordinates are the brane (embedding) functions. In ${\cal M}$-space we
can define the distance, metric, connection, covariant derivative,
curvature, etc. We show that the brane dynamics can be derived from the
principle of minimal length in ${\cal M}$-space; a brane follows a geodetic
path in ${\cal M}$. The situation is analogous to the free fall of an
ordinary point particle as described by general relativity. Instead of keeping
the metric fixed, we then add to the action a kinetic term for the metric of
${\cal M}$-space and so we obtain a background independent brane
theory in which there is no preexisting space.

\section{Brane space ${\cal M}$ (brane kinematics)}

We will first treat the brane kinematics, and only later we will introduce
a brane dynamics. We assume that the basic kinematically possible objects
are $n$-dimen\-sional, arbitrarily deformable branes ${\cal V}_n$ living
in an $N$-dimensional embedding (target) space $V_N$. Tangential deformations
are also allowed. This is illustrated in Fig.\,2. Imagine a rubber sheet
on which we paint a grid of lines. Then we deform the sheet in such
a way that mathematically the surface remains the same, nevertheless
the deformed object is physically different from the original object.

\begin{figure}[ht]
\centering
\includegraphics[width=60mm]{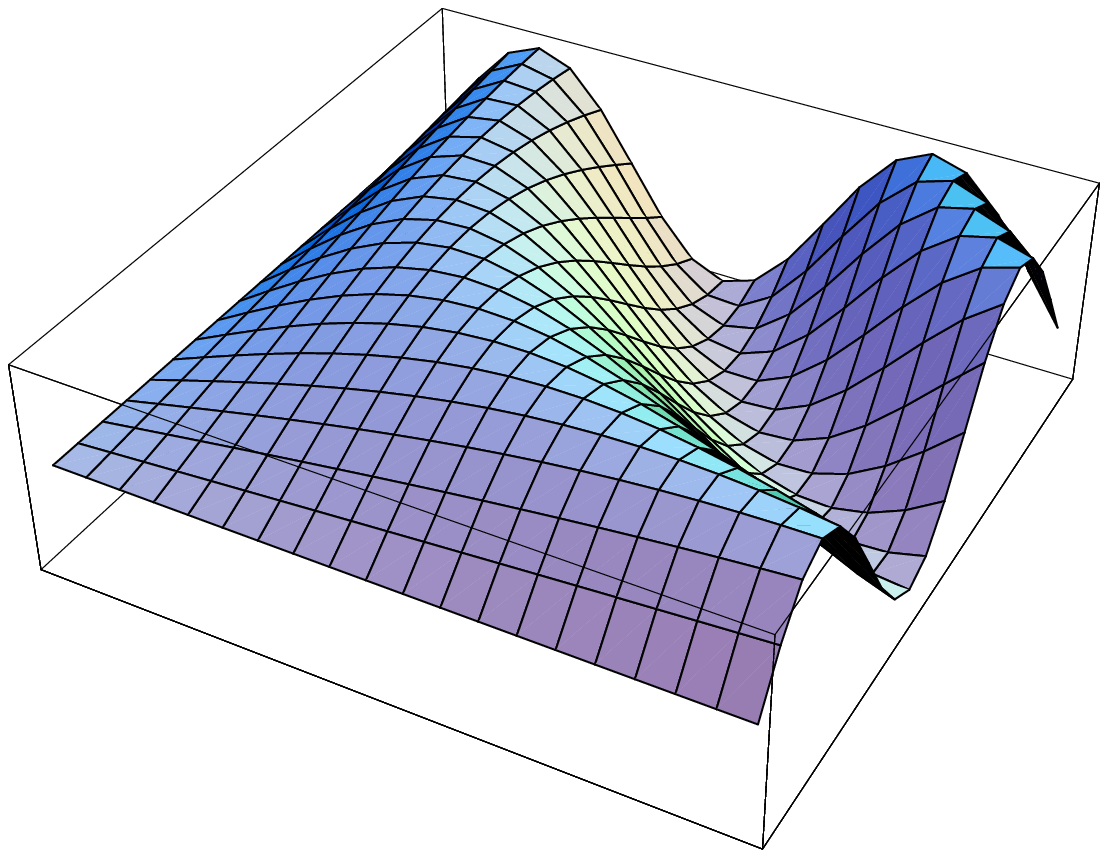}\hspace{-1cm}
\includegraphics[width=60mm]{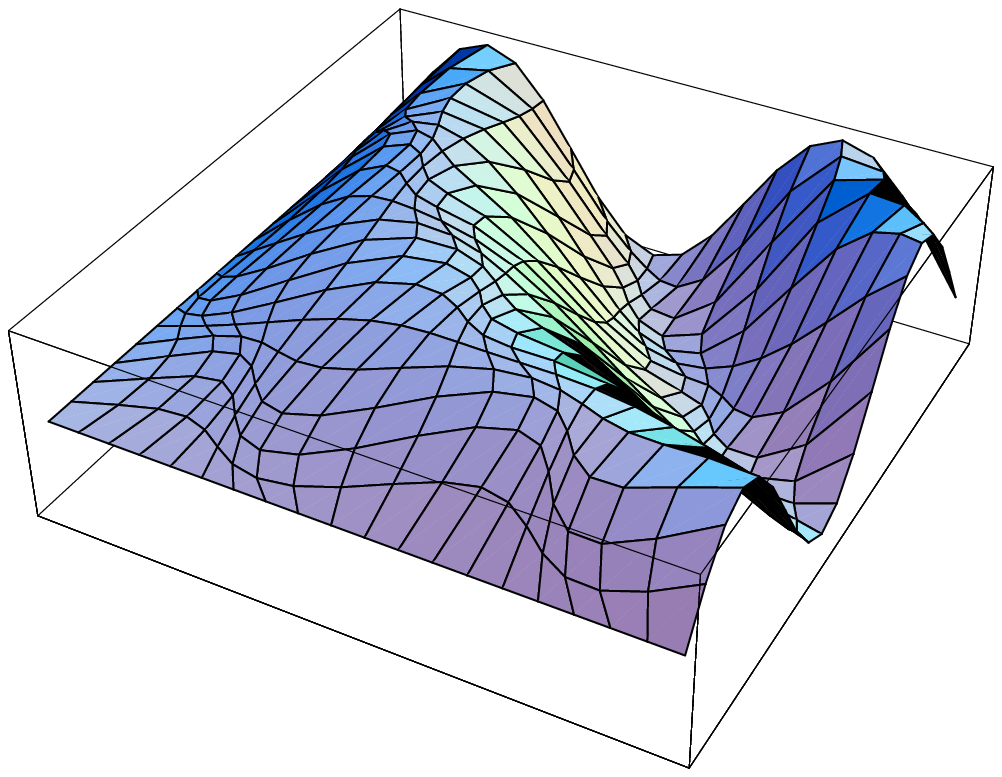}
\caption{%
Examples of tangentially deformed membranes. 
Mathematically the surface on the left and is the same as the surface
on the right. Physically the two surfaces are different.}
\end{figure} 

We represent ${\cal V}_n$ by functions $X^\mu (\xi^a)~, ~~\mu = 0,1,...,N-1$,
where $\xi^a,~a=0,1,2,...,n-1$ are parameters on ${\cal V}_n$.
According the assumed interpretation, different functions $X^\mu (\xi^a)$
can represent physically different branes. That is, if we perform
an {\it active diffeomorphism} $\xi^a \rightarrow \xi'^a = f^a (\xi)$, then
the new functions $X^\mu (f^a (\xi)) = X'^\mu (\xi)$ represent a physically
different brane ${\cal V}'_n$. For a more complete and detailed discussion
see ref. \cite{book}.

The set of all possible
${\cal V}_n$ forms {\it the brane space} ${\cal M}$. A brane ${\cal V}_n$
can be considered as a point in ${\cal M}$ parametrized by coordinates
$X^\mu (\xi^a) \equiv X^{\mu (\xi)}$ which bear a discrete index $\mu$ and
$n$ continuous indices $\xi^a$. That is,
$\mu(\xi)$ as superscript or
subscript denotes a single index which consists of the discrete part $\mu$
and the continuous part $(\xi)$.

In analogy with the finite-dimensional case we can introduce the {\it distance}
$\MPdd \ell$ in the infinite-dimensional space ${\cal M}$:
\begin{equation}
    {\MPdd} {\ell}^2 = \int {\MPdd} \xi \, {\MPdd} \zeta \, {\rho}_{\mu \nu}
    (\xi,\zeta) \,
    {\MPdd} X^{\mu} (\xi) \, {\MPdd} X^{\nu} (\zeta) \nonumber \\
     = {\rho}_{\mu(\xi )
    \nu(\zeta)} \, 
    {\MPdd} X^{\mu (\xi)} \, {\MPdd} X^{\nu (\zeta)}  ,
\label{mp1}
\end{equation}
where ${\rho}_{\mu \nu} (\xi ,\zeta) \equiv {\rho}_{\mu(\xi)\nu(\zeta)}$
is the metric in
${\cal M}$. Let us consider a particular choice of metric
\begin{equation}
       {\rho}_{\mu(\xi) \nu (\zeta)} = \sqrt{|f|} \, \alpha \, g_{\mu \nu}
       \delta (\xi - \zeta) ,
\label{mp2}
\end{equation}
where $f \equiv {\rm det} \, f_{ab}$ is the determinant of the
induced metric $f_{ab} \equiv {\partial}_a X^{\alpha} {\partial}_b X^{\beta} \, 
g_{\alpha \beta}$
on the sheet $V_n$, whilst $g_{\mu \nu}$ is the metric
tensor of the embedding space $V_N$, and $\alpha$ an arbitrary function
of $\xi^a$ or, in particular, a constant. Then the line element (\ref{mp1})
becomes
\begin{equation}
    {\MPdd} {\ell}^2 = \int  {\MPdd} \xi \sqrt{|f|} \, \alpha \, 
    g_{\mu \nu} \, {\MPdd} X^\mu (\xi) {\MPdd} X^\nu (\xi) .
\label{mp3}
\end{equation}

The invariant volume (measure) in ${\cal M}$ is
\begin{equation}
   \sqrt{|\rho|} {\cal D} X = {\left ( \mbox{\rm Det} \, {\rho}_{\mu \nu} (\xi,\zeta) 
    \right )}
    ^{1/2} \prod_{\xi,\mu} {\MPdd} X^{\mu} (\xi) .
\label{mp4}
\end{equation}
Here Det denotes a continuum determinant taken over $\xi ,\zeta$
as well as over $\mu, \nu$. In the case of the diagonal metric (\ref{mp2})
we have
\begin{equation}
     \sqrt{|\rho|}  {\cal D} X = \prod_{\xi ,\mu} \left ( \sqrt{|f|} \, \alpha 
       \, |g|\right ) ^{1/2}
       {\MPdd} X^{\mu} (\xi)
\label{mp5}
\end{equation}

Tensor calculus in ${\cal M}$-space is analogous to that in a finite dimensional
space. The differential of coordinates ${\MPdd} X^{\mu} (\xi)
\equiv {\MPdd} X^{\mu(\xi)}$ is a vector in ${\cal M}$. The coordinates
$X^{\mu(\xi)}$ can be transformed into new coordinates
${X '}^{\mu(\xi)}$ which are functionals of $X^{\mu(\xi)}$ :
\begin{equation}
       {X'}^{\mu (\xi)} = F^{\mu (\xi)} [X] .
\label{mp6}
\end{equation}
If functions $X^{\mu} (\xi)$ represent a brane ${\cal V}_n$, then
functions $X'^{\mu} (\xi)$ obtained from $X^{\mu} (\xi)$
by a functional transformation represent the same (kinematically possible)
brane. 

Under a general coordinate transformation (\ref{mp6}) a generic
vector $A^{\mu(\xi)} \equiv A^{\mu} (\xi)$ transforms as\footnote{
A similar formalism, but for a specific type of the functional
transformations, namely the reparametrizations which
functionally depend on string coordinates, was developed by
Bardakci \cite{Bardakci} }
\begin{equation}
      A^{\mu(\xi)} = {{\partial {X'}^{\mu(\xi)}} \over {\partial X^{\nu(\zeta)}}}
     A^{\nu (\zeta)} \equiv \int {\MPdd} \zeta 
     {{\delta {X'}^{\mu} (\xi)} \over {\delta X^{\nu} (\zeta)}} 
     A^{\nu} (\zeta)\, ,
\label{mp7}
\end{equation}
where $\delta/\delta X^{\mu} (\xi)$ denotes
the functional derivative. 

Similar transformations hold for a covariant vector $A_{\mu(\xi)}$,
a tensor $B_{\mu(\xi)\nu(\zeta)}$, etc.. Indices are
lowered and raised, respectively, by ${\rho}_{\mu(\xi)\nu(\zeta)}$
and ${\rho}^{\mu(\xi)\nu(\zeta)}$, the latter being
the inverse metric tensor satisfying
\begin{equation}
     {\rho}^{\mu(\xi) \alpha (\eta)} {\rho}_{\alpha (\eta) \nu (\zeta)} =
     {{\delta}^{\mu(\xi)}}_{\nu(\zeta)} .
\label{mp8}
\end{equation}

As can be done in a finite-dimensional space, we can also 
define the {\it covariant
derivative} in ${\cal M}$. When acting on a {\it scalar} $A[X (\xi)]$
the covariant derivative coincides with the ordinary
functional derivative:
\begin{equation}
    A_{;\mu(\xi)} = {{\delta A} \over {\delta X^{\mu} (\xi)}}
    \equiv A_{,\mu(\xi)} .
\label{9}
\end{equation}
But in general a geometric object in ${\cal M}$ is a tensor of 
arbitrary rank, 
$${A^{{\mu}_1 (\xi_1) \mu_2 (\xi_2)...}}
_{\nu_1 (\zeta_1) \nu_2 (\zeta_2)...},$$
which is a functional of $X^{\mu} (\xi)$, and its covariant derivative
contains the affinity ${\Gamma}_{\nu(\zeta)\sigma(\eta)}^{\mu(\xi)}$
composed of the metric $\rho_{\mu (\xi) \nu (\xi')}$
\cite{Pavsic1}. For instance, when acting
on a vector the covariant derivative gives
\begin{equation} 
    {A^{{\mu}(\xi)}}_{; \nu(\zeta)} = {A^{\mu(\xi)}}_{, \nu(\zeta)}
    + {\Gamma}_{\nu(\zeta)\sigma(\eta)}^{\mu(\xi)}
    A^{\sigma(\eta)}
\label{10}
\end{equation}
In a similar way we can write the covariant derivative acting on a tensor
of arbitrary rank.

In analogy to the notation as employed in the finite dimensional
tensor calculus we can use the following variants of notation for the
ordinary and covariant derivative:
\begin{eqnarray}
     &&{{\delta } \over {\delta X^{\mu} (\xi)}} \equiv
     {{\partial} \over {\partial X^{\mu (\xi)}}} \equiv {\partial}_{\mu(\xi)} \quad 
     \mbox{for functional derivative} \nonumber \\
     &&{{\MPDD} \over {{\MPDD} X^{\mu} (\xi)}} \equiv
     {{\MPDD} \over {{\MPDD} X^{\mu (\xi)}}} \equiv {\MPDD}_{\mu(\xi)} \quad
     \mbox{for covariant derivative in ${\cal M}$}
\label{11}
\end{eqnarray}
Such  shorthand notations for functional derivative is very effective.
\section{Brane dynamics: brane theory as free fall in ${\cal M}$-space}

So far we have considered kinematically possible branes as the points in
the brane space ${\cal M}$. Instead of one brane we can consider a one
parameter family of branes $X^\mu (\tau,\xi^a) \equiv X^{\mu (\xi)} (\tau)$,
i.e., a curve (or trajectory) in ${\cal M}$. Every trajectory is kinematically
possible in principle. A particular dynamical theory then
selects which amongst those kinematically possible branes and 
trajectories are also dynamically possible. We will assume that dynamically
possible trajectories are {\it geodesics} in ${\cal M}$ described by
the minimal length action \cite{book}:
\begin{equation}
     I[X^{\alpha (\xi)}] = \int {\MPdd} \tau' \left (\rho_{\alpha (\xi')
     \beta (\xi'')} {\dot X}^{\alpha (\xi')} {\dot X}^{\beta (\xi'')} 
     \right )^{1/2} .
\label{12}
\end{equation}

Let us introduce the shorthand notation
\begin{equation}
          \mu \equiv  \rho_{\alpha (\xi') \beta (\xi'')}
          {\dot X}^{\alpha (\xi')} {\dot X}^{\beta (\xi'')}
\label{13}
\end{equation}
and vary the action (\ref{12}) with respect to $X^{\alpha (\xi)} (\tau)$. 
If the expression for the metric $\rho_{\alpha (\xi') \beta (\xi'')}$ does not
contain the velocity ${\dot X}^{\mu}$ we obtain
\begin{equation}
     {1\over \mu^{1/2}}{{\MPdd} \over {{\MPdd} \tau}} 
     \left ({{{\dot X}^{\mu (\xi)}}\over \mu^{1/2}} \right ) + 
     {\Gamma^{\mu (\xi)}}_{\alpha (\xi') \beta (\xi'')}
     {{{\dot X}^{\alpha (\xi')} {\dot X}^{\beta (\xi'')}}\over \mu} = 0
\label{16}
\end{equation}
which is a straightforward generalization of the usual geodesic equation from a
finite-dimensional space to an infinite-dimensional ${\cal M}$-space.

Let us now consider a particular choice of the ${\cal M}$-space metric:
\begin{equation}
     \rho_{\alpha (\xi') \beta (\xi'')} = \kappa {{\sqrt{|f(\xi')|}} \over
     {\sqrt{{\dot X}^2 (\xi')}} } \, \delta (\xi' - \xi'') \eta_{\alpha \beta}
\label{17}
\end{equation}
where ${\dot X}^2 \equiv g_{\mu \nu} {\dot X}^\mu {\dot X}^\nu$ is the
square of velocity ${\dot X}^\mu$. Therefore, the metric (\ref{17})
depends on velocity. If we insert it into the action (\ref{12}), then
after performing the functional derivatives and the integrations over
$\tau$ and $\xi^a$ (implied in the repeated indexes 
$\alpha (\xi')$, $\beta (\xi'')$) we obtain the following equations of motion:
\begin{equation}
    {{\MPdd}\over {{\MPdd} \tau}} \left ({1\over {\mu^{1/2}}} {{\sqrt{|f|}}\over
    {\sqrt{{\dot X}^2}}} \, {\dot X}_{\mu} \right ) + {1\over {\mu^{1/2}}}
    \partial_a \left ( \sqrt{|f|} \sqrt{{\dot X}^2} \partial^a X_{\mu} \right ) = 0
\label{18}
\end{equation}
If we take into account the relations
\begin{equation}
    {{{\MPdd} \sqrt{|f|}}\over {{\MPdd} \tau}} = {{\partial \sqrt{|f|} } \over {\partial f_{ab}}}\,
    {\dot f}_{ab} = \sqrt{|f|} \, f^{ab} \partial_a {\dot X}^{\mu} \partial_b X_{\mu} =
    \sqrt{|f|} \, \partial^a X_{\mu} \partial_a {\dot X}^{\mu}
\label{19}
\end{equation}
and
\begin{equation}
     {{{\dot X}_{\mu}} \over {\sqrt{{\dot X}^2}}}             
    {{{\dot X}^{\mu}} \over {\sqrt{{\dot X}^2}}} = 1 \quad \Rightarrow \quad
    {{\MPdd}\over {{\MPdd} \tau}} \left ( {{{\dot X}_{\mu}} \over {\sqrt{{\dot X}^2}}}
    \right ) {\dot X}^{\mu} = 0 
\label{20}
\end{equation}
it is not difficult to find that
\begin{equation}
{{{\MPdd} \mu} \over {{\MPdd} \tau}} = 0
\label{21}
\end{equation}
Therefore, instead of (\ref{18}) we can write
\begin{equation}
    {{\MPdd} \over {{\MPdd} \tau}} \left ( {{\sqrt{|f|}}\over {\sqrt{{\dot X}^2}}} \, 
    {\dot X}_{\mu} \right ) +  
    \partial_a \left ( \sqrt{|f|} \sqrt{{\dot X}^2} \partial^a X_{\mu} \right ) = 0 .
\label{22}
\end{equation} 
This are precisely the equation of motion for the Dirac-Nambu-Goto brane,
written in a particular gauge.

The action (\ref{12}) is by definition invariant under reparametrizations
of $\xi^a$. In general, it is not invariant under reparametrization of
the parameter $\tau$. If the expression for the metric
$\rho_{\alpha (\xi') \beta(\xi'')}$ does not contain the velocity
${\dot X}^{\mu}$, then the action (\ref{12}) is invariant under
reparametrizations of $\tau$. This is no longer true if $\rho_{\alpha (\xi') \beta(\xi'')}$
contains ${\dot X}^{\mu}$. Then the action (\ref{12}) is not
invariant under reparametrizations of $\tau$.

In particular, if metric is given by eq.\,(\ref{17}), then the action becomes
explicitly
\begin{equation}
     I[X^\mu (\xi)] = \int \MPdd \tau \, \left ( 
     \MPdd \xi \, \kappa \, \sqrt{|f|}\, 
     \sqrt{{\dot X}^2} \right )^{1/2}
\label{23}
\end{equation}
and the equations of motion (\ref{18}), as we have seen,  automatically
contain the relation
\begin{equation}
     {{\MPdd}\over{{\MPdd} \tau}} \left ( {\dot X}^{\mu (\xi)} {\dot X}_{\mu (\xi)}
     \right )
     \equiv {{\MPdd}\over{{\MPdd} \tau}} \int {\MPdd} \xi \, \kappa \sqrt{|f|}
     \sqrt{{\dot X}^2} = 0 .
\label{24}
\end{equation}
The latter relation is nothing but {\it a gauge fixing relation},
where by ``gauge" we mean here a choice of parameter $\tau$. The action
(\ref{12}), which in the case of the metric (\ref{17}) is not
reparametrization invariant, contains the gauge fixing term.

In general the exponent in the Lagrangian is not necessarily
${1\over 2}$, but can be arbitrary:
\begin{equation}
     I[X^{\alpha (\xi)}] = \int {\MPdd} \tau \, \left (
     \rho_{\alpha (\xi') \beta(\xi'')} {\dot X}^{\alpha (\xi')}
     {\dot X}^{\beta (\xi'')} \right )^a .
\label{25}
\end{equation}
For the metric (\ref{17}) we have explicitly
\begin{equation}
     I[X^\mu (\xi)] = \int \MPdd \tau \, \left ( 
     \MPdd \xi \, \kappa \, \sqrt{|f|}\, 
     \sqrt{{\dot X}^2} \right )^a
\label{26}
\end{equation}
The corresponding equations of motion are
\begin{equation}
    {{\MPdd}\over{{\MPdd} \tau}} \left ( a \mu^{a-1} {{\kappa \sqrt{|f|}}\over
    {\sqrt{{\dot X}^2}}} \, {\dot X}_{\mu} \right ) + a \mu^{a - 1} \partial_a
    \left ( \kappa \sqrt{|f|} \sqrt{{\dot X}^2} \partial^a X_{\mu} \right ) = 0 .
\label{27}
\end{equation}

We distinguish two cases: 

\ (i) $a \neq 1$. Then the action 
 is {\it not} invariant under reparametrizations of
$\tau$. The equations of motion (\ref{27}) for $a\neq 1$ imply
the gauge fixing relation $\MPdd \mu/\MPdd \tau = 0$, that is, the relation
(\ref{24}). 

(ii) $a=1$. Then the action (\ref{26}) is invariant under 
reparametrizations of $\tau$. The equations of motion for $a=1$ contain 
no gauge fixing term. In both cases, (i) and (ii), we obtain the same
equations of motion
(\ref{22}).

Let us focus our attention to the action with $a=1$:
\begin{equation}
     I[X^{\alpha (\xi)}] = \int {\MPdd} \tau \, \left (
     \rho_{\alpha (\xi') \beta(\xi'')} {\dot X}^{\alpha (\xi')}
     {\dot X}^{\beta (\xi'')} \right ) = 
     \int \MPdd \tau \, \MPdd \xi \, \kappa \, \sqrt{|f|}\, 
     \sqrt{{\dot X}^2}
\label{28}
\end{equation}
It is invariant under the transformations
\begin{eqnarray}
     &&\tau \rightarrow \tau' = \tau' (\tau)  \label{29} \\
      &&\xi^a \rightarrow \xi'^a = \xi'^a (\xi^a) \label{30}
\end{eqnarray}
in which $\tau$ and $\xi^a$ do not mix.

Invariance of the action (\ref{28}) under reparametrizations (\ref{29})
of the evolution parameter $\tau$ implies the existence of a constraint
among the canonical momenta $p_{\mu (\xi)}$ and coordinates $X^{\mu (\xi)}$.
Momenta are given by
\begin{eqnarray}
    p_{\mu (\xi)} &=& {{\partial L}\over {\partial {\dot X}^{\mu (\xi)}}} = 
    2 \rho_{\mu (\xi)
    \nu (\xi')} {\dot X}^{\nu (\xi')} + 
    {{\partial \rho_{\alpha (\xi') \beta (\xi'')}} \over
    {\partial {\dot X}^{\mu (\xi)}}} {\dot X}^{\alpha (\xi')} {\dot X}^{\beta 
    (\xi'')} \nonumber \\    
     &=& {{\kappa \sqrt{|f|}}\over {\sqrt{{\dot X}^2}}} \, {\dot X}_{\mu} .
\label{31}
\end{eqnarray}
By distinguishing covariant and contravariant components one finds
\begin{equation}
    p_{\mu (\xi)} = {\dot X}_{\mu (\xi)} = \rho_{\mu (\xi) \nu (\xi')}
    {\dot X}^{\nu (\xi')} \; , \quad 
    p^{\mu (\xi)} = {\dot X}^{\mu (\xi)}  .
\label{32}
\end{equation}
We define $p_{\mu (\xi)} \equiv p_{\mu} (\xi) \equiv p_{\mu} \; ,
\quad {\dot X}^{\mu (\xi)} \equiv {\dot X}^{\mu} (\xi) \equiv {\dot X}^{\mu}$.
Here $p_{\mu}$ and ${\dot X}^{\mu}$ have the meaning of the usual finite
dimensional vectors whose components are lowered and raised by
the finite-dimensional metric tensor $g_{\mu \nu}$ and its inverse
$g^{\mu \nu}$: $p^{\mu} = g^{\mu \nu} p_{\nu} \; , 
\quad {\dot X}_{\mu} = g_{\mu \nu} {\dot X}^{\nu}$.

The {\it Hamiltonian} belonging to the action (\ref{28}) is
\begin{equation}
   H = p_{\mu (\xi)} {\dot X}^{\mu (\xi)} - L = 
  \int \MPdd \xi \, {{\sqrt{{\dot X}^2}}\over {\kappa \sqrt{|f|} }} \,
    (p^{\mu} p_{\mu} - \kappa^2 |f|) = p_{\mu (\xi)} p^{\mu (\xi)} - K = 0
\label{35}
\end{equation}
where $K = K[X^{\mu (\xi)}] = \int \MPdd \xi \, \kappa \, \sqrt{|f|}\, 
    \sqrt{{\dot X}^2} = L$.
It is identically zero. The ${\dot X}^2$ entering the integral for $H$ is 
arbitrary due to arbitrary reparametrizations of $\tau$ (which change
${\dot X}^2$). Consequently, $H$ vanishes when the following expression
under the integral vanishes:
\begin{equation}
      p^{\mu} p_{\mu} - \kappa^2 |f| = 0
\label{37}
\end{equation}
Expression (\ref{37}) is the usual constraint
for the Dirac-Nambu-Goto brane ($p$-brane).
It is satisfied at every $\xi^a$.

In ref.\,\cite{book} it is shown that the constraint is conserved in $\tau$ and
that as a consequence we have
\begin{equation}
     p_{\mu} \partial_a X^{\mu} = 0 .
\label{38}
\end{equation}
The latter equation is yet another set of constraints\footnote{
Something similar happens in canonical gravity. Moncrief and Teitelboim
\cite{Moncrief}
have
shown that if one imposes the Hamiltonian constraint on the Hamilton
functional then the momentum constraints are automatically satisfied.}
which are satisfied at any point $\xi^a$ of
the brane world manifold $V_{n+1}$.

Both kinds of constraints  are thus automatically implied by the
action (\ref{28}) in which the choice  (\ref{17}) of ${\cal M}$-space
metric tensor has been taken.

Introducing a more compact notation $\phi^A = (\tau,\xi^a)$ and
$X^{\mu (\xi)} (\tau) \equiv X^\mu (\phi^A) \equiv X^{\mu (\phi)}$
we can write
\begin{equation}
     I[X^{\mu (\phi)}] = \rho_{\mu (\phi) \nu (\phi')} {\dot X}^{\mu (\phi)}
     {\dot X}^{\nu (\phi')} = \int \MPdd^{n+1} \phi \, \sqrt{|f|} \, 
     \sqrt{{\dot X}^2}
\label{39}
\end{equation}
where
\begin{equation}
    \rho_{\mu (\phi') \nu (\phi'')} = \kappa {{\sqrt{|f(\xi')|}} \over
     {\sqrt{{\dot X}^2 (\xi')}} } \, \delta (\xi' - \xi'') 
     \delta (\tau' - \tau'') \eta_{\mu \nu} 
\label{40}
\end{equation}

Variation of the action (\ref{39}) with respect to $X^{\mu (\phi)}$ gives
\begin{equation}
{{{\MPdd} {\dot X}^{\mu (\phi)}} \over {\MPdd \tau}} + \Gamma_{\alpha (\phi')
    \beta (\phi'')}^{\mu (\phi)}
     {\dot X}^{\alpha (\phi')} {\dot X}^{\beta (\phi'')} = 0
\label{41}
\end{equation}
which is the geodesic equation in the space ${\cal M}_{V_{n+1}}$
of brane world
manifolds $V_{n+1}$ described by $X^{\mu (\phi)}$. For simplicity we will
omit the subscript and call the latter space ${\cal M}$-space as well.

Once we have the constraints we can write the first order or phace space
action
\begin{equation}
    I[X^{\mu},p_{\mu},\lambda,\lambda^a] = \int {\MPdd} \tau \, {\MPdd} \xi \,
    \left ( p_{\mu} {\dot X}^{\mu} - {\lambda \over {2 \kappa \sqrt{|f|}}}
    (p^{\mu} p_{\mu} - \kappa^2 |f|) - \lambda^a p_{\mu} \partial_a X^{\mu}
    \right ) ,
\label{42}
\end{equation}
where $\lambda$ and $\lambda^a$ are Lagrange multipliers. It is classically
equivalent to the {\it minimal surface action} for the $(n+1)$-dimensional
world manifold $V_{n+1}$
\begin{equation}
      I[X^{\mu}] = \kappa \int {\MPdd}^{n+1} \phi \, ({\rm det} \, 
      \partial_A X^{\mu} \partial_B X_{\mu})^{1/2} .
\label{43}
\end{equation}
This is the conventional Dirac--Nambu--Goto action, invariant under
reparametrizations of $\phi^A$.

\section{Dynamical metric field in ${\cal M}$-space}

Let us now ascribe the dynamical role to the ${\cal M}$-space metric.
>From ${\cal M}$-space perspective we have motion of a point ``particle"
in the presence of a metric field $\rho_{\mu (\phi) \nu (\phi')}$ which is
itself dynamical.

As a model let us consider the action
\begin{equation}
    I[\rho] = \int {\cal D} X \sqrt{|\rho |} \, \left 
    ( \rho_{\mu (\phi) \nu (\phi')} 
    {\dot X}^{\mu (\phi)} {\dot X}^{\nu (\phi')} +
    {\epsilon \over {16 \pi}} {\cal R} \right )  .
\label{45}
\end{equation}
where $\rho$ is the determinant of the metric $\rho_{\mu (\phi) \nu (\phi')}$
and $\epsilon$ a constant.
Here ${\cal R}$ is the Ricci scalar in ${\cal M}$-space, defined according to
${\cal R} = \rho^{\mu (\phi) \nu (\phi')} {\cal R}_{\mu (\phi) \nu (\phi')}$,
where ${\cal R}_{\mu (\phi) \nu (\phi')}$ is the Ricci tensor in 
${\cal M}$-space \cite{book}.

Variation of the action (\ref{45}) with respect to $X^{\mu (\phi)}$ and
$\rho_{\mu (\phi) \nu (\phi')}$ leads to (see ref.\cite{book}) the {\it geodesic
equation} (\ref{41})
and to the {\it Einstein equations} in ${\cal M}$-space
\begin{equation}
    {\dot X}^{\mu (\phi)} {\dot X}^{\nu (\phi)} + {\epsilon \over {16 \pi}}
    {\cal R}^{\mu (\phi) \nu (\phi')} = 0
\label{48}
\end{equation}
In fact, after performing the variation we had a term with ${\cal R}$ and
a term with ${\dot X}^{\mu (\phi)} {\dot X}_{\mu (\phi)}$ in the Einstein equations.
But, after performing the contraction with the metric, we find that the two
terms cancel each other resulting in the simplified equations (\ref{48})
(see ref.\cite{book}).

The metric $\rho_{\mu (\phi) \nu (\phi')}$ is a functional of the
variables $X^{\mu (\phi)}$ and in eqs.\,(\ref{41}),(\ref{48}) we have a
system of functional differential equations which determine the set
of possible solutions for $X^{\mu (\phi)}$ and 
$\rho_{\mu (\phi) \nu (\phi')}$. Our brane model (including strings) is
background independent: there is no preexisting space with a preexisting
metric, neither curved nor flat.

We can imagine a model universe consisting of a single brane. Although we
started from a brane embedded in a higher dimensional finite space, we
have subsequently arrived at the action(\ref{45}) in which the dynamical
variables $X^{\mu (\phi)}$ and $\rho_{\mu (\phi) \nu (\phi')}$ are defined
in ${\cal M}$-space.  
In the latter model the concept of an underlying
finite dimensional space, into which the brane is embedded, is in fact
abolished. We keep on talking about ``branes" for convenience reasons,
but actually there is no embedding space in this model. 
The metric $\rho_{\mu (\phi) \nu (\phi')} [X]$ is defined only on the brane.
There is no metric of a space into which the brane is embedded. Actually,
there is no embedding space. All what exists is a brane configuration
$X^{\mu (\phi)}$ and the corresponding metric $\rho_{\mu (\phi) \nu (\phi')}$
in ${\cal M}$-space.

\paragraph{A system of branes (a brane configuration)}
Instead of a single brane we can consider a system of branes described
by coordinates $X^{\mu (\phi,k)}$, where $k$ is a discrete index
that labels the branes (Fig.\,3). If we replace $(\phi)$ with $(\phi,k)$, or,
alternatively, if we interpret $(\phi)$ to include the index $k$, then
the previous action (\ref{45}) and equations of motion (\ref{41}),(\ref{48})
are also valid for a system of branes.

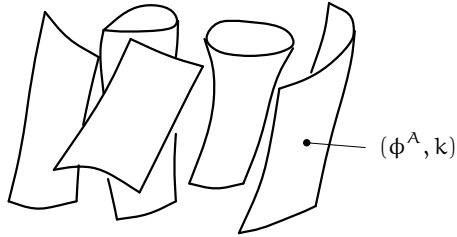
\begin{figure}[ht]
\setlength{\unitlength}{.6mm}
\begin{picture}(120,50)(-36,15)

\put(80,35){\circle*{1.5}}
\put(80,35){\line(12,-1){13}}
\put(96,32.6){$(\phi^A,k)$}

\thicklines
\spline(14,22)(20,20)(25,21)(29,22)
\spline(29,22)(29.5,30)
\spline(30.5,39)(32,46)(34,54)
\spline(34,54)(27,58)(22,64)
\spline(22,64)(19,50)(17,35)(14,22)

\spline(24,29)(30,30)(35,29)(40,26)(43,24)
\spline(43,24)(46,30)(52,42)(57,52)
\spline(57,52)(50,54)(41,58)
\spline(41,58)(35,45)(24,29)

\spline(35,60)(40,59.5)(50,60)(52,62)(50,65)(45,66)(40,64.5)(35,60)
\spline(35,60)(34,55)(35,46)
\spline(36,27)(37,23)(38,18)
\spline(38,18)(45,19)(51,22)
\spline(51,22)(50.5,30)(51,37)
\spline(51,54)(51.5,58)(51.5,62)

\spline(65,62)(60,62)(57,58)(60,55)(65,54)(70,54)(75,55)(78,57)(75,61)(65,62)
\spline(57.4,58)(60,50)(59,39)(54,26)
\spline(54,26)(60,24.5)(66,26)
\spline(66,26)(70,36)(73,49)(77,56)

\spline(74,47)(80,49)(85,52)(90,57)(91,60)(90,64)(85,66)
\spline(85,66)(84,59)(81,50.6)
\spline(74,47)(72,35)(70,26)(65,15)
\spline(65,15)(70,16)(75,18)(80,20)(82,21)
\spline(82,21)(85,35)(90,52)(90.7,59.8)

\end{picture}

\caption{%
The system of branes is represented as being 
embedded in a finite-dimensional
space $V_N$. The concept of a continuous 
embedding space is only an approximation
which, when there are many branes, becomes good 
at large scales (i.e., at the
``macroscopic" level). The metric is defined 
only at the points $(\phi ,k)$ situated on
the branes. At large scales (or equivalently, 
when the branes are ``small"
and densely packed together) the set of all the 
points $(\phi ,k)$ is a good approximation
to a continuous metric space $V_N$.}

\end{figure}

A brane configuration is all what exists in such a model. It is identified
with the embedding space\footnote{Other authors also considered a class of
brane theories in which the
embedding space has no prior existence, but is instead coded completely
in the degrees of freedom that reside on the branes. They take for
granted that, as the background is not assumed to exist, there are no
embedding coordinates (see e.g., \cite{MPSmolin}). This seems to be
consistent with our usage of $X^{\mu (\phi)}$ which, at the fundamental
level, are not considered as the embedding coordinates, but as the
${\cal M}$-space coordinates. Points of ${\cal M}$-space are
described by coordinates $X^{\mu (\phi)}$, and the distance between
the points is determined by the metric $\rho_{\mu (\phi) \nu (\phi')}$,
which is dynamical.. In the limit of infinitely many densely packed
branes, the set of points $(\phi^A,k)$ is supposed to become
a continuous, finite dimensional
metric space $V_N$.}.

\paragraph{From ${\cal M}$-space to spacetime}
We now define ${\cal M}$-space as the space of all possible brane
configurations. Each brane configuration is considered as a point
in ${\cal M}$-space described by coordinates $X^{\mu (\phi,k)}$.
The metric $\rho_{\mu (\phi,k) \nu (\phi',k')}$ determines the {\it distance}
between two points {\it belonging to two different brane configurations}:
\begin{equation}
     \MPdd \ell^2 = \rho_{\mu (\phi,k) \nu (\phi',k')}
     \MPdd X^{\mu (\phi,k)} \MPdd X^{\nu (\phi',k')}
\label{49}
\end{equation}
where
\begin{equation}
       \MPdd X^{\mu (\phi,k)} = X'^{\mu (\phi ,k)} - X^{\mu (\phi,k)} .
\label{50}
\end{equation}

Let us now introduce another quantity which connects two different 
points, in the usual sense of the word, 
{\it within the same brane configuration}:
\begin{equation}
    {\widetilde \Delta} X^{\mu} (\phi ,k) \equiv X^{\mu (\phi' , k')} - 
    X^{\mu (\phi ,k)} .
\label{51}
\end{equation}
and define
\begin{equation}
        \Delta s^2 = \rho_{\mu (\phi ,k) \nu (\phi' , k')} 
        {\widetilde \Delta} X^{\mu}
    (\phi ,k) {\widetilde \Delta} X^{\nu} (\phi' , k') .
\label{52}
\end{equation}
In the above formula summation over the repeated indices $\mu$ and $\nu$ is
assumed, but no integration over $\phi$, $\phi'$ and no summation over 
$k$, $k'$.

Eq.(\ref{52}) denotes the distance between the points within a given
brane configuration. This is the quadratic form in the skeleton
space $S$. The metric $\rho$ in the skeleton space $S$ is the prototype of
the metric in target space $V_N$ (the embedding space). A brane
configuration is a skeleton $S$ of a target space $V_N$.

\section{Conclusion}

We have taken the brane space ${\cal M}$ seriously as an arena for
physics. The arena itself is also a part of the dynamical system, it is not
prescribed in advance. The theory is thus background independent. It is
based on a geometric principle which has its roots in the brane space
${\cal M}$. We can thus complete the picture that occurred in the
introduction:

\begin{figure}[ht]
\setlength{\unitlength}{.4mm}
\begin{picture}(120,90)(-80,15)

\put(40,105){\line(0,-1){88}}
\put(53,78){$I[g_{\mu \nu}] = \int  \MPdd^4 x \, \sqrt{-g}\, R $}
\put(53,36){$I[\rho_{\mu (\phi) \nu (\phi')}] = \int {\cal D} X \, 
       \sqrt{|\rho|} \, {\cal R}$ }

\thicklines

\spline(14,67)(12,73)(15,82)(17,93)(15,101)

\spline(9,22)(7,28)(10,37)(12,48)(10,56)
\spline(19,22)(17,28)(19,37)(22,48)(20,56)
\closecurve(10,56, 12,54.5, 15,54.5, 17,55, 20,56, 17,57.5, 
        15,57.9, 12,57.5, 10,56)
\spline(9,22)(12,20.5)(14,20.3)(16,20.5)(19,22)

\end{picture}
\caption{%
Brane theory is formulated in ${\cal M}$-space. The action
is given in terms of the ${\cal M}$-space curvature scalar ${\cal R}$.}
\end{figure}

We have formulated a theory in which an embedding space {\it per se}
does not exist, but is intimately connected to the existence of branes
(including strings). Without branes there is no embedding space. There is
no preexisting space and metric: they appear dynamically as solutions
to the equations of motion. Therefore the model is background independent.

All this was just an introduction into a generalized theory of branes.
Much more can be found in a book \cite{book} where the description with a metric
tensor has been surpassed. Very promising is the description in terms
of the Clifford algebra equivalent of the tetrad which simplifies
calculations significantly. The relevance of
the concept of Clifford space for physics is discussed in refs.\,\cite{book}, 
\cite{Castro}--\cite{CliffordMankoc}).

There are possible connections to other topics. 
The system, or condensate of branes (which, in particular, may be so dense
that the corresponding points form a continuum), represents a 
{\it reference system}
or {\it reference fluid} with respect to which 
the points of the target space are
defined. Such a system was postulated by DeWitt \cite{DeWittReference},
and recently 
reconsidered by Rovelli \cite{RovelliReference} in relation to the
famous Einstein's `hole argument' according to which the
points of spacetime cannot be identified. The brane model
presented here can also be related to
the {\it Mach principle} according to which the motion of 
matter at a given location 
is determined by the contribution of all the matter in the universe and this
provides an explanation for inertia (and inertial mass). 
Such a situation is implemented
in the model of a universe consisting of a system of branes described
by eqs.\,(\ref{41}),(\ref{48}): 
the motion of a $k$-th
brane, including its inertia (metric), is determined by the presence of
all the other branes.

\section*{Acknowledgement}
This work has been supported by the Ministry
of Education, Science and Sport of Slovenia under the contract PO-0517.

\title*{Cosmological Neutrinos}
\author{Gianpiero Mangano}
\institute{%
INFN, Sezione di Napoli, \\
Dipartimento di Scienze Fisiche, Universita' di Napoli Federico
II, Italy}

\titlerunning{Cosmological Neutrinos}
\authorrunning{Gianpiero Mangano}
\maketitle

\begin{abstract}
We review present information on cosmological neutrinos, and more
generally on relativistic degrees of freedom at the Cosmic
Microwave Background formation epoch, in view of the recent
results of WMAP collaborations on temperature anisotropies of the
CMB, as well as of recent detailed analysis of Primordial
Nucleosynthesis.
\end{abstract}

\section{Introduction}
\setcounter{equation}0 \noindent We are pretty confident that our
Universe is presently filled with quite a large number of
neutrinos, of the order of $100$ $cm^{-3}$ for each flavor,
despite of the fact that there are no direct evidences for this
claim and, more sadly, it will be also very hard to achieve this
goal in the future. However several stages of the evolution of the
Universe have been influenced by neutrinos, and their $silent$
contribution has been first communicated to other species via weak
interactions, and eventually through their coupling with gravity.
In fact, Big Bang Nucleosynthesis (BBN), the Cosmic Microwave
Background (CMB) and the spectrum of Large Scale Structure (LSS)
keep traces of their presence, so that by observing the power
spectrum $P(k)$, the photon temperature-temperature angular
correlation, and primordial abundances of light nuclei, we can
obtain important pieces of information on several features of the
neutrino background, as well as on some fundamental parameters,
such as their mass scale. It is astonishing, at least for all
those of the elementary particle community who moved to
"astroparticle" physics, to see that in fact the present bound on
the neutrino mass , order 1 $eV$, obtained by studying their
effect on suppressing structure formation at small scales, is
already stronger than the limit obtained in terrestrial
measurement from $^3H$ decay.

In this short lecture I briefly review some of the cosmological observables
which indeed lead to relevant information on both dynamical (number
density, chemical potential) and kinematical (masses) neutrino properties,
as well as on extra weakly coupled light species.

\section{Cosmological neutrinos: standard features}
\setcounter{equation}0
\noindent

For large temperatures neutrinos are kept in thermodynamical
equilibrium with other species, namely $e^--e^+$ and nucleons,
which in turn share the very same temperature of photons because
of electromagnetic interactions. The key phenomenon for
cosmological neutrinos is that for temperatures of the order of
$T_d \sim 1 MeV$ weak interactions become unable to sustain
equilibrium, since the corresponding effective rate $\Gamma_w$
(cross-section $\sigma_w$ times electron number density $n_e$)
falls below the expansion rate $H$, the Hubble parameter. We can
in fact estimate $\sigma_w \sim G_F^2 T^2$ and $n_e \sim T^3$, so
that $\Gamma_w =G_F^2 T^5$, and since for a radiation dominated
universe $H \sim \sqrt{G_N} T^2$, with $G_N$ the Newton
constant\footnote{I adopt the standard unit system $\hbar=c=k=1$}
it is straightforward to get \begin{equation} T_d \sim
\frac{G_N^{1/6}}{G_F^{2/3}} \sim 1 MeV \label{Td} \end{equation} From this
epoch on, neutrinos freely stream with an (almost) perfect
Fermi-Dirac distribution, the one they had at decoupling, while
momentum red-shifts as expansion goes on. In terms of the comoving
momentum $y=k a$, with $a$ the scale factor, \begin{equation} dn_\nu =
a^{-3}\frac{1}{e^{y}+1} \frac{d^3y}{(2 \pi)^3} \label{fnu} \end{equation}
Actually neutrinos are slightly heated up during the $e^--e^+$
annihilation phase, which takes place at temperatures of the order
of the electron mass and release entropy mainly to photons, but
also to neutrinos. This is because the neutrino decoupling is not
an instantaneous phenomenon, but it partially overlaps the
$e^--e^+$ annihilation phase. A detailed analysis, which also
takes into account QED plasma effects on the $e^--e^+$ pairs
\cite{mmpp} shows that neutrino distribution is slightly different
than a pure black body distribution, and the corresponding energy
density differ from the instantaneous decoupling result at the
level of percent.

It is customary to parameterize the contribution $\rho_R$ of
relativistic species to the expansion rate of the universe in
terms of the $effective$ neutrino number $N_{eff}$ defined as
follows \begin{equation} \rho_R = \rho_\gamma+\rho_\nu +\rho_X = \left[ 1+
\frac{7}{8}\left(\frac{4}{11}\right)^{4/3} N_{eff}\right]
\rho_\gamma \label{neff} \end{equation} with $\rho_\gamma$, $\rho_\nu$ and
$\rho_X$ the energy density of photons, neutrinos and of extra
(unknown) light species, respectively. The two factors $7/8$ and
$(4/11)^{4/3}$ are due to the difference in the form of
equilibrium distribution (Bose-Einstein for photons, Fermi-Dirac
for neutrinos) and to the different temperature of photons and
neutrinos after $e^+-e^-$ pair annihilation. With this definition,
three massless neutrinos with a pure equilibrium distribution and
zero chemical potential give $N_{eff}=3$. In view of the partial
neutrino reheating from $e^+-e^-$ the actual value is slightly
larger $N_{eff}=3.04$. The role of this parameter is crucial in
our understanding of fundamental physics. Any result in favor of a
larger (or a smaller) value for $N_{eff}$ would imply some exotic
non standard physics at work in the early universe. In the
following Sections we will see how this parameter is in fact
constrained by some crucial cosmological observables, such as CMB
or BBN.

\section{$\Omega_b h^2$ and $N_{eff}$ after WMAP}
\setcounter{equation}0
\noindent

The peak structure of the CMB power spectrum has been beautifully
confirmed by a series of experiment in the past few years
(BOOMERanG, MAXIMA, DASI, CBI, ACBAR) and more recently by WMAP
collaboration \cite{GPMwmap} with a very high accuracy. The
improvement in angular resolution from the $7$ degrees across the
sky of COBE to the order 0.5 degrees of WMAP, allows us to have a
better understanding of several features of our Universe and in
particular of its matter-energy content in terms of cosmological
constant, dark matter and baryons.

The role of relativistic species at the CMB formation, at
redshifts of the order of $z \sim 1100$, is mainly to shift the
matter radiation equality time, which results in both shifting the
peak location in angular scale and changing the power around the
first acoustic peak. This is basically due to a change in the
early integrated Sachs-Wolfe effect. Several groups have studied
this topics, obtaining comparable bounds on $N_{eff}$  but using
different priors \cite{crotty}-\cite{cuoco}. For example, in our
analysis \cite{cuoco}, $N_{eff} = 2.6^{+3.5}_{-2.0}$, using WMAP
data only and weak prior on the value of the Hubble parameter,
$h=0.7 \pm 0.2$. The reason for such a wide range for $N_{eff}$ is
ultimately due to the many unknown cosmological parameters which
determine the power spectrum, and in particular to the presence of
several degeneracies, i.e. the fact that different choices for
some of these parameters produce the very same power spectrum. As
an example if we increase both $N_{eff}$ and the amount of dark
matter $\Omega_{cdm}$ we can obtain the same power spectrum
provided we do not change the radiation-matter equality. This is
shown in Fig.1, a plot of the bi-dimensional likelihood contours
in the $N_{eff}-\Omega_{cdm}$ plane

\begin{figure}[t]
\begin{center}
\includegraphics[width=7cm]{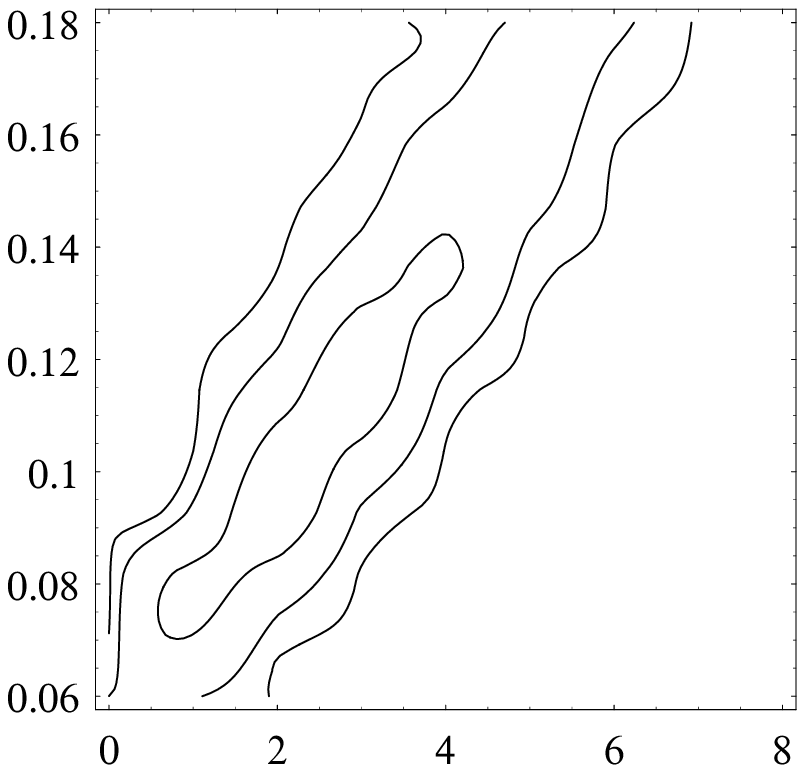}
\end{center}
\caption{Likelihood contours from WMAP data in the ${N_{eff}}$
(x-axis) - $\Omega_{cdm}h^2$ (y-axis) plane \cite{cuoco}.}
\label{fig:nnuomegacdm}
\end{figure}

\begin{figure}[t]
\begin{center}
\includegraphics[width=7cm]{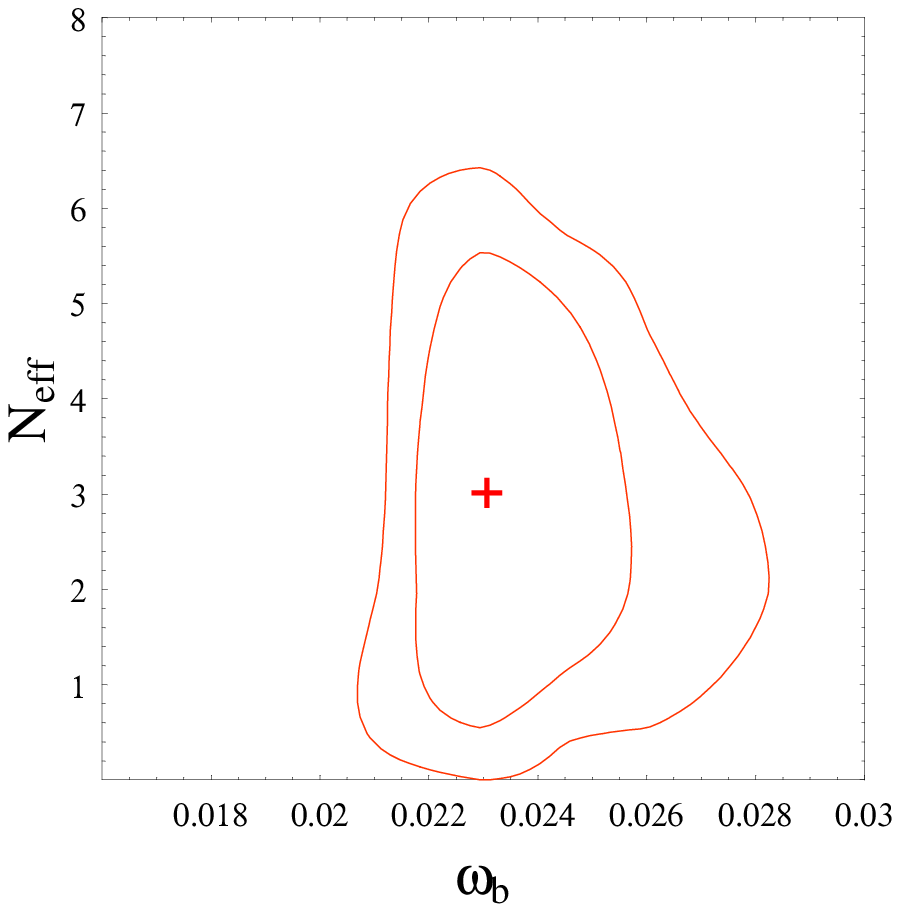}
\end{center}
\caption{The 68 and 95\% C.L. likelihood contours from WMAP data
in the ${N_{eff}}-{\omega_b}$ plane.} \label{fig:cmb}
\end{figure}

The baryon density parameter $\Omega_b h^2$ can be much more
severely constrained from the power spectrum. Increasing the
baryons in the plasma enhances the effective mass of the fluid and
this leads to more pronounced compression peaks. By a likelihood
analysis the bound obtained in \cite{cuoco} is $\Omega_b h^2 =
0.023 \pm 0.002$ ($1-\sigma$ error), fully compatible with the one
quoted by the WMAP Collaboration, $\Omega_b h^2 = 0.022 \pm
0.001$, \cite{GPMwmap}. In Fig.2 we show the likelihood contours in
the $N_{eff}-\Omega_b h^2 $ plane. This result is extremely
important. The WMAP data tell us the value of baryon density with
a better accuracy than BBN, so we can test the standard scenario
of light nuclei formation with basically no free parameters but
the value of $N_{eff}$.

\section{$\Omega_b h^2$ and $N_{eff}$ and Big Bang Nuclesynthesis}
\setcounter{equation}0 \noindent

The primordial production of light nuclei, mainly $^4He$, $D$ and
$^7Li$, takes place when the temperature of the electromagnetic
plasma is in the range $1 \div 0.01 MeV$, and is strongly
influenced by the two parameters $\Omega_b h^2$ and ${N_{eff}}$.
Increasing the value of the baryon to photon number density
enhances the fusion mechanism, so it leads to a larger eventual
amount of $^4He$, the most tightly bound light nucleus ($^4He$
binding energy per nucleon is of the order of $7 MeV$). On the
other hand, $D$ rapidly decreases with $\Omega_b h^2$, so the
experimental result on this species is a very sensitive measure of
baryons in the universe. The contribution of relativistic degrees
of freedom to the expansion rate, parameterized by ${N_{eff}}$ affects
instead the decoupling temperature of weak reaction which keep in
chemical equilibrium protons and neutrons. For large temperatures
in fact the ratio of their densities is given by equilibrium
conditions, $n/p = \exp(-(m_n-m_p)/T)$, therefore if weak
interactions were efficient down to very low temperatures, much
smaller than the neutron-proton mass difference, neutrons would
completely disappear. We mentioned however that the rate of these
processes indeed becomes smaller than the expansion rate $H$ for
temperatures of the order of $T_D \sim 1 MeV$, so that the $n/p$
ratio freezes-out at the value $n/p = \exp(-(m_n-m_p)/T_D)$. Since
almost all neutrons are eventually bound in $^4He$ nuclei it is
then straightforward to get for the Helium mass fraction \begin{equation} Y_p
\equiv \frac{4 n_{He}}{n_b} \sim 2 \frac{n/p}{1+n/p} = 2
\frac{1}{e^{(m_n-m_p)/T_D}+1} \label{np} \end{equation} When correcting this
result for neutron spontaneous decay one gets $Y_p \sim 0.25$,
already an excellent estimate compared with the result of detailed
numerical calculations. Changing ${N_{eff}}$ affect the decoupling
temperature $T_D$ and so the amount of primordial Helium.

An accurate analysis of BBN can be only achieved by numerically
solving a set of coupled differential equations, taking into
account quite a complicated network of nuclear reactions. Some of
these reactions are still affected by large uncertainties, which
therefore introduce an error in the theoretical prediction for,
mainly, $D$ and $^7Li$ abundances. As we mentioned Helium
prediction is mainly influenced by $n \leftrightarrow p$
processes, which are presently known at a high level of accuracy
\cite{lopez}-\cite{emmp}. Quite recently a big effort has been
devoted in trying to quantify the role of each nuclear reaction to
the uncertainties on nuclei abundances, using either Monte Carlo
\cite{krauss} or linear propagation \cite{lisi} techniques. The
most recent analysis \cite{cuoco}, \cite{burles}, \cite{vangioni}
have benefited from the NACRE nuclear reaction catalogue
\cite{nacre}, as well as of very recent results, as for example
the LUNA Collaboration measurement of the $D(p,\gamma)^3He$
\cite{luna}. We report here the results obtained in \cite{cuoco}
for the total relative theoretical uncertainties $\sigma_i^{th}$
on $Y_p$ and $D$ and $^7Li$ number fractions $X_i=n_i/n_b$ \begin{equation}
\frac{\sigma_D}{X_D} \sim 10 \%, \,\,\,\frac{\sigma_{He}}{Y_p}
\sim 0.1 \% ,\,\,\, \frac{\sigma_{Li}}{X_{Li}} \sim 25 \%
\label{uncert} \end{equation} The large error on $^7Li$ is mainly due to the
uncertainty on the rate for the process $^4He(^3He,\gamma)^7Be$, a process
which is also of great interest for the determination of both
$^7Be$ and $^8B$ neutrino fluxes from the sun. Hopefully it will
be studied at low energies in the near future.

The experimental determination of primordial abundances is really
a challenging task. The strategy is to identify metal poor
environment, which are not been severely polluted by star
contamination in their light nuclei content, and possibly to
correct the observations for the effect of galactic evolution.

The $^4He$ mass fraction is determined by regression to zero
metallicity of the values obtained by observing $HeII \rightarrow
HeI$ recombination lines in extragalactic ionized gas. There are
still quite different results (see e.g. \cite{sarkar} for a review
and references), a $low$ one, $Y_p = 0.234 \pm 0.003$, and a
$high$ value, $Y_p =0.244 \pm 0.002$. In the following we also use
a conservative estimate, $Y_p = 0.239 \pm 0.008$.

The best estimate of Primordial $D$ comes from observations of
absorption lines in gas clouds in the line of sight between the
earth and Quasars at very high redshift ($z \sim 2 - 3$), which
give $X_D= (2.78^{+0.44}_{-0.38}){\cdot} 10^{-5}$ \cite{tytler}.
Finally $^7Li$ is measured via observation of absorption lines in
spectra of POP II halo stars, which show a saturation of $^7Li$
abundance at low metallicity (Spite plateau).

\begin{figure}[t]
\begin{center}
\includegraphics[width=8cm]{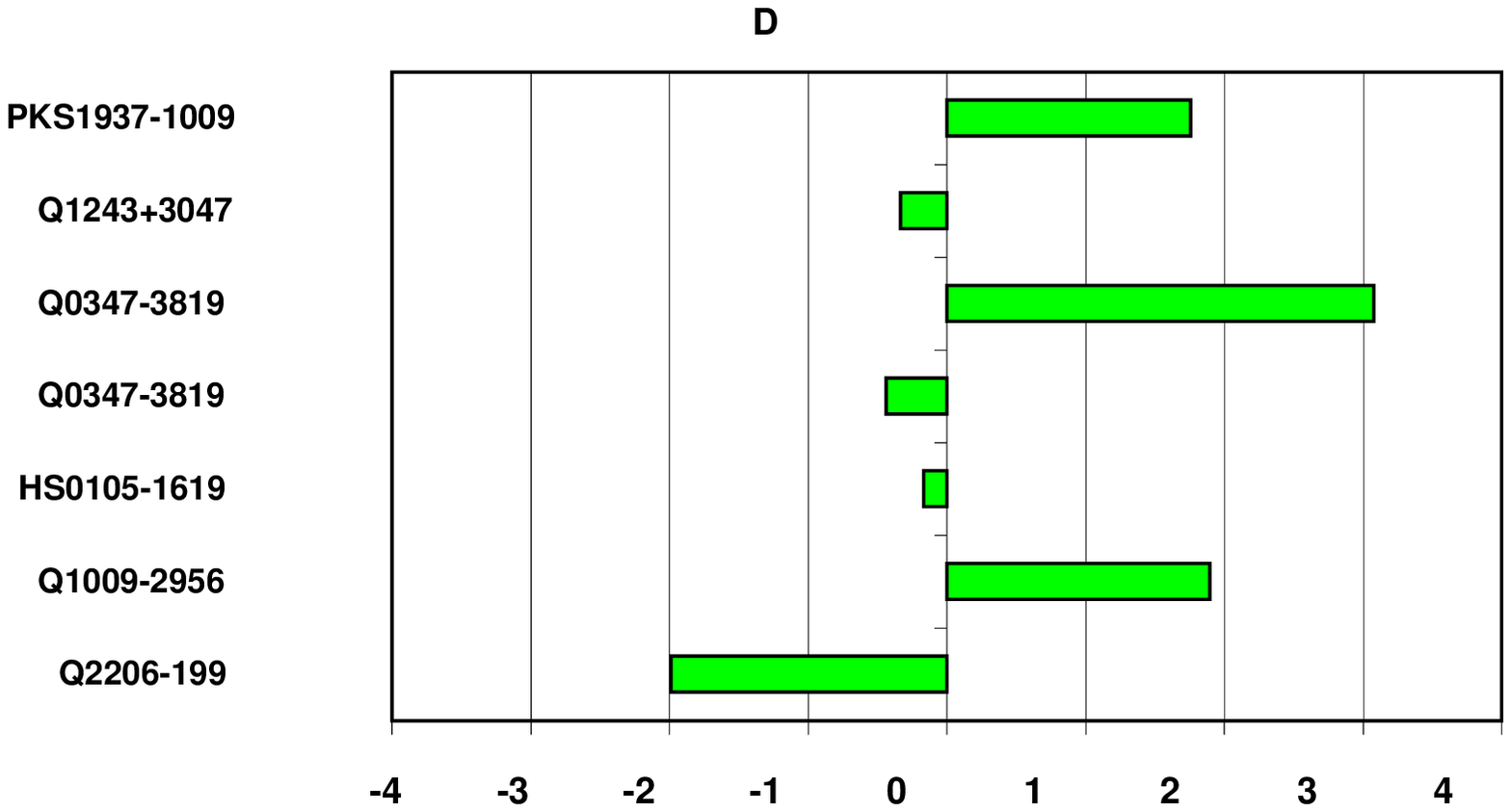}
\end{center}
\caption{The pulls of QSO D measurements with respect to the
theoretical prediction for ${N_{eff}}=3.04$ and ${\omega_b}=0.023$, in units
of $\left((\sigma^{th}_{DD})^2 + (\sigma^{exp}_D)^2
\right)^{1/2}$.} \label{fig:pulld}
\end{figure}

\begin{figure}[t]
\begin{center}
\includegraphics[width=8cm]{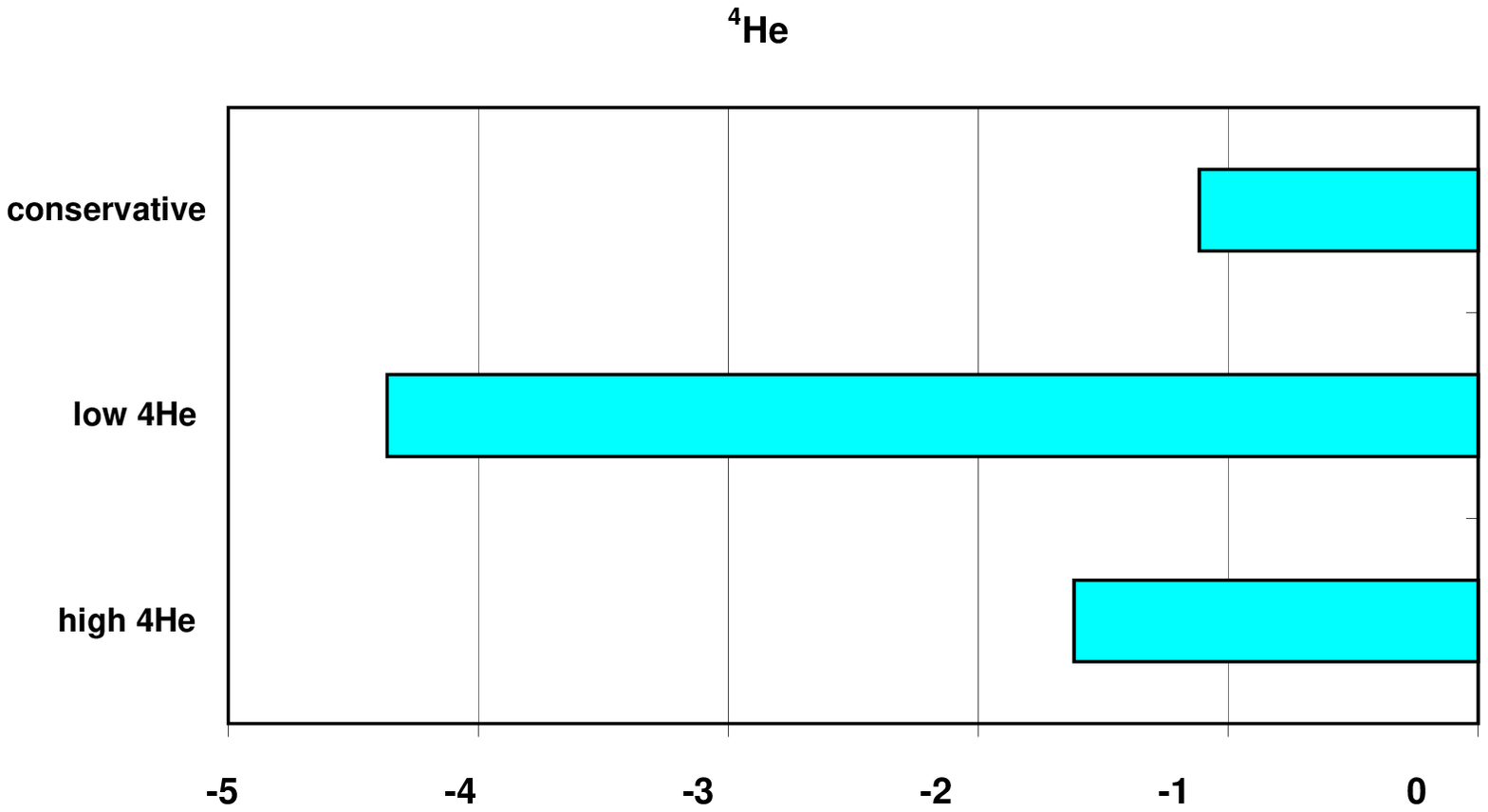}
\end{center}
\caption{The pulls of $Y_p$ measurements with respect to the
theoretical prediction for ${N_{eff}}=3.04$ and ${\omega_b}=0.023$, in units
of $\left( (\sigma^{th}_{44})^2 + (\sigma^{exp}_4)^2
\right)^{1/2}$.} \label{fig:pullhe}
\end{figure}

\begin{figure}[t]
\begin{center}
\includegraphics[width=8cm]{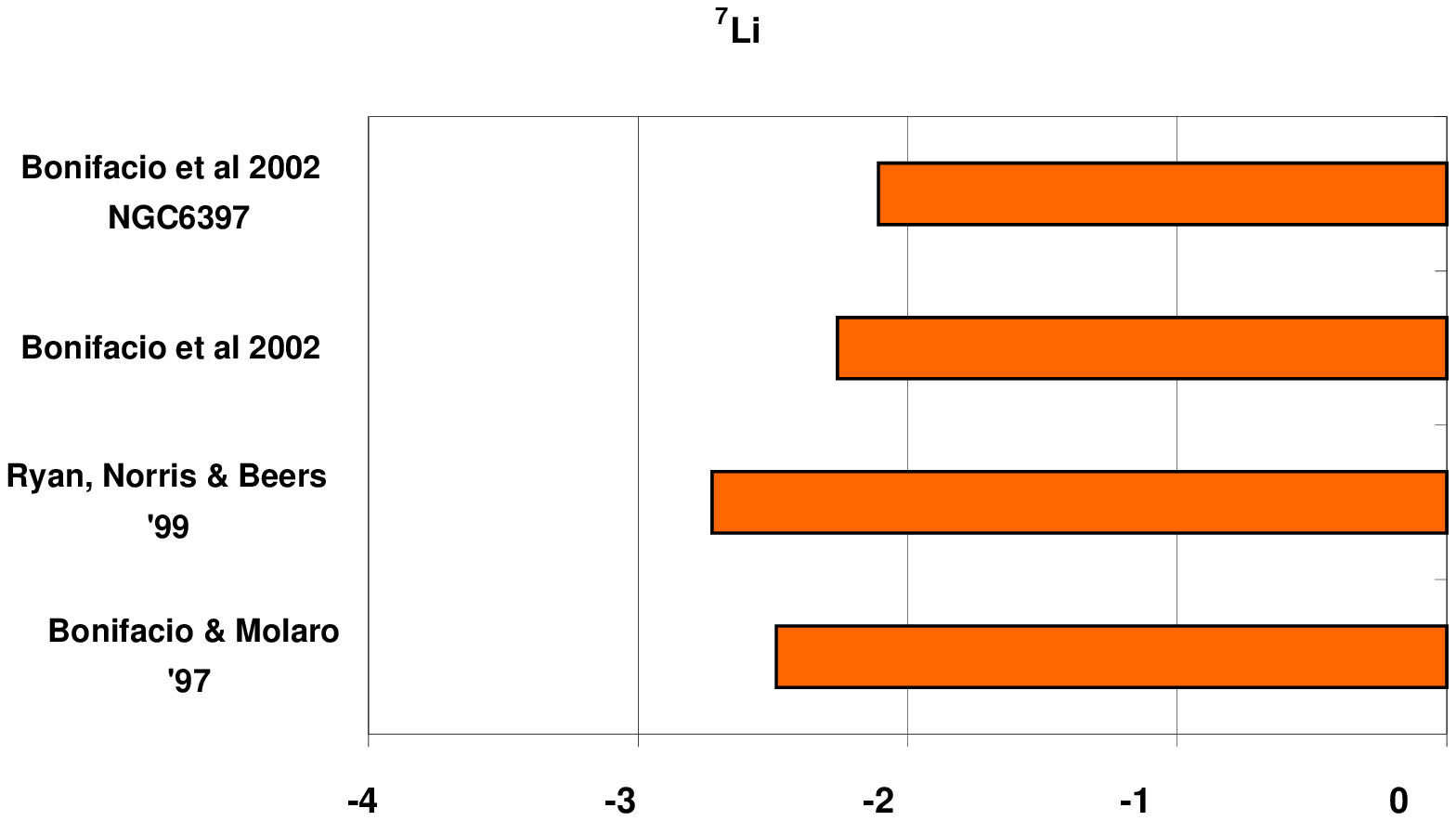}
\end{center}
\caption{The pulls of $X_{7Li}$ measurements with respect to the
theoretical prediction for ${N_{eff}}=3.04$ and ${\omega_b}=0.023$, in units
of $\left((\sigma^{th}_{77})^2 + (\sigma^{exp}_7)^2
\right)^{1/2}$.} \label{fig:pullli}
\end{figure}

The present status of BBN, in the standard scenario, using the
value of baryon density as determined by WMAP and ${N_{eff}}=3.04$ is
summarized in Fig.s 3-5. Here I report the difference between the
theoretical and the experimental determination, normalized to the
total uncertainty, theoretical and experimental, summed in
quadrature. The average of the several $D$ measurements, reported
above, is indeed in very good agreement with theory. This is a
very crucial result since, as we said already, $D$ is strongly
influenced by $\Omega_b h^2$, which is now fixed by WMAP. In Fig.
6 I show the combined likelihood contours at $2 \sigma$ in the
${N_{eff}}-\Omega_b h^2$ plane obtained when using the WMAP result
$and$ $D$ measurement only (colored area) and the $D+^4He$ results
using the conservative $Y_p$ shown before.
\begin{figure}[t]
\begin{center}
\includegraphics[width=8cm]{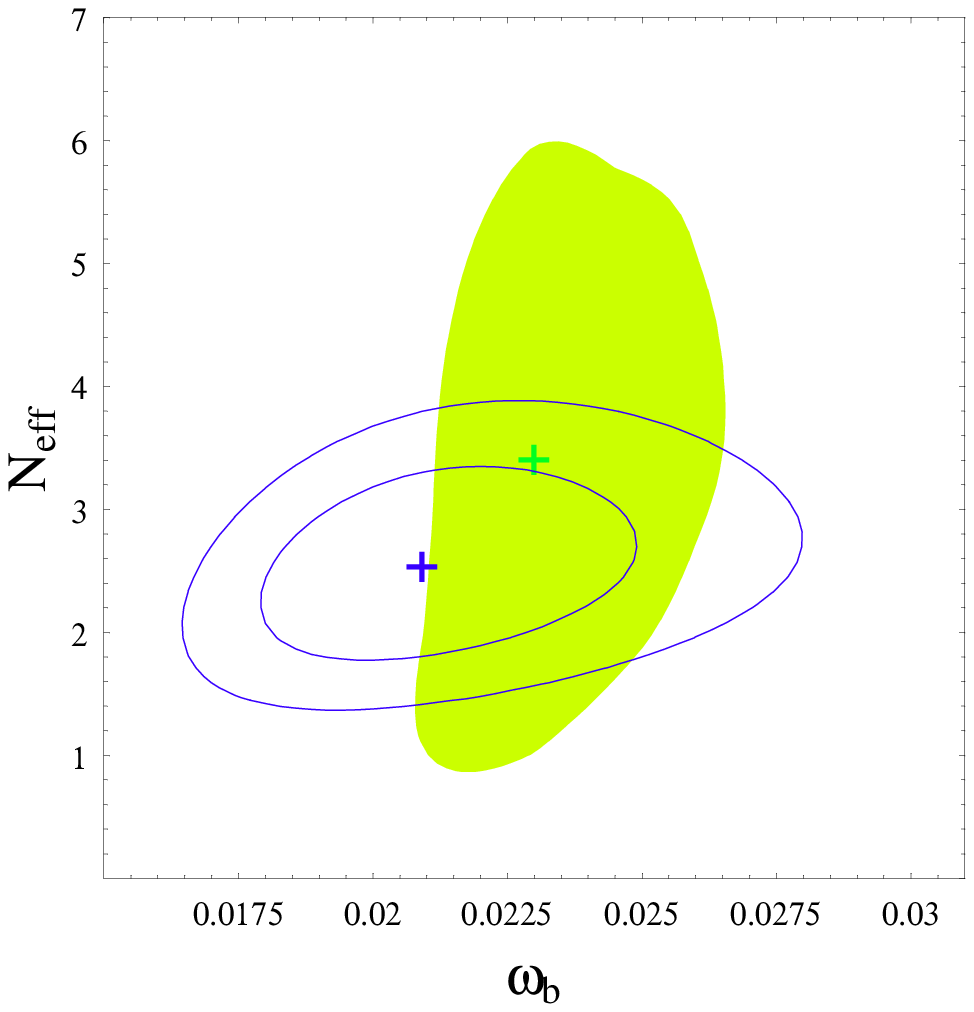}
\end{center}
\caption{The 68 and $95\%$ C.L. contours for the $D+^4He$
likelihood function in the ${\omega_b}-{N_{eff}}$ plane (${\omega_b}=\Omega_b
h^2$). We also show the result of the $CMB+D$ analysis (colored
area).} \label{fig:likelihoodbbn}
\end{figure}

It is evident that the effect of $^4He$ is to shift the values of
both $\Omega_b h^2$ and ${N_{eff}}$ towards smaller values, which
produces a smaller theoretical value for $Y_p$. Though this may be
seen as a (weak) indication of the fact that a slightly lower
value for ${N_{eff}}$ is preferred, I would more conservatively say
that, waiting for a more clear understanding of possible
systematics in $Y_p$ experimental determination, the standard
scenario for BBN is in reasonable good shape. An open problem is
however still represented by the evidence for $^7Li$ depletion,
which is not fully understood (see Fig. 5). The theoretical result
for $X_{Li}$ is in fact a factor 2-3 larger than the present
experimental determination.

\section{Neutrino-antineutrino asymmetry}
\setcounter{equation}0
\noindent

While the electron-positron asymmetry density is severely
constrained, of the order of $10^{-10}$ in unit of the photon
density, we have no bounds at all on neutrino asymmetry from
charge neutrality of the universe. Defining $\xi_x=\mu_x/T_x$,
with $\mu_x$ the chemical potential for the $\nu_x$ species, with
$x=e,\mu,\tau$, we recall that for a Fermi-Dirac distribution the
particle-antiparticle asymmetry is simply related to $\xi_x$ (I
assume here for simplicity massless neutrinos) \begin{equation}
n(\nu_x)-n(\overline{\nu}_x) = \frac{T_x^3}{6} \left( \xi_x +
\frac{\xi_x^3}{\pi^2} \right) \label{asym} \end{equation} since neutrinos
decoupled as hot relics starting from a chemical equilibrium
condition with $e^\pm$, so that $\mu_x\equiv \mu(\nu_x) =
-\mu(\overline{\nu}_x)$. Non vanishing values for $\xi_x$ affects
very weakly CMB, while it is much more constrained by BBN. In fact
any asymmetry in the neutrino sector contribute to the Hubble
expansion rate, i.e. to ${N_{eff}}$ \begin{equation} {N_{eff}} \rightarrow {N_{eff}} +
\sum_x \left[ \frac{30}{7} \left(\frac{\xi_x}{\pi}\right)^2 +
\frac{15}{7} \left(\frac{\xi_x}{\pi}\right)^4 \right]
\label{neffdeg} \end{equation} In addition the asymmetry in the electron
neutrino sector directly affects the $n/p$ value at the freeze-out
of weak interactions, since they directly enter in the processes
governing this phenomenon, namely $n+\nu_e \leftrightarrow p+e^-$,
$n \leftrightarrow p+e^- + \overline{\nu}_e$ and $n+e^+
\leftrightarrow p+ \overline{\nu}_e$.

\begin{figure}[t]
\begin{center}
\includegraphics[width=7cm]{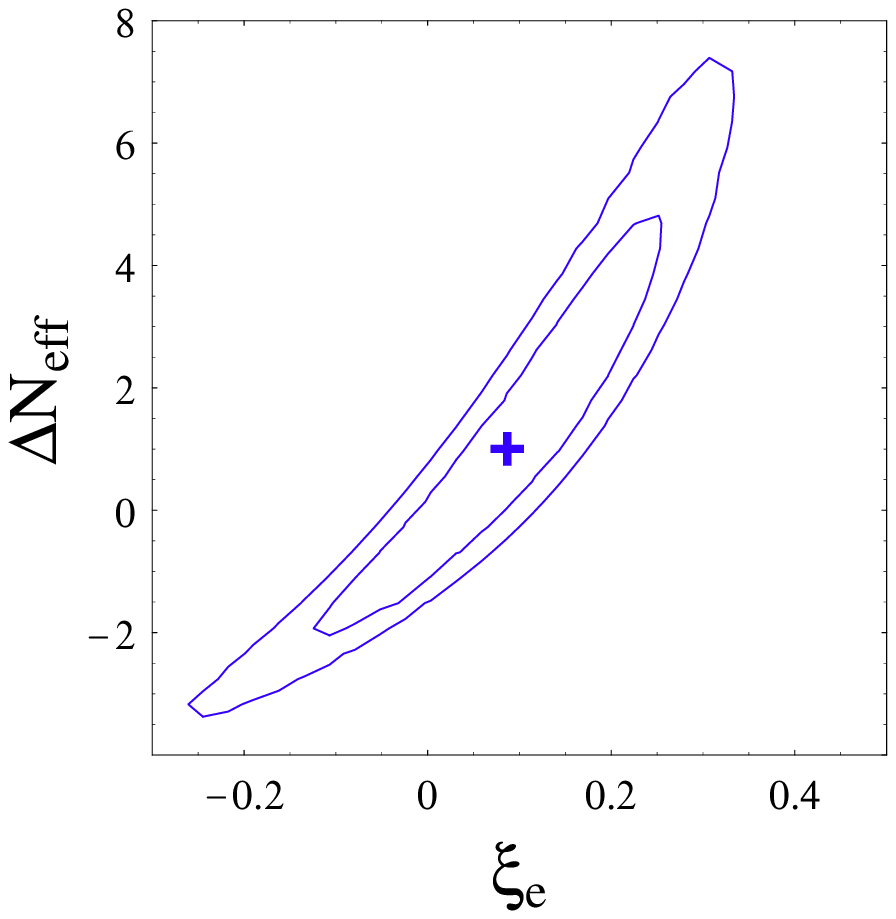}
\end{center} \caption{The 68 and $95\%$ C.L. likelihood contours in the
 $\xi_e-\Delta{N_{eff}}$ plane from BBN, with $\Delta {N_{eff}}={N_{eff}}-3.04$.} \label{fig:likelihoode}
\end{figure}

It was recently realized \cite{petcov} that indeed, because of
flavor oscillation, using present determination of mass
differences and mixing angles from atmospheric and solar
neutrinos, the three $\xi_x$ should be very close each other, so
the bound on their (common) value $\xi$ mainly come from the fact
that $\xi_e$ should be quite small $\xi \leq 0.1$, to give a value
for the $n/p$ ratio (and so for $^4He$) in agreement with data. In
Fig. 7 I show the likelihood contour obtained in the $\xi-{N_{eff}}$
plane \cite{cuoco}. Though the standard BBN is preferred, there is
still room for very exotic scenarios, with larger neutrino
degeneracies and even very large (or very small) ${N_{eff}}$.

\section{Cosmological bounds on neutrino mass scale}
\setcounter{equation}0
\noindent

Despite of the fact that we presently know neutrino mass
differences from oscillation effects in atmospheric and solar
neutrino fluxes, there is still quite a wide range for their
absolute mass scale, spanning several order of magnitude, from few
$eV$ down to $10^{-2}$ $eV$. Terrestrial bounds come from Tritium
decay experiments \cite{tritium}, which presently give $m(\nu_e)
\leq 2.2 eV$. This result will be greatly improved by next
generation experiment KATRIN, which should reach a sensitivity
after three years of running of the order of $0.35$ $eV$
\cite{katrin}

An independent source of information will be provided by
neutrinoless beta decay, which is sensitive to the effective
$\nu_e$ mass \begin{equation} <m_e> = |U_{e1}|^2 m_1 + |U_{e2}|^2 e^{i \phi_2}
m_2 + |U_{e3}|^2 e^{i \phi_3} m_3 \end{equation} with $U_{ei}$ the electron
neutrino projection onto mass eigenstates with mass $m_i$, and
$\phi_i$ CP violating Majorana phases. Planned experiments CUORE
\cite{cuore} and GENIUS \cite{genius} will have a sensitivity on
this parameter of the order of $10^{-1}-10^{-2}$ $eV$.

Interestingly, quite severe constraints on neutrino masses come
from cosmology. Massive neutrinos in fact contribute to the
present total energy density of the Universe as $m_\nu n_\nu$ so
we get \begin{equation} \Omega_\nu h^2 = \frac{\sum_x m(\nu_x)}{92.5 \,eV} \end{equation}
which gives a generous bound when imposing $\Omega_\nu h^2<1$.

Neutrino masses however also enter in the way structures grow for
gravitational instability from the initial seed likely given by
adiabatic perturbations produced during the inflationary phase. In
fact neutrinos free stream and tend to suppress structure
formation on all scales unless they are massive. In this case they
can only affect scales smaller than the Hubble horizon when they
eventually become non relativistic, the ones with a corresponding
wave number larger than \cite{hu} \begin{equation} k_{nr} \sim 0.026 \left(
\frac{m_\nu}{1 eV} \right) ^{1/2} \Omega_m ^{1/2} h Mpc^{-1} \end{equation}
Several authors have recently considered this issue in details
\cite{wmap2}- \cite{hann}, combining WMAP data and the results of
the 2dFGRS survey \cite{2df}. Actually the result also depends on
the specific value of ${N_{eff}}$. A conservative value is given by
$\sum_x m(\nu_x) \leq 2 \, eV$.

\section{Acknowledgments}

It is a pleasure to thank Norma Mankoc Borstnik for organizing the
Portoroz meeting and for her warm hospitality. I also thank my
collaborators at Naples University A. Cuoco, F. Iocco, G. Miele,
O. Pisanti and P. Serpico.

\makeatletter
\def\gsim{\compoundrel>\over\sim}
\def\CDFlsim{\compoundrel<\over\sim}
\def\compoundrel#1\over#2{\mathpalette\compoundreL{{#1}\over{#2}}}
\def\compoundreL#1#2{\compoundREL#1#2}
\def\compoundREL#1#2\over#3{\mathrel
	   {\vcenter{\hbox{$\m@th\buildrel{#1#2}\over{#1#3}$}}}}
\makeatother
\def\CDFsleq{\raisebox{-.6ex}{${\textstyle\stackrel{<}{\sim}}$}}
\def\CDFsVEV#1{\left\langle #1\right\rangle}
\title*{The Problem of Mass}
\author{Colin D. Froggatt\thanks{E-mail:
c.froggatt@physics.gla.ac.uk}}
\institute{%
Department of Physics and Astronomy, Glasgow University,\\
Glasgow G12 8QQ, Scotland, UK}

\titlerunning{The Problem of Mass}
\authorrunning{Colin D. Froggatt}
\maketitle

\begin{abstract}
The quark-lepton mass problem and the ideas of mass protection are
reviewed. The Multiple Point Principle is introduced and used
within the Standard Model to predict the top quark and Higgs
particle masses. We discuss the lightest family mass generation
model, in which all the quark mixing angles are successfully
expressed in terms of simple expressions involving quark mass
ratios. The chiral flavour symmetry of the family replicated gauge
group model is shown to provide the mass protection needed to
generate the hierarchical structure of the quark-lepton mass
matrices.
\end{abstract}


\section{Introduction}
\label{introduction}

The most important unresolved problem in particle physics is the
understanding of flavour and the fermion mass spectrum. The observed
values of the fermion masses and mixing angles constitute the bulk of
the Standard Model (SM) parameters and provide our main experimental clues
to the underlying flavour dynamics. In particular the non-vanishing
neutrino masses and mixings provide direct evidence for physics beyond
the SM.

The charged lepton masses can be directly measured and correspond to
the poles in their propagators:
\begin{equation}
M_e = 0.511 \ \makebox{MeV} \qquad M_{\mu} = 106 \ \makebox{MeV}
\qquad M_{\tau} = 1.78 \ \makebox{GeV}
\end{equation}
However the quark masses have to be extracted from the properties of
hadrons and are usually quoted as running masses $m_q(\mu)$ evaluated at
some renormalisation scale $\mu$, which are related to the  propagator
pole masses $M_q$ by
\begin{equation}
M_q = m_q(\mu=m_q)\left[ 1+ \frac{4}{3}\alpha_3(m_q) \right]
\end{equation}
to leading order in QCD. The light $u$, $d$ and $s$ quark masses
are usually normalised to the scale $\mu=1$ GeV (or $\mu=2$ GeV
for lattice measurements) and to the quark mass itself for the
heavy $c$, $b$ and $t$ quarks. They are typically given
\cite{pdg} as follows\footnote{Note that the top quark mass,
$M_t = 174 \pm 5$ GeV, measured at FermiLab is interpreted as
the pole mass.}:
\begin{eqnarray}
m_u(1 \ \makebox{GeV})  = 4.5 \pm 1 \ \makebox{MeV} &\qquad &
m_d(1 \ \makebox{GeV})  = 8 \pm 2 \ \makebox{MeV} \nonumber \\
m_c(m_c)  =  1.25 \pm 0.15 \ \makebox{GeV} &\qquad &
m_s(1 \ \makebox{GeV}) =  150 \pm 50 \ \makebox{MeV} \nonumber \\
m_t(m_t)  =  166 \pm 5 \ \makebox{GeV} &\qquad &
m_b(m_b)  =  4.25 \pm 0.15 \ \makebox{GeV}
\end{eqnarray}
However we only have an upper limit on the neutrino masses of
$m_{\nu_i} \CDFlsim 1$ eV from tritium beta decay and from cosmology,
and measurements of the neutrino mass squared differences:
\begin{equation}
 \Delta m_{21}^2 \sim 5 \times 10^{-5} \makebox{eV}^2 \qquad
 \Delta m_{32}^2 \sim 3 \times 10^{-3} \makebox{eV}^2
 \label{dm2}
\end{equation}
from solar and atmospheric neutrino oscillation data \cite{garcia}.

The magnitudes of the quark mixing matrix $V_{CKM}$ are well measured
\begin{equation}
 |V_{CKM}| = \begin{pmatrix}
0.9734 \pm 0.0008 & 0.2196 \pm 0.0020 & 0.0036 \pm 0.0007 \cr
0.224 \pm 0.016   & 0.996 \pm 0.013   & 0.0412 \pm 0.002 \cr
0.0077 \pm 0.0014 & 0.0397 \pm 0.0033 & 0.9992 \pm 0.0002 \cr
\end{pmatrix}
\end{equation}
and a CP violating phase of order unity:
\begin{equation}
\sin^2\delta_{CP} \sim 1
\end{equation}
can reproduce all the CP violation data. Neutrino
oscillation data constrain the magnitudes of the lepton
mixing matrix elements to lie in the following $3\sigma$ ranges
\cite{garcia}:
\begin{equation}
 |U_{MNS}| = \begin{pmatrix}
 0.73-0.89 & 0.45-0.66 & <0.24     \cr
 0.23-0.66  & 0.24-0.75 & 0.52-0.87 \cr
 0.06-0.57  & 0.40-0.82 & 0.48-0.85 \cr
 \end{pmatrix}
 \label{mns}
\end{equation}
Due to the Majorana nature of the neutrino mass matrix, there are
three unknown CP violating phases $\delta$, $\alpha_1$ and
$\alpha_2$ in this case \cite{garcia}.

The charged fermion masses range over five orders of magnitude,
whereas there seems to be a relatively mild neutrino mass hierarchy.
The absolute neutrino mass scale ($m_{\nu} < 1$
eV) suggests a new physics mass scale -- the
so-called see-saw scale $\Lambda_{seesaw} \sim 10^{15}$ GeV.
The quark mixing matrix $V_{CKM}$ is also hierarchical, with small
off-diagonal elements. However the elements of $U_{MNS}$ are all
of the same order of magnitude except for $|U_{e3}|<0.24$,
corresponding to two leptonic mixing angles being close to
maximal ($\theta_{atmospheric} \simeq \pi/4$ and $\theta_{solar}
\simeq \pi/6$).

We introduce the mechanism of mass protection by approximately
conserved chiral charges in section \ref{protection}. The top
quark mass is the dominant term in the SM fermion mass matrix, so
it is likely that its value will be understood dynamically before
those of the other fermions. In section \ref{top} we discuss the
connection between the top quark and Higgs masses and how they can
be determined from the so-called Multiple Point Principle. We
present the lightest family mass generation model in section
\ref{ansatze}, which provides an ansatz for the texture of fermion
mass matrices and expresses all the quark mixing angles
successfully in terms of simple expressions involving quark mass
ratios. The family replicated gauge group model is presented in
section \ref{origin}, as an example of a model whose gauge group
naturally provides the mass protecting quantum numbers needed to
generate the required texture for the fermion mass matrices.
Finally we present a brief conclusion in section \ref{conclusion}.

\section{Fermion Mass and Mass Protection}
\label{protection}

A fermion mass term
\begin{equation}
{\cal{L}}_{mass} =  -m \overline{\psi}_L \psi_R + h. c.
\end{equation}
couples together a left-handed Weyl field $\psi_L$ and a
right-handed Weyl field $\psi_R$, which then satisfy the Dirac
equation
\begin{equation}
 i\gamma^{\mu} \partial_{\mu} \psi_L = m \psi_R
\end{equation}
If the two Weyl fields are not charge conjugates $\psi_L \neq
(\psi_R)^c$ we have a Dirac mass term and the two fields $\psi_L$
and $\psi_R$ together correspond to a Dirac spinor. However if the
two Weyl fields are charge conjugates $\psi_L = (\psi_R)^c$ we
have a Majorana mass term and the corresponding four component
Majorana spinor has only two degrees of freedom. Particles
carrying an exactly conserved charge, like the electron carries
electric charge, must be distinct from their anti-particles and
can only have Dirac masses with $\psi_L$ and $\psi_R$ having equal
conserved charges. However a neutrino could be a Majorana
particle.

If $\psi_L$ and $\psi_R$ have different quantum numbers,
i.e.~belong to inequivalent representations of a symmetry group
$G$ ($G$ is then called a chiral symmetry), a Dirac mass term is
forbidden in the limit of an exact $G$ symmetry and they represent
two massless Weyl particles. Thus the $G$ symmetry ``protects'' the
fermion from gaining a mass. Such a fermion can only gain a mass
when $G$ is spontaneously broken.

The left-handed and right-handed top quark, $t_L$ and $t_R$, carry
unequal Standard Model $SU(2) \times U(1)$ gauge charges
$\vec{Q}$:
\begin{equation}
 \vec{Q}_L \neq \vec{Q}_R \qquad \mathrm{(Chiral\  charges)}
\end{equation}
Hence electroweak gauge invariance protects the quarks and leptons
from gaining a fundamental mass term ($\overline{t}_L t_R$ is not
gauge invariant). This {\em mass protection} mechanism is of
course broken by the Higgs effect, when the vacuum expectation
value of the Weinberg-Salam Higgs field
\begin{equation}
 <\phi_{WS}> = \sqrt{2} v = 246 \ GeV
\end{equation}
breaks the gauge symmetry and the SM gauge invariant Yukawa
couplings $\frac{y_i}{\sqrt{2}}$ generate the running quark masses
$m_i = y_i v  =  174 \, y_i \ \makebox{GeV}$. In this way a top
quark mass of the same order of magnitude as the SM Higgs field
vacuum expectation value (VEV) is naturally generated (with $y_t$
unsuppressed). Thus the Higgs mechanism explains why the top quark
mass is suppressed, relative to the fundamental (Planck, GUT...)
mass scale of the physics beyond the SM, down to the scale of
electroweak gauge symmetry breaking. However the further
suppression of the other quark-lepton masses ($y_b$, $y_c$, $y_s$,
$y_u$, $y_d$ $\ll$ 1) remains a mystery, which it is natural to
attribute to mass protection by other approximately conserved
chiral gauge charges beyond the SM, as discussed in section
\ref{origin} for the family replicated gauge group model.

Fermions which are vector-like under the SM gauge group
($\vec{Q}_L = \vec{Q}_R$) are not mass protected and are expected
to have a large mass associated with new (grand unified,
string,..) physics. The Higgs particle, being a scalar, is not
mass protected and {\em a priori} would also be expected to have a
large mass; this is the well-known gauge hierarchy problem
discussed at Portoroz by Holger Nielsen \cite{holger}.

\section{Top Quark and Higgs Masses from the Multiple Point Principle}
\label{top}

\begin{figure}
\centering
\includegraphics[width=5cm]{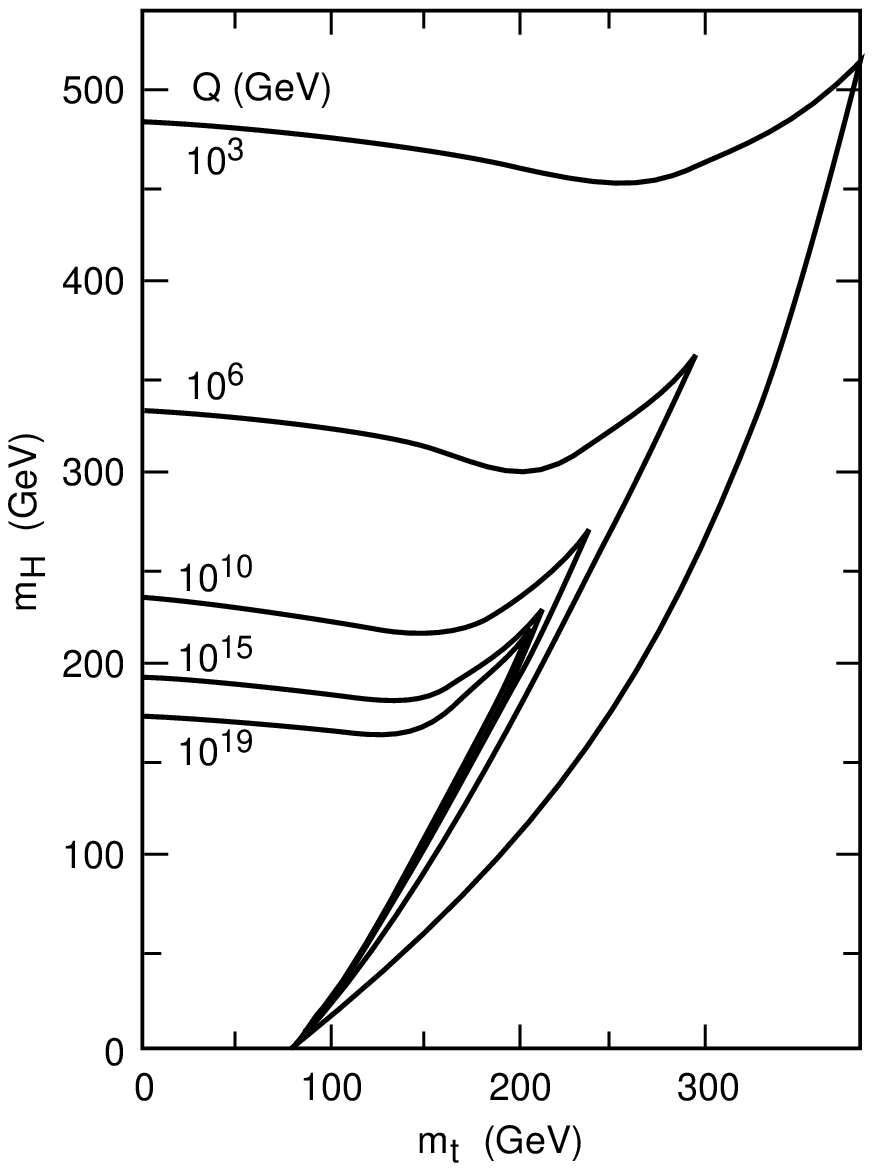} 
\caption{SM bounds in the ($m_t$,$m_H$) plane for  various values
of $\Lambda = Q$, the scale at which new physics enters.}
\label{fig:Maiani}
\end{figure}
It is well-known \cite{maiani} that the self-consistency of the
pure SM up to some physical cut-off scale $\Lambda$ imposes
constraints on the top quark and Higgs boson masses. The first
constraint is the so-called triviality bound: the running Higgs
coupling constant $\lambda(\mu)$ should not develop a Landau pole
for $\mu < \Lambda$. The second is the vacuum stability bound: the
running Higgs coupling constant $\lambda(\mu)$ should not become
negative leading to the instability of the usual SM vacuum. These
bounds are illustrated \cite{lindner} in Fig. \ref{fig:Maiani},
where the combined triviality and vacuum stability bounds for the
SM are shown for different values of the high energy cut-off
$\Lambda$. The allowed region is the area around the origin
bounded by the co-ordinate axes and the solid curve labelled by
the appropriate value of $\Lambda$.  The upper part of each curve
corresponds to the triviality bound. The lower part of each curve
coincides with the vacuum stability bound and the point in the top
right hand corner, where it meets the triviality bound curve, is
the infra-red quasi-fixed point for that value of $\Lambda$. Here
the vacuum stability curve, for a large cut-off of order the
Planck scale $\Lambda_{Planck} \simeq 10^{19}$ GeV, is important
for the discussion of the values of the top quark and Higgs boson
masses predicted from the Multiple Point Principle.

\begin{figure}
\centering
\includegraphics*[width=7cm,bbllx=95pt,bblly=130pt,bburx=510pt,bbury=555pt,angle=270]{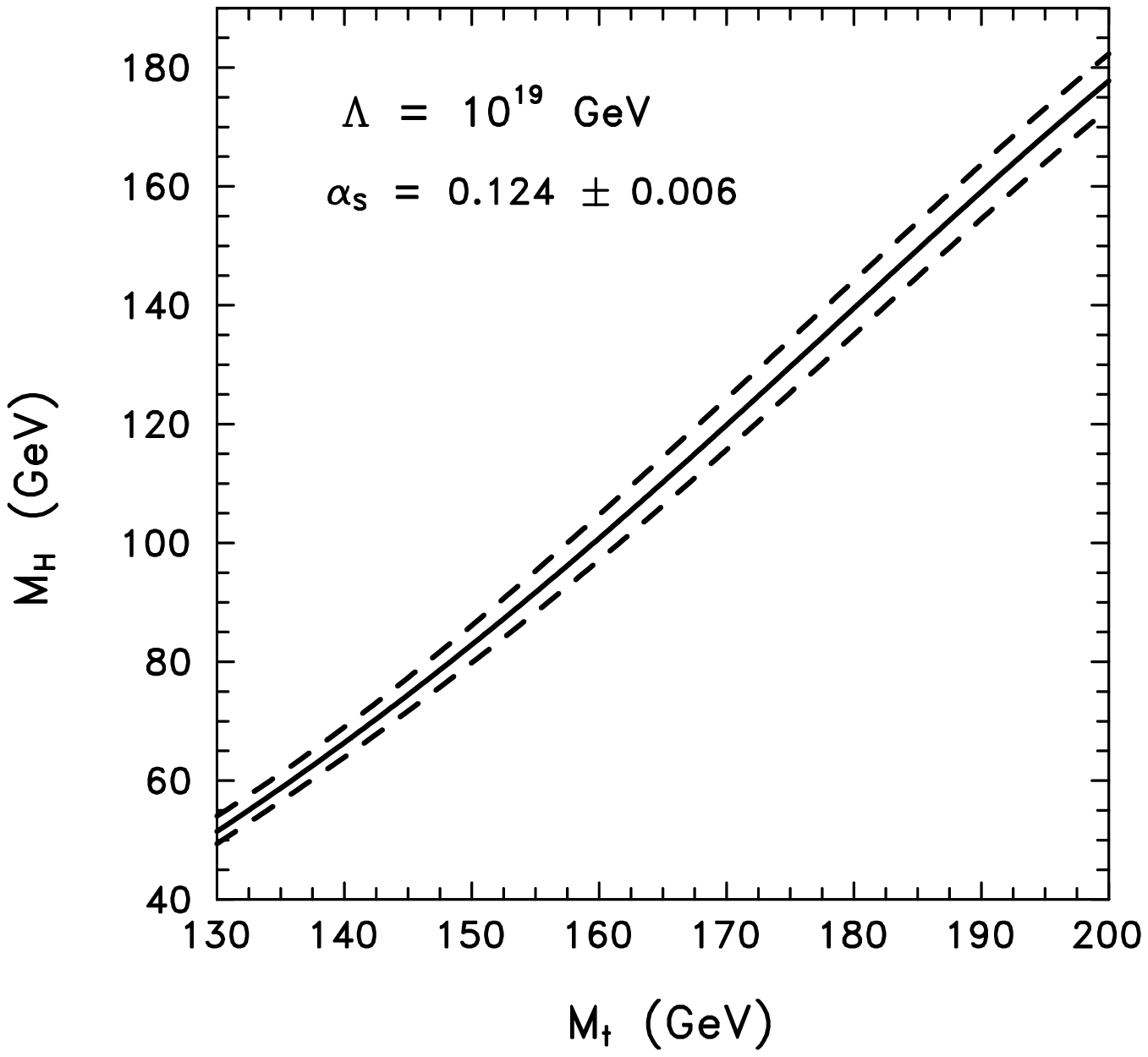}
\caption{SM vacuum stability curve for $\Lambda
= 10^{19}$ GeV and $\alpha_s = 0.124$ (solid line), $\alpha_s =
0.118$ (upper dashed line), $\alpha_s = 0.130$ (lower dashed
line).} \label{fig:vacstab}
\end{figure}
According to the Multiple Point Principle (MPP), Nature chooses
coupling constant values such that a number of vacuum states have
the same energy density (cosmological constant). This fine-tuning
of the coupling constants is similar to that of temperature for a
mixture of co-existing phases such as ice and water. We have
previously argued \cite{glasgowbrioni} that baby-universe like
theories \cite{baby}, having a mild breaking of locality and
causality, may contain the underlying physical explanation of the
MPP, but it really has the status of a postulated new principle.
Here we apply it to the pure Standard Model \cite{fn2}, which we
assume valid up close to $\Lambda_{Planck}$. So we shall postulate
that the effective potential $V_{eff}(\phi)$ for the SM Higgs
field $\phi$ should have a second minimum, at $<\phi> =
\phi_{vac\; 2}$, degenerate with the
well-known first minimum at the electroweak scale $<\phi> =
\phi_{vac\; 1} = 246$ GeV:
\begin{equation}
V_{eff}(\phi_{vac\; 1}) = V_{eff}(\phi_{vac\; 2}) \label{eqdeg}
\end{equation}
Thus we predict that our vacuum is barely stable and we just lie
on the vacuum stability curve in the top quark, Higgs particle
(pole) mass ($M_t$, $M_H$) plane, shown \cite{casas} in Fig.
\ref{fig:vacstab} for a cut-off $\Lambda = 10^{19}$ GeV.
Furthermore we expect the second minimum to be within an order of
magnitude or so of the fundamental scale, i.e. $\phi_{vac\; 2}
\simeq \Lambda_{Planck}$. In this way, we essentially select a
particular point on the SM vacuum stability curve and hence the
MPP condition predicts precise values for $M_t$ and $M_H$.
\begin{figure}
\centering
\includegraphics[width=6.8cm]{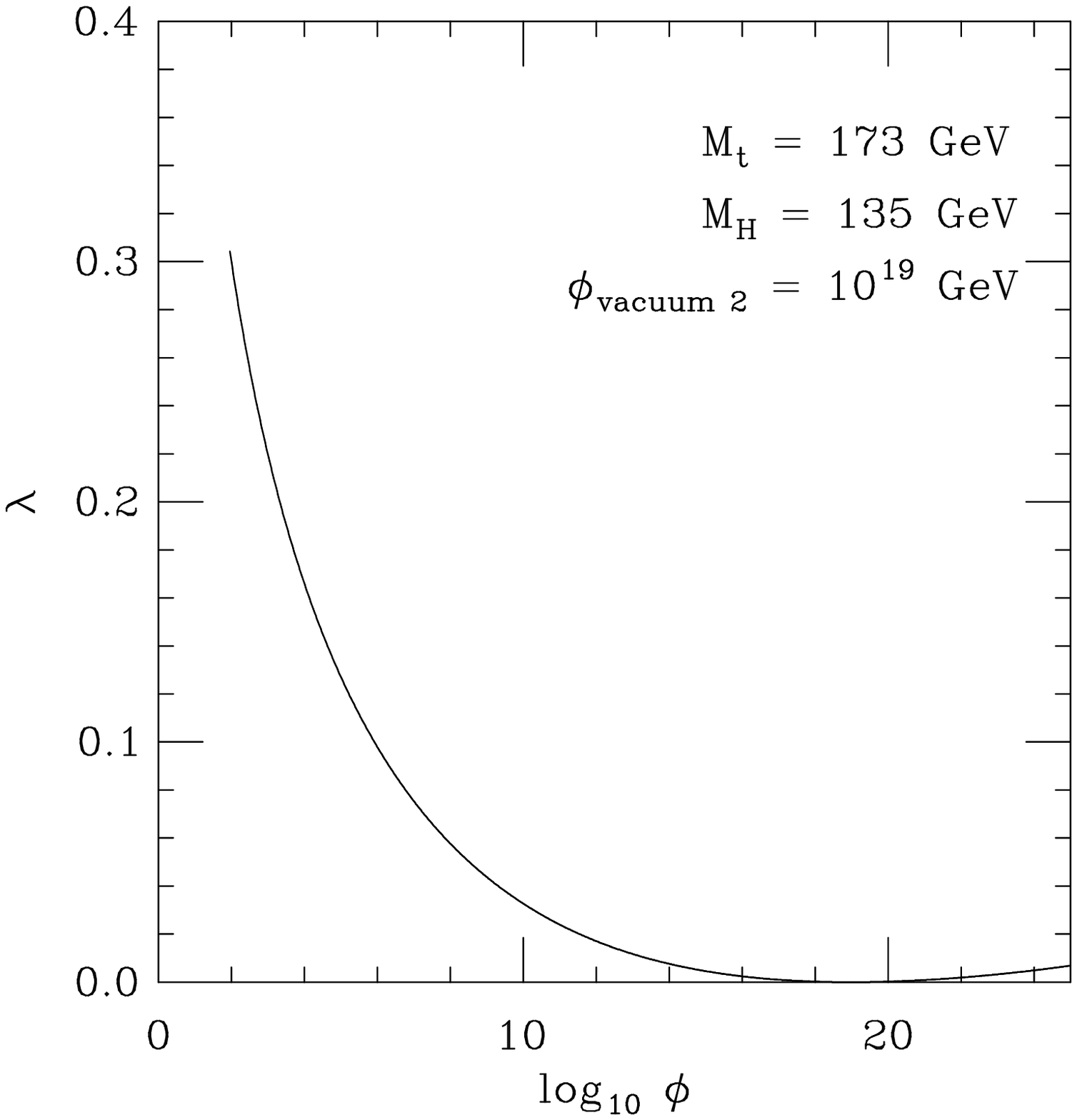}
\hspace{-0.6cm} 
\includegraphics[width=6.8cm]{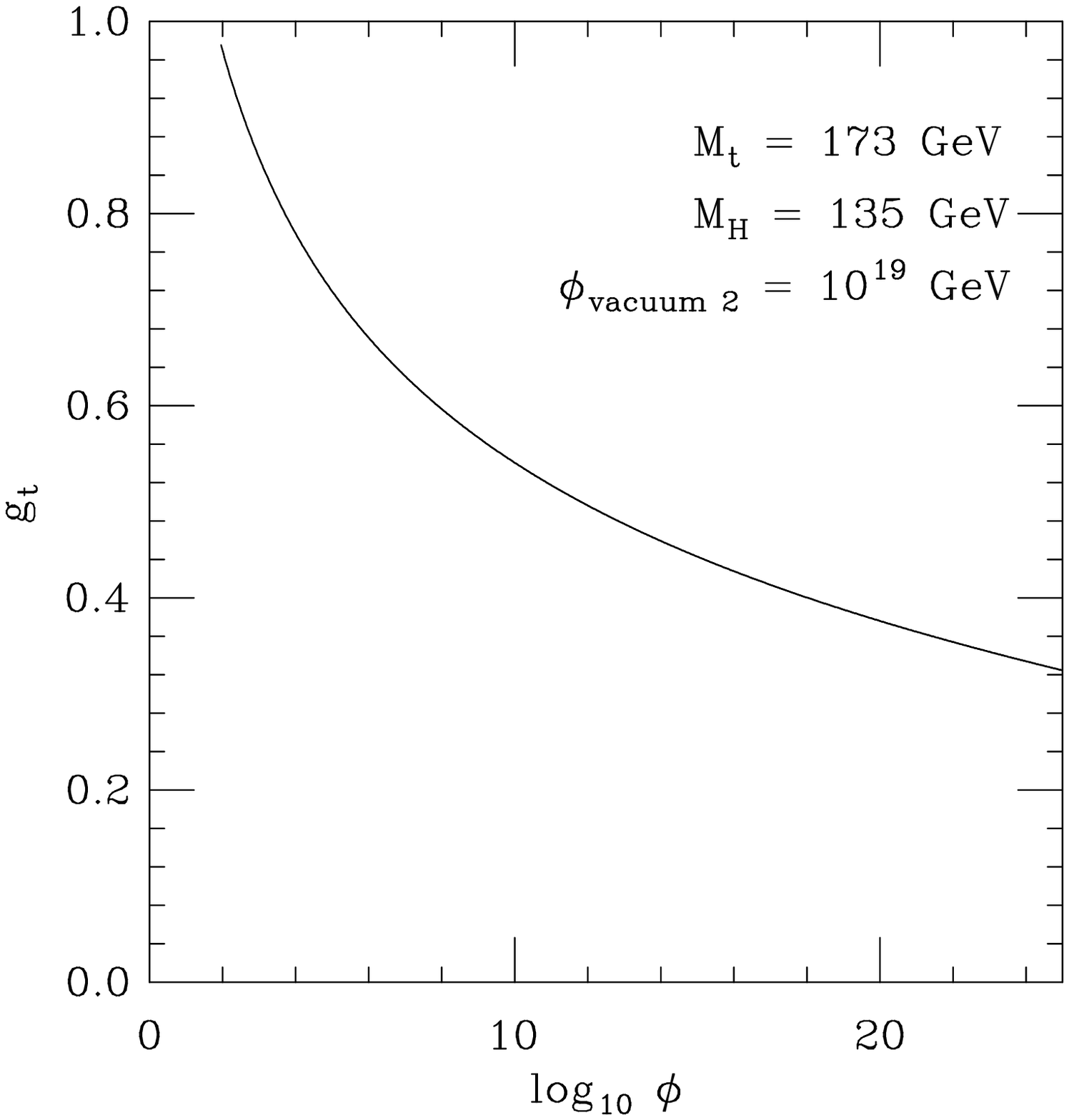}
\caption{Plots of $\lambda$ and $g_t$ as functions
of the scale of the Higgs field $\phi$ for degenerate vacua with
the second Higgs VEV at the Planck scale $\phi_{vac\;2}=10^{19}$
GeV. We formally apply the second order SM renormalisation group
equations up to a scale of $10^{25}$ GeV.} \label{fig:lam19}
\end{figure}

For large values of the SM Higgs field $\phi >> \phi_{vac\; 1}$,
the renormalisation group improved tree level effective potential
is very well approximated by 
$$V_{eff}(\phi) \simeq
\frac{1}{8}\lambda (\mu = |\phi | ) |\phi |^4$$ 
and the degeneracy
condition, eq.~(\ref{eqdeg}), means that $\lambda(\phi_{vac\; 2})$
should vanish to high accuracy. The derivative of the effective
potential $V_{eff}(\phi)$ should also be zero at $\phi_{vac\; 2}$,
because it has a minimum there. Thus at the second minimum of the
effective potential the beta function $\beta_{\lambda}$ vanishes
as well:
\begin{equation}
\beta_{\lambda}(\mu = \phi_{vac\; 2}) = \lambda(\phi_{vac\; 2}) = 0
\end{equation}
which gives to leading order the relationship:
\begin{equation}
\frac{9}{4}g_2^4 + \frac{3}{2}g_2^2g_1^2 + \frac{3}{4}g_1^4 -
12g_t^4 = 0
\end{equation}
between the top quark Yukawa coupling $g_t(\mu)$ and the electroweak
gauge coupling constants $g_1(\mu)$ and $g_2(\mu)$ at the scale
$\mu = \phi_{vac\; 2} \simeq \Lambda_{Planck}$. We use the
renormalisation group equations to relate the couplings at the
Planck scale to their values at the electroweak scale. Figure
\ref{fig:lam19} shows the running coupling constants
$\lambda(\phi)$ and $g_t(\phi)$ as functions of $\log(\phi)$.
Their values at the electroweak scale give our predicted
combination of pole masses \cite{fn2}:
\begin{equation}
M_{t} = 173 \pm 5\ \mbox{GeV} \quad M_{H} = 135 \pm 9\ \mbox{GeV}
\end{equation}

We have also considered \cite{metaMPP} a slightly modified version
of MPP, according to which the two vacua are approximately
degenerate in such a way that they should both be physically
realised over comparable amounts of space-time four volume. This
modified MPP corresponds to the Higgs mass lying on the vacuum
{\em metastability} curve rather than on the vacuum stability
curve, giving a Higgs mass prediction of $122 \pm 11$ GeV. We
should presumably not really take the MPP predictions to be more
accurate than to the order of magnitude of the variation between
the metastability and stability bounds. However we definitely
predict a light Higgs mass in this range, as seems to be in
agreement with indirect estimates of the SM Higgs mass from
precision data \cite{pdg}.

This application of the MPP assumes the existence of the hierarchy
$$v/\Lambda_{Planck} \sim 10^{-17}.$$ 
Recently we have speculated
\cite{itepfn} that this huge scale ratio is a consequence of the
existence of yet another vacuum in the SM, at the electroweak
scale and degenerate with the two vacua discussed above. The two
SM vacua at the electroweak scale are postulated to differ by the
condensation of an S-wave bound state formed from 6 top and 6
anti-top quarks mainly due to Higgs boson exchange forces. This
scenario is discussed in more detail in Holger Nielsen's talk
\cite{holger}.

\section{Lightest Family Mass Generation Model}
\label{ansatze}

Motivated by the famous Fritzsch ansatz \cite{fritzsch} for the two
generation quark mass matrices:
\begin{equation}
M_U =\begin{pmatrix}
          0      & B\cr
          B^{\ast}   & A\cr
     \end{pmatrix}
\qquad
M_D =\begin{pmatrix}
          0          & B^\prime\cr
          B^{\prime\ast}  & A^\prime\cr
     \end{pmatrix}
\label{fritzsch_ansatz}
\end{equation}
several ans\"{a}tze have been proposed for the fermion mass
matrices---for example, see
\cite{rrr} for a systematic analysis of symmetric quark mass
matrices with texture zeros at the SUSY-GUT scale.
Here I will concentrate on the lightest family mass generation
model \cite{lfm}. It successfully generalizes the well-known
formula
\begin{equation}
\left| V_{us} \right| \simeq
\left|\sqrt{\frac{m_d}{m_s}} -
e^{i\phi} \sqrt{\frac{m_u}{m_c}} \right|
\label{fritzsch1}
\end{equation}
for the Cabibbo angle derived from the above ansatz,
eq.~(\ref{fritzsch_ansatz}), to simple working formulae for all
the quark mixing angles in terms of quark mass ratios. According
to this model the flavour mixing for quarks is basically
determined by the mechanism responsible for generating the
physical masses of the up and down quarks, $m_{u}$ and $m_{d}$
respectively. So, in the chiral symmetry limit, when $m_{u}$ and
$m_{d}$ vanish, all the quark mixing angles vanish. Therefore we
are led to consider an ansatz in which the diagonal mass matrix
elements for the second and third generations are practically the
same in the gauge (unrotated) and physical bases.

The mass matrix for the down quarks ($D$ = $d$, $s$, $b$) is taken
to be hermitian with three texture zeros of the following form:
\begin{equation}
M_{D}=\begin{pmatrix} 0 & a_D & 0 \cr a_D^{\ast} & A_D & b_D \cr 0 &
b_D^{\ast} & B_D \cr\end{pmatrix}
\label{LFM1}
\end{equation}
where
\begin{equation}
B_{D}=m_{b}+\delta_{D} \qquad A_{D}= m_{s} + \delta_{D}^{\prime }
\qquad |\delta_D |\ll m_{s} \qquad |\delta _{D}^{\prime }|\ll m_{d}
\label{BA}
\end{equation}
It is, of course, necessary to assume some hierarchy between the
elements, which we take to be: $B_{D}\gg A_{D}\sim \left|
b_{D}\right| \gg \left| a_{D}\right| $. The zero in the $\left(
M_{D}\right) _{11}$ element corresponds to the commonly accepted
conjecture that the lightest family masses appear as a direct
result of flavour mixings. The zero in $\left( M_{D}\right) _{13}$
means that only minimal ``nearest neighbour'' interactions occur,
giving a tridiagonal matrix structure. Since the trace and
determinant of the hermitian matrix $M_{D}$ gives the sum and
product of its eigenvalues, it follows that
\begin{equation}
\delta _{D}\simeq - m_{d}  \label{del}
\end{equation}
while $\delta _{D}^{\prime }$ is vanishingly small and can be neglected
in further considerations.

It may easily be shown that equations (\ref{LFM1} - \ref{del}) are
entirely equivalent to the condition that the diagonal
elements ($A_{D}$, $B_{D}$) of $M_{D}$ are proportional
to the modulus square of the off-diagonal elements ($a_{D}$, $b_{D}$):
\begin{equation}
\frac{A_{D}}{B_{D}}=\left| \frac{a_{D}}{b_{D}}\right| ^{2}
\label{ABab}
\end{equation}
Using the conservation of the trace, determinant and sum of
principal minors of the hermitian matrix $M_{D}$ under unitary
transformations, we are led to a complete determination of the
moduli of all its elements, which can be expressed to high
accuracy as follows:
\begin{equation}
\left| M_{D} \right| = \begin{pmatrix} 0 & \sqrt{m_d m_s} & 0 \cr
\sqrt{m_d m_s} & m_s & \sqrt{m_d m_b} \cr 0 &
\sqrt{m_d m_b} & m_b - m_d \cr\end{pmatrix}  \label{LFM1A}
\end{equation}

The mass matrix for the up quarks is taken to be of the following
hermitian form:
\begin{equation}
M_{U}=\begin{pmatrix} 0 & 0 & c_U \cr 0 & A_U & 0 \cr c_U^{\ast} & 0 & B_ U \cr\end{pmatrix}
\label{LFM2}
\end{equation}
The moduli of all the elements of $M_{U}$ can also be readily
determined in terms of the physical masses as follows:
\begin{equation}
\left| M_{U} \right| = \begin{pmatrix} 0 & 0 & \sqrt{m_u m_t} \cr
0 & m_c & 0 \cr \sqrt{m_u m_t} & 0 & m_t - m_u \cr\end{pmatrix}
\label{LFM2A}
\end{equation}

The CKM quark mixing matrix elements can now be readily calculated
by diagonalising the mass matrices $M_D$ and $M_U$. They are
given  in terms of quark mass ratios as follows:
\begin{eqnarray}
\left| V_{us}\right| = \sqrt{\frac{m_{d}}{m_{s}}} = 0.222 \pm 0.004
\qquad \left| V_{us}\right|_{exp} = 0.221 \pm 0.003 \\
\left|V_{cb}\right| = \sqrt{\frac{m_{d}}{m_{b}}} = 0.038 \pm 0.004
\qquad \left|V_{cb}\right|_{exp} = 0.039 \pm 0.003 \\
\left|V_{ub}\right| = \sqrt{\frac{m_{u}}{m_{t}}} = 0.0036 \pm 0.0006
\qquad \left|V_{ub}\right|_{exp} = 0.0036 \pm 0.0006  \\
\left|V_{td}\right| = \left|V_{us}V_{cb}-V_{ub}\right|
= 0.009 \pm 0.002
\qquad \left|V_{ub}\right|_{exp} = 0.0077 \pm 0.0014
\label{angles}
\end{eqnarray}
As can be seen, they are in impressive agreement with the experimental
values. The MNS lepton mixing matrix can also be fitted, if the texture
of eq.~(\ref{LFM1}) is extended to the Dirac and Majorana right-handed
neutrino mass matrices \cite{matsuda}.

The proportionality condition, eq.~(\ref{ABab}), is not so easy to
generate from an underlying symmetry beyond the Standard Model,
but it is possible to realise it in a local chiral $SU(3)$ family
symmetry\footnote{See ref.~\cite{SU3ross} for a local chiral
SU(3) family model with an alternative texture.} model \cite{SU3}.

\section{Family Replicated Gauge Group Model}
\label{origin}

As pointed out in section \ref{protection}, a natural explanation
of the charged fermion mass hierarchy would be mass protection due
to the existence of some approximately conserved chiral charges
beyond the SM. An attractive possibility is that these chiral
charges arise as a natural feature of the gauge symmetry group of
the fundamental theory beyond the SM. This is the case in the
family replicated gauge group model (also called the anti-grand
unification model) \cite{smg3m,fnt}. The new chiral charges
provide selection rules forbidding the transitions between the
various left-handed and right-handed quark-lepton states, except
for the top quark. In order to generate mass terms for the other
fermion states, we have to introduce new Higgs fields, which break
the symmetry group $G$ of the fundamental theory down to the SM
group. We also need suitable intermediate fermion states to
mediate the forbidden transitions, which we take to be vector-like
Dirac fermions with a mass of order the fundamental scale $M_F$ of
the theory. In this way effective SM Yukawa coupling constants are
generated \cite{fn1}, which are suppressed by the appropriate
product of Higgs field VEVs measured in units of $M_F$. We assume
that all the couplings in the fundamental theory are unsuppressed,
i.e.~they are all naturally of order unity.

The family replicated gauge group model is based on a non-simple
non-supersymmetric extension of the SM with three copies of the SM
gauge group---one for each family or generation. With the
inclusion of three right-handed neutrinos, the gauge group becomes
$G = (SMG \times U(1)_{B-L})^3$, where the three copies of the SM
gauge group are supplemented by an abelian $(B-L)$ (= baryon
number minus lepton number) gauge group for each
family\footnote{The family replicated gauge groups $(SO(10))^3$
and $(E_6)^3$ have recently been considered by Ling and Ramond
\cite{ling}.}. The gauge group $G$ is the largest anomaly free
group, transforming the known 45 Weyl fermions plus the three
right-handed neutrinos into each other unitarily, which does {\em
not} unify the irreducible representations under the SM gauge
group. It is supposed to be effective at energies near to the
Planck scale, $M_F = \Lambda_{Planck}$, where the $i$'th
proto-family couples to just the $i$'th group factor $SMG_i\times
U(1)_{B_i-L_i}$. The gauge group $G$ is broken down by four Higgs
fields $W$, $T$, $\rho$ and $\omega$, having VEVs about one order
of magnitude lower than the Planck scale, to its diagonal
subgroup:
\begin{equation}
(SMG \times U(1)_{B-L})^3 \rightarrow SMG\times U(1)_{B-L}
\end{equation}
The diagonal $U(1)_{B-L}$ is broken down at the see-saw scale, by
another Higgs field $\phi_{SS}$, and the diagonal $SMG$ is broken
down to $SU(3) \times U(1)_{em}$ by the Weinberg-Salam Higgs field
$\phi_{WS}$.

\begin{table}[!th]
\caption{All $U(1)$ quantum charges of the Higgs fields in the
$(SMG \times U(1)_{B-L})^3$ model.} \vspace{3mm} \label{qc}
\begin{center}
\begin{tabular}{|c||c|c|c|c|c|c|} \hline
& $y_1/2$& $y_2/2$ & $y_3/2$ & $(B-L)_1$ & $(B-L)_2$ & $(B-L)_3$
\\ \hline\hline
$\omega$ & $\frac{1}{6}$ & $-\frac{1}{6}$ & $0$ & $0$ & $0$ & $0$\\
$\rho$ & $0$ & $0$ & $0$ & $-\frac{1}{3}$ & $\frac{1}{3}$ & $0$\\
$W$ & $0$ & $-\frac{1}{2}$ & $\frac{1}{2}$ & $0$ & $-\frac{1}{3}$
& $\frac{1}{3}$ \\
$T$ & $0$ & $-\frac{1}{6}$ & $\frac{1}{6}$ & $0$ & $0$ & $0$\\
$\phi_{\scriptscriptstyle WS}$ & $0$ & $\frac{2}{3}$ & $-\frac{1}{6}$ & $0$
& $\frac{1}{3}$ & $-\frac{1}{3}$ \\
$\phi_{\scriptscriptstyle SS}$ & $0$ & $1$ & $-1$ & $0$ & $2$ & $0$ \\
\hline
\end{tabular}
\end{center}
\end{table}
The $(SMG \times U(1)_{B-L})^3$ gauge quantum numbers of the
quarks and leptons are uniquely determined by the structure of the
model and they include 6 chiral abelian charges---the weak
hypercharge $y_i/2$ and $(B-L)_i$ quantum number for each of the
three families, $i=1,2,3$. With the choice of the abelian charges
in Table \ref{qc} for the Higgs fields, it is possible to generate
a good order of magnitude fit to the SM fermion masses, with VEVs
of order $M_F/10$. In this fit, we do not attempt to guess the
spectrum of superheavy fermions at the Planck scale, but simply
assume a sufficiently rich spectrum to mediate all of the symmetry
breaking transitions in the various mass matrix elements. Then,
using the quantum numbers of Table \ref{qc}, the suppression
factors are readily calculated as products of Higgs field VEVs
measured in Planck units for all the fermion Dirac mass matrix
elements\footnote{For clarity we distinguish between Higgs fields
and their hermitian conjugates.}, giving for example:
\begin{eqnarray}
M_{\scriptscriptstyle U} \simeq \frac{\CDFsVEV{(\phi_{\scriptscriptstyle\rm WS})^\dagger}}
{\sqrt{2}}\hspace{-0.1cm} \left(\!\begin{array}{ccc}
        (\omega^\dagger)^3 W^\dagger T^2
        & \omega \rho^\dagger W^\dagger T^2
        & \omega \rho^\dagger (W^\dagger)^2 T\\
        (\omega^\dagger)^4 \rho W^\dagger T^2
        &  W^\dagger T^2
        & (W^\dagger)^2 T\\
        (\omega^\dagger)^4 \rho
        & 1
        & W^\dagger T^\dagger
\end{array} \!\right)\label{M_U}
\end{eqnarray}
for the up quarks. Similarly the right-handed neutrino Majorana
mass matrix is of order:
\begin{eqnarray}
M_R \simeq \CDFsVEV{\phi_{\scriptscriptstyle\rm SS}}\hspace{-0.1cm} \left
(\hspace{-0.1 cm}\begin{array}{ccc} (\rho^\dagger)^6 T^6 &
(\rho^\dagger)^3 T^6
& (\rho^\dagger)^3 W^3 (T^\dagger)^3 \\
(\rho^\dagger)^3 T^6
& T^6 & W^3 (T^\dagger)^3 \\
(\rho^\dagger)^3 W^3 (T^\dagger)^3 & W^3 (T^\dagger)^3 & W^6
(T^\dagger)^{12}
\end{array} \hspace{-0.1 cm}\right ) \label{Mmajo}
\end{eqnarray}
and the effective light neutrino mass matrix can be calculated
from the Dirac neutrino mass matrix $M_N$ and $M_R$ using the
see-saw formula \cite{seesaw}:
\begin{equation}
M_{\nu} = M_N M_R^{-1} M_N^T
\end{equation}
In this way we obtain a good 5 parameter fit to the orders of
magnitude of all the quark-lepton masses and mixing angles, as
given in Table \ref{convbestfit}, actually even with the expected
accuracy \cite{douglas}.
\begin{table}[!t]
\caption{Best fit to quark-lepton mass spectrum. All masses are
running masses at $1~\mbox{\rm GeV}$ except the top quark mass which is the
pole mass.}
\begin{displaymath}
\begin{array}{|c|c|c|}
\hline\hline
 & {\rm Fitted} & {\rm Experimental} \\ \hline
m_u & 4.4~\mbox{\rm MeV} & 4~\mbox{\rm MeV} \\
m_d & 4.3~\mbox{\rm MeV} & 9~\mbox{\rm MeV} \\
m_e & 1.6~\mbox{\rm MeV} & 0.5~\mbox{\rm MeV} \\
m_c & 0.64~\mbox{\rm GeV} & 1.4~\mbox{\rm GeV} \\
m_s & 295~\mbox{\rm MeV} & 200~\mbox{\rm MeV} \\
m_{\mu} & 111~\mbox{\rm MeV} & 105~\mbox{\rm MeV} \\
M_t & 202~\mbox{\rm GeV} & 180~\mbox{\rm GeV} \\
m_b & 5.7~\mbox{\rm GeV} & 6.3~\mbox{\rm GeV} \\
m_{\tau} & 1.46~\mbox{\rm GeV} & 1.78~\mbox{\rm GeV} \\
V_{us} & 0.11 & 0.22 \\
V_{cb} & 0.026 & 0.041 \\
V_{ub} & 0.0027 & 0.0035 \\ \hline
\Delta m^2_{\odot} & 9.0 \times 10^{-5}~\mbox{\rm eV}^2 &  5.0 \times 10^{-5}~\mbox{\rm eV}^2 \\
\Delta m^2_{\rm atm} & 1.7 \times 10^{-3}~\mbox{\rm eV}^2 &  2.5 \times 10^{-3}~\mbox{\rm eV}^2\\
\tan^2\theta_{\odot} &0.26 & 0.34\\
\tan^2\theta_{\rm atm}& 0.65 & 1.0\\
\tan^2\theta_{\rm chooz}  & 2.9 \times 10^{-2} & \CDFsleq~2.6 \times 10^{-2}\\
\hline\hline
\end{array}
\end{displaymath}
\label{convbestfit}
\end{table}

\section{Conclusion}
\label{conclusion}

The hierarchical structure of the quark-lepton spectrum was
emphasized and interpreted as due to the existence of a mass
protection mechanism, controlled by approximately conserved chiral
flavour quantum numbers beyond the SM. The family replicated gauge
group model assigns a unique set of anomaly free gauge charges to
the quarks and leptons. With an appropriate choice of quantum
numbers for the Higgs fields, these chiral charges naturally
generate a realistic set of quark-lepton masses and mixing angles.
The top quark dominates the fermion mass matrices and we showed
how the Multiple Point Principle can be used to predict the top
quark and SM Higgs boson masses. We also discussed the lightest
family mass generation model, which gives simple and compact
formulae for all the CKM mixing angles in terms of the quark
masses.

\section*{Acknowledgements}
I should like to thank my collaborators Jon Chkareuli, Holger Bech
Nielsen and Yasutaka Takanishi for many discussions.


\title*{How to Approach Quantum Gravity --
Background Independence in $1+1$ Dimensions}
\author{Daniel~Grumiller\footnote{grumil@hep.itp.tuwien.ac.at} {} 
and Wolfgang~Kummer\footnote{wkummer@tph.tuwien.ac.at}}
\institute{%
Institute f. Theor. Physics\\
Vienna University of Technology\\
Wiedner Hauptstr. 8-10, A-1040 Vienna, Austria
}

\titlerunning{How to Approach Quantum Gravity}
\authorrunning{Daniel~Grumiller and Wolfgang~Kummer}
\maketitle

\begin{abstract}

The application of quantum theory to gravity is beset with
many technical and conceptual problems. After a short tour d'horizon
of recent attempts to master those problems by the introduction of
new approaches, we show that the aim, a background independent quantum
theory of gravity, can be reached in a particular area, 2d dilaton quantum
gravity, {\it without} any assumptions beyond standard
quantum field theory.

\end{abstract}

\section{Introduction}\label{se:1}

It has been realized for some time that a merging
of quantum theory with Einstein's theory of general relativity\footnote{Several reviews on quantum gravity have emerged at the turn of the millennium, cf.\ e.g.\  \cite{Horowitz:1996qd,Carlip:2001wq}.} (GR)
is necessitated by consistency arguments. In {\it Gedankenexperimenten}
the interaction of a classical gravitational wave with a quantum system
inevitably leads to contradictions \cite{Eppley:1977}. Arguments of this type are important because no relevant experimental data are available -- we are very far from the quantum gravity analogue of the Balmer series.

On the other hand, when a quantum theory (QT) of gravity is
developed along usual lines, one is confronted with a fundamental
problem, from which many other (secondary) difficulties can be traced.
The crucial difference to quantum field theory
(QFT) in flat space is the fact that the variables of gravity exhibit
a dual role, they are fields living on a manifold which is determined
by themselves, {}``stage'' and {}``actors'' coincide. But there exist also
numerous other problems: 
the time variable, an object with special properties already
in QT, in GR appears on an equal footing with the space coordinates 
(``problem of time'' which manifests itself in many disguises); the information paradox
\cite{Banks:1995ph}; perturbative non-renormalizability \cite{'tHooft:1974bx} etc.

In section \ref{se:2} we discuss some key-points regarding the definition
of physical observables in QFT and the ensuing ones in quantum gravity
(QGR). Then we critically mention some {}``old'' and {}``new''
approaches to QGR (section \ref{se:3}) from a strictly quantum field theorist's
point of view. Finally we give some highlights on the ``Vienna approach''
to 2d dilaton quantum gravity with matter, including a new result (within that approach) on entropy corrections which is in agreement with the one found in literature (section \ref{se:4}). In that area which contains also
models with physical relevance (e.g.\ spherically reduced gravity)
the application of just the usual concepts of (even nonpertubative!)
QFT lead to very interesting consequences \cite{Grumiller:2002nm} which allow
physical interpretations in terms of {}``solid'' traditional QFT
observables.

\section{Observables}\label{se:2}

\subsection{Cartan variables in GR}

Physical observables in the sense used here are certain functionals
of the field variables which are directly accessible to experimental
measurements.

The metric $g$ in GR can be considered as a {}``derived'' field variable
\begin{equation}
g=e^{a}\otimes e^{b}\eta_{ab},\label{eq:1}\end{equation}
because it is the direct product of the dual basis
one forms\footnote{For details on gravity in the Cartan formulation we refer to the mathematical literature, e.g.\ \cite{nakaharageometry}} $e^{a}=e_{\mu}^{a}\, dx^{\mu}$ contracted with the flat local Lorentz
metric $\eta_{ab}$ which is used to raise and lower ``flat indices'' denoted by Latin letters ($\eta={\rm diag}(1,-1,-1,-1,....),\,\, x^{\mu}=\left\{ x^{0},x^{i}\right\}$).
Local Lorentz invariance leads to the ``covariant derivative'' $D^{a}{}_b=\delta_{b}^{a}d+\omega^{a}{}_b$
with a spin connection 1-form $\omega^{a}{}_b$
as a gauge field. Its antisymmetry $\omega^{ab}=-\omega^{ba}$ implies metricity. Thanks to the Bianchi identities all covariant tensors relevant for constructing actions in even dimensions 
can be expressed in terms of $e^{a}$, the curvature 2-form
$R^{ab}=(D\omega)^{ab}$ and the torsion 2-form $T^{a}=(De)^{a}$.
For nonvanishing torsion the affine connection $\Gamma_{\mu\nu}$$\,^{\rho}=E_{a}^{\rho}\,\left(D_{\mu}e\right)_{\nu}^{a}$,
expressed in terms of components $e_{\mu}^{a}$ and of its inverse
$E_{a}^{\rho}$, besides the usual Christoffel symbols also contains
a contorsion term in $\Gamma_{\left(\mu\nu\right)}\,^{\rho}$, whereas
$\Gamma_{\left[\mu\nu\right]}\,^{\rho}$ are the components of torsion.
Einstein gravity in d=4 dimensions postulates vanishing
torsion $T^{a}=0$ so that $\omega=\omega(e)$. This theory can be
derived from the Hilbert action ($G_{N}$ is Newton's constant; dS space results
for nonvanishing cosmological constant $\Lambda$ from the replacement $R^{ab}\rightarrow R^{ab}-\frac{4}{3}\Lambda e^{a}\wedge e^{b}$)
\begin{equation}
L_{\left(H\right)}=\frac{1}{16\pi G_{n}}\int_{\mathcal{M}_{4}}\, R^{ab}\wedge e^{c}\wedge e^{d}\epsilon_{abcd}+L_{\left({\rm matter}\right)}.\label{eq:2}\end{equation}
Because of the {}``Palatini mystery'', independent
variation of $\delta\omega$ yields $T^{a}=0,$ whereas $\delta e$
produces the Einstein equations.

Instead of working with the metric (\ref{eq:1}) the {}``new''
approaches \cite{Sen:1982qb} are based upon a gauge field related to
$\omega^{ab}$
\begin{equation}
A^{ab}=\frac{1}{2}\left(\omega^{ab}-\frac{\gamma}{2}\epsilon^{ab}{}_{cd}\,\omega^{cd}\right).\label{eq:3}\end{equation}
The Barbero-Immirzi parameter $\gamma$ \cite{Barbero:1995ap} is an arbitrary constant. 
The extension to complex gravity
$\left(\gamma=i\right)$ makes $A^{a}$ a self-adjoint field and transforms
the Einstein theory into the one of an $SU\left(2\right)$ gauge field
\begin{equation}
A_{i}^{\underline{a}}=\epsilon^{0\,\underline{a}}\,_{\underline{b}\,\underline{c}}\, A_{i}^{\,\underline{b}\,\underline{c}},\label{eq:4}\end{equation}
where the index $\underline{a}=1,2,3.$ This formulation
is the basis of loop quantum gravity and spin foam models (see below).

\subsection{Observables in classical GR}

The exploration of the global properties of a certain solution
of (\ref{eq:2}), its singularity structure etc., is only possible
by means of the introduction of an additional test field, most simply
a test particle with action
\begin{eqnarray}
L_{\left({\rm test}\right)} & = & -m_{0}\int\left|ds\right|,\nonumber \\
ds^{2} & = & g_{\mu\nu}\left(x\left(\tau\right)\right)\frac{dx}{d\tau}^{\mu}\frac{\, dx\,^{\nu}}{d\tau},\label{eq:5}\end{eqnarray}
which is another way to incorporate Einstein's old
proposal \cite{Einstein:1916vd} of a {}``net of geodesics''. The
path $x^{\mu}\left(\tau\right)$ is parameterized by the affine parameter
$\tau$ (actually only timelike or lightlike $ds^{2}\geq 0$ describes the paths of
a physical particle).

It is not appreciated always that the global properties of
a manifold are \textit{\large defined} in terms of a specific
device like (\ref{eq:5}). Whereas the usual geodesics derived from
(\ref{eq:5}) depend on $g_{\mu\nu}$ through the Christoffel symbols
only, e.g.\ in the case of torsion also the contorsion may contribute
({}``autoparallels'') in the affine connection; spinning particles
{}``feel'' the gravimagnetic effect etc. As a consequence, when
a field dependent transformation of the gravity variables is performed
(e.g.\ a conformal transformation from a {}``Jordan frame'' to an
{}``Einstein frame'' in Jordan-Brans-Dicke \cite{Fierz:1956} theory)
the action of the device must be transformed in the same way.

\subsection{Observables in QFT}

In flat QFT one starts from a Schr\"odinger
equation, dependent on field operators and, proceeding
through Hamiltonian quantization to the path integral, the experimentally
accessible observables are the elements of the S-matrix, or quantities
expressible by those.\footnote{Note that ordinary quantum mechanics and its Schr\"odinger equation
appear as the nonrelativistic, weak coupling limit of the Bethe-Salpeter
equation of QFT \cite{Salpeter:1951sz}. Useful notions like eigenvalues of Hermitian operators,
collapse of wave functions etc.\ are not basic concepts in this more
general frame (cf.\ footnote 2 in ref.\ \cite{Kummer:2001ip}).}
It should be recalled that the properly defined renormalized
S-matrix element obtains by amputation of external propagators in
the related Green function, multiplication with polarizations and
with a square root of the wave function renormalization constant,
taking the mass-shell limit.

In gauge theories  one
encounters the additional problem of gauge-de\-pen\-dence, i.e.\ the dependence
on some gauge parameter $\beta$ introduced by generic gauge fixing.
Clearly the S-matrix elements must be and indeed are \cite{Kummer:2001ip}
independent of $\beta$. But other objects, in particular matrix-elements
of gauge invariant operators $\mathcal{O}_{A}$, depend on $\beta$.
In addition, under renormalization they mix with operators $\tilde{\mathcal{O}}_{\tilde{A}}$
of the same {}``twist'' (dimension minus spin) which depend on Faddeev-Popov
ghosts \cite{Dixon:1974ss} and are not gauge-invariant:
\begin{eqnarray}
&& \mathcal{O}_{A}^{\left(ren\right)} = Z_{AB}\mathcal{O}_{B}+ Z_{A\tilde{B}}\tilde{\mathcal{O}}_{\tilde{B}}\nonumber \\
&& \tilde{\mathcal{O}}_{\tilde{A}}^{\left(ren\right)} =  \phantom{Z_{AB}\mathcal{O}_{B}+\ } Z_{\tilde{A}\tilde{B}}\tilde{\mathcal{O}}_{\tilde{B}}\label{eq:6}\end{eqnarray}
The contribution of such operators to the S-matrix element
(sic!) of e.g.\ the scaling limit for deep inelastic scattering \cite{Friedman:1991} of leptons on protons \cite{Gross:1973ju} occurs only through the anomalous dimensions 
($\propto\partial Z_{AB}/\partial\Lambda$ for a regularisation
cut-off $\Lambda$). And those objects, also thanks to the triangular form of
(\ref{eq:6}), are gauge-independent!

In flat QFT, as well as in QGR, the (gauge invariant) {}``Wilson
loop''
\begin{equation}
W_{\left(\mathcal{C}\right)}={\rm Tr}\, P\exp{\big(i\oint\limits _{\mathcal{C}}A_{\mu\, dx^{\mu}}\big)},\label{eq:7}\end{equation}
parameterized by a path ordered closed curve $\mathcal{C},$
often is assumed to play an important role. In covariant gauges it
is multiplicatively renormalizable with the renormalization constant
depending on the length of $\mathcal{C}$, the UV cut-off and eventual
cusp-angles in $\mathcal{C}$ \cite{Polyakov:1980ca}. Still the relation
to experimentally observable quantities (should one simply drop the renormalization constant or proceed \cite{Kummer:2001ip} as for an S-matrix?) is unclear. Worse, for lightlike axial gauges $\left(nA\right)=0$ $\left(n^{2}=0\right)$ multiplicative renormalization is not
applicable \cite{Andrasi:1999hu}. Then, only for a matrix element of (\ref{eq:7}) between
{}``on-shell gluons'', this type of renormalization is restored.
Still the renormalization constant is different from covariant gauge,
except for the anomalous dimension derived from it (cf.\ precisely that
feature of operators in deep inelastic scattering).

\section{Approaches to QGR}\label{se:3}

``Old'' QGR worked with a separation of the two aspects
of gravity variables by the decomposition of the metric
\begin{equation}
g_{\mu\nu}=g_{\mu\nu}^{\left(0\right)}+h_{\mu\nu},\label{eq:8}
\end{equation}
which consists of a (fixed) classical background $g_{\mu\nu}^{\left(0\right)}$
({}``stage'') with small quantum fluctuations $h_{\mu\nu}$ ({}``actors'').
The {}``observable'' (to be tested by a classical device) would
be the effective matrix $g_{\mu\nu}^{\left({\rm eff}\right)}=g_{\mu\nu}^{\left(0\right)}+<h_{\mu\nu}>.$
Starting computations from the action (\ref{eq:2}) one finds that an ever increasing number of counter-terms is necessary. They are different from the terms in
the Lagrangian $\mathcal{L}=\sqrt{-g}\, R/\left(16\pi G_{N}\right)$
in (\ref{eq:2}). This is the reason why QGR is called (perturbatively) {}``nonrenormalizable''
\cite{'tHooft:1974bx}. Still, at energies E $\ll\left(G_{N}\right)^{-1/2}$,
i.e.\ much below the Planck mass scale $m_{\rm Pl}\propto\left(G_{N}\right)^{-1/2}$,
such calculations can be meaningful in the sense of an {}``effective
low energy field theory'' \cite{Donoghue:1994dn}, irrespective of the fact that (perhaps
by embedding gravity into string theory) by inclusion of further fields
at higher energy scales (Planck scale), QGR may become renormalizable.
Of course, such an approach even when it is modified by iterative
inclusion of $<h_{\mu\nu}>$ into $g_{\mu\nu}^{\left(0\right)}$ etc.\
-- which is technically quite hopeless -- completely misses inherent
background independent effects, i.e.\ effects when $g_{\mu\nu}^{\left(0\right)}=0$.

One could think also of applying nonperturbative methods developed
in numerical lattice calculations for QCD. However, there are problems
to define the Euclidean path integral for that, because the Euclidean
action is not bounded from below (as it is the case in QCD) \cite{Gibbons:1994cg}.

The quantization of gravity which -- at least in principle
-- avoids background dependence is based upon the ADM approach to the
Dirac quantization of the Hamiltonian \cite{Arnowitt:1962}. Space-time is foliated
by a sequence of three dimensional space-like manifolds $\sum_{3}$
upon which the variables $g_{ij}=q_{ij}$ and associated canonical
momenta $\pi_{ij}$ live. The constraints associated to the further
variables lapse $\left(N_{0}\right)$ and shift $\left(N_{i}\right)$
in the Hamiltonian density
\begin{equation}
\mathcal{H}=N_{0}\, H^{0}\left(q,\pi\right)+N_{i}\, H^{i}\left(q,\pi\right)\label{eq:9}\end{equation}
are primary ones. The Poisson brackets of the secondary constraints
$H^{\mu}$ closes. $H^{i}$ generates diffeomorphisms on $\sum_{3}$.
In the quantum version of (\ref{eq:9}) the solutions of the Wheeler-deWitt
equation involving the Hamiltonian constraint
\begin{equation}
\int\limits _{\sum_{3}}\, H^{0}\,\left(q,\,\frac{\delta}{i\delta q}\right)\mid\psi>=0\label{eq:10}\end{equation}
 formally would correspond to a nonperturbative QGR. Apart
from the fact that it is extremely difficult to find a general solution
to (\ref{eq:10}) there are several basic problems with a quantum
theory based upon that equation (e.g.\ no Hilbert space $\mid\psi>$
can be constructed, no preferred time foliation exists with ensuing inequivalent
quantum evolutions \cite{Torre:1998eq}, problems with usual {}``quantum
causality'' exist, the {}``axiom'' that fields should commute at
space like distances does not hold etc.). A restriction to a finite
number of degrees of freedom ({}``mini superspace'') \cite{Dewitt:1967yk} or infinite number of degrees of freedom (but still less than the original theory -- so-called ``midi superspace'') \cite{Kuchar:1971xm}
has been found to miss essential features.

As all physical states $\mid\psi>$ must be annihilated by
the constraints $H^{\mu}$, a naive Schr\"odinger equation involving
the Hamiltonian constraint $H^{0}$,
\begin{equation}
i\hbar\frac{\partial\mid\psi>}{\partial t}=H^{0}\mid\psi>=0,\label{eq:11}
\end{equation}
cannot contain a time variable ({}``problem of time'').
A kind of Schr\"odinger equation can be produced by the definition of
a {}``time-function'' $T\left(q,\pi,x\right)$, at the price of
an even more complicated formalism \cite{Bergmann:1962} with quite ambiguous
results -- and the problem, how to connect those with {}``genuine''
observables. All these problems are aggravated, when one tries to
first eliminate constraints by solving them explicitly before quantization.
In this way, clearly part of the quantum fluctuations are eliminated
from the start. As a consequence different quantum theories, constructed
in this way, are not equivalent.

The {}``new'' gravities (loop quantum gravity, spin foam models) reformulate the quantum theory of space-time
by the introduction of novel variables, based upon the concept of
Wilson loops (\ref{eq:7}) applied to the gauge-field (\ref{eq:4}).
The operator
\begin{equation}
U\left(s_{1},s_{2}\right)={\rm Tr}\, P\, \exp{\left( i\int\limits _{s_{1}}^{s_{2}}ds\frac{dx^{i}}{ds}\, A_{i}\right)} \label{eq:12}
\end{equation}
defines a holonomy. It is generalized by inserting further
invariant operators at intermediate points between $s_{1}$ and $s_{2}$
. From such holonomies a spin network can be created which represents spacetime
(in the path integral it is dubbed ``spin foam'').

These approaches claim several successes \cite{Carlip:2001wq}. Introducing
as a basis diffeomorphism equivalence classes of {}``labeled graphs''
a finite Hilbert space can be constructed and some solutions of the
Wheeler-deWitt equation (\ref{eq:10}) have been obtained. The methods
introduce a {}``natural'' coarse graining of space-time which implies
a $UV$ cutoff. {}``Small'' gravity around certain states leads
in those cases to corresponding linearized Einstein gravity.

However, despite of very active research in this field a number
of very serious open questions persists: The Hamiltonian constructed
from spin networks does not lead to massless excitations (gravitons)
in the classical limit. The Barbero-Immirzi parameter $\gamma$ has to be fixed by
the requirement of a ``correct'' Bekenstein-Hawking entropy for
the Black Hole. The most severe problem, however, is the one of observables.
By some researchers in this field it has been claimed that by {}``proper
gauge fixing'' (!) area and volume can be obtained as quantized {}``observables'',
which is a contradiction in itself from the point of view of QFT.
We must emphasize too that also in an inherently $UV$ regularized
theory (finite) renormalization remains an issue to be dealt with
properly. Also the fate of S-matrix elements, which play such a central
role as the proper observables in QFT, is completely unclear in these
setups.

Embedding QGR into (super-)string theory \cite{Polchinski:1998rq} does not remove the
key problems related to the dual role of the metric. Gravity may well be a string excitation
in a string/brane world of 10-11 dimensions, possibly a finite
theory of everything. Nevertheless, at low energies Einstein gravity (eventually
plus an antisymmetric $B$-field) remains the theory for which computations
must be performed.\footnote{It should be noted that the now widely confirmed astronomical
observations of a positive cosmological constant \cite{Riess:2001gk}
(if it is a constant and not a {}``quintessence'' field in a theory
of type \cite{Fierz:1956}) precludes immediate application of supersymmetry
(supergravity) in string theory, because only AdS space
is compatible with supergravity \cite{VanNieuwenhuizen:1981ae}.}
Unfortunately, the proper choice (let alone the derivation)
of a string vacuum in our d=4 space-time is an unsolved problem.


Many other approaches exist, including noncommutative geometry, twis\-tors, causal sets, 3d approaches, dynamical triangulations, Regge calculus etc., each of which has certain attractive features and difficulties (cf.\ e.g.\ \cite{Carlip:2001wq} and refs.\ therein).

To us all these {}``new'' approaches appear as -- very ingenious
-- attempts to bypass the technical problems of directly applying standard
QFT to gravity -- without a comprehensive solution of the main problems of QGR being in sight. Thus the
main points of a {}``minimal'' QFT for gravity should be based upon
``proven concepts'' of QFT with a point of departure characterizing QGR
as follows:
\begin{itemize}
\item[(a)] QGR is an {}``effective'' low energy theory and therefore need not
be renormalizable to all orders.
\item[(b)] QGR is based upon classical Einstein (-dS) gravity with usual
variables (metric or Cartan variables).
\item[(c)] At least the quantization of geometry must be performed in a background independent (nonperturbative) way.
\item[(d)] Absolutely {}``safe'' quantum observables are only the S-matrix
elements of QFT ${<f\mid S\mid i>}$, where initial state ${\mid i>}$
and final state ${<f\mid}$ are defined only when those states are
realized as Fock states of particles in a (at least approximate) flat
space environment. In certain cases it is permissible to employ a
semi-classical approach: expectation values of quantum corrections
may be added to classical geometric variables, and a classical computation
is then based on the effective variables, obtained in this way.
\end{itemize}
Clearly item (d) by construction
excludes any application to quantum cosmology, where $\mid i>$ would
be the (probably nonexistent) infinite past before the Big Bang.

Obviously the most difficult issue is (c). We describe in the following
section how gravity models in d=2 (e.g.\ spherically reduced gravity)
permit a solution of just that crucial point, leading to novel results.

\section{``Minimal'' QGR in 1+1 dimensions}\label{se:4}

\subsection{Classical theory: first order formulation}

In the 1990s the interest in dilaton gravity in d=2 was rekindled by results from string theory \cite{Mandal:1991tz}, but it existed as a field on its own more or less since the 1980s \cite{Barbashov:1979}. For a review on dilaton gravity ref.\ \cite{Grumiller:2002nm} may be consulted. For sake of self-containment the study of dilaton gravity will be motivated briefly from a purely geometrical point of view.

The notation of ref.\ \cite{Grumiller:2002nm} is used: $e^a$ is the
zweibein one-form, $\epsilon = e^+\wedge e^-$ is the volume two-form. The one-form
$\omega$ represents the  spin-connection $\omega^a{}_b=\varepsilon^a{}_b\omega$
with  the totally antisymmetric Levi-Civit{\'a} symbol $\varepsilon_{ab}$ ($\varepsilon_{01}=+1$). With the
flat metric $\eta_{ab}$ in light-cone coordinates
($\eta_{+-}=1=\eta_{-+}$, $\eta_{++}=0=\eta_{--}$) the torsion 2-form reads
$T^\pm=(d\pm\omega)\wedge e^\pm$. The curvature 2-form $R^a{}_b$ can be presented by the 2-form $R$ defined by 
$R^a{}_b=\varepsilon^a{}_b R$, $R=d\wedge\omega$.
Signs and factors of the Hodge-$\ast$ operation are defined by $\ast\epsilon=1$. 

Since the Hilbert action $\int_{\mathcal{M}_2}  R\propto(1-g)$ yields just the Euler number for a surface with genus $g$ one has to generalize it appropriately. The simplest idea is to introduce a Lagrange multiplier for curvature, $X$, also known as ``dilaton field'', and an arbitrary potential thereof, $V(X)$, in the action $\int_{\mathcal{M}_2}  \left(XR+\epsilon V(X)\right)$. In particular, for $V\propto X$ the Jackiw-Teitelboim model emerges \cite{Barbashov:1979}. Having introduced curvature it is natural to consider torsion as well. By analogy the first order gravity action \cite{Ikeda:1993aj}
\begin{equation}
L^{(1)}=\int_{\mathcal{M}_2}  \left(X_aT^a+XR+\epsilon\mathcal{V} (X^aX_a,X)\right)
\label{eq:FOG}
\end{equation}
can be motivated where $X_a$ are the Lagrange multipliers for torsion. It encompasses essentially all known dilaton theories in 2d, also known as Generalized Dilaton Theories (GDT). Spherically reduced gravity (SRG) from d=4 corresponds to $\mathcal{V} = -X^+X^-/(2X)-{\rm const}$.

Without matter there are no physical propagating degrees of freedom, which is advantageous mathematically but not very attractive from a physical point of view. Thus, in order to describe scattering processes matter has to be added. The simplest way is to consider a massless Klein-Gordon field $\phi$,
\begin{equation}
L^{(m)} = \frac{1}{2} \; \int_{\mathcal{M}_2}\;
F(X)\, d \phi  \wedge \ast d \phi\,, 
\label{eq:matter}
\end{equation}
with a coupling function $F(X)$ depending on the dilaton (for dimensionally reduced theories typically $F\propto X$ holds). 

\subsection{Quantum theory: Virtual Black Holes}

It turned out that even in the presence of matter an exact path integration 
of all geometric quantities is possible for all GDTs, proceeding along well 
established paths of QFT\footnote{We mention just a few technical details: no ordering ambiguities arise, the (nilpotent) BRST charge is essentially the same as for Yang-Mills theory (despite of the appearance of nonlinearities in the algebra of the first class secondary constraints), the gauge fixing fermion is chosen such that ``temporal'' gauge is obtained, the Faddeev-Popov determinant cancels after integrating out the ``unphysical'' sector, and ``natural'' boundary conditions cannot be imposed on the fields, so one has to be careful with the proper treatment of the boundary.} \cite{Kummer:1992rt}. 

The effective theory obtained in 
this way solely depends on the matter fields in which it is nonlocal and 
non-polynomial. Already at the level of the (nonlocal) vertices of matter 
fields, to be used in a systematic perturbative expansion in terms of Newton's 
constant, a highly nontrivial and physically intriguing phenomenon can be 
observed, namely the so-called  ``virtual black hole'' (VBH). This notion 
originally has been introduced by S. Hawking \cite{Hawking:1996ag}, but in our 
recent approach the VBH for SRG emerges naturally
in Minkowski signature space-time, without the necessity of additional {\em ad 
hoc} assumptions. 

For non-minimally coupled scalars the 
lowest order S-matrix indeed exhibited interesting
features: forward scattering poles, monomial scaling with energy, CPT 
invariance, and pseudo-self-similarity in its kinematic sector 
\cite{Grumiller:2000ah}.

\begin{figure}
\center
\includegraphics[width=80pt]{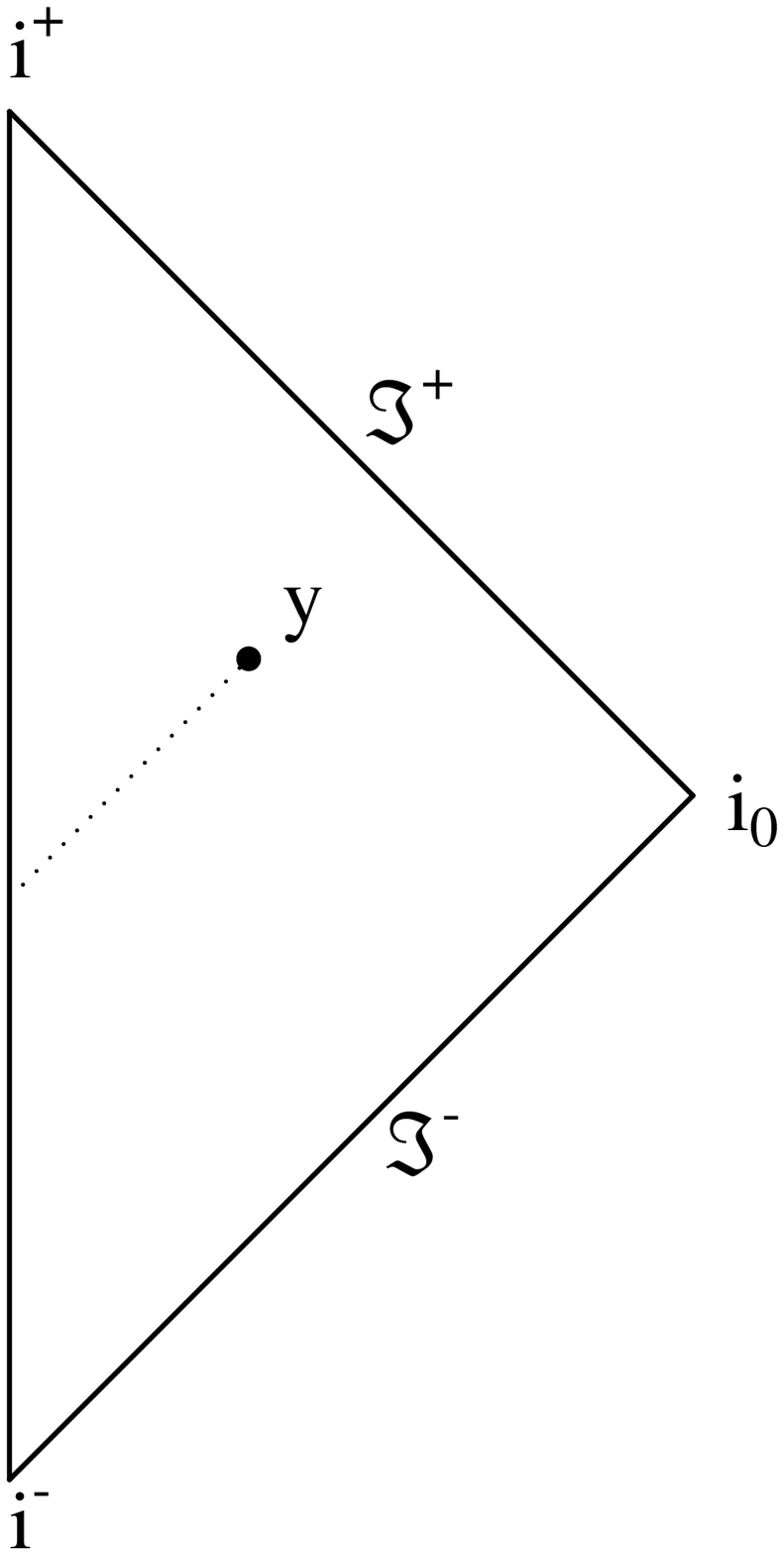}
\caption{CP diagram of the VBH}
\label{fig:cp}
\end{figure}
It was possible to reconstruct geometry self-consistently from a 
(perturbative or, if available, exact) solution of the effective theory. For
the simplest case of four-point tree-graph scattering the corresponding 
Carter-Penrose (CP) diagram is presented in Fig. \ref{fig:cp}. It is non-local 
in the sense that it depends not only on one set of coordinates but on two. 
This was a consequence of integrating out geometry non-perturbatively. For 
each choice of $y$ (one of the two sets of coordinates) it is possible to draw 
an ordinary CP-diagram. The non-trivial part of our effective geometry (i.e.\
the VBH) is concentrated on the light-like cut. For SRG the ensuing 
line-element has Sachs-Bondi form
\begin{equation}
(ds)^2 = 2 dr du + \left(1 - \frac{2m(u,r)}{r} - a(u,r) r + d(u,r)\right) (du)^2\,,
\end{equation}
with $m$, $a$ and $d$ being localized\footnote{The localization of ``mass'' 
and ``Rindler acceleration'' on a light-like cut  
is not an artifact of an accidental gauge choice, but has a physical 
interpretation in terms of the Ricci-scalar. Certain parallels to Hawking's
Euclidean VBHs can be observed, but also essential differences. The main one 
is our Minkowski signature which we deem to be a positive feature.} 
on the cut $u=u_0$ with compact support
$r<r_0$. These quantities depend on the second set of coordinates $u_0$, $r_0$.

One should not take the effective geometry at face value -- this would be like 
over-interpreting the role of virtual particles in a loop diagram. It is a 
nonlocal entity and one still has to ``sum'' (read: integrate) over all 
possible geometries of this type in order to obtain the nonlocal vertices and 
the scattering amplitude. Nonetheless, the simplicity of this geometry and the 
fact that all possible configurations are summed over are nice features of 
this picture. Moreover, all VBH geometries coincide asymptotically and differ only very little from each other in the asymptotic region. This observation allows for the following interpretation: the boundaries of the diagram, $\mathcal{I}^\pm$ and $i^0$, behave in a classical way\footnote{Clearly the imposed boundary conditions play a crucial role in this context. They produce effectively a fixed background, but only at the boundary.} (thus enabling one to construct an ordinary Fock space like in fixed background QFT), but the more one zooms into the geometry the less classical it becomes. The situation seems to be quite contrary to Kucha\v{r}'s proposal of geometrodynamics\footnote{This approach considers only the matterless case and thus a full comparison to our results is not possible.} of BHs: while we have fixed boundary conditions for the target space coordinates (and hence a fixed ADM mass) but a ``smeared geometry'' (in the sense that a continuous spectrum of asymptotically equivalent VBHs contributes to the S-matrix), Kucha\v{r} encountered a ``smeared mass'' (obeying a Schr\"odinger equation) but an otherwise fixed geometry \cite{Kuchar:1994zk}.

Qualitatively it is clear what has to be done in order to obtain the 
S-matrix\footnote{The idea that BHs must be considered in the S-matrix 
together with 
elementary matter fields has been put forward some time ago 
\cite{'tHooft:1996tq}. The approach \cite{Grumiller:2000ah} 
reviewed here, for the first time allowed to derive (rather than to 
conjecture) the appearance of the BH states in the quantum scattering matrix 
of gravity.}: 
Take all possible VBHs of Fig.\ \ref{fig:cp} and sum them coherently with proper
weight factors and suitably attached external legs of scalar fields. This had
been done quantitatively in a straightforward but rather lengthy calculation
for gravitational scattering of s-waves in the framework of SRG, the result of 
which yielded the lowest order tree-graph S-matrix for ingoing modes with 
momenta $q,q'$ and outgoing ones $k,k'$,
\begin{equation}
T(q, q'; k, k') = -\frac{i\kappa\delta\left(k+k'-q-q'\right)}{2(4\pi)^4 
|kk'qq'|^{3/2}} E^3 \tilde{T}\,,
\end{equation}
with the total energy $E=q+q'$, $\kappa=8\pi G_N$, 
\begin{eqnarray}
\tilde{T} (q, q'; k, k') := \frac{1}{E^3}{\Bigg [}\Pi \ln{\frac{\Pi^2}{E^6}}
+ \frac{1} {\Pi} \sum_{p \in \left\{k,k',q,q'\right\}}  
\nonumber \\
p^2 \ln{\frac{p^2}{E^2}} {\Bigg (}3 kk'qq'-\frac{1}{2}
\sum_{r\neq p} \sum_{s \neq r,p}\left(r^2s^2\right){\Bigg )} {\Bigg ]}\,,
\end{eqnarray}
and the momentum transfer function $\Pi = (k+k')(k-q)(k'-q)$. The interesting 
part of the scattering amplitude is encoded in the scale independent factor 
$\tilde{T}$. The forward scattering poles occurring for $\Pi=0$ should be noted.

It is possible to generalize the VBH phenomenon to arbitrary GDTs with
matter as well as most of its properties (for instance, the CP-diagram, CPT 
invariance and the role played in the S-matrix) \cite{Grumiller:2002dm}.

\subsection{New results and outlook}

Recently quantum corrections to the specific heat of the dilaton BH have been calculated by applying the quantization method discussed above \cite{Grumiller:2003mc}. The result is $C_s:=dM/dT=96\pi^2M^2/\lambda^2$, where $\lambda$ is the scale parameter of the theory. Thus, in that particular case quantum corrections lead to a stabilization of the system. The mass of the BH is found to be decreasing according to
\begin{equation}
M(u)\approx M_0-\frac{\pi}{6} (T_H^0)^2(u-u_0)-\frac{\lambda}{24\pi}\ln{\frac{M(u)}{M_0}} + {\mathcal O}\left(\frac{\lambda}{M(u)}\right)\,.
\label{referee:2}
\end{equation}
The first term is the ADM mass, the second term corresponds to a linear decrease due to the (in leading order) constant Hawking flux and the third term provides the first nontrivial correction.

Applying simple thermodynamical methods\footnote{The review \cite{Wald:1999vt} may be consulted in this context.} ($dS=C_sdT/T$) and exploiting the quantum corrected mass/temperature relation $T/T_0=1-\lambda/(48\pi M)$ it is possible to calculate also entropy corrections:
\begin{equation}
S = S_0 - \frac{1}{24} \ln{S_0} + {\mathcal O}(1)\,,\quad S_0:=\frac{2\pi M}{\lambda}=\left. 2\pi X\right|_{\rm horizon}
\label{eq:entropy}
\end{equation}
The logarithmic behavior is in qualitative agreement with the one found in the literature by various methods \cite{Mann:1998hm}; the prefactor $1/24$ coincides with \cite{Zaslavskii:1996dg}.  

An extension of the results obtained in the first order formulation to dilaton supergravity is straightforward in principle but somewhat tedious in detail. It permitted, among other results, to obtain for the first time a full solution of dilaton supergravity \cite{Bergamin:2003am}.

All these exciting applications indicate that the strict application of standard QFT concepts to gravity (at least in d=2 or in models dimensionally reduced to d=2) shows great promise.

\section*{Acknowledgement}
 
This work has been supported by project P-14650-TPH of the Austrian Science Foundation (FWF) and by EURESCO. We are grateful to the organizers of the workshop ``What comes beyond the Standard Model?'' for an enjoyable meeting in Slovenia. We thank M.\ Bojowald and O.\ Zaslavskii for useful discussions at the Erwin Schr\"odinger Institut in Vienna, L.\ Bergamin for collaboration on 2d dilaton supergravity and D.\ Vassilevich for a long time collaboration on 2d dilaton gravity.


\newcommand{\SO}{{\rm SO}}
\newcommand{\SL}{{\rm SL}}
\newcommand{\Sp}{{\rm Sp}}
\newcommand{\ISO}{{\rm ISO}}
\newcommand{\SU}{{\rm SU}}
\newcommand{\U}{{\rm U}}
\newcommand{\Spin}{{\rm Spin}}
\newcommand{\MDfft}[2]{{\frac{#1}{#2}}}
\newcommand{\MDft}[2]{{\textstyle{\frac{#1}{#2}}}}
\ifx\ltimes\undefined
\newcommand{\ltimes}{{\kern3pt\hbox{\vrule width 0.4pt height 5.30pt depth
.0pt}\kern-1.76pt\times\kern1pt}}
\fi

\title*{Hidden Spacetime Symmetries and Generalized Holonomy in 
M-theory\thanks{MCTP-03-11 and hep-th/0303140}${}^,$
\thanks{Research supported in part by DOE Grant DE-FG02-95ER40899.}}
\author{Michael~J.~Duff\thanks{mduff@umich.edu} and
James T.~Liu\thanks{jimliu@umich.edu}}

\institute{%
Michigan Center for Theoretical Physics\\
Randall Laboratory, Department of Physics, University of Michigan\\
Ann Arbor, MI 48109--1120, USA}

\titlerunning{Hidden Spacetime Symmetries and Generalized Holonomy in 
M-theory}
\authorrunning{Michael~J.~Duff and James T.~Liu}
\maketitle

\begin{abstract}
In M-theory vacua with vanishing 4-form $F_{(4)}$, one can invoke the
ordinary Riemannian holonomy $H \subset \SO(10,1)$ to account for
unbroken supersymmetries $n=1$, 2, 3, 4, 6, 8, 16, 32.  However, the
generalized holonomy conjecture, valid for non-zero $F_{(4)}$, can
account for more exotic fractions of supersymmetry, in particular
$16<n<32$.  The conjectured holonomies are given by ${\cal H} \subset
{\cal G}$ where ${\cal G}$ are the generalized structure groups ${\cal
G}=\SO(d-1,1) \times G(spacelike)$, ${\cal G}= \ISO(d-1) \times G(null)$
and ${\cal G}=\SO(d) \times G(timelike)$ with $1\leq d<11$.  For example,
$G(spacelike)=\SO(16)$, $G(null)=[\SU(8) \times \U(1)]\ltimes {\mathbb R}^{56}$
and $G(timelike)=\SO^*(16)$ when $d=3$.  Although extending spacetime
symmetries, there is no conflict with the Coleman-Mandula theorem.
The holonomy conjecture rules out certain vacua which are otherwise
permitted by the supersymmetry algebra.
\end{abstract}


\section{Introduction}
\label{Physics}

M-theory not only provides a non-perturbative unification of the five
consistent superstring theories, but also embraces earlier work on
supermembranes and eleven-dimensional supergravity \cite{World}.  It
is regarded by many as the dreamed-of final theory and has accordingly
received an enormous amount of attention.  It is curious, therefore,
that two of the most basic questions of M-theory have until now remained
unanswered:

i) {\it What are the symmetries of M-theory?}

ii) {\it How many supersymmetries can vacua of M-theory preserve?}

The first purpose of this paper is to argue that M-theory possesses
previously unidentified hidden spacetime (timelike and null) symmetries in
addition to the well-known hidden internal (spacelike) symmetries.  These
take the form of generalized structure groups ${\cal G}$ that replace
the Lorentz group $\SO(10,1)$.

The second purpose is to argue that the number of supersymmetries
preserved by an M-theory vacuum is given by the number of singlets
appearing in the decomposition of the 32-dimensional representation of
${\cal G}$ under ${\cal G} \supset {\cal H}$ where ${\cal H}$ are
generalized holonomy groups.

The equations of M-theory display the
maximum number of supersymmetries $N$=32, and so $n$, the number of
supersymmetries preserved by a particular vacuum, must be some integer
between 0 and 32.  But are some values of $n$ forbidden and, if so,
which ones?  For quite some time it was widely believed that, aside
from the maximal $n=32$, $n$ is restricted to $0\leq n\leq 16$ with
$n=16$ being realized by the fundamental BPS objects of M-theory: the
M2-brane, the M5-brane, the M-wave and the M-monopole.  The subsequent
discovery of intersecting brane configurations with
$n=0$, 1, 2, 3, 4, 5, 6, 8, 16 lent credence to this argument.  In
\cite{Gauntletthull1}, on the other hand, it was shown that all values
$0\leq n \leq 32$ are allowed by the M-theory algebra \cite{Townsend},
and examples of vacua with $16< n < 32$ have indeed since been found.
Following \cite{Duffstelle} and \cite{MDDuff}, we here
put forward a {\it generalized holonomy conjecture} according to
which the answer lies somewhere in between.  Evidence in favor of
this conjecture includes the observations that there are no known
counterexamples and that a previously undiscovered example predicted in
\cite{MDDuff}, namely $n$=14, has recently been found \cite{Harmark}.

As we shall see, these conjectures are based on a group-theoretical argument
which applies to the fully-fledged M-theory.  To get the ball rolling,
however, we begin with the low energy limit of M-theory, namely $D=11$
supergravity.  The unique $D=11$ supermultiplet is comprised of
a graviton $g_{MN}$, a gravitino $\Psi_M$ and $3$-form gauge field
$A_{MNP}$, where $M=0, 1, \ldots 10$, with $44$, $128$ and $84$
physical degrees of freedom, respectively. In section \ref{hidden}, we
conjecture that the supergravity equations of motion for this set of
fields admit hidden timelike and null symmetries (in addition to
previously demonstrated hidden spacelike ones).  Then in section
\ref{holonomy} we propose that, so long as the $D=11$ Killing spinor
equation has such hidden symmetries, we may enlarge the tangent space
group into a generalized structure group.  This allows us to analyze
the number of supersymmetries based on a generalized holonomy conjecture.
Partial justification for this conjecture is presented in section
\ref{sec:dimred} in the context of a dimensionally reduced theory.  In
section~\ref{N} we discuss some consequences of generalized holonomy
for classifying supersymmetric vacua, and finally conclude in
section~\ref{M}.


\section{Hidden spacetime symmetries of D=11 supergravity}
\label{hidden}

Long ago, Cremmer and Julia \cite{Cremmerjulia}
pointed out that, when dimensionally reduced to $d$ dimensions, $D=11$
supergravity exhibits hidden symmetries.  For example ${\rm E}_7(global)
\times \SU(8)(local)$ when $d=4$ and ${\rm E}_8(global) \times \SO(16)(local)$
when $d=3$.  The question was then posed \cite{Fradkin}: do these
symmetries appear magically only after dimensional reduction, or were
they already present in the full uncompactified and untruncated $D=11$
theory?  The question was answered by de Wit and Nicolai
\cite{Dewitnicolai2,Nicolai} who made a $d/(11-d)$ split and fixed the
gauge by setting to zero the off-diagonal components of the elfbein.
They showed that in the resulting field equations the local symmetries
are indeed already present, but the global symmetries are not.  For
example, after making the split $\SO(10,1) \supset \SO(3,1) \times
\SO(7)$, we find the enlarged symmetry $\SO(3,1) \times \SU(8)$.  There
is no global ${\rm E}_7$ invariance (although the 70 internal components of
the metric and 3-form may nevertheless be assigned to an ${\rm E}_7/\SU(8)$
coset).  Similar results were found for other values of $d$: in each
case the internal subgroup $\SO(11-d)$ gets enlarged to some compact
group $G(spacelike)$ while the spacetime subgroup $\SO(d-1,1)$ remains
intact%
\footnote{We keep the terminology ``spacetime'' and ``internal''
even though no compactification or dimensional reduction is implied.}.
In this paper we ask instead whether there are hidden {\it spacetime}
symmetries.  This is a question that could have been asked long ago,
but we suspect that people may have been inhibited by the
Coleman-Mandula theorem which forbids combining spacetime and internal
symmetries \cite{MDColeman}.  However, this is a statement about
Poincare symmetries of the S-matrix and here we are concerned
with Lorentz symmetries of the equations of motion, so there will be no
conflict.

The explicit demonstration of $G(spacelike)$ invariance by de Wit and
Nicolai is very involved, to say the least.  However, the result is
quite simple: one finds the same $G(spacelike)$ in the full
uncompactified $D=11$ theory as was already found in the spacelike
dimensional reduction of Cremmer and Julia.  Here we content ourselves
with the educated guess that the same logic applies to $G(timelike)$
and $G(null)$: they are the same as what one finds by timelike and
null reduction, respectively. So we propose
that, after making a $d/(11-d)$ split, the Lorentz subgroup
$G=\SO(d-1,1) \times \SO(11-d)$ can be enlarged to the
generalized structure groups ${\cal G}=\SO(d-1,1) \times
G(spacelike)$, ${\cal G}= \ISO(d-1) \times G(null)$ and ${\cal G}= \SO(d)
\times G(timelike)$ as shown in Tables~\ref{sgen}, \ref{ngen}
and \ref{tgen}.

\begin{table}[t]
\begin{center}
\begin{tabular}{c|lc}
$d/(11-d)$&${\cal G}= \SO(d-1,1) \times G(spacelike)$&$\epsilon$ representation\\
\hline
10/1&$\SO(9,1) \times \{1\}$&$16+\overline{16}$\\
9/2&$\SO(8,1) \times \SO(2)$&$16_{\pm1/2}$\\
8/3&$\SO(7,1) \times \SO(3) \times \SO(2)$&$(8_s,2)_{1/2}+(8_c,2)_{-1/2}$\\
7/4&$\SO(6,1) \times \SO(5)$&$(8,4)$\\
6/5&$\SO(5,1) \times \SO(5) \times \SO(5)$&$(4,4,1)+(\overline{4},1,4)$\\
5/6&$\SO(4,1) \times {\rm USp}(8)$&$(4,8)$\\
4/7&$\SO(3,1) \times \SU(8)$&$(2,1,8)+(1,2,\overline{8})$\\
3/8&$\SO(2,1) \times \SO(16)$&$(2,16)$\\
\hline
2/9&$\SO(1,1) \times \SO(16) \times \SO(16)$&$(16,1)_{1/2}+(1,16)_{-1/2}$\\
1/10&$\{1\}   \times \SO(32)$&$32$
\end{tabular}
\end{center}
\caption{Generalized structure groups: spacelike case.  The last column
denotes the representation of $\epsilon$ under ${\cal G}$.}
\label{sgen}
\end{table}

\begin{table}[t]
\begin{center}
\begin{tabular}{c|lc}
$d/(11-d)$&${\cal G}=\ISO(d-1) \times G(null)$&$\epsilon$ representation\\
\hline
10/1&$\ISO(9)$&$16+16$\\
9/2&$\ISO(8) \times {\mathbb R}$&$8_s+8_s+8_c+8_c$\\
8/3&$\ISO(7) \times \ISO(2) \times {\mathbb R}$&$8_{\pm1/2}+8_{\pm1/2}$\\
7/4&$\ISO(6) \times [\SO(3) \times \SO(2)] \ltimes
{\mathbb R}^6_{(3,2)}$&$(4,2)_{\pm_1/2}+(\overline{4},2)_{\pm1/2}$\\
6/5&$\ISO(5) \times \SO(5) \ltimes {\mathbb R}^{10}_{(10)}$&$(4,4)+(4,4)$\\
5/6&$\ISO(4) \times [\SO(5) \times \SO(5)] \ltimes
{\mathbb R}^{16}_{(4,4)}$&$(2,1,4,1)+(2,1,1,4)$\\
&&\qquad$+(1,2,4,1)+(1,2,1,4)$\\
4/7&$\ISO(3) \times {\rm USp}(8)\ltimes {\mathbb R}^{27}_{(27)}$&$(2,8)+(2,8)$\\
3/8&$\ISO(2) \times [\SU(8) \times \U(1)]\ltimes
                        {\mathbb R}^{56}_{(28_{1/2},\overline{28}_{-1/2})}$&$
(8_{1/2})_{\pm1/2}+(\overline{8}_{-1/2})_{\pm1/2}$\\
\hline
2/9&${\mathbb R}      \times \SO(16) \ltimes {\mathbb R}^{120}_{(120)}$&$16+16$\\
1/10&$\{1\}  \times [\SO(16)\times\SO(16)]\ltimes{\mathbb R}^{256}_{(16,16)}$&$(16,1)
+(1,16)$
\end{tabular}
\end{center}
\caption{Generalized structure groups: null case.  The last column
denotes the representation of $\epsilon$ under the maximum compact
subgroup of ${\cal G}$.}
\label{ngen}
\end{table}

\begin{table}[t]
\begin{center}
\begin{tabular}{c|lc}
$d/(11-d)$&${\cal G}= \SO(d) \times G(timelike)$&$\epsilon$ representation\\
\hline
10/1&$\SO(10) \times \{1\}$&$16+\overline{16}$\\
9/2&$\SO(9) \times \SO(1,1)$&$16_{\pm1/2}$\\
8/3&$\SO(8) \times \SO(2,1) \times \SO(1,1)$&$(8_s,2)_{1/2}+(8_c,2)_{-1/2}$\\
7/4&$\SO(7) \times \SO(3,2)$&$(8,4)$\\
6/5&$\SO(6) \times \SO(5,{\mathbb C})$&$(4,4)+(\overline{4},\overline{4})$\\
5/6&$\SO(5) \times {\rm USp}(4,4)$&$(4,8)$\\
4/7&$\SO(4) \times \SU^*(8)$&$(2,1,8)+(1,2,\overline{8})$\\
3/8&$\SO(3) \times \SO^*(16)$&$(2,16)$\\
\hline
2/9&$\SO(2) \times \SO(16,{\mathbb C})$&$16_{1/2}+\overline{16}_{-1/2}$\\
1/10&$\{1\} \times \SO(16,16)$&$32$
\end{tabular}
\end{center}
\caption{Generalized structure groups: timelike case.  The last column
denotes the representation of $\epsilon$ under ${\cal G}$.}
\label{tgen}
\end{table}

Some of the noncompact groups appearing in the Tables may be unfamiliar,
but a nice discussion of their properties may be found in \cite{Gilmore}.
For $d>2$ the groups $G(spacelike)$, $G(timelike)$ and $G(null)$ are
the same as those obtained from the spacelike dimensional reductions
of Cremmer and Julia \cite{Cremmerjulia}, the timelike reductions of
Hull and Julia \cite{Hulljulia}%
\footnote{Actually, for the 8/3 split, we have the factor $\SO(1,1)$
instead of their $\SO(2)$.},
and the null
reduction of section \ref{genhol}, respectively. For our purposes,
however, their physical interpretation is very different.  They are
here proposed as symmetries of the full $D=11$ equations of motion;
there is no compactification involved, whether toroidal or otherwise.
This conjecture that these symmetries are present in the full theory
and not merely in its dimensional reductions may be put to the test,
however, as we shall later describe.  For $d\leq 2$ it is less clear whether
these generalized structure groups are actually hidden symmetries.  See
the caveats of section~\ref{sec:dimred}. The $\SO(16) \times \SO(16)$ for
$d=2$ is also discussed by Nicolai \cite{Nicolai:1991tt}.


\section{Hidden Symmetries and Generalized Holonomy}
\label{holonomy}

We begin by reviewing the connection between holonomy and the number of
preserved supersymmetries, $n$, of supergravity vacua.  This also serves
to define our notation.  Subsequently, we introduce a generalized holonomy
which involves the hidden symmetries conjectured in the previous
section.

\subsection{Riemannian Holonomy}

We are interested in solutions of the bosonic field equations
\begin{equation}
R_{MN}=\frac{1}{12}\left(F_{MPQR}F_{N}{}^{PQR}-\frac{1}{12}g_{MN}
F^{PQRS}F_{PQRS}\right)
\end{equation}
and
\begin{equation}
d*\!F_{(4)}+\MDfft12F_{(4)} \wedge F_{(4)}=0,
\end{equation}
where $F_{(4)}=dA_{(3)}$.  The supersymmetry transformation rule of the
gravitino reduces in a purely bosonic background to
\begin{equation}
\delta \Psi_{M}={\tilde D}_{M} \epsilon,
\label{eq:11gto}
\end{equation}
where the parameter $\epsilon$ is a 32-component anticommuting spinor,
and where
\begin{equation}
\label{covariant}
{\tilde D}_{M}=D_{M}-
\frac{1}{288}(\Gamma_M{}^{NPQR}-8\delta_M^N\Gamma^{PQR})F_{NPQR},
\end{equation}
where $\Gamma^{A}$ are the $D=11$ Dirac matrices.  Here $D_{M}$ is the
usual Riemannian covariant derivative involving the connection $\omega_{M}$
of the usual structure group $\Spin(10,1)$, the double cover of $\SO(10,1)$,
\begin{equation}
D_{M}=\partial_{M}+\frac{1}{4}\omega_{M}{}^{AB}\Gamma_{AB}.
\end{equation}
The number of supersymmetries preserved by an M-theory background depends
on the number of covariantly constant spinors,
\begin{equation}
{\tilde D}_{M}\epsilon=0,
\end{equation}
called {\it Killing} spinors.  It is the presence of the terms involving
the 4-form $F_{(4)}$ in (\ref{covariant}) that makes this counting difficult.
So let us first examine the simpler vacua for which $F_{(4)}$ vanishes.
Killing spinors then satisfy the integrability condition
\begin{equation}
[{D}_{M}, {D}_{N}] \epsilon=\frac{1}{4}R_{MN}{}^{AB}\Gamma_{AB}\epsilon=0,
\label{integrability}
\end{equation}
where $R_{MN}{}^{AB}$ is the Riemann tensor.  The subgroup of
$\Spin(10,1)$ generated by this linear combination of $\Spin(10,1)$ generators
$\Gamma_{AB}$ corresponds to the ${\it holonomy}$ group ${H}$ of the
connection $\omega_{M}$.  The number of supersymmetries, $n$, is then
given by the number of singlets appearing in the decomposition of the
$32$ of $\Spin(10,1)$ under ${H}$.  In Euclidean signature, connections
satisfying (\ref{integrability}) are automatically Ricci-flat and
hence solve field equations when $F_{(4)}=0$.  In Lorentzian signature,
however, they need only be Ricci-null \cite{Fig} so Ricci-flatness has
to be imposed as an extra condition.  In Euclidean signature, the
holonomy groups have been classified \cite{Berger}.  In Lorentzian
signature, much less is known but the question of which subgroups ${H}$
of $\Spin(10,1)$ leave a spinor invariant has been answered in
\cite{Bryant}.  There are two sequences according as the vector
$v_{A}=\overline{\epsilon}\,\Gamma_{A}\epsilon$ is timelike or null, as shown
in Tables~\ref{static} and \ref{wave}. Since $v^{2} \leq 0$, the
spacelike $v_A$ case does not arise. The timelike $v_A$ case corresponds
to static vacua, where ${H} \subset \Spin(10) \subset \Spin(10,1)$ while
the null case to non-static vacua where ${H} \subset \ISO(9) \subset
\Spin(10,1)$.  It is then possible to determine the possible $n$-values
\cite{Acharya:1998yv,Acharya:1998st} and one finds $n=2$, 4, 6, 8,
16, 32 for static vacua, as shown in Table~\ref{static}, and $n=1$,
2, 3, 4, 8, 16, 32 for non-static vacua, as shown in
Table~\ref{wave}.

\begin{table}[t]
\begin{center}
\begin{tabular}{c|cc}
$d/(11-d)$&$H \subset \SO(11-d)\subset\Spin(10)$& $n$\\
\hline
7/4&$\SU(2) \cong \Sp(2)$& $16$\\
5/6&$\SU(3)$& $8$\\
4/7&${\rm G}_2$& $4$\\
3/8&$\SU(2) \times \SU(2)$& $8$\\
&$\Sp(4)$& $6$\\
&$\SU(4)$& $4$\\
&$\Spin(7)$& $2$\\
1/10&$\SU(2) \times \SU(3)$& $4$\\
&$\SU(5)$& $2$
\end{tabular}
\end{center}
\caption{Holonomy of static M-theory vacua with $F_{(4)}=0$ and their
supersymmetries.}
\label{static}
\end{table}

\begin{table}[t]
\begin{center}
\begin{tabular}{c|cc}
$d/(11-d)$&$H \subset \ISO(d-1)\times\ISO(10-d)\subset\Spin(10,1)$& $n$\\
\hline
10/1&${\mathbb R}^9$ & $16$\\
6/5&${\mathbb R}^5\times(\SU(2) \ltimes {\mathbb R}^4)$ & $8$\\
4/7&${\mathbb R}^3\times(\SU(3) \ltimes {\mathbb R}^6)$ & $4$\\
3/8&${\mathbb R}^2\times({\rm G}_2 \ltimes {\mathbb R}^7)$ & $2$\\
2/9&${\mathbb R}\times(\SU(2) \ltimes {\mathbb R}^4) \times (\SU(2) \ltimes {\mathbb R}^4)$ & $4$\\
&${\mathbb R}\times(\Sp(4) \ltimes {\mathbb R}^8)$ & $3$\\
&${\mathbb R}\times(\SU(4) \ltimes {\mathbb R}^8)$ & $2$\\
&${\mathbb R}\times(\Spin(7) \ltimes {\mathbb R}^8)$ & $1$
\end{tabular}
\end{center}
\caption{Holonomy of non-static M-theory vacua with $F_{(4)}=0$ and their
supersymmetries.}
\label{wave}
\end{table}

\subsection{Generalized holonomy}
\label{genhol}

When we want to include vacua with $F_{(4)}\neq 0$ we face the problem
that the connection in (\ref{covariant}) is no longer the spin
connection to which the bulk of the mathematical literature on
holonomy groups is devoted.  In addition to the $\Spin (10,1)$
generators $\Gamma_{AB}$, it is apparent from (\ref{covariant})
that there are terms involving $\Gamma_{ABC}$ and $\Gamma_{ABCDE}$.
As a result, the connection takes its values in the full $D=11$
Clifford algebra.  Moreover, this connection can preserve exotic
fractions of supersymmetry forbidden by the Riemannian connection.
For example, the M-branes at angles in \cite{Ohta} include $n$=5, the
11-dimensional pp-waves in
\cite{Michelson,Cvetic:2002si,Gauntletthull2,Bena} include $n=18$, 20,
22, 24, 26 (and $n=28$ for Type IIB), the squashed $N(1,1)$ spaces in
\cite{Page} and the M5-branes in a pp-wave background in \cite{Singh}
include $n$=12 and the G\"{o}del universes in \cite{Gauntlett:2002nw}
include $n=18$, 20, 22, 24.

However, we can attempt to quantify this in terms of generalized
holonomy groups ${\cal H} \subset {\cal G}$ where ${\cal G}$ are
the generalized structure groups discussed in section \ref{hidden}.
The generalized holonomy conjecture
\cite{Duffstelle,MDDuff} states that one can assign a holonomy ${\cal H}
\subset {\cal G}$ to the generalized connection%
\footnote{A related conjecture was made in \cite{Berkooz}, where the
generalized holonomy could be any subgroup of $\SO(16,16)$.  This also
appears in our conjectured hidden structure groups under the 1/10 split,
though only in the timelike case ${\cal G}(timelike)$.}
appearing in the supercovariant
derivative (\ref{covariant}). Here we propose that, after making
a $d/(11-d)$ split, the Lorentz subgroup $G=\SO(d-1,1) \times \SO(11-d)$
can be enlarged to the generalized structure groups ${\cal
G}=\SO(d-1,1) \times G(spacelike)$, ${\cal G}= \ISO(d-1) \times
G(null)$ and ${\cal G}= \SO(d) \times G(timelike)$ as shown in
Tables~\ref{sgen}, \ref{ngen} and \ref{tgen}.
Note that in the right hand column of the tables we have listed the
corresponding ${\cal G}$ representations under which the 32 supersymmetry
parameters $\epsilon$ transform.  The number of supersymmetries preserved
by an M-theory vacuum is then given by the number of singlets
appearing in the decomposition of these representations under ${\cal G}
\supset {\cal H}$.


\section{Structure groups from dimensional reduction}
\label{sec:dimred}

In this section we provide partial justification for the conjectured
hidden symmetries by demonstrating their presence in the gravitino variation
of the dimensionally
reduced theory.  In particular, we consider a spacelike dimensional
reduction corresponding to a $d/(11-d)$ split.  Turning on only
$d$-dimensional scalars, the reduction ansatz is particularly simple
\begin{equation}
g^{(11)}_{MN}=\begin{pmatrix}\Delta^{-1/(d-2)}g_{\mu\nu}&0\cr0&g_{ij}\end{pmatrix},\qquad
A^{(11)}_{ijk}=\phi_{ijk},
\end{equation}
where $\Delta=\det{g_{ij}}$.  For $d\le5$, we must also consider the
possibility dualizing either $F_{(4)}$ components or (for $d=3$)
Kaluza-Klein vectors to scalars.  We will return to such possibilities
below.  But for now we focus on $d\ge6$.  In this case, a standard
dimensional reduction of the $D=11$ gravitino transformation,
(\ref{eq:11gto}), yields the $d$-dimensional gravitino transformation
\begin{equation}
\delta\psi_\mu=[D_\mu+\MDft14Q_\mu{}^{ab}\Gamma_{ab}+\MDft1{24}\partial_\mu
\phi_{ijk}\Gamma^{ijk}]\epsilon.
\label{eq:dgto}
\end{equation}
For completeness, we also note that the $d$-dimensional dilatinos
transform according to
\begin{equation}
\delta\lambda_i=-\MDft12\gamma^\mu[P_{\mu\,ij}\Gamma^j
-\MDft1{36}(\Gamma_i{}^{jkl}-6\delta_i^j\Gamma^{kl})\partial_\mu\phi_{jkl}]
\epsilon.
\end{equation}
In the above, the lower dimensional quantities are related to their $D=11$
counterparts through
\begin{eqnarray}
&&\psi_\mu=\Delta^{\MDfft1{4(d-2)}}\left(\Psi^{(11)}_\mu+\MDfft1{d-2}\gamma_\mu
\Gamma^i\Psi^{(11)}_i\right),\qquad
\lambda_i=\Delta^{\MDfft1{4(d-2)}}\Psi^{(11)}_i,\nonumber\\
&&\epsilon=\Delta^{\MDfft1{4(d-2)}}\epsilon^{(11)},\nonumber\\
&&Q^{ab}_\mu=e^{i[a}\partial_\mu e_i{}^{b]},\qquad
P_{\mu\,ij}=e_{(i}^a\partial_\mu e_{j)\,a}.
\end{eqnarray}

We now see that the lower dimensional gravitino transformation,
(\ref{eq:dgto}), may be written in terms of a covariant derivative
under a generalized connection
\begin{equation}
\delta\psi_\mu=\hat D_\mu\epsilon,\qquad
\hat D_\mu=\partial_\mu+\MDft14\Omega_\mu,
\end{equation}
where
\begin{equation}
\Omega_\mu=\omega_\mu{}^{\alpha\beta}\gamma_{\alpha\beta}
+Q_\mu{}^{ab}\Gamma_{ab}+\MDft1{3!}e^{ia}e^{jb}e^{kc}\partial_\mu\phi_{ijk}
\Gamma_{abc}.
\label{eq:gencon}
\end{equation}
Here $\gamma_\alpha$ are $\SO(d-1,1)$ Dirac matrices, while $\Gamma_a$
are $\SO(11-d)$ Dirac matrices.  This decomposition is suggestive of a
generalized structure group with connection given by $\Omega_\mu$.
However one additional requirement is necessary before declaring this an
enlargement of $\SO(d-1,1)\times \SO(11-d)$, and that is to ensure that the
algebra generated by $\Gamma_{ab}$ and $\Gamma_{abc}$ closes within
itself.  Along this line, we note that the commutators of these internal
Dirac matrices have the schematic structure
\begin{equation}
[\Gamma^{(2)},\Gamma^{(2)}]=\Gamma^{(2)},\qquad
[\Gamma^{(2)},\Gamma^{(3)}]=\Gamma^{(3)},\qquad
[\Gamma^{(3)},\Gamma^{(3)}]=\Gamma^{(6)}+\Gamma^{(2)}.
\label{eq:diralg}
\end{equation}
Here the notation $\Gamma^{(n)}$ indicates the antisymmetric product of
$n$ Dirac matrices, and the right hand sides of the commutators only
indicate what possible terms may show up.  The first commutator above
merely indicates that the $\Gamma_{ab}$ matrices provide a
representation of the Riemannian $\SO(11-d)$ structure group.

For $d\ge6$, the internal space is restricted to five or fewer
dimensions.  In this case, the antisymmetric product $\Gamma^{(6)}$
cannot show up, and the algebra clearly closes on $\Gamma^{(2)}$ and
$\Gamma^{(3)}$.  Working out the extended structure groups for these
cases results in the expected Cremmer and Julia groups listed in the
first four lines of Table~\ref{sgen}.  A similar analysis follows for
$d\le5$.  However, in this case, we must also dualize an additional set
of fields to see the hidden symmetries.  For $d=5$, an additional scalar
arises from the dual of $F_{\mu\nu\rho\sigma}$; this yields an addition
to (\ref{eq:gencon}) of the form $\Omega_\mu^{\rm additional}=\MDfft1{4!}
\epsilon_\mu{}^{\nu\rho\sigma\lambda} F_{\mu\nu\rho\sigma}\Gamma_{123456}$.
This $\Gamma^{(6)}$ term is precisely what is necessary for the closure
of the algebra of (\ref{eq:diralg}).  Of course, in this case, we must
also make note of the additional commutators
\begin{equation}
[\Gamma^{(2)},\Gamma^{(6)}]=\Gamma^{(6)},\qquad
[\Gamma^{(3)},\Gamma^{(6)}]=\Gamma^{(7)}+\Gamma^{(3)},\qquad
[\Gamma^{(6)},\Gamma^{(6)}]=\Gamma^{(10)}+\Gamma^{(6)}+\Gamma^{(2)}.
\label{eq:gam6com}
\end{equation}
However neither $\Gamma^{(7)}$ nor $\Gamma^{(10)}$ may show up in $d=5$ for
dimensional reasons.

The analysis for $d=4$ is similar; however here
$$\Omega_\mu^{\rm
additional}=\MDfft1{3!}\epsilon_\mu{}^{\nu\rho\sigma}e^{ia}F_{\nu\rho\sigma i}
\Gamma_a\Gamma_{1234567}.$$  Closure of the algebra on $\Gamma^{(2)}$,
$\Gamma^{(3)}$ and $\Gamma^{(6)}$ then follows because, while $\Gamma^{(7)}$
may in principle arise in the middle commutator of (\ref{eq:gam6com}),
it turns out to be kinematically forbidden.  For $d=3$, on the other
hand, in additional to a contribution $\Omega_\mu^{\rm additional}
=\MDfft1{2!\cdot2!}\epsilon_\mu{}^{\nu\rho}e^{ia}e^{jb}F_{\nu\rho ij}
\Gamma_{ab}\Gamma_{12345678}$, one must also dualize the Kaluza-Klein
vectors $g_\mu{}^i$.  Doing so gives rise to a $\Gamma^{(7)}$ in the
generalized connection which, in addition to the previously identified
terms, completes the internal structure group to $\SO(16)$.

The remaining two cases, namely $d=2$ and $d=1$, fall somewhat outside the
framework presented above.  This is because in these low dimensions the
generalized connections $\Omega_\mu$ derived via reduction are partially
incomplete.  For $d=2$, we find
\begin{equation}
\Omega_\mu^{(d=2)}=\omega_\mu{}^{\alpha\beta}
\gamma_{\alpha\beta}+Q_\mu{}^{ab}\Gamma_{ab}+\MDft19(\delta_\mu^\nu-
\MDft12\gamma_\mu{}^\nu)e^{ia}e^{jb}e^{kc}\partial_\nu\phi_{ijk}\Gamma_{abc},
\label{eq:d2omega}
\end{equation}
where $\gamma_{\mu\nu}=-\MDfft12\epsilon_{\mu\nu}
(\epsilon^{\alpha\beta}\gamma_{\alpha\beta})$ is necessarily proportional
to the two-dimensional chirality matrix.  Hence from a two-dimensional
point of view, the scalars from the metric enter non-chirally, while the
scalars from $F_{(4)}$ enter chirally.  Taken together, the generalized
connection (\ref{eq:d2omega}) takes values in $\SO(16)_+\times
\SO(16)_-$, which we regard as the enlarged structure group.  However not
all generators are present because of lack of chirality in
the term proportional to $Q_\mu{}^{ab}$.  Thus at this point the
generalized structure group deviates from the hidden symmetry group,
which would be an infinite dimensional subgroup of affine ${\rm E}_8$.
Similarly, for $d=1$, closure of the connection $\Omega_\mu^{(d=1)}$
results in an enlarged $\SO(32)$ structure group.  However this is not
obviously related to any actual hidden symmetry of the $1/10$ split.

Until now, we have considered the spacelike reductions leading to the
generalized structure groups of Table~\ref{sgen}.  For a timelike
reduction, we simply interchange a time and a space direction in the
above analysis%
\footnote{By postulating that the generalized structure groups survive as
hidden symmetries of the full uncompactified theory, we avoid the undesirable
features associated with compactifications including a timelike direction
such as closed timelike curves.}.
This results in an internal Clifford algebra with signature $(10-d,1)$,
and yields the extended symmetry groups indicated in Table~\ref{tgen}.
Turning finally to the null case, we may replace one of the internal
Dirac matrices with $\Gamma_+$ (where $+$, $-$ denote light-cone
directions).  Since $(\Gamma_+)^2=0$, this indicates that the extended
structure groups for the null case are contractions the corresponding
spacelike (or timelike) groups.  In addition, by removing $\Gamma_+$
from the set of Dirac matrices, we essentially end up in the case of one
fewer compactified dimensions.  As a result, the $G(null)$ group in
$d$-dimensions must have a semi-direct product structure involving the
$G(spacelike)$ group in $(d+1)$-dimensions.  Of course, these groups
also contain the original $\ISO(10-d)$ structure group as a subgroup.
The resulting generalized structure groups are given in
Table~\ref{ngen}%
\footnote{The reduction of $D$-dimensional pure gravity along a single
null direction was analyzed by Julia and Nicolai \cite{Julianicolai}.}.
%


\section{Counting supersymmetries}
\label{N}

Having defined a generalized holonomy for vacua with $F_{(4)}\ne0$, we
now turn to some elementary examples.  For the basic objects of
M-theory, the M2-brane configuration may be placed under the 3/8 (spacelike)
classification, as it has three longitudinal and eight transverse directions.
Focusing on the transverse directions (which is the analog of looking at
$\hat D_\mu$), the M2-brane has generalized holonomy $\SO(8)$ contained in
$\SO(2,1)\times\SO(16)$ \cite{Duffstelle}.  In this case, the spinor
decomposes as
$(2,16)=2(8)+16(1)$, indicating the expected presence of 16 singlets.
For the M5-brane with 6/5 (spacelike) split, the generalized $\hat D_\mu$
holonomy is
given by $\SO(5)_+\subset \SO(5,1)\times\SO(5)_+\times\SO(5)_-$, with
the spinor decomposition $(4,4,1)+(\overline{4},1,4)=4(4)+16(1)$.  Since
the wave solution depends on nine space-like coordinates, we may regard
it as a 1/10 (null) split.  In this case, it has generalized $\tilde D_M$
holonomy
${\mathbb R}^{9}\subset [\SO(16)\times\SO(16)]\ltimes {\mathbb R}^{256}_{(16,16)}$.  The
spinor again decomposes into 16 singlets.  Note, however, that since
the wave is pure geometry, it could equally well be categorized under a
10/1 split as ${\mathbb R}^9\subset \ISO(9)$.  Finally, the KK monopole
is described by a 7/4 (spacelike) split, and has $\hat D_\mu$ holonomy
$\SU(2)_+ \subset \SO(6,1)\times\SO(5)$, where the spinor decomposes as
$(8,4)=8(2)+16(1)$.  In all four cases, the individual objects
preserve exactly half of the 32 supersymmetries.  However each object
is associated with its own unique generalized holonomy, namely
$\SO(8)$, $\SO(5)$, ${\mathbb R}^{9}$ and $\SU(2)$ for the M2, M5, MW and MK,
respectively.

The supersymmetry of intersecting brane configurations may be understood
in a similar manner based on generalized holonomy.  For example, for a
M5 and MK configuration sharing six longitudinal directions, we may
choose a 6/5 split.  In this case, the structure group is
$\SO(5,1)\times \SO(5)_+\times \SO(5)_-$, and the $\hat D_\mu$ holonomies
of the
individual objects are $\SO(5)_+$ and $\SU(2)\subset\SO(5)_{\rm diag}$,
respectively.  The holonomy for the combined configuration turns out to
be $\SO(5)_+\times\SU(2)_-$, with the spinor decomposing as
$(4,4,1)+(\overline{4},1,4)=4(4,1)+4(1,2)+8(1,1)$.  The resulting eight
singlets then signify the presence of a $1/4$ supersymmetric configuration.
In principle, this analysis may be applied to more general brane
configurations.  However one goal of understanding enlarged holonomy is
to obtain a classification of allowed holonomy groups and, as a result, to
obtain a unified treatment of counting supersymmetries.  We now provide
some observations along this direction.

We first note the elementary fact that a $p$-dimensional representation
can decompose into any number of singlets between 0 and $p$, {\it
except} $(p-1)$, since if we have $(p-1)$ singlets, we must have $p$.
It follows that in theories with $N$ supersymmetries, $n=N-1$
is ruled out, even though it is permitted by the supersymmetry algebra.

In some cases, additional restrictions on $n$ may be obtained.  For example,
if the supersymmetry charge transforms as the $(2,16)$ representation of
${\cal G}$ when $d=3$,
then $n$ is restricted to 0, 2, 4, 6, 8, 10, 12, 14, 16, 18, 20, 22, 24,
26, 28, 32 as first noted in \cite{MDDuff}.  No new values of $n$ are generated
by $d>3$ reps.  For example, the 4 of $\SO(5)$ can decompose only into
0, 2 or 4 singlets but not 1.

We note that all the even values of $n$ discussed so far appear in the
list and that $n=30$ is absent.  This is consistent with the presence of
pp-waves with $n=16$, 18, 20, 22, 24, 26 (and $n=28$ for Type IIB) but the
absence of $n$=30 noted in
\cite{Bena,Michelson,Cvetic:2002si,Gauntletthull2}.  Of course a good
conjecture should not only account for the existing data but should go
on to predict something new.  For example, Gell-Mann's flavor $\SU(3)$
not only accounted for the nine known members of the baryon decuplet
but went on to predict the existence of the $\Omega^{-}$, which was
subsequently discovered experimentally.  For M-theory supersymmetries,
the role of the $\Omega^{-}$ is played by $n=14$ which at the time of
its prediction had not been discovered ``experimentally''.  We note
with satisfaction, therefore, that this missing member has recently been
found in the form of a G\"{o}del universe \cite{Harmark}.

The $d=2$ and $d=1$ cases are more problematic since $\SO(16)\times\SO(16)$
and $\SO(32)$ in principle allow any $n$ except $n=31$.  So more work is
required to explain the presence of M-branes at angles with $n=0$, 1,
2, 3, 4, 5, 6, 8, 16 but the absence of $n=7$ noted in \cite{Ohta}.
Presumably, a more detailed analysis will show that only those
subgroups compatible with these allowed values of $n$ actually appear as
generalized holonomy groups.  The beginnings of a classification of
all supersymmetric $D=11$ solutions may be found in \cite{Gauntlett:2002fz}.

We can apply similar logic to theories with fewer than 32
supersymmetries.  Of course, if M-theory really underlies all
supersymmetric theories then the corresponding vacua will all be
special cases of the above.  However, it is sometimes useful to focus
on such a sub-theory, for example the Type I and heterotic strings
with $N=16$.  Here $G(spacelike)= \SO(d) \times \SO(d)$, $G(null)=
\ISO(d-1) \times \ISO(d-1)$ and $G(timelike)=\SO(d-1,1) \times
\SO(d-1,1)$.  If the supersymmetry charge transforms as a $(2,8)$
representation of the generalized structure group when $d=3$, then $n$
is restricted to 0, 2, 4, 6, 8, 10, 12, 16.  No new values of $n$ are
generated from other $d>4$ reps. Once again, the $d=2$ and $d=1$ cases
require a more detailed analysis.


\section{The full M-theory}
\label{M}

We have focused on the low energy limit of M-theory, but
since the reasoning that led to the conjecture is based just on group
theory, it seems reasonable to promote it to the full
M-theory%
\footnote{Similar conjectures can be applied to M-theory in signatures
(9,2) and (6,5) \cite{Blencowe:1988sk}, the so-called M$^\prime$ and M$^*$
theories \cite{Hull:1998ym}, but the groups will be different.}.
When counting the $n$ value of a particular vacuum, however, we should be
careful to note the phenomenon of {\it supersymmetry without
supersymmetry}, where the supergravity approximation may fail to
capture the full supersymmetry of an M-theory vacuum. For example,
vacua related by T-duality and S-duality must, by definition, have the
same $n$ values. Yet they can appear to be different in
supergravity \cite{DLP1,DLP2}, if one fails to take into account winding
modes and non-perturbative solitons. So more work is needed to verify
that the $n$ values found so far in $D=11$ supergravity exhaust those of
M-theory, and to prove or disprove the conjecture.


\section*{Notes added}

After this paper was posted on the archive, a very interesting paper by
Hull appeared \cite{Hull:2003mf} which generalizes and extends the
present theme.  Hull conjectures that the hidden symmetry of M-theory is
as large as $\SL(32,{\mathbb R})$ and that this is necessary in order to accommodate
all possible generalized holonomy groups.  We here make some remarks in
the light of Hull's paper:

{\it Hidden symmetries}:

Hull stresses that, as a candidate hidden symmetry, $\SL(32,{\mathbb R})$
is background independent.  However, the hidden symmetries displayed in
Tables~\ref{sgen}, \ref{ngen} and \ref{tgen} are also background
independent.  They depend only on the choice of non-covariant split and gauge
in which to write the field equations.  Hull's proposal is nevertheless
very attractive since $\SL(32,{\mathbb R})$ contains all the groups in
Tables~\ref{sgen}, \ref{ngen} and \ref{tgen} as subgroups and would
thus answer the question of whether all these symmetries are present
at the same time.

One can accommodate
$\SL(32,{\mathbb R})$ by extending the $d/(11-d)$ split to include the $d=0$
case.  Then the same $\SL(32,{\mathbb R})$ would appear in all three tables.  At
the other end, one could also include the $d=11$ case.  Then the same
$\SO(10,1)$ would appear in all three tables.  Our reason for not
including the $d=0$ case stems from the apparent need to make a
non-covariant split and to make the corresponding gauge choice before
the hidden symmetries become apparent \cite{Dewitnicolai2,Nicolai}.
Moreover, from the point of view of guessing the hidden symmetries
from the dimensional reduction, the $d=0$ case would be subject to the
same caveats as the $d=1$ and $d=2$ cases: not all group generators
are present in the covariant derivative.  $\SL(32,{\mathbb R})$ requires
$\{\Gamma^{(1)},\Gamma^{(2)},
\Gamma^{(3)},\Gamma^{(4)},\Gamma^{(5)}\}$ whereas only
$\{\Gamma^{(2)}, \Gamma^{(3)},\Gamma^{(5)}\}$ appear in the covariant
derivative.  This is an important issue deserving of further study.
That M-theory could involve a ${\rm GL}(32,{\mathbb R})$ has also been conjectured by
Barwald and West \cite{Barwald:1999is}.

{\it Generalized holonomy}:

Hull goes on to stress the importance of $\SL(32,{\mathbb R})$ by finding solutions
whose holonomy is contained in $\SL(32,{\mathbb R})$ but not in Tables~\ref{sgen},
\ref{ngen} and \ref{tgen}. Although not all generators are present in the
covariant derivative, they are all present in the commutator. So we agree
with Hull that $\SL(32,{\mathbb R})$ is necessary if one wants to embrace all
possible generalized holonomies.

Indeed, since the basic objects of M-theory discussed in section \ref{N}
involve warping by a harmonic function, the $\hat D$ holonomy is smaller
than the $\tilde D$ holonomy, which requires extra ${\mathbb R}^n$ factors.
Interestingly enough, the $\hat D$ holonomy nevertheless yields the
correct counting of supersymmetries.

Hull points out that, in contrast to the groups appearing in Tables~\ref{sgen},
\ref{ngen} and \ref{tgen}, $\SL(32,{\mathbb R})$ does not obey the $n \neq N-1$
rule of section \ref{N}, and hence M-theory vacua with $n=31$ are in
principle possible%
\footnote{The case for $n=31$ has also been made by Bandos {\it et al.}
\cite{Bandos:2001pu} in the different context of hypothetical preons of
M-theory preserving 31 out of 32 supersymmetries.}.
Of course we do not yet know whether the required ${\mathbb R}^{31}$
holonomy actually appears.  To settle the issue of which $n$ values
are allowed, it would be valuable to do for supergravity what Berger
\cite{Berger} did for gravity and have a complete classification of
all possible generalized holonomy groups. But this may prove quite
difficult.

So we remain open-minded about a formulation of M-theory with
$\SL(32,{\mathbb R})$ symmetry, but acknowledge the need for $\SL(32,{\mathbb R})$ from
the point of view of generalized holonomy.


\section*{Acknowledgments}

We have enjoyed useful conversations with Hisham Sati.


\def\JMt#1{\tilde #1}
\def\JMb#1{{\mathbb #1}}
\def\JMc#1{{\cal #1}}
\def\JMkbar {{\mathchar'26\mkern-9muk}}
\def\JMtfrac #1#2{\textstyle{\frac{#1}{#2}}}

\title*{On the Resolution of Space-Time Singularities II}
\author{Marco Maceda${}^1$ and John Madore${}^{1,2}$}
\institute{%
        Laboratoire de Physique Th\'eorique\\ Universit\'e
        de Paris-Sud, B\^atiment 211, F-91405 Orsay \and
        Max-Planck-Institut f\"ur Physik\\ F\"ohringer
        Ring 6, D-80805 M\"unchen}
\titlerunning{On the Resolution of Space-Time Singularities II}
\authorrunning{Marco Maceda and John Madore}
\maketitle

\begin{abstract}
In previous articles it has been argued that a differential
  calculus over a noncommutative algebra uniquely determines a
  gravitational field in the commutative limit and that there is a
  unique metric which remains as a commutative `shadow'. Some examples
  were given of metrics which resulted from a given algebra and given
  differential calculus. Here we aboard the inverse problem, that of
  constructing the algebra and the differential calculus from the
  commutative metric.  As an example a noncommutative version of the
  Kasner metric is proposed which is periodic.

\end{abstract}

\section{Motivation}

A definition has been given~\cite{DubMadMasMou96b} of a torsion-free
metric-compatible linear connection on a differential calculus
$\Omega^*(\JMc{A})$ over an algebra $\JMc{A}$ which has certain rigidity
properties provided that the center $\JMc{Z}(\JMc{A})$ of $\JMc{A}$ is
trivial. It was argued from simple examples that a differential
calculus over a noncommutative algebra uniquely determines a
gravitational field in the commutative limit. Some examples have been
given~\cite{DubMadMasMou95,ChoMadPar98,FioMad99} of metrics which
resulted from a given algebra and given differential calculus. Here we
aboard the inverse problem, that of constructing the algebra and the
differential calculus from the commutative metric.  As an example we
construct noncommutative versions of the Kasner metric and we show
that it is possible to choose an algebra such that the metric is
nonsingular before taking the commutative limit.  The `II' on the
title alludes to a preliminary version given at the Torino
Euroconference~\cite{Mad00b} on noncommutative
geometry~\cite{Torino99}.

The physical idea we have in mind is that the description of
space-time using a set of commuting coordinates is only valid at
length scales greater than some fundamental length. At smaller scales
it is impossible to localize a point and a new geometry must be used.
We can use a solid-state analogy and think of the ordinary Minkowski
coordinates as macroscopic order parameters obtained by
`course-graining' over regions whose size is determined by a
fundamental area scale $\JMkbar $, which is presumably, but not
necessarily, of the order of the Planck area $G\hbar$. They break down
and must be replaced by elements of a noncommutative algebra when one
considers phenomena on smaller scales.  A simple visualization is
afforded by the orientation order parameter of nematic liquid
crystals. The commutative free energy is singular in the core region
of a disclination. There is of course no physical singularity; the
core region can simply not be studied using the commutative order
parameter.  

As a concrete example we have chosen, for historical
reasons, the Kasner metric; we show that its singularity can be
resolved into an essentially noncommutative structure.  We do not
however claim that an arbitrary singularity in a metric on an
arbitrary smooth manifold can be resolved using a noncommutative
structure. From the point of view we are adopting a commutative
geometry is a rather singular limit. The close relation between the
differential calculus and the metric can at most be satisfied when the
center is trivial. This manifests itself in the fact that on an
ordinary manifold one can put any metric with any singularity. We
argue only that those metrics which are `physical' in some sense, for
example are Ricci flat, can have resolvable singularities. 

There is a similarity of the method we use to resolve the singularity
with the method known in algebraic geometry as `blowing up' a
singularity~\cite{Abb98} as well as with the method used by 't~Hooft and
Polyakov to resolve the monopole singularity. The regular solution
found in this case can in fact be considered as the Dirac monopole
solution on a noncommutative geometry which contains the $2\times 2$
matrix algebra as extra factor. Since we are now dealing with a
dynamical field configuration it is improbable that the singularity
will admit being blown up by a finite-dimensional algebra. Our
solutions offer evidence however in favor of this possibility with an
algebra of infinite dimension.

In previous articles the algebra and the differential calculus were
given and the linear connection and metric were constructed.  It was
argued~\cite{DimMad96,MadMou98} that given the algebra $\JMc{A}$ the
structure of $\Omega^*(\JMc{A})$ is intimately connected with the
gravitational field which remains on $V$ as shadow in the commutative
limit $\JMkbar \rightarrow 0$.  Within the general framework which we
here consider, the principal difference between the commutative and
noncommutative cases lies in the spectrum of the operators which we
use to generate the noncommutative algebra which replaces the algebra
of functions. This in turn depends not only on the structure of this
algebra as abstract algebra but on the representation of it which we
choose to consider. Here we attempt the inverse problem, that of
constructing the algebra {\it and} the differential calculus from the
commutative linear connection. We cannot claim that the procedure is
in any way unique.

For a discussion of the relation of noncommutative geometry to the
problem of space-time singularities from rather different points of
view from the one we adopt we refer, for example, to Heller \&
Sasin~\cite{HelSas96}, to Hawkins~\cite{Haw02} or to Lizzi
{\it et al.}\ ~\cite{LizManMiePel02}. For a recent discussion of diffeomorphism
invariance within the context of commutative gravity we refer to Gaul
\& Rovelli~\cite{GauRov00}. For a discription of a noncommutative
approach to gravity using a choice of metric which does not fulfill
the criteria which we use we refer to Aschieri \&
Castellani~\cite{AscCas93}. There seems to be a relation between the
quadratic momentum algebra we use and non-linear representations of
momentum operators considered recently. We refer to
Chakrabarti~\cite{Cha02} for review of this and references to the
previous literature.  We refer elsewhere for a description of
the same `quantization' applied to the $PP$ wave~\cite{MacMadRob02} and
for a possible cosmological application~\cite{MacMadManZou03}.

In the next sections we introduce the general formalism of
noncommutative geometry which we use and we make some general remarks
concerning the problem of `quantization' of space-time.  In Section~7
we recall the commutative Kasner metric. In Sections~8 and~9 we study
the structure of the algebra we associate to the metric; we make also
some remarks concerning perturbative approximations to noncommutative
geometry and present the Kasner solution as a perturbative solution in
$\JMkbar $. The key formulae are~(\ref{duality0}) and~(\ref{j1}). They
can be used to construct other examples without knowledge of the
preceding material.

There is much that needs further work. We have not explicitly
constructed the differential calculus nor have we examined in detail
the complex structure of the algebra.  There seems to be evidence of a
cosmological constant associated to the noncommutativity but this
remains elusive. All we can affirm is that a noncommutative structure
is similar in certain aspects to extra dimensions and so could be
expected to yield an effective cosmological constant in dimension four
in the same way that extra (Ricci non-flat) dimensions give rise to
one.

Greek indices take values from 0 to 3; the first half of the alphabet
is used to index (moving) frames and the second half to index
generators. Latin indices $a$, $b$, {\it etc.} take values from 1 to
$3$ and the indices $i$, $j$, {\it etc.} values from 0 to $n-1$.

\section{The general formalism}

The notation is the same as that of a previous article~\cite{Mad00b} on
the symplectic structure of space-time and is based on a
noncommutative generalization~\cite{Mad89c,DimMad96,CerFioMad00c} of the
Cartan moving-frame formalism .  Let $\JMc{A} = \JMc{C}(V)$ be the algebra
of smooth real-valued functions on a space-time $V$ which for
simplicity we shall suppose parallelizable and with a metric and
linear connection defined in terms of a globally defined moving frame
$\theta^\alpha$. Let $\Omega^*(\JMc{A})$ be the algebra of de~Rham
differential forms. The space $\Omega^1(\JMc{A})$ of 1-forms is free of
rank 4 as a $\JMc{A}$-module.  According to the general idea outlined
above a singularity in the metric is due to the use of commuting
coordinates beyond their natural domain of definition into a region
where they are physically inappropriate. From this point of view the
space-time $V$ should be more properly described `near the
singularity' by a noncommutative $*$-algebra $\JMc{A}$ over the complex
numbers with four hermitian generators $x^\lambda$.  The observables
will be some subset of the hermitian elements of $\JMc{A}$.  We shall
not discuss this problem here; we shall implicitly suppose that all
hermitian elements of $\JMc{A}$ are observables, including the
`coordinates'.  We shall not however have occasion to use explicitly
this fact.

We introduce 6 additional elements $J^{\mu\nu}$ of $\JMc{A}$ by the
relations
\begin{equation}
[x^\mu, x^\nu] = i \JMkbar J^{\mu\nu}.                      \label{xx}
\end{equation}
The details of the structure of $\JMc{A}$ will be contained for example
in the commutation relations $[x^\lambda, J^{\mu\nu}]$.  
One can define recursively an infinite sequence of elements
by setting for $p \geq 1$
\begin{equation}
[x^\lambda, J^{\mu_1 \cdots \mu_p}] = 
i \JMkbar J^{\lambda\mu_1 \cdots \mu_p}.                    \label{x-x}
\end{equation}
We shall assume that for the description of a generic (strong)
gravitational field the appropriate algebra $\JMc{A}$ has a trivial
center $\JMc{Z}(\JMc{A})$:
\begin{equation}
\JMc{Z}(\JMc{A}) = \JMb{C}.                              \label{2.3}
\end{equation}
The only argument we have in favour of this assumption is the fact
that it would be difficult to interpret the meaning of the center.
The $x^\mu$ will be referred to as `position generators'.  We
shall suppose also that there is a set of $n(=4)$ antihermitian
`momentum generators' $\lambda_\alpha$ and a `Fourier transform'
$$
F: x^\mu \longrightarrow \lambda_\alpha = F_\alpha (x^\mu)
$$
which takes the position generators to the momentum generators.

Let $\rho$ be a representation of $\JMc{A}$ as an algebra of linear
operators on some Hilbert space.  For every $k_\mu \in \JMb{R}^4$ one
can construct a unitary element $u(k) = e^{ik_\mu x^\mu}$ of $\JMc{A}$
and one can consider the weakly closed algebra $\JMc{A}_\rho$ generated
by the image of the $u(k)$ under $\rho$.  The momentum operators
$\lambda_\alpha$ are also unbounded but using them one can construct
also a set of `translation' operators $\hat u(\xi) = e^{\xi^\alpha
\lambda_\alpha}$ whose image under $\rho$ belongs also to
$\JMc{A}_\rho$. In general $\hat u u \neq u \hat u$; if the metric which
we introduce is the flat metric then we shall see that
$[\lambda_\alpha, x^\mu] = \delta^\mu_\alpha$ and in this case we can
write the commutation relations $\hat u u = q u \hat u$ with $q = e^{i
k_\mu \xi^\mu}$; the `Fourier transform' is the simple linear
transformation
\begin{equation}
\lambda_\alpha = 
\frac{1}{i\JMkbar }J^{-1}_{\alpha\mu} x^\mu        \label{Four}
\end{equation}
for some symplectic structure $J^{\alpha\mu}$. If the
structure is degenerate then it is no longer evident that the algebra
can be generated by either the position generators or the momentum
generators alone. In such cases we define the algebra $\JMc{A}$ to be
the one generated by both sets. The derivations could be considered as
outer derivations of the smaller algebra generated by the $x^\mu$;
they become inner in the extended algebra,

We shall suppose that $\JMc{A}$ has a commutative limit which is an
algebra $\JMc{C}(V)$ of smooth functions on a space-time $V$ endowed
with a globally defined moving frame $\theta^\alpha$ and thus a
metric.  By parallelizable we mean that the module $\Omega^1(\JMc{A})$
has a basis $\theta^\alpha$ which commutes with the elements of
$\JMc{A}$. For all $f \in \JMc{A}$
\begin{equation}
f\theta^\alpha = \theta^\alpha f.  \label{f-theta}
\end{equation}
We shall see that this implies that the metric components must be
constants, a condition usually imposed on a moving frame. It also
means that we have `frozen out' local Lorentz transformations since
they do not leave this condition invariant.  The frame $\theta^\alpha$
allows one~\cite{Mad00c} to construct a representation of the
differential algebra from that of $\JMc{A}$. Following strictly what one
does in ordinary geometry, we shall introduce the set of derivations
$e_\alpha$ to be dual to the frame $\theta^\alpha$, that is with
\begin{equation}
\theta^\alpha(e_\beta) = \delta^\alpha_\beta.     \label{e-theta}
\end{equation}
We define the differential exactly as did E.~Cartan in the commutative
case. If $e_\alpha$ is a derivation of $\JMc{A}$ then for every element
$f \in \JMc{A}$ we define $df$ by the constraint $df(e_\alpha) =
e_\alpha f$. The differential calculus is defined as the largest one
consistent with the module structure of the 1-forms so constructed.
One can at this point take the classical limit to obtain four
functions $\JMt{\lambda}_\alpha(\JMt{x}^\mu)$ which satisfy the equations
$$
\{\JMt{\lambda}_\alpha, \JMt{x}^\mu\} = \JMt{e}^\mu_\alpha.
$$
This defines a Poisson structure directly from which one can calculate
the $\{\JMt{x}^\mu, \JMt{x}^\nu\}$. In this way only at the last moment
does one pass to a noncommutative algebra and most of the problem
remains within the category of smooth manifolds.

It follows from general arguments that the momenta $\lambda_\alpha$
must satisfy the consistency condition
\begin{equation}
2 \lambda_\gamma \lambda_\delta P^{\gamma\delta}{}_{\alpha\beta} -
\lambda_\gamma F^\gamma{}_{\alpha\beta} - 
K_{\alpha\beta} = 0.                                  \label{consis}
\end{equation}
The $P^{\gamma\delta}{}_{\alpha\beta}$ define the product $\pi$ in the
algebra of forms:
\begin{equation}
\theta^\alpha \theta^\beta = P^{\alpha\beta}{}_{\gamma\delta}
\theta^\gamma \otimes \theta^\delta.  \label{struc}
\end{equation}
This product is defined to be the one with the least relations which
is consistent with the module structure of the 1-forms.  The
$F^\gamma{}_{\alpha\beta}$ are related to the 2-form $d\theta^\alpha$
through the structure equations:
\begin{equation}
d\theta^\alpha = - {1\over 2}C^\alpha{}_{\beta\gamma} \theta^\beta
\theta^\gamma.                                        \label{se}
\end{equation}
In the noncommutative case the structure elements are defined as
\begin{equation}
C^\alpha{}_{\beta\gamma} = 
F^\alpha{}_{\beta\gamma} - 2 \lambda_\delta
P^{(\alpha\delta)}{}_{\beta\gamma}.                     \label{s-e}
\end{equation}
It follows that
\begin{equation}
e_\alpha C^\alpha{}_{\beta\gamma} = 0.                  \label{fix}
\end{equation}
This must be imposed then at the classical level and can be used as a
gauge-fixing condition.  We impose also, without loss of generality 
the conditions
$$
F^\eta{}_{\alpha\beta} P^{\alpha\beta}{}_{\gamma\delta} =
F^\eta{}_{\gamma\delta},\qquad
K_{\alpha\beta} P^{\alpha\beta}{}_{\gamma\delta} =
K_{\gamma\delta}.
$$
There follows a similar relation for the $F^\gamma{}_{\alpha\beta}$.

{F}inally, to complete the definition of the coefficients of the
consistency condition~(\ref{consis}) we introduce the special 1-form
$\theta = -\lambda_\alpha\theta^\alpha$. In the commutative, flat
limit
$$
\theta \rightarrow i\partial_\alpha dx^\alpha.
$$
As an (antihermitian) 1-form $\theta$ defines a covariant
derivative on an associated $\JMc{A}$-module with local gauge
transformations given by the unitary elements of $\JMc{A}$.  The
$K_{\alpha\beta}$ are related to the curvature of $\theta$:
$$
d\theta + \theta^2 = K, \qquad 
K = - {1\over 2} K_{\alpha\beta} \theta^\alpha \theta^\beta.
$$
All the coefficients lie in the center $\JMc{Z}(\JMc{A})$ of the
algebra. 

The condition~(\ref{consis}) can be expressed
also in terms of a twisted commutator
$$
[\lambda_\alpha,\lambda_\beta]_P = 2
P^{\gamma\delta}{}_{\alpha\beta} \lambda_\gamma \lambda_\delta
$$
as
$$
[\lambda_\alpha, \lambda_\beta]_P = \lambda_\gamma
F^\gamma{}_{\alpha\beta} + K_{\alpha\beta}.
$$
It is also connected with the condition that $d^2 f = 0$.
The differential $df$ of an element $f\in \JMc{A}$ is given by 
$df = e_\alpha f \theta^\alpha$. Since, in particular
$$
d^2 \lambda_\gamma = 
d( [\lambda_\beta, \lambda_\gamma] \theta^\beta)
= ([\lambda_\alpha, [\lambda_\beta, \lambda_\gamma]]
- \JMtfrac 12 [\lambda_\mu,\lambda_\gamma] C^\mu{}_{\alpha\beta})
\theta^\alpha \theta^\beta
$$
it follows that
$$
P^{\alpha\beta}{}_{\gamma\delta} e_\alpha e_\beta -
C^\gamma{}_{\alpha\beta} e_\gamma = 0.
$$
This is the same as Equation~(\ref{consis}).

We must now compare in some way the commutators 
$[\lambda_\alpha, \lambda_\beta]$ and $[e_\alpha, e_\beta]$. Consider 
$C^\alpha{}_{\beta\gamma}$ as defined by the structure equation~(\ref{se}).
Suppose further that for some numbers $c_1$ and $c_2$ the relations
$$
[\lambda_\alpha, \lambda_\beta] = 
c_1 C^\gamma{}_{\alpha\beta} \lambda_\gamma,\qquad
[e_\alpha, e_\beta] = c_2 C^\gamma{}_{\alpha\beta} e_\gamma
$$
hold. The identities
\begin{equation}
([e_\alpha, e_\beta] - C^\gamma{}_{\alpha\beta} e_\gamma) x^\mu =
(c_2 - c_1) C^\gamma{}_{\alpha\beta} e_\gamma x^\mu  +
[C^\gamma{}_{\alpha\beta}, x^\mu] \lambda_\gamma       \label{comp}
\end{equation}
place restrictions on the coefficients. In the commutative limit one
must have $c_1 = c_2 = 1$.  In a general noncommutative geometry there
is no relation between the 2-form $d\theta^\alpha$ and the commutator
$[e_\alpha, e_\beta]$. Such a relation would fix the value of $c_2$.
In the formalism we are here considering the
$C^\gamma{}_{\alpha\beta}$ are linear functionals of the momentum
generators. (The theory has that is only four degrees of freedom and
not the ten of general relativity. There are six which have been fixed
by the choice of frame but they are not to be identified with the
missing six.) It follows that
$$
[C^\gamma{}_{\alpha\beta}, x^\mu] \lambda_\gamma = 
C^\gamma{}_{\alpha\beta} e_\gamma x^\mu.
$$
Because of the Leibniz rule then one must have in general the
relation $c_2 = 2 c_1$. We shall choose $c_2=1$ so that the relation
of the commutative limit is satisfied in general. We are forced then
in general to choose $c_1 = \JMtfrac 12$.  We shall assume that the
commutator and the gravitational field are two aspects of the same
phenomenon which we call gravity. We therefore suppose that in a
`realistic' situation both are present and one should take into
account the relation found between $c_1$ and $c_2$.

\section{The algebra}

Equation~(\ref{s-e}) is the correspondence principle which associates
a differential calculus to a metric.  On the left in fact the quantity
$C^\alpha{}_{\beta\gamma}$ determines a moving frame, which in turn
fixes a metric; on the right are the elements of the algebra which fix
to a large extent the differential calculus. A `blurring' of a
geometry proceeds via this correspondence. It is evident that in the
presence of curvature the 1-forms cease to anticommute. On the other
hand it is possible for flat `space' to be described by `coordinates'
which do not commute.  The correspondence principle between the
classical and noncommutative geometries can be also described as the
map
\begin{equation}
\JMt{\theta^\alpha} \mapsto \theta^\alpha                   \label{cp}
\end{equation}
with the product satisfying the condition
$$
\JMt{\theta^\alpha} \JMt{\theta^\beta} \mapsto
P^{\alpha\beta}{}_{\gamma\delta} \theta^\gamma \theta^\delta.
$$
The tilde on the left is to indicate that it is the classical form.
The condition can be written also as
$$
\JMt{C}^\alpha{}_{\beta\gamma} \mapsto C^\alpha{}_{\eta\zeta}
P^{\eta\zeta}{}_{\beta\gamma}
$$
or as
\begin{equation}
\lim_{\JMkbar \to 0}C^\alpha{}_{\beta\gamma} =
\JMt{C}^\alpha{}_{\beta\gamma}.                         \label{cl}
\end{equation}
A solution to these equations would be a solution to the problem we
have set. It would be however unsatisfactory in that no smoothness
condition has been imposed. This can at best be done using the inner
derivations. We shall construct therefore the set of momentum
generators. The procedure we shall follow is not always valid; a
counter example has been constructed~\cite{MadSae98} for the flat
metric on the torus. The correspondence principle which in fact we
shall actually use is a modified version of the map
$$
\JMt{e}_\alpha \mapsto \lambda_\alpha
$$
which is the inverse of that introduced by von~Neumann to represent
the Heisenberg algebra. 

We introduce an involution~\cite{Con95} on the algebra of forms
using~\cite{FioMad98a} a reality condition on derivations, a procedure
which is more or less a straightforward generalization of that which
is used in the case of ordinary differential manifolds. The involution
depends on the form of the product projection $\pi$.
For general $\xi, \eta \in \Omega^1(\JMc{A})$ it follows that
$$
(\xi \eta)^* = - \eta^* \xi^*.                    
$$
In particular
$$
(\theta^\alpha \theta^\beta)^* = - \theta^\beta \theta^\alpha.
$$
The product of two frame elements is hermitian then if and only if
they anticommute. Recall that the product of two hermitian elements
$f$ and $g$ of the algebra is hermitian if and only if they commute.
When the frame exists one has necessarily also the relations
$$
(f \xi \eta)^* = (\xi \eta)^* f^*, \qquad (f \xi \otimes \eta)^* =
(\xi \otimes \eta)^* f^*
$$
for arbitrary $f \in \JMc{A}$.

We write $P^{\alpha\beta}{}_{\gamma\delta}$ in the form
\begin{equation}
P^{\alpha\beta}{}^{\phantom{\alpha\beta}}_{\gamma\delta} = \JMtfrac 12
\delta^{[\alpha}_\gamma \delta^{\beta]}_\delta + 
i\JMkbar Q^{\alpha\beta}{}^{\phantom{\alpha\beta}}_{\gamma\delta}   \label{P}
\end{equation}
of a standard projector plus a perturbation and
we decompose $Q^{\alpha\beta}{}_{\gamma\delta}$ as the sum
of two terms
$$
Q^{\alpha\beta}{}_{\gamma\delta} =
Q_-^{\alpha\beta}{}_{\gamma\delta} +
Q_+^{\alpha\beta}{}_{\gamma\delta}
$$
symmetric (antisymmetric) and antisymmetric (symmetric) with
respect to the upper (lower) indices. The condition that
$P^{\alpha\beta}{}_{\gamma\delta}$ be a projector is satisfied to
first order in $\JMkbar $ because of the property that
$$
Q^{\alpha\beta}{}_{\gamma\delta} = P^{\alpha\beta}{}_{\zeta\eta}
Q^{\zeta\eta}{}_{\gamma\delta} + Q^{\alpha\beta}{}_{\zeta\eta}
P^{\zeta\eta}{}_{\gamma\delta}.
$$
The compatibility condition with the product
$$
(P^{\alpha\beta}{}_{\zeta\eta})^* P^{\eta\zeta}{}_{\gamma\delta} =
P^{\beta\alpha}{}_{\gamma\delta}
$$
is satisfied provided $Q^{\alpha\beta}{}_{\gamma\delta}$ is real.

To simplify the formulae we introduce the notation
\begin{equation}
[\lambda_\alpha, \lambda_\beta] = \Lambda_{\alpha\beta}.     \label{rhs}
\end{equation}
We can then write~(\ref{consis}) in the form
\begin{equation}
\Lambda_{\alpha\beta} + 2i\JMkbar \Lambda_{\gamma\delta}\, 
Q_-^{\gamma\delta}{}_{\alpha\beta} = K_{\alpha\beta}
+ \lambda_\gamma (F^\gamma{}_{\alpha\beta} - 2i\JMkbar \lambda_\delta
Q_-^{\gamma\delta}{}_{\alpha\beta}).                            \label{Gab2}
\end{equation}
This implies that to lowest order we can rewrite~(\ref{consis}) as two
independent equations
\begin{eqnarray}
&&
\Lambda_{\alpha\beta}\phantom{0} = 
K_{-\alpha\beta} + \lambda_\gamma
F_-^\gamma{}_{\alpha\beta} - 
2i\JMkbar \lambda_\gamma\lambda_\delta
Q_-^{\gamma\delta}{}_{\alpha\beta},                           \label{Gab1a}
\\[4pt]
&&
\phantom{\Lambda_{\alpha\beta}}0 = 
K_{+\alpha\beta} + \lambda_\gamma
F_+^\gamma{}_{\alpha\beta} - 
2i\JMkbar \lambda_\gamma\lambda_\delta
Q_+^{\gamma\delta}{}_{\alpha\beta}.                           \label{Gab1b}
\end{eqnarray}
This is the form which we shall use. If Equation~(\ref{Gab1b}) is
non-trivial then one can substitute into the third term on the
right-hand side the expression~(\ref{Gab1a}) for the commutator:
$$
2i\JMkbar \lambda_\gamma \lambda_\delta Q^{\gamma\delta}{}_{\alpha\beta}
=i\JMkbar \Lambda_{\gamma\delta} Q^{\gamma\delta}{}_{\alpha\beta}.
$$
From the definition~(\ref{s-e})  it follows that
\begin{equation}
C^\gamma{}_{\alpha\beta} = 
F^\gamma{}_{\alpha\beta} - 
4i\JMkbar \lambda_\delta Q_-^{\gamma\delta}{}_{\alpha\beta}.      \label{s-e2}
\end{equation}
We must choose the $\Lambda_{\alpha\beta}$ so that for arbitrary 
$f \in \JMc{A}$ in the classical limit when $f \to \JMt{f}$
$$
[\JMt{e}_\alpha, \JMt{e}_\beta] f =  \JMt{C}^\gamma{}_{\alpha\beta} \JMt{e}_\gamma f.
$$
From the general considerations of Section~2 and in particular
Equation~(\ref{Gab2}) we have
$$
[\Lambda_{\alpha\beta}, f] = F^\gamma{}_{\alpha\beta} [\lambda_\gamma , f]
- 2i\JMkbar Q^{\gamma\delta}{}_{\alpha\beta} 
[\lambda_\gamma \lambda_\delta, f]
$$
which we rewrite as
\begin{equation}
[e_\alpha, e_\beta] f = 
\JMtfrac 12 C^{\gamma}{}_{\alpha\beta} e_\gamma f
+ \JMtfrac 12  e_\gamma f C^{\gamma}{}_{\alpha\beta}.          \label{ee-cr}
\end{equation}
We shall assume that
$$
F^\gamma{}_{\alpha\beta} = 0.
$$
In the classical limit we have
$$
C^\alpha{}_{\beta\gamma} \to \JMt{\theta}^\alpha_\mu \JMt{e}_{[\beta} 
\JMt{e}^\mu_{\gamma]}.
$$
The gauge-fixing condition can be written to the classical approximation
$$
e_\gamma ( \theta^\gamma_\mu e_{[\alpha} e^\mu_{\beta]}) = 0.
$$
We assume that the noncommutative expression is the same to within
a reordering of the factors. To determine the order we consider the
Jacobi identity
$$
[[\lambda_\alpha, \lambda_\beta], x^\mu] +
[[\lambda_\beta, x^\mu], \lambda_\alpha] +
[[x^\mu, \lambda_\alpha], \lambda_\beta] = 0.
$$
Using~(\ref{ee-cr}) we can rewrite this as 
$$
\JMtfrac 12 C^\gamma{}_{\alpha\beta}e^\mu_\gamma +
\JMtfrac 12 e^\mu_\gamma C^\gamma{}_{\alpha\beta}
= e_{[\alpha}e^\mu_{\beta]}.
$$
We define then $ C^\gamma{}_{\alpha\beta}$ to be a solution of this
equation. Because of the standard ordering problems familiar from
quantum mechanics the solution will not be unique.

To lowest order one can write
$$
C^\gamma{}_{\alpha\beta} = 
\theta^\gamma_\mu e_{[\alpha}e^\mu_{\beta]} - 
\JMtfrac 12 [\theta^\gamma_\mu, e_{[\alpha}e^\mu_{\beta]}] =
\JMtfrac 12 (\theta^\gamma_\mu e_{[\alpha}e^\mu_{\beta]}+
e_{[\alpha}e^\mu_{\beta]}\theta^\gamma_\mu ).
$$

\section{The connection}

It is necessary~\cite{DubMadMasMou96b} to introduce a flip operation
$$
\sigma:\;\Omega^1(\JMc{A}) \otimes \Omega^1(\JMc{A}) \rightarrow
\Omega^1(\JMc{A}) \otimes \Omega^1(\JMc{A})
$$
to define the reality condition and the Leibniz rules.  If we write
$$
S^{\alpha\beta}{}_{\gamma\delta} = \delta^\beta_\gamma
\delta^\alpha_\delta + i\JMkbar T^{\alpha\beta}{}_{\gamma\delta}
$$
we find that a choice~\cite{DimMad96} of connection which is
torsion-free, and satisfies all Leibniz rules is given by
\begin{equation}
\omega^\alpha{}_{\beta} = \JMtfrac 12 F^\alpha{}_{\gamma\beta}
\theta^\gamma + i\JMkbar \lambda_\gamma
T^{\alpha\gamma}{}_{\delta\beta}\theta^\delta.      \label{t-free}
\end{equation}
The relation
$$
\pi \circ (1 + \sigma) = 0
$$
must hold~\cite{DimMad96,MadMou98} to assure that the torsion be a
bilinear map. 

To all orders one has
\begin{equation}
\omega^\alpha{}_{\eta\zeta}P^{\eta\zeta}{}_{\beta\gamma}
 = \JMtfrac 12 C^\alpha{}_{\beta\gamma}.                    \label{T0}
\end{equation}
This is the usual relation between the Ricci-rotation coefficients and
the Levi-Civita connection. Using it one can deduce~(\ref{s-e})
from~(\ref{t-free}). One can also write~(\ref{t-free}) in the form
$$
2\omega^\alpha{}_{\delta\beta} = C^\alpha{}_{\delta\beta} + i\JMkbar
\lambda_\gamma T^{\alpha\gamma}{}_{(\delta\beta)} - i\JMkbar
\lambda_\gamma T^{\alpha\gamma}{}_{(\eta\zeta)}
Q^{\eta\zeta}{}_{\delta\beta}                    
$$
which includes the part symmetric in the second two indices and
which must be determined by the condition that the connection be
metric.
In terms of the coefficients
$P^{\alpha\beta}{}_{\gamma\delta}$ the relation can be written in the
form
\begin{equation}
T^{\alpha\beta}{}_{\eta\zeta} P^{\eta\zeta}{}_{\gamma\delta} +
Q^{(\alpha\beta)}{}_{\gamma\delta} = 0.                 \label{t-q}
\end{equation}
To lowest order this becomes
\begin{eqnarray}
&&Q_-^{\alpha\beta}{}_{\gamma\delta} = 
- \JMtfrac 14 T^{\alpha\beta}{}_{[\gamma\delta]}, \label{QT1}\\[4pt]
&&T^{\alpha\beta}{}_{\gamma\delta} = 
- 2(Q_-^{\alpha\beta}{}_{\gamma\delta} -
Q_{+(\gamma}{}^\beta{}_{\delta)}{}^\alpha).              \label{QT2}
\end{eqnarray}
The symmetric part of $T^{\alpha\beta}{}_{\gamma\delta}$ has been here
fixed by the condition that the connection be metric.  

Under a change of frame basis the coefficients of the spin connection
also change. We mention only the linear approximation. If
$$
\theta^{\prime\alpha} = \theta^\alpha - H^\alpha{}_\beta\theta^\beta
$$
then
$$
C^{\prime\alpha}{}_{\beta\gamma} = C^{\alpha}{}_{\beta\gamma} +
D_{[\beta} H^\alpha{}_{\gamma]}.
$$
The only restriction on $H^\alpha{}_\beta$, apart from the
condition that it be small and antisymmetric, is that it must leave the
condition~(\ref{fix}) invariant or impose it if it is not satisfied.
If we treat the $\lambda_\alpha$ as the components of a 1-form
$-\theta$, which they are, and take their covariant derivative as if
they formed an ordinary covector, which they do not, then we find that
$$
D_{[\alpha} \lambda_{\beta]} - 2 [\lambda_\alpha, \lambda_\beta]
- C^\gamma{}_{\alpha\beta} \lambda_\gamma = 2K_{\alpha\beta}.
$$
It is difficult to interpret this equation.

\section{The metric}

We shall suppose that $\JMc{A}$ has a metric
\begin{equation}
g:\; \Omega^1(\JMc{A}) \otimes \Omega^1(\JMc{A}) \rightarrow \JMc{A}.
\label{metric}
\end{equation} 
In terms of the frame one can define the metric by the condition that
\begin{equation}
g(\theta^\alpha \otimes \theta^\beta)= g^{\alpha\beta}.  \label{f-m}
\end{equation}
The $g^{\alpha\beta}$ are taken to form an arbitrary complex matrix
which satisfies~\cite{FioMad98a} the symmetry condition
\begin{equation}
P^{\alpha\beta}{}_{\gamma\delta} g^{\gamma\delta} = 0 \label{symmet}
\end{equation}
as well as the reality condition
$$
g^{\beta\alpha} + i\JMkbar T^{\alpha\beta}{}_{\gamma\delta}
g^{\gamma\delta} = (g^{\beta\alpha})^*.
$$
When constructing an algebra from a given classical geometry
usually, but not necessarily, one starts with the matrix
$g^{\alpha\beta}$ of standard Minkowski or euclidian metric
components. One must read the symmetry and reality conditions then as
conditions on the maps $\pi$ and $\sigma$.

If we write $g^{\alpha\beta}=\eta^{\alpha\beta}+i\JMkbar h^{\alpha\beta}$
then we find that to first order the symmetry and reality become 
respectively
$$
h^{\alpha\beta} = 
-  Q_+^{\alpha\beta}{}_{\gamma\delta} \eta^{\gamma\delta} = 
\JMtfrac 12  T^{\alpha\beta}{}_{\gamma\delta} \eta^{\gamma\delta}.
$$
If we require that the flip be an involution then we find to first
order that
$$
T^{\alpha\beta}{}_{\gamma\delta} + 
2 Q^{\alpha\beta}{}_{\gamma\delta} = 0.
$$
The various reality conditions~\cite{Con95,FioMad98a} imply also
that
$$
(Q^{\alpha\beta}{}_{\gamma\delta})^* =
Q^{\alpha\beta}{}_{\gamma\delta} + o(i\JMkbar ), \qquad
(T^{\alpha\beta}{}_{\gamma\delta})^* =
T^{\alpha\beta}{}_{\gamma\delta} + o(i\JMkbar ).
$$
The sum of two idempotents is also an idempotent if the two terms
are orthogonal.  

The condition that the product be a projector implies that it be
hermitian with respect to the usual inner product on the tensor
product:
$$
g^{\alpha\gamma} g^{\beta\delta} = (\theta^\alpha \otimes
\theta^\beta, \theta^\gamma \otimes \theta^\delta).
$$
Therefore we have the condition
\begin{equation}
P^{\alpha\beta}{}_{\gamma\delta} g^{\gamma\zeta} g^{\delta\eta} =
P^{\zeta\eta}{}_{\gamma\delta} g^{\gamma\alpha} g^{\delta\beta}  
                                                         \label{her-c}
\end{equation}
for $P^{\alpha\beta}{}_{\gamma\delta}$ and the metric.  
A weaker condition is the orthogonality condition
$$
P^{\alpha\beta}{}_{\gamma\delta} g^{\gamma\zeta} g^{\delta\eta}
(\delta^\mu_\zeta \delta^\nu_\eta + S^{\mu\nu}{}_{\zeta\eta}) = 0.
$$
With the Ansatz we shall use this second condition is an identity.

The connection is compatible with the metric if
\begin{equation}
T^{\alpha\gamma}{}_{\delta\epsilon} g^{\epsilon\beta} +
T^{\beta\gamma}{}_{\delta\epsilon} g^{\alpha\epsilon} + i\JMkbar
T^{\beta\gamma}{}_{\epsilon\zeta} g^{\eta\zeta}
T^{\alpha\epsilon}{}_{\delta\eta} = 0.                       \label{m-c3}
\end{equation}
To first order this simplifies to the usual condition
\begin{equation}
T^{(\alpha\gamma}{}_{\delta}{}^{\beta)} = o(i\JMkbar ).         \label{m-c2}
\end{equation}
The index was lifted here with the lowest-order, symmetric part of the
metric.

\section{Speculations}

It is tempting to suppose that to lowest order at least, in a
semi-classical approximation, there is an analogue of Darboux's lemma
and that it is always possible to choose generators which satisfy
commutation relations of the form~(\ref{x-x}) with the right-hand in
the center.  However the example we shall examine in detail shows that
this is not always the case. Having fixed the generators, the
manifestations of curvature would be found then in the form of the
frame. The two sets of generators $x^\mu$ and $\lambda_\alpha$
satisfy, under the assumptions we make, three sets of equations.  The
commutation relations~(\ref{x-x}) for the position generators $x^\mu$
and the associated Jacobi identities permit one definition of the
algebra. The commutation relations for the momentum generators permit
a second definition. The conjugacy relations assure that the two
descriptions concern the same algebra.  We shall analyze these
identities later using the example to show that they have interesting
non-trivial solutions.

The problem of gauge invariance and the algebra of observables is a
touchy one upon which we shall not dwell. It is obvious that not all
of the elements of $\JMc{A}$ are gauge invariant but not that all
observables are gauge-invariant.  One of the principles of the theory
of general relativity is that all (regular) coordinates systems or
frames are equal. In the noncommutative case one finds that some are
more equal than others. If one quantize a space-time using two
different moving frames one will obtain two different differential
calculi, although the two underlying algebras might be the same. This
is equivalent to the fact that the canonical transformations of a
commutative phase space are a very special set of phase-space
coordinate transformations. It can also be expressed as the fact that
the Poisson structure which remains on space-time as the commutative
limit of the commutation relations breaks Lorenz invariance.  In the
special case where the $H^\alpha_\beta$ are constants then the two
quantized frames will be also equivalent. Since we have decided to
work only with algebras whose centers are trivial the converse will
also be true. For a discussion of Poisson structures on curved
manifolds we refer to Fedosov~\cite{Fed96}. Since we are interested in
finding the `simplest' differential calculus, one of the aspects of
the problem is the choice of `correct' moving frame to start with.

One possible method of looking for a solution is to consider a
manifold $V$ embedded in $\JMb{R}^d$ for some $d$ with the commutation
relations
$$
[y^i, y^j] = i \JMkbar J^{ij}, \qquad J^{ij} \in \JMb{R}.
$$
This will induce a symplectic structure on $V$ which is intimately
related to the one we shall exhibit in the following sections. The
details of this have yet to be investigated. Let the larger algebra be
$\JMc{B}$. It has a natural differential calculus defined by imposing
the condition $[y^i, dy^j] = 0$ that the differentials of the
generators be a frame. It follows that the associated metric is
flat. The projection
$$
\Omega^*(\JMc{B}) \longrightarrow \Omega^*(\JMc{A})
$$
would yield a solution to the problem but it is not necessarily
easier to find. In fact a similar situation arises in one of the
possible definitions of a differential calculus as a quotient of the
universal differential calculus by a differential ideal. In that case
the projection is strictly equivalent to the calculus.  One could
also consider the problem of finding the metric as an evolution
equation in field theory in the sense that one can pass from the
Schr\"odinger picture to the Heisenberg picture with the help of an
evolution hamiltonian.

It is interesting to notice how the old Kaluza-Klein idea of gauge
transformations as coordinate transformations appears here. Gauge
transformations are inner automorphisms of the algebra with respect to
some unitary (pseudo-)group $\JMc{G}_G \subset \JMc{A}$ of elements; the
complete dynamical evolution of the system can be described as an
involution with respect to one unitary element $U=e^{iHt}$ of a
(pseudo-)group $\JMc{G}_H \subset \JMc{A}$ of elements of $\JMc{A}$, just as
in quantum field theory. The difference lies in the `size' of the
subalgebra $\JMc{A}_G$ in which $\JMc{G}$ takes its values, as can be
measured for example by the dimension of the commutant of the
subalgebra generated by it; whereas in general
$\mbox{dim}(\JMc{A}^\prime_G) = \mbox{dim}(\JMc{A}^\prime)$, since gauge
transformations are relatively unimportant, in general
$\mbox{dim}(\JMc{A}^\prime_H) = 0$. A topological field theory has
$\mbox{dim}(\JMc{A}^\prime_G) = \mbox{dim}(\JMc{A}^\prime_H)$.

A Riemann-flat solution to the problem is given by choosing
$$
e^\mu_\alpha = \delta^\mu_\alpha, \qquad K_{\alpha\beta} = - \frac
1{i\JMkbar } J^{-1}_{\alpha\beta} \in \JMc{Z}(\JMc{A}).
$$
We have introduced the inverse matrix $J^{-1}_{\alpha\beta}$
of $J^{\alpha\beta}$; we must suppose the Poisson structure to be
non-degenerate: $\det J^{\alpha\beta} \neq 0$.  The relations can
be written in the form
\begin{equation}
\lambda_\alpha = - K_{\alpha\mu} x^\mu, \qquad [\lambda_\alpha,
\lambda_\beta] = K_{\alpha\beta}.  \label{flat}
\end{equation}
This structure is flat according to our definitions.

The most natural Ansatz for the coefficient array $Q$ would seem to be
of the form
$$
Q_-^{\alpha\beta}{}_{\gamma\delta} = 
\JMtfrac 14 k^{(\alpha} Q^{\beta)}_{[\gamma}k_{\delta]}
$$
with $k_\alpha$ a principle null vector and $Q^{\beta}_{\gamma}$
a matrix. Because of the symmetry and reality conditions on the metric
one  must suppose that
$$
g_{\alpha\beta}Q_-^{\alpha\beta}{}_{\gamma\delta} = 0.
$$
This equation will be satisfied if $k^\alpha$ is an  eigenvector of  
$Q^\beta_\alpha$. We set
$$
Q^\beta_\alpha k^\alpha = q k^\beta
$$
and we conclude that 
$$
k_\alpha Q_-^{\alpha\beta}{}_{\gamma\delta} = 0.
$$
From~(\ref{s-e2}) we obtain the expression
$$
C^\alpha{}_{\beta\gamma} \lambda_\alpha = 
F^\alpha{}_{\beta\gamma} \lambda_\alpha -
i\JMkbar\mu^2 \lambda_\alpha \lambda_\delta 
k^{(\alpha} Q^{\delta)}_{[\beta}k_{\gamma]}
$$
This leads to an expression for $\tau$ which is a constant
multiple of $\lambda_\alpha k^\alpha$:
$$
\tau = - 2i\JMkbar\mu^2 \lambda_\alpha k^\alpha.
$$
We would like $\tau$ to play the role of time and so must impose the
condition that $e_a \tau = 0$ and $e_0 \tau \neq 0$. One sees, with 
$F^\alpha{}_{\beta\gamma} =  0$ and to lowest order, that
$$
e_\alpha \tau \propto k^\beta [\lambda_\alpha,\lambda_\beta] =
k^\beta K_{\alpha\beta}.
$$
The null vector must be chosen to have the correct relation also with
the symplectic form. It seems difficult however to work with this
Ansatz  and so we shall opt for a simpler if less elegant one.

We shall find it convenient to consider a curved geometry as a
perturbation of a noncommutative flat geometry. The measure of
noncommutativity is the parameter $\JMkbar $; the measure of curvature
is the quantity $\mu^2$. There are two special interesting limits.  If
we keep $\JMkbar \mu^2$ small but fixed then we can let $\JMkbar \to 0$ or
$\JMkbar \to \infty$. The former (latter) corresponds to a `small'
(`large') universe filled with `small' (`large') cells. The number of
cells is given by $(\JMkbar \mu^2)^{-1}$. We can assume the flat-space
limit to have commutation relations of the form~(\ref{xx}) with
$$
J^{\mu\nu} = J_{(0)}^{\mu\nu} ( 1 + o(i\JMkbar \mu^2)).
$$
Finally we mention that according to the standard definition of
curvature, even corrected to account for the bimodule structure of the
module of 1-forms, all the geometries we consider here have constant
curvature~\cite{FioMad98a}. One of the motivations for considering the
examples is the hope that they will lead to a definition of curvature
which is both bilinear as a map and also in some sense time-dependent.

\section{The Kasner geometry}                      \label{classical}

The Kasner metric is of Petrov type~I and has four distinct principal
null vectors. The limiting Poisson structure defines an additional two
principle null vectors. We must also choose the frame so that it is in
some way adapted to these vectors.  A major problem is to possess a
criterium by which one can decide if the frame is well-chosen. One
obvious condition is that the frame components of all principal null
vectors lie in the center of the algebra.

Choose a symmetric matrix $P = (P^a_b)$ of real
numbers. A moving frame for the Kasner metric is given by
\begin{equation}
\JMt{\theta}^0 = d\JMt{t}, \qquad 
\JMt{\theta}^a = d\JMt{x}^a - P^a_b \JMt{x}^b  \JMt{t}^{-1} d\JMt{t}.   \label{3.1}
\end{equation}
The 1-forms $\JMt{\theta}^\alpha$ are dual to the derivations
$$
\JMt{e}_0 = \JMt{\partial}_0 + P^i_j \JMt{x}^j  \JMt{t}^{-1}\JMt{\partial}_i, \qquad 
\JMt{e}_a = \JMt{\partial}_a
$$
of the algebra $\JMc{A}$. The space $\JMc{X}$ of all derivations is
free of rank 4 as an $\JMc{A}$-module and the $\JMt{e}_\alpha$ form a basis.
The Lie-algebra structure of $\JMc{X}$ is given by the commutation
relations
\begin{equation}
[\JMt{e}_a, \JMt{e}_0] = \JMt{C}^b{}_{a0} \JMt{e}_b, \qquad 
[\JMt{e}_a, \JMt{e}_b] = 0                                      \label{ccr}
\end{equation}
with
$$
\JMt{C}^b{}_{a0} = P^b_a \JMt{t}^{-1}.
$$
For fixed time it is a solvable Lie algebra which is not nilpotent.
We have written the frame in coordinates which are adapted to the
asymptotic condition.  There is a second set which will
be convenient with space coordinates $x^{\prime a}$ given in matrix
notation by
$$
x^{\prime a} = (t^{-P} x)^a.
$$
The frame can be then written, again in matrix notation, with space components 
in the form
$$
\theta^a = (t^P d (t^{-P} x))^a = (t^P d x^{\prime})^a.
$$
This transformation can be considered as an inner automorphism of the
differential 
$$
d \mapsto d^\prime = t^P \circ d \circ t^{-P}
$$
and it would transform the differential calculus into a new one with a
flat metric,

The expression for $\JMt{C}^b{}_{a0}$ contains no parameters with
dimension but it has the correct physical dimensions. Let $G_N$ be
Newton's constant and $\mu$ a mass such that $G_N \mu$ is a length
scale of cosmological order of magnitude. As a first guess we would
like to identify the length scale determined by $\JMkbar$ with the
Planck scale: $\hbar G_N \sim \JMkbar$ and so we have $\JMkbar \sim
10^{-87} \mbox{sec}^2$ and since $\mu^{-1}$ is the age of the universe
we have $\mu \sim 10^{-17} \mbox{sec}^{-1}$. The dimensionless
quantity $\JMkbar\mu^2$ is given by $\JMkbar\mu^2 \sim 10^{-120}$.  In the
Kasner case the role of $\mu$ is played by $\JMt{t}^{-1}$ at a given
epoch $\JMt{t}_0$.

We saw, and we shall see below, that the spectrum of the commutator
of two momenta is the sum of a constant term of order $\JMkbar^{-1}$ and
a `gravitational' term of order $\mu \JMt{t}^{-1} = \JMkbar^{-1} \times
(\JMkbar \mu) \JMt{t}^{-1}$. So the gravitational term in the units we are
using is relatively important for $\JMt{t} \lesssim \JMkbar \mu$. The
existence of the constant term implies that the gravitational field is
not to be identified with the noncommutativity {\it per se} but rather
with its variation in space and time. 

The components of the curvature form are given by
\begin{eqnarray}
&&\JMt{\Omega}^a{}_0 = (P^2 -P)^a_b  
\JMt{t}^{-2} \JMt{\theta}^0 \JMt{\theta}^b,                  \label{3.6a}
\\[6pt]
&&\JMt{\Omega}^a{}_b = - \JMtfrac 12 P^a_{[c} P_{d]b} 
 \JMt{t}^{-2} \JMt{\theta}^c \JMt{\theta}^d.                    \label{3.6b}
\end{eqnarray}                
The curvature form is invariant under a uniform scaling of all
coordinates. The Riemann tensor has components
$$
\JMt{R}^a{}_{0c0} = (P^2 - P)^a_b \JMt{t}^{-2}, \qquad 
\JMt{R}^a{}_{bcd} = P^a_{[c} P_{d]b} \JMt{t}^{-2}.
$$
The vacuum field equations reduce to the equations
$$
\tr (P) = 1, \qquad \tr (P^2) = 1.                     
$$
If $p_a$ are the eigenvalues of the matrix $P^a_b$ there is
a 1-parameter family of solutions given by
\begin{equation}
p_a =\frac{1}{1+\omega+\omega^2} 
\left(1+\omega,\;\omega(1+\omega),\; -\omega\right).      \label{pa}
\end{equation}
The most interesting value is $\omega=1$ in which case
$$
p_a = \JMtfrac 13 (2,\;2,\;-1).
$$
The curvature invariants are proportional to $\JMt{t}^{-2}$; they are
singular at $\JMt{t} = 0$ and vanish as $\JMt{t} \rightarrow \infty$.  

The values $p_a = c$ for the three parameters are also of interest.
The Einstein tensor is given by
$$
\JMt{G}^0_{0} = - 3 c^2  \JMt{t}^{-2} , \qquad 
\JMt{G}^a_{b} = - c (3c - 2)\delta^a_b \JMt{t}^{-2}.
$$
For the value $c=2/3$ the space is a flat FRW with a
dust source given by
$$
\JMt{T}_{00} = - \frac 1{8\pi G_N}\JMt{G}_{00} = \frac 1{6\pi G_N}.
$$
For $c = 1/3$ the space is Einstein with a time-dependent
cosmological `constant' which we interpret as due to the presence of
the supplementary noncommutative dimensions. We start then with a
Kasner solution in dimension 4 and we add a fuzzy structure which
forces us out of the 1-parameter family of vacuum solutions into a
family of solutions with similar properties except for the existence
of a non-zero value for $\Lambda_c$. We interpret the time dependence
of the `constant' as due to the time variation of the internal
structure which gave rise to it but we cannot write explicitly a
formulae as one does in the case of an finite-dimensional manifold as
internal structure or in the case~\cite{Mad00c} of a
finite-dimensional noncommutative structure. We have in the present
case an infinite-dimensional noncommutative algebra and we would
expect rather $\Lambda_c$ to yield information about the algebra than
the inverse.

To illustrate the notation one can analyze the non-commutative version
of the region in parameter space around the flat solution given by
$\omega = -1$.  The local Lorentz rotation which makes this explicit
is equivalent to a change of coordinates. If we choose the
$\JMt{z}$-axis along the vector with the non-vanishing eigenvalue then
the transformation is given by
$$
\JMt{t}^\prime = \JMt{z} \sinh (\JMt{t}/\JMt{z}), \qquad 
\JMt{z}^\prime = \JMt{z} \cosh (\JMt{t}/\JMt{z}).
$$
It follows that $\JMt{t}^2 = \JMt{t}^{\prime 2} -  \JMt{z}^{\prime 2}$; the
origin of the Kasner time coordinate, exactly at the flat-space values
of the parameters and because of the singular nature of the
transformation, becomes a null surface.

\section{The momentum generators}                    

There are four sets of Jacobi identities to satisfy, depending on how
many momentum factors are present. We shall start with those depending
only on the momenta since they were analyzed in detail in the general
discussion. If we assume that $F^\alpha{}_{\beta\gamma} = 0$ and that the
noncommutative extension has the same form as the commutative limit
then from the results of Section~2, in the Kasner case, we find 
that the commutator $\Lambda_{\alpha\beta}$ is of the form
\begin{eqnarray}
&&\Lambda_{ab} = K_{ab},                                   \label{Va1}\\[4pt]
&&\Lambda_{0a} = K_{0a} + \JMtfrac 12 C^b{}_{0a} \lambda_b.  \label{Va2}
\end{eqnarray}
The second term on the right-hand side of the second equation can also
be written as
$$
\JMtfrac 12 C^b{}_{0a} \lambda_b = 
- 2i \JMkbar \lambda_b \lambda_c Q^{bc}{}_{0a}.
$$
We shall suppose that for some matrix $P^a_b$
$$
P^{bc}{}_{a0} = \JMtfrac 14 \mu^2 P^{(b}_a P^{c)}_d k^{\prime d}
$$
so that
$$
C^b{}_{a0} \lambda_b = 
- i \JMkbar \mu^2 \lambda_{(b} \lambda_{c)} P^{b}_a P^{c}_d k^{\prime d}.
$$
From the classical limit we can infer therefore that for some $\tau$
$$
P^b_{a} (\tau \lambda_b + \lambda_b \tau) =
- 2i \JMkbar \lambda_{(b} \lambda_{c)} P^{b}_a P^{c}_d k^{\prime d}.
$$
We have argued in fact that in general the commutative
limit should be flat and so this relation should be expected to be
satisfied only in the limiting case $\omega \to 1$. We define
\begin{equation}
\tau = -2i\JMkbar\mu^2 k^{\prime a} P^b_a \lambda_b                  \label{tx}
\end{equation}
to obtain a consistent relation. 

We recall that the right-hand side of (\ref{Va1}) as well as the
first term on the right-hand side of (\ref{Va2}2) diverge as 
$\JMkbar \to 0$.  To stress this fact we write
$$
K_{0a} = (i\JMkbar)^{-1} l_a, \qquad 
K_{ab} = (i\JMkbar)^{-1} \epsilon_{abc} k^c
$$
with two space-like vectors $l_a$ and $k^a$.  These constitute a
sort of `vacuum-energy density'; they imply that in the flat limit the
commutation relations are not necessarily trivial. We shall require
that $\tau$ depend only on time, that
$$
e_a \tau = 0.
$$
From the definition it follows that
$$
e_a \tau = -2i\JMkbar \mu^2 k^{\prime c} P^b_c K_{ab}
$$
and therefore that
$$
k^{\prime c} P^b_c K_{ab} = 0.
$$
We choose accordingly
$$
k^a = P^a_b k^{\prime b}, \qquad  
\tau = -2i\JMkbar\mu^2 k^{a} \lambda_a.
$$
It will be convenient to introduce the matrix
$$
W(t) = \exp{-\JMtfrac 12 P \int_1^t \tau dt}
$$
solution to the differential equation
$$
\dot W + \JMtfrac 12 P\tau W = 0.
$$
The three eigenvalues of this matrix are the integrating factors for the 
frame. Once we have found a particular solution $\lambda_a$ to
Equation~(\ref{Va2}) then the most general solution is obtained by adding
to it a term of the form $W^b_a \mu_{b}$, with $\mu_{b}$
a triplet of operators which commute with $t$.

A natural connection between the symplectic and metric structures is
that the vector $k^a$ dual to $K_{ab}$ be an eigenvector of $P^a_b$.
Let $p$ be the corresponding eigenvalue.  It follows that
if we multiply both sides of~(\ref{Va2}) by $-2i \mu^2 k^a$ we
find then for $\tau$ the equation
\begin{equation}
\dot \tau + \JMtfrac 12 p\tau^2 + 2 \mu^2 l_a k^a = 0.          \label{de0}
\end{equation}
This equation transforms under the action
of the `duality' transformation
$$
\tau \mapsto -\mu^2 \tau^{-1}
$$
into 
\begin{equation}
\dot \tau + 2 l_ak^a \tau^2 + \JMtfrac 12 p\mu^2 = 0.             \label{de0dual}
\end{equation}
If one neglect the higher-order terms the behaviour of these solutions
near the singularity is respectively
$$
\tau \simeq \frac{2}{pt}, \qquad \tau \simeq \frac{1}{2l_ak^at}.
$$
In the commutative limit $W \to \JMt{W}$ with
$$
\JMt{W} = \JMt{t}^{-P/p}.
$$ 
It would appear here that the classical limit is not a smooth
one. This point is not clear. We cannot expect the two solutions to
behave similarly near the singularity; indeed we hope to eventually
find a `smooth' noncommutative analog. Therefore $\tau - t^{-1}$ should
diverge near the singularity. On the other hand in the asymptotic
region the noncommutative extension has a cosmological constant and is
thus not comparable with the classical solution either. The best we
can claim here according to Equation~(\ref{de000}) (with the plus
sign) is that near the flat solution with $p \simeq 1$ and in the
intermediate region with $\tau \simeq \mu$ the noncommutative solution
behaves like the classical analog.
 
It will be convenient to impose the condition
$$
4 l_a k^a = \pm p
$$
and write Equation~(\ref{de0}) as
\begin{equation}
\dot \tau + \JMtfrac 12 p (\tau^2 \pm  \mu^2) =  0.              \label{de000}
\end{equation}
Equation~(\ref{de000}) is a degenerate form of the Riccati equation and
can be completely solved.  With the plus sign the function
\begin{equation}
\tau = \mu\cot (p \mu t/2)                 \label{sol11}
\end{equation}
is a solution.  With this solution the expression for $W$ becomes
$$
W = \sin^{-P/p}  (p \mu t/2).
$$
Using the Jacobi identity as a Leibniz rule, we find the differential
equation
\begin{equation}
\dot \Lambda_{ab} = - \JMtfrac 12 P^{c}_{[a} \Lambda_{cb]} \label{cd}
\end{equation}
for the commutators $\Lambda_{ab}$. This equation can solved to yield
the expression
$$
\Lambda_{ab} = W^c_a W^d_b K_{cd} = 
(\sin^{-P/p})^c_a(\sin^{-P/p})^d_b K_{cd}.
$$
In the preferred coordinate system, with the $z$-axis along
the direction of the distinguished eigenvector $k^a$, the only
nonvanishing element is
$$
\Lambda_{12} = \sin^{(p-1)/p} (p \mu t/2) K_{12}.
$$
We shall suppose that $g^{ab}l_b$ and $k^a$ are parallel. We can
solve then the system of equations~(\ref{Va2}). We find that for $a = 1,2$
\begin{equation}
\lambda_a = W^b_a \mu_{b}                       \label{Dress}
\end{equation}
with $\mu_{b}$ a doublet of operators which commute with
$t$. The third element $\lambda_3$ is a constant multiple of $\tau$.

To clarify this a bit one can consider a perturbative solution near
flat space, that is with the parameter $\omega$ approximately equal to
the flat-space value $\omega = -1$.  Since we have claimed that
non-trivial commutation relations and a curved metric are in fact two
aspects of the same reality, we allow ourselves the freedom to
identify the difference $\omega + 1$ with the parameter $\JMkbar \mu^2$.
We write then $\omega = -1 + \JMkbar\mu^2$ and expand $P$ in a
power series:
$$
P = P_0 +\JMkbar\mu^2 P_1 + (\JMkbar\mu^2)^2 P_2 + \cdots.
$$
The coefficients can be written in the form
$$
P_0 = \left(\begin{array}{ccc}0 &0 &0 \\ 0 &0  &0 \\ 0 &0 &1
\end{array}\right), \qquad
P_1 = \left(\begin{array}{ccc}1 &0 &0 \\ 0 &-1 &0 \\ 0 &0 &0
\end{array}\right), \qquad
P_2 = \left(\begin{array}{ccc}1 &0 &0 \\ 0 &0  &0 \\ 0 &0 &-1
\end{array}\right).
$$
To first order the expansion of the right-hand side of~(\ref{Va2})
is given by
\begin{eqnarray}
&&[\lambda_0, \lambda_1] = - \JMtfrac 12\JMkbar\mu^2 \tau \lambda_1, \\[4pt]
&&[\lambda_0, \lambda_2] = \JMtfrac 12\JMkbar\mu^2 \tau \lambda_2, \\[4pt]
&&[\lambda_0, \lambda_3] = \frac 1{i\JMkbar} l_3 - \JMtfrac 12 \tau \lambda_3.
\end{eqnarray}
The structure of the classical algebra of derivations is given
by~(\ref{ccr}); the noncommutative generalization is an algebra
defined by the relations~(\ref{Va1}) and~(\ref{Va2}) with however a
different right-hand side for the former.

It is possible to represent $\JMc{A}$ as a tensor product of two
Heisenberg algebras. We introduce $\mu_\alpha$ with
$$
[\mu_\alpha, \mu_\beta] = K_{\alpha\beta}
$$
and we define $\JMc{A}_{12}$ ($\JMc{A}_{30}$) to be the algebra 
generated by $(\mu_1,\mu_2)$ ($(\mu_3,\mu_0)$). We define an embedding
$$
0 \rightarrow \JMc{A}_{30} \buildrel \phi \over \rightarrow \JMc{A}
$$
by setting
$$
\lambda_0 = \phi(\mu_0) = \mu_0,\qquad 
\lambda_3 = \phi(\mu_3) = 
-(2i\JMkbar\mu)^{-1} \cot(\JMtfrac 12 p \mu \mu_3).
$$
The image of $\JMc{A}_{30}$ is an 2-sided ideal of $\JMc{A}$ and the
projection
$$
\JMc{A}_{12} \otimes  \JMc{A}_{30}
\buildrel \phi \over \rightarrow \JMc{A}_{12} \rightarrow 0
$$
is defined by setting, for $a=1,2$
\begin{equation}
\lambda_a = \phi_a(\mu_a,\mu_3) = W_a^b \mu_b.             \label{lm}
\end{equation}
We have represented $\JMc{A}$ as the direct product of
$\JMc{A}_{12}$ and $\JMc{A}_{30}$.
The `canonical transformation' $\mu_\alpha \mapsto \lambda_\alpha$
is partially given by a change in momenta defined by a transendental
function; it is highly nonlocal.  There is evidence of what could
be considered a noncommutative Darboux theorem. In the formal analogy
with quantum field theory the $\mu_\alpha$ could be perhaps considered
`bare' momenta and the $\lambda_\alpha$ the corresponding `dressed'
momenta. There has however been no use made of the latter and we shall
wait until we have introduced the position generators and can speak of
`bare' and `dressed' fields to pursue this analogy.

\section{The position generators}                    

The conjugacy relations which define the Kasner metric are most easily
defined using the second set $x^{\prime a}$ of coordinates introduced
in Section~\ref{classical}. They are given by
\begin{equation}
\begin{array}{ll}
[\lambda_0, t^\prime] = 1, &[\lambda_0, x^{\prime j}] = 0,\\[4pt]
[\lambda_a, t^\prime] = 0, &[\lambda_a, x^{\prime j}] = (W^{-2})^j_a.
\end{array}                                             \label{duality0}
\end{equation}
This set of equations constitutes a relation between the coordinate
generators and the momenta generators.  Using them one can
find differential equations for the $J^{\mu\nu}$:
\begin{equation}
\begin{array}{ll}
[\lambda_0, J^{\prime ij}] = 0,
&[\lambda_a, J^{\prime ij}] = 
(PW^{-2})^{[i}_a \tau J^{\prime 0j]},                  \\[8pt]
[\lambda_0, J^{\prime 0j}] = 0,
&[\lambda_a, J^{\prime 0j}] = 0.   
\end{array}                                          \label{j1}
\end{equation}
These can be solved immediately to yield the expresssions
\begin{equation}
J^{\prime ij} = S_{(0)}^{\prime ij} + L^{\prime ij}, \qquad 
L^{\prime ij} = P^{[i}_k x^{\prime k} \tau J^{\prime 0j]}, \qquad 
J^{\prime 0i} = J^{\prime 0j}_{(0)},                         \label{tom}
\end{equation}
for the commutator as a sum of a constant `spin' $S^{\prime ij}$ and `orbital
momentum' $L^{\prime ij}$. They can also be written in terms of the original
coordinates $x^i$, without due regard to hermiticity, as
$$
J^{0i} = [t, (W^{2} x^\prime)^{ i}] =  
(W^{2} J^\prime)^{0i} = (W^{2})^i_j J^{\prime 0j}_{(0)}.      \label{Marco}
$$
and a rather more complicated expression
$$
J^{ij} = [ (W^{2} x^\prime)^{i}, (W^{2} x^\prime)^{j}]
$$
for the angular momentum. The components of $L^{\prime ij}$ behave as
$t^{-2}$

If we choose the $z$-axis so that $i\JMkbar J^{\prime 0i} = (0,0,ih)$ and neglect
the `spin' then we find the Lie algebra structure
$$
[x^\prime,y^\prime] = 0, \qquad 
[y^\prime,z^\prime] = ih y\prime, \qquad 
[z^\prime,x^\prime] = - ih x^\prime 
$$
of the solvable subgroup of $SL(2,\JMb{R})$. This algebra appears also
in a noncommutative version of the Lobachevski plane. We refer 
elsewhere~\cite{Mad00c} for a discussion of this as well as for the
reference to the original literature.

It is difficult to appreciate the significance of these components
since at the origin the coordinate system becomes singular.  
The frame components
$$
I^{\alpha\beta} = \theta^\alpha_\mu \theta^\beta_\nu J^{\mu\nu}
$$
might be considered more significant. One finds that
$$
I^{0a} = (W^{2}J_{(0)})^{0a}, \qquad I^{ab} = S^{ab} = (W^{4}S_{(0)})^{ab}
$$
provided one uses the `frame components' of the coordinate (generator)
$$
x^a = (W^{2})^a_i x^{\prime i} = x^i.
$$
We can choose as only non-vanishing components $S_{(0)}^{12}$ and
$J^{03}$ in which case near the singularity we find that
$$
I^{a3} \sim t^{-2}.
$$
The `spin' has to lowest order the same time dependence as the curvature.
the singularity is infinitely fuzzy. In this sense
we have resolved the point singularity.

We have noticed that when $p_a = 2/3(1,1,1)$ the space becomes the
flat FRW solution with a pressure-free dust as source. The matrix $W$
is a multiple of the unit matrix and we find for the
momenta commutators
$$
\Lambda_{ab} \sim t^{-2}
$$
The standard coordinates one uses are the analogs of the 
$x^{\prime i}$ coordinates introduced in Section~\ref{classical} for
the Kasner metric. In this case the position generators have
commutators which near the origin behave as
$$
L^{\prime ij} = \JMtfrac 23 x^{\prime [i} t^{-1} J^{\prime 0j]}.
$$
The covariant derivatives $D_\gamma I^{\alpha\beta}$ of the `spin tensor'
are given as
$$
\begin{array}{ll}
D_0 I^{0a} = e_0  I^{0a} = P^a_b \tau J^{b0},
&D_b I^{0a} = 0,\\[6pt]
D_0 I^{ab} = e_0 I^{ab} =  \tau P^{[a}_{c} I^{bc]},
&D_c I^{ab} = e_c I^{ab} + \omega^{[b}{}_{a0} J^{0c]}
= \tau P^{[a}_{c} I^{0b]}.
\end{array}
$$
We find then a `Maxwell field strength' 
$F_{\alpha\beta} = \mu^2 I_{\alpha\beta}$ obeying `Maxwell's
equations' with a current $j_E$ defined by
$$
j_E^0 = 0, \qquad 
j_E^a = D_\alpha I^{\alpha a} = \tau \mu^2 I^{0a}.
$$
There is also a `magnetic monopole' density given by
$$
j_M^a = \JMtfrac 12 P^{[a}_b \tau I^{0b]} = 
\JMtfrac 12 P^{a}_b j_E^b - j_E^a.
$$
This relation recalls somewhat the relation between an magnetic
field and an electric field in a uniformly moving Lorentz frame.  If
$B_a = \epsilon_{abc} v^b E^c$ depends only on time then
$$
j_{Ma} = \dot B_a = \epsilon_{abc} v^b \dot E^c =
\epsilon_{abc} v^b j_E^c.
$$
One obtains the former from the latter by the replacement
$$
\epsilon_{abc} v^b \mapsto \JMtfrac 12 P_{ab} - g_{ab}
$$
of an antisymmetric tensor by a symmetric one.

It is interesting to note than one can consider Equation~(\ref{cd})
as the equation
$$
\nabla_\alpha \theta^\beta_\nu = 0
$$
which permits one to pass from a coordinate frame to an orthonormal
frame. The covariant derivative here is with respect to the complete
set of indices. This equation yields the relation between the
Christophel symbols and the Ricci rotation coefficients.

\section*{Discusssion}
 
In a subsequent publication we shall discuss the differential
calculus.  There is a well-defined if perhaps complicated algorithm to
construct the frame starting with the momenta $\lambda_\alpha$.
Having `blurred' the Kasner metric and deformed the resulting algebra
we can now take the `sharp' limit and see what we obtain.  With the
form of $P^a_b$ we have the metric cannot be Ricci-flat but has an
induced cosmological constant due to the
noncommutativity~\cite{MacMadManZou03}.  The theory we are
investigating has certain similarities with theories of the type
called Kaluza-Klein. That is, the additional noncommutative structure
can perhaps at least to a certain extent be assimilated to an
effective commutative theory in higher dimensions.  This means that
even if one could define a curvature tensor in a satisfactory manner
there is no reason to expect the Ricci tensor to vanish. We shall
assume that to the lowest approximation the Ricci tensor of the total
structure does vanish and we shall use the Ricci tensor of the four
dimensions to elucidate the structure of the hidden dimensions.

\section*{Acknowledgments} One of the authors (JM) would like to thank 
M.~Buri\'c, S.~Cho, G~Fiore, E.~Floratos, H.~Grosse, J.~Mourad, A.~Sykora,
S.~Theisen, P.~Tod and G.~Zoupanos for interesting conversations.
Preliminary work was to a large extent carried out while he was
visiting the MPI in M\"unchen.  He is grateful to Julius Wess for
financial support during this period. The results included here were
presented by him at the II Summer School in Modern Mathematical
Physics, Kopaonik, the XXI Workshop on Geometric Methods in
Physics, Bialowieza, the Erwin Schr\"odinger Institute workshop
on Noncommutative Geometry, Feynman Diagrams and Quantum Field Theory
and the Euresco meeting on physics beyond the standard model in
Portoroz, Slovenia.


\providecommand{\href}[2]{#2}\begingroup\raggedright\endgroup


\title*{The Multiple Point Principle: Realized Vacuum 
in Nature is Maximally Degenerate}
\author{Donald~L.~Bennett\thanks{E-mail:bennett@alf.nbi.dk}${}^1$ and 
Holger Bech Nielsen\thanks{E-mail:hbech@alf.nbi.dk}${}^2$}
\institute{%
${}^1$~Brookes Institute for Advanced Studies, 
B\o gevej 6, 
2900 Hellerup, Denmark\\
${}^2$~The Niels Bohr Institute, Blegdamsvej 17, 
2100 Copenhagen {\O}, Denmark}

\titlerunning{The Multiple Point Principle}
\authorrunning{Donald~L.~Bennett and Holger Bech Nielsen}
\maketitle

\begin{abstract}
We put forward the multiple point principle as a 
fine-tuning mechanism
that can explain why some of the parameters of the standard model have the 
values observed experimentally. The principle states that the parameter
values realized in Nature coincide with the surface (e.g. the point) in 
the action parameter space that lies in the boundary that separates the maximum
number of regulator-induced phases (e.g., the lattice artifact phases of
a lattice gauge theory). We argue that a mild form of non-locality - namely
that embodied in allowing diffeomorphism invariant contributions to the 
action - seems to be needed for some fine-tuning problems. 
We demonstrate that the
multiple point principle solution to fine-tuning has the  very 
special property of avoiding the paradoxes that can arise in the
presence of non-locality. The non-renormalizability of gravity suggests ---
in a manner reminiscent of baby universe theory --- the presence of non-local
effects without which the phenomenologically observed  high degree of flatness 
of spacetime would seem mysterious. In our picture, different vacuum states 
are realized in different spacetime regions of the cosmological history.
\end{abstract}

\section{Introduction}

Except for providing explanations for neutrino-oscillation masses of neutrinos, 
dark matter, the baryon assymmetry and inflation, the standard model serves 
physicists extremely well for the moment. In our view the major motivation for 
seekng a theory beyond the standard model is the need for a theory that predicts 
the
20 or so free parameters of the standard model. Our by now old proposal of an 
assumption about the parameters, couplings and masses that can be appended to 
any proposed quantum field theory provides predictions for these free 
parameters. We call this assumption the Multiple Point Principle (MPP) 
\cite{bennett1,bennett2,bennett3}. The
multiple point is a point in the action parameter space of a theory that is 
special in a way that is analogous to the way that the triple point of water
is the special point in the phase diagram spanned by temperature and presure
at which the solid, liquid and vapor phases of water coexist.    
The MPP states that fundamental physical
parameters assume values that correspond to having a maximal number of
different coexisting ``phases'' for the physically realized vacuum.   
There is phenomenological evidence 
suggesting that some or all of the about 20 parameters in the Standard
Model (SM) that are not predicted within the framework of the SM correspond
to the MPP values of these parameters. 
That these parameters take on special
values (i.e., the multiple point values) poses from one viewpoint a fine 
tuning problem (why do constants of Nature take the MPP values). 
From another viewpoint, assuming the MPP as a law of 
Nature leads to a mechanism for fine-tuning. It is the latter viewpoint that
is developed here. Moreover, we shall argue that a mild form of non locality
is inherant to fine-tuning problems in general. We therefore develope a model
for the relationship between fine-tuning, non-locality and the MPP.    

\section{Arguments for non-locality}\label{nonloc}

\subsection{From cosmological constant fine-tuning}
 
Explaining the (dressed) value of the cosmological constant is an example of a 
fine 
tuning problem that would seem to require
the breakdown of locality at least in a mild sense. As with any fine-tuning 
problem, the cosmological constant problem calls for a way to make the coupling
dynamical in such a way that the values of such couplings are
maintained at constant values (required for translational invariance).
But this leads to a problem: if a coupling (e.g., the cosmological constant)
is dynamical, the demands of a strictly local theory would be that the bare 
coupling
can only depend on the spacetime point in question and indirectly on the 
past but certainly not on the future. However, if the bare cosmological constant
(that is to be dynamically maintained at a constant value) immediately following
the big bang is to already have its value fine-tuned once and for all - to
say 120 decimal places - to the value that makes the dressed cosmological
constant so small as suggested phenomenologically, we definitely have a
problem with locality.

The problem is that the bare cosmological constant is relateable to the
value of the dressed cosmological constant only if the details of the dressed
cosmological constant (that did not exist when the bare value was already 
tuned 
to the
valued required for the dressed vacuum) that will evolve in the future are
known at the time of big bang\cite{bennett4}.
We are forced to conclude that a strict principle of locality is not allowed
if we want to have a dynamically maintained bare coupling and renormalization 
group corrections of a quantum field theory with a well-defined vacuum.

This suggests models with a mild form of non-locality consisting of an 
interaction that is
the same between any pair of points in spacetime independent of the distance 
between these points. Assuming that this sort of non-locality is manifested
through a non-local action $\hat{S}_{nl}$, this symmetry between 
any pair of space time points (i.e., identical interaction regardless of 
separation) 
is insured by requiring the invariance of $\hat{S}_{nl}$ under
diffeomorphisms (reparameterization invariance). The non-local action 
$\hat{S}_{nl}$ is
a function of functionals $I_{f_j}[\phi(x)]$: 
$\hat{S}_{nl}=\hat{S}_{nl}(\{I_{f_j}[\phi(x)\})$ where 
$I_{f_j}[\phi(x)]\stackrel{def}{=}\int dx^4\sqrt{g(x)}f_j(\phi(x))$ 
and $f_j(\phi)$ might 
typically be a Lagrange density e.g.,  $f_j(\phi)={\cal 
L}_j(\phi(x),\partial_{\mu}\phi(x))$.
The symbol $\phi(x)$ stands for all the fields (and derivatives of same) 
of the theory.

An example of a nonlocal action would be any nonlinear function
of the (reparameterization invariant) functionals $I_{f_i},I_{f_j},\cdots$; 
e.g., a term $$\int d^4x \int d^4y \sqrt{g(x)g(y)}\phi^2(x)\phi^4(y).$$
Another example of a non-local (and nonlinear) action term more relevant 
to this paper is associated with having  fixed values 
$I_{fixed\;f_j}$ (fixed in the sense of being a law of Nature)
of some extensive quantities $I_{f_j}[\phi]$. This 
amounts to having a $\delta$-function term 
$\exp(S_{nl}(\{I_{f_j}\})=\prod_j\delta(I_{f_j}[\phi]-I_{fixed\;f_j})$ in 
the functional integration measure and results in the nonlocality that, 
strictly speaking, is inherant to any microcanonical ensemble (but which 
often is ``approximated away'' by using a canonical ensemble when phase 
space volume (or functional integration measure) is a sufficiently 
rapidly varying 
function of the extensive quantities).

An extensive quantity $I_{f_j}[\phi(x)]$ has a value for each imaginable 
Feynman path intgeral history of the Universe as it evolves from Big Bang 
to Big Crunch. The value $I_{fixed\;f_j}$ is by assumption ``frozen in''  
and cannot change during the lifetime of the Universe. This unchangeable 
``choice'' $I_{fixed\;f_j}$ then singles out a subset of all possible
Feynman path integral histories that is consistent with the spacetime evoluion 
of our actually realized Universe having $I_{f_j}[\phi]=I_{fixed\;f_j}$.

An interaction that is the same between the
fields at {\it any} pair of spacetime points - regardless of separation - 
would not
likely be perceived as a non-local interaction. Rather such spacetime
omnipresent fields - a sort of background that is forever everywhere the 
same - would likely be interpreted as simply constants of Nature. This feature
is reminiscent of baby universe theory the essence of which is that a physical
constant can depend on something and still be a constant as a function of 
spacetime.

\subsection{Arguments from short distance flucuation cancellation}

Given that the prevailing feature of spacetime foam --- the term coined by Wheeler 
to conjure up a picture of the Planck scale structure of spacetime --- is a 
multitude of topologically nontrivial structures in a high curvature spacetime, 
it is natural to wonder how it comes about that spacetime at human distance 
scales of one meter say only deviate from being completely flat (i.e., ordinary 
Euclidean geometry) by tiny gravitational field effects that are almost 
negligible. 

It is, however, not so trivial to see that short distance fluctuations will sum 
up to make large distance fluctuations in the curvature of spacetime. If we
consider the parallel transport of a little vector from a genuine point in 
spacetime around a closed curve and back it is of course impossible that the 
fluctuation in angle of rotation caused by parallel transport along a long 
closed curve could be smaller than around a small closed curves of Planck size.  

However, if we instead define an averaged effective geometry and consider a 
locally smeared way of parallel transporting and then parallel transport a 
vector defined in this smeared way along a long closed curve, it is no longer so 
that the fluctuation of the vector rotation after parallel transporting cannot
be much smaller that the small (Planck) scale local fluctuations. However it is 
not immediately obvious what it means to define such an average of the geometry 
over large volumes in order to average out small distance fluctuations. For 
instance it is not trivial to construct an approximately parallel vector field 
on a highly curved space.

\section{The Multiple Point Principle (MPP)}

The MPP was originally put forward in connection
with theoretical predictions for the values of the three gauge coupling 
constants \cite{bennett1,bennett2}. In addition to the assumption of the MPP, we 
also assumed in this
first application of MPP our so-called Anti-GUT gauge group 
$G_{Anti-GUT}$ which consists of the 3-fold replication
of the Standard Model Group (SMG): 
$G_{Anti-GUT}=SMG \otimes SMG \otimes SMG \stackrel{def}{=}
SMG^3$ (in the extended version: $(SMG\times U(1))^3$ )
having one $SMG$ factor for each generation of fermions
and gauge bosons. We postulate that $G_{Anti-GUT}$ is broken to
the diagonal subgroup (i.e., the usual SMG) at roughly the Planck scale.

In the original context of predicting the standard model gauge couplings
using MPP (originally referred to as the principle of multiple point
criticality), the principle asserts that the Planck scale values of the
standard model gauge group couplings coincide with the multiple point, i.e.,  
the point that lies in the boundary separating the maximum number of phases in 
the action parameter space corresponding to the gauge group
$G_{Anti-GUT}$. The (Planck scale) 
predictions for the gauge couplings are subsequently   will
sum 
identified with the
parameter values at the point in the action parameter space for the diagonal 
subgroup of $G_{Anti-GUT}$ that is
inherited from the multiple point for $G_{Anti-GUT}$ after the Planck
scale breakdown of the latter.

The idea was developed in the context of lattice gauge theory and the
phases to which we refer are usually dismissed as lattice artifacts.
(e.g., a Higgsed phase, a confined or Coulomb-like phase).
Such phases have been studied extensively in the litterature for simple
gauge groups and semi simple gauge groups with discrete subgroups (e.g.
$SU(2)$ and $SU(3)$). One typically finds first order phase transitions
between confined and Coulomb-like phases at critical values of the action
parameters.    

Taking such lattice artifact phases as physical reflects our 
suspicion that such phases are inherant to having a regulator. As a
regulator in some form (be it a lattice, strings or whatever) is always
needed for the consistency of any quantum field theory, it is consistent to 
assume the existence of a fundamental regulator.
The ``artifact'' phases that arise in a theory with such a fundamental 
regulator (that we have chosen to implement as
a fundamental lattice) are accordngly taken as ontological phases that have
physical significance at the scale of the fundamental regulator (e.g.,
lattice). The assumption of an ontological fundamental regulator implies
the existence of monopoles in terms of which the regulator induced phase
can also be studied\cite{bennett5}.

Finding the multiple point in an action parameter space corresponding to
the gauge group $G_{Anti-GUT}$ is more complicated than for a single
$SU(2)$ or $SU(3)$ say. The boundaries between phases in the action parameter
space (i.e., the phase diagram) must be sought in a high dimensional parameter 
space essentially because $G_{Anti-GUT}$ being a non-simple group has many
subgroups and invariant subgroups.

In fact there is a distinct phase for each subgroup pair $(K,H)$ 
where $K$ is a 
subgroup and $H$ is an invariant subgroup such that 
$H \triangleleft K \subseteq G_{Anti-GUT}$. An
element $U \in G_{Anti-GUT}$ can be parameterized as $U=U(g,k,h)$
where the Higgsed (gauge) degrees of freedom 
are elements $g$ of the homogeneous space $G_{Anti-GUT}/K$. 
The (un-Higgsed) Coulomb-like and confined degrees of freedom
are respectively the elements $k$ of the factor group $K/H$ and the 
elements $h \in H$. 
    
\section{A Familiar Analogy to the MPP as a Fine-
Tuning Mechanism} 

Some important features of the MPP as a fine-tuning mechanism can be
illustrated using an analogy to the familiar system in which the solid,
liquid and vapour phases of water coexist. This occurs at the 
``triple point'' of water, i.e., at the ``triple point'' values of
temperature and presure. Because the transitions between these three phases
are all first order, there is a whole range of combinations of the extensive 
variables energy and volume for which the system can only be
realized by having the
coexistence of the ice, liquid and vapour phases. But these three phases
coexist only for the triple point values of temperature and presure,
so there is a whole range of combinations of 
energy and volume that map onto the triple point values of the
conjugate intensive variables temperature and presure with the result that these 
variables 
are fine-tuned to the
triple point values. In this illustrative analogy, the triple point of
water in the phase diagram spanned by the intensive parameters temperature 
and presure is analogous to the multiple point. As already stated, the 
multiple point is the (or a) point in the phase diagram that ``touches'' the
maximun number of phases. In a phase diagram spanned by $D$ intensive parameters
(couplings), a generic multiple point can be in contact with up to $D+1$ 
phases (in the illustrative example, $D=2$ and the triple point is in contact 
with the $D+1=3$ phases ice, liquid and vapour). In a non-generic situation,
the multiple point can be in contact with more than $D+1$ phases (e.g., 
accidently or due to symmetries).

For ease of illustration, consider now the even simpler system
consisting of $n_{H_2O}$ moles of $H_2O$ in which just the ice and liquid 
phases 
coexist (at constant presure). 
Such a system is unavoidably 
realized (and the temperature fine-tuned to $0^oC=273.15^oK$) for {\it any}
value of the energy density $\rho_E=E/V_{n_{H_2O}}$ ($E$ and $V_{n_{H_2O}}$
are respectively the energy and volume of the $n_{H_2O}$ moles of $H_2O$)
in the {\it finite} interval
\begin{eqnarray}
\frac{n_{H_2O}}{V_{n_{H_2O}}}\int^{273^o\;K}_{0^o\;K}C_{p,ice}(T)dT 
 &<& \rho_E <
\frac{n_{H_2O}}{V_{n_{H_2O}}}\biggl(\;\int^{273^o\;K}_{0^o\;K}\;C_{p,ice}(T)dT
  \biggr. \nonumber\\
 & & \qquad\qquad +\biggl.\  
      (molar\;heat\; of\;melting)\biggr) \label{interval}
\end{eqnarray}
\[ \mbox{($C_{p,ice}$ is the molar heat capacity of ice at constant
pressure (e.g., 1 atm.)).}\]
For any $\rho_E$ in this interval, the system cannot be realized as a single 
phase
but rather only as an equilibriated mixture of ice and liquid water. 
Even choosing  $\rho_E$ at random there is a finite chance of
landing in this interval in which case the temperature will be fine-tuned
to $273.15^oK$.

\section{The History of Our Universe as a Fine Tuner}\label{ouruniverse}

Consider an analogy between the (3-dimensional) 
ice-water system with $\rho_E$
in the interval of Eqn.(\ref{interval}) and our 4-dimensional
universe with the value of an extensive variable 
$I_{f_j}[\phi(x)]\stackrel{def}{=}\int dx^4 
\sqrt{g(x)}f_j(\phi(x))$ 
(with $f_j$ any function of $\phi$ - see also Sec.~\ref{nonloc}
for notation)
primordially fixed  at a value $I_{fixed\;f_j}$ that
can only be realized as a combination of  
two (for the sake of example - really there could be more than two) 
coexisting phases i.e., two degenerate vacuum states at field values that
we denote as $\phi_{us}$ and $\phi_{other}$ where we take 
$\phi_{us}<\phi_{other}$. Here we are anticipating the introduction of an
effective potential $V_{eff}$ that has relative minima at the field values 
$\phi_{us}$ and $\phi_{other}$. 
In 4-space, one generic possibility for having coexistent phases would be to 
have a phase with $\phi_{us}$ in an early epoch including say the universe 
as we know it and a phase with $\phi_{other}$ in a later epoch:  
\begin{equation}
I_{fixed\;f_j}=
f_j(\phi_{us})(t_{ignit}-t_{BB})V_3+ f_j(\phi_{other})(t_{BC}-t_{ignit})V_3
\label{coexist}\end{equation}
where $t_{ignit}$ is the ``ignition'' time (in the future) at which there is a 
first
order phase transition from the vacuum at $\phi_{us}$ to the later
vacuum at $\phi_{other}$. $V_3$ is the 3-volume of the universe.
The value of the ``coupling constant'' conjugate to $I_{fixed\;f_j}$ gets fine 
tuned (unavoidably by assumption
of the coexistence of the two phases separated by a first
order transition) by a mechanism that also depends on a phase that will 
first be realized in the future (at $t_{ignit}$). Such a mechanism is  
non-local. Note in particular that the right hand side of Eqn.~\ref{coexist} 
depends on $t_{ignit}$.  

In order to formally define a ``coupling constant'' (intensive quantity)
conjugate to some extensive quantity (e.g., $I_{fixed\;f_j}$), we introduce
non-loality more abstractly. Let us restrict the non-local 
action $\hat{S}_{nl}=\hat{S}_{nl}(\{I_j[\phi(x)]\})$ to being a (also 
reparameterization invariant) non-local
potential $V_{nl}$ that is a function of (not necessarily independent) 
functionals
\[ V_{nl}=
\stackrel{def}{=} V_{nl}(I_{f_i}[\phi],I_{f_j}[\phi],\cdots). \]
Define now an effective potential $V_{eff}$ such that
\begin{equation}
\frac{\partial V_{eff}(\phi(x))}{\partial \phi(x)} \stackrel{def}{=}
\frac{\delta V_{nl}(\{I_{f_j}[\phi]\})}{\delta \phi(x)}
\left|_{near\;\;min.} \right.=
\sum_i\left(\frac{\partial V_{nl}(\{I_{f_j}\})}{\partial I_{f_i}}\frac{\delta
I_{f_i}[\phi]}
{\delta \phi(x)}\right)\left|_{near\;\;min.} \right. \label{eq11}
\end{equation}
\[ =\sum_i\frac{\partial V_{nl}(\{I_{f_j}\})}{\partial I_{f_i}}
\left|_{near\;\;min.} \right. f_i^{\prime}(\phi(x))  \]
The subscript ``near min'' 
denotes  the approximate ground state of the whole universe, up to deviations
of $\phi(x)$ from  its vacuum value (or vacuum values for a multi-phase vacuum)
by any amount in relatively small spacetime regions.
The solution to Eq.~(\ref{eq11}) is
\begin{equation}
V_{eff}(\phi)=\sum_i \frac{\partial V_{nl}(\{I_{f_j}\})}{\partial 
I_{f_i}}f_i(\phi)
\label{eq12} \end{equation}
We can identify the
$\frac{\partial V_{nl}(\{I_{f_j}\})}{\partial I_{f_i}}$ as intensive quantities
conjugate to the $I_{f_i}$.

Consider now the effective potential
(\ref{eq12}) in the special case that
$V_{nl}(\{I_{f_j}\})=V_{nl}(I_2,I_4)\stackrel{def}{=}
V_{nl}(\int d^4x\sqrt{g(x)}\phi^2(x),\int d^4y\sqrt{g(y)}\phi^4(y))$
in which case, (\ref{eq12}) becomes
\begin{equation} V_{eff}=\frac{\partial V_{nl}(I_2,I_4)}{\partial 
I_2}\phi^2(x)+
\frac{\partial V_{nl}(I_2,I_4)}{\partial I_4}\phi^4(x)\stackrel{def}{=}
\frac{1}{2}m^2_{Higgs}\phi^2(x)+
\frac{1}{4}\lambda\phi^4(x) \label{msqh}\end{equation}
where the right hand side of this equation, which also defines 
the (intensive) couplings $m^2_{Higgs}$ and
$\lambda$, is recognised as a prototype scalar potential
at the tree level.
Of course the
form of $V_{nl}$ is, at least {\em a priori}, completely unknown to us,
so - for example - the coupling
constant $m^2_{Higgs}$ cannot be calculated from Eqn.~\ref{msqh}.
The potential of Eqn.~\ref{msqh}
with $m_{Higgs}^2<0$ has an asymmetric minimun --- at, say, the value $\phi_{us}$ 
resulting in 
spontaneous symmetry breakdown 
in the familiar way. This is just standard physics (without non-locality).

Actually we want to consider the  potential $V_{eff}$ having the two relative 
minima 
$\phi_{us}$ and $\phi_{other}$ - both at nonvanishing values of $\phi$ - 
alluded to at the beginning of this section.
The second minimum comes about at a value $\phi_{other}>\phi_{us}$ when 
radiative corrections to (\ref{msqh}) are taken into 
account and the top quark mass is not too 
large\cite{bennett6,bennett7,bennett3}.
Which of these vacua - the one at $\phi_{us}$ or $\phi_{other}$ - 
would be the stable one in this 
two-minima Standard Model effective Higgs field potential depends on
the value of $m^2_{Higgs}$. 
Since $I_2$ and $I_4$ are  functions of $t_{ignit}$ (as seen from 
Eqn.~\ref{coexist} with $f_j=\phi^2$ or $\phi^4$),
$m^2_{Higgs}\stackrel{def}{=}
\frac{\partial V_{nl}(\{I_2,I_4\})}{\partial I_2}$ is also a function of
$t_{ignit}$.

\begin{figure}
\centering
\includegraphics*[width=10cm,bbllx=10pt,bblly=40pt,bburx=230pt,bbury=270pt]{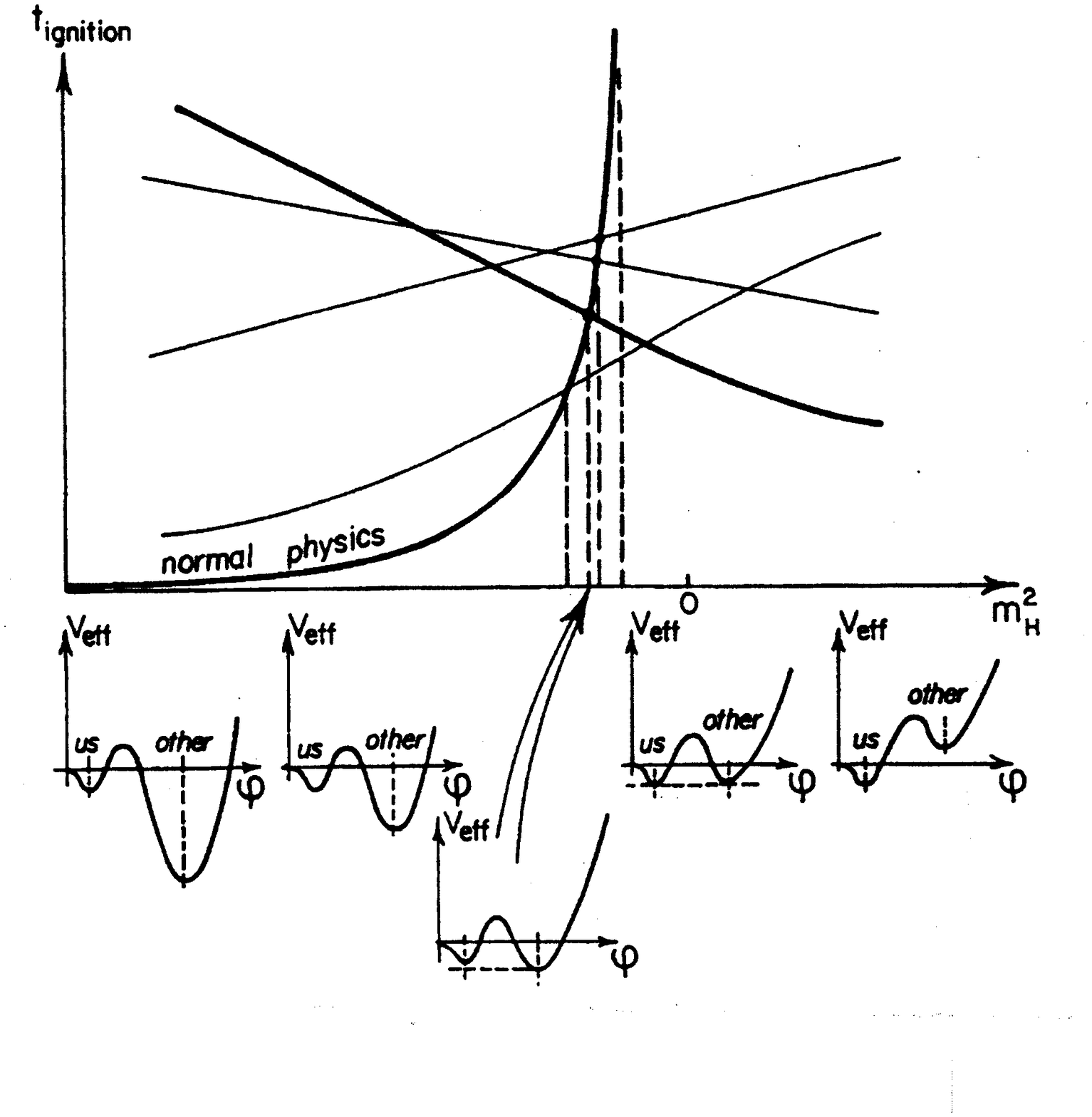}
\caption{\label{fignonloc}
The development of the double well
potential and $m_{Higgs}$ as a function of $t_{ignit}$. Note that
all the more or less randomly drawn non-locality curves intersect the
``normal physics'' curve
near where the vacua are degenerate (i.e., the MPP solution).}
\end{figure}

\begin{figure}
\centering
\includegraphics*[width=10cm,bbllx=250pt,bblly=15pt,bburx=480pt,bbury=320pt]{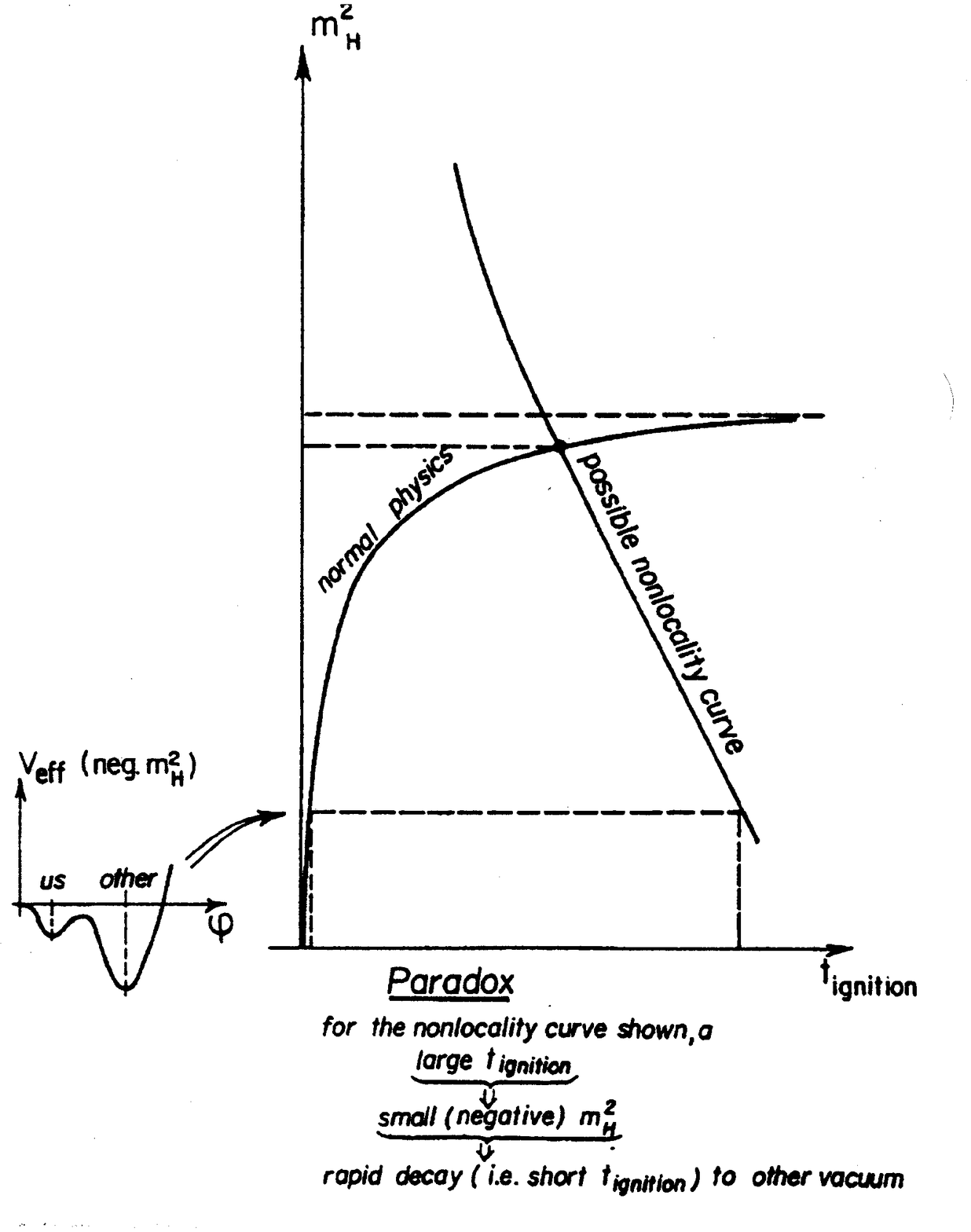}
\caption{\label{paradox} 
Many non-locality curves could lead to
paradoxes similar to the ``matricide'' paradox. Such paradoxes are avoided if
the value of $m_{Higgs}$ is fine-tuned to the multiple point critical value.
This corresponds to the intersection of the ``normal physics'' curve with the 
``possible nonlocality'' curve.}
\end{figure}

Let us first use  ``normal physics'' to see how the relative depths of the
two minima of the double well are related to $m^2_{Higgs}$ and to
$t_{ignit}$. It can be deduced from\cite{bennett7} that a large
negative value of $m^2_{Higgs}$ corresponds to the relative minimum 
$V_{eff}(\phi_{other})$ being deeper than $V_{eff}(\phi_{us})$
(in which by assumption the Universe starts off following Big Bang)
than for less negative values
of $m^2_{Higgs}$ (see Fig.~\ref{fignonloc}). It can also be argued quite
plausibly that a minimum in $V_{eff}$ at $\phi_{other}$ much deeper
than that at $\phi_{us}$ would correspond to an
early (small) $t_{ignit}$ inasmuch as the ``false'' vacuum at
$\phi_{us}$ would be very unstable. However, as the value
of the potential at $\phi_{other}$ approaches that at
$\phi_{us}$, $t_{ignit}$ becomes longer and longer
and approaches infinity as the values of $V_{eff}$ at
$\phi_{us}$ and $\phi_{other}$
become
the same. 
The development of the double well potential and
$m^2_{Higgs}$ as a function of $t_{ignit}$ is illustrated in 
Fig.~\ref{fignonloc}.
Note that the larger the difference
$|V_{eff}(\phi_{other})-V_{eff}(\phi_{us})|$
the more the realization of say $I_{fixed\;2}$ will in general depend on 
$t_{ignit}$. 
If $V_{eff}(\phi_{us})=V_{eff}(\phi_{other})$, 
$t_{ignit}$ plays no role in realizing e.g. $I_{fixed\;2}$ and the value
of $m^2_{Higgs}$ becomes independent of $t_{ignit}$.

\section{Avoiding paradoxes arising from non-locality}

In general the presence of non-locality leads to paradoxes. While
the form that the non-local action (or potential $V_{nl}$ in this discussion)
is unknown to us, we make the 4 generically representative guesses 
portrayed as the 4 non-locality curves 
in Fig.~\ref{fignonloc}. 
In particular, non-locality curves having a negative slope as a function
of $t_{ignit}$ lead to paradoxes in the following manner.
Consider the non-locality curve in Fig.~\ref{fignonloc} drawn with bold line
that is redrawn in a rotated position in Fig.~\ref{paradox}. Let us make the
assumption that  $t_{ignit}$ is large and see that this leads to a
contradiction. Assuming that $t_{ignit}$ is large, it is seen from
the non-locality function in Fig. \ref{paradox} 
(call it $m^2_{Higgs\;nl}(t_{ignit})$ to distinguish it from the 
``normal physics'' $m_{Higgs}^2(t_{ignit})$) that this
implies that the ``normal physics'' $m^2_{Higgs}$ has a large negative value.
But a large negative value of $m^2_{Higgs}$ corresponds in ``normal physics''to
a (false) vacuum at $\phi_{us}$ that is very
unstable and therefore to a very short $t_{ignit}$ corresponding to a rapid
decay to the stable vacuum at $\phi_{other}$. So the
paradox
appears: the assumption of a {\em large} $t_{ignit}$ implies a
{\em small} $t_{ignit}$. This happens because in general 
$m^2_{Higgs}(t_{ignit})\neq m^2_{Higgs\;nl}(t_{ignit})$ and is akin 
to the ``matricide''
paradox encountered for example when dealing with ``time machines''.
It is well known\cite{bennett8,bennett9,bennett10} that Nature avoids such 
paradoxes by
choosing a very clever solution in situations where these paradoxes lure.

In the case of the paradoxes that can come about due to non-locality of the
type considered here, a clever solution that avoids paradoxes is available to 
Nature in the form of the Multiple Point Principle (MPP).
The MPP solution corresponds
to the intersection of the ``normal physics'' curve and the ``non-locality
curve'' in Fig.~\ref{paradox}. 
because here the vacua at $\phi_{us}$ and $\phi_{other}$ are (essentially) 
degenerate. But at this intersection point, $m_{Higgs}^2(t_{ignit})=
m^2_{Higgs\;nl}(t_{ignit})$
so the paradox is avoided. So the paradox is avoided at the multiple point.
But at the multiple point, an intensive
parameter has its value fine-tuned for a wide range of values of the 
conjugate extensive quantity.
Fine-tuning can therefore be understood as a
consequence of Nature's way of avoiding paradoxes that can come about due to
non-locality.

It should be pointed out that the paradox-free solution corresponding to the
intersection of the two curves in Fig.~\ref{paradox} occurs for a value of
$m^2_{Higgs}$ corresponding to ``our'' vacuum at $\phi_{us}$
being very slightly unstable. The value of $m^2_{Higgs}$ corresponding to the
vacua at $\phi_{us}$ and $\phi_{other}$
being (precisely) degenerate is slightly less negative than that corresponding
to the multiple point value of $m^2_{Higgs}$ at the intersection of the
curves.
Note that 
the multiple point value of $m^2_{Higgs}$ is very insensitive to which
``guess''
we use for the non-local action. Indeed all the ``non-locality'' curves in
Fig.~\ref{fignonloc} intersect the ``normal physics'' curve at values of
$m^2_{Higgs}$ that are tightly nested together. The reason for this is
that $m^2_{Higgs}$ is a very slowly varying function of
$t_{ignit}$ as $m_{Higgs}^2(t_{ignit})$ approaches the value
corresponding to degenerate vacua. The more nearly parallel the ``normal
physics'' and the
``non-locality'' curves at the point of intersection, the less are the
(paradoxical) effects of non-locality.
For a point of intersection at values of $t_{ignit}$ sufficiently large
that (the ``normal physics'') 
$m^2_{Higgs}(t_{ignit})\approx m^2_{Higgs}(\infty)$,
the non-locality effects disappear as the curves become parallel since
both curves become independent of $t_{ignit}$.
If the curves were parallel, there would  also be the possibility  that these
do not
intersect in which case there would be  no ``miraculous solution'' that could 
avoid
the paradoxes imbued in having non-locality.

If the interval $|\phi^2_{other}$-$\phi^2_{us}|$
is large (e.g. of the order of the largest physically conceivable
scale (Planck?) if tuning is to be maximally effective)
and if $I_{fixed\;2}$ falls not too close to the ends of this interval, 
then
$t_{ignit}$ will be something of the order of half the life of the universe.
Actually, the approximate degeneracy of the vacua
$V_{eff}(\phi_{us})\approx V_{eff}(\phi_{other})$
may be characteristic of the
temperature of the post-Big Bang universe in the present epoch and
{\em not} characteristic of the high temperature that prevailed immediately
following the Big Bang.
At high temperatures,
the free energy is considerably less than the total energy if the entropy
is large enough. A phase with a large number of light particles - for
example a Coulomb-like vacuum such as the ``us'' phase in which we live -
could very plausibly be so strongly favoured at high temperatures that other
phases - for example the ``other'' vacuum - simply disappeared at the high
temperature of the universe immediately following the Big Bang.

If this
were to have depleted the universe of the phase having
$\phi_{other}$ at high temperatures,
it would indeed be difficult to
re-establish
it in a lower temperature universe even if the vacuum at
$\phi_{us}$ were to be only meta-stable and the vacuum at
$\phi_{other}$ 
were the true vacuum at the lower temperature.
Such an exchange of the true vacuum is indeed a possibility in
going to lower temperatures inasmuch as the
difference between the total energy and the free energy decreases in going to
lower temperatures. Accordingly this difference becomes less effective in
favouring a Coulomb-like phase at the expense of a phase with heavier particles.

At this point we point out that when we say the ``vacuum at $\phi_{us}$''
and ``vacuum at $\phi_{other}$'' we are thinking of the approximation 
$\phi=\phi_{us}$ and $\phi=\phi_{other}$ almost everywhere and at 
all times in respectively the early and late epochs of the universe in
our discussion. More correctly we should talk about
vacuum densities $\langle \phi(x^0,\vec{x})\rangle_{us}$ and
$\langle \phi(x^0,\vec{x})\rangle_{other}$ where

\[ \langle \phi(x^0,\vec{x})\rangle_{us}\stackrel{def}{=}\frac{1}{V_{us}}
\int_{t_{BB}}^{t_{ignit}}dx^0\int d^3\vec{x}\sqrt{g(x)}\phi(x^0,\vec{x})\]
where $V_{us}$ denotes the the 4-volume of the universe in the first epoch.
Density $\langle \phi(x^0,\vec{x})\rangle_{other}$ is defined analogously. 
The more correct 
$\langle \phi(x^0,\vec{x})\rangle_{us}$ is mentioned here so as not to 
confuse the reader
when we talk about the changes in $V_{eff}(\phi_{us}(x^0,\vec{x}))$ as the 
universe cools 
following Big Bang.
 
Recall now that the value of say
of $I_{fixed\;2}$ can easily
(i.e. as a generic possibility) assume a value that requires that
the universe to be in the ``phase'' with 
$\langle \phi(x^0,\vec{x})\rangle_{other}$
 during a sizeable part of its life if the universe is to
have multiple point parameters in the course of its evolution (as required
for avoiding the paradoxes that accompany non-locality). How can Nature
overcome the energy barrier that must be surmounted in order to bring about
the decay of the slightly unstable (false) vacuum 
with $\langle \phi(x^0,\vec{x})\rangle_{us}$ to the vacuum with
$\langle \phi(x^0,\vec{x})\rangle_{other} $? Even  producing just
a tiny ``seed'' of the ``true'' vacuum having 
$\langle \phi(x^0,\vec{x})\rangle_{other}$ would
be very difficult.
What miraculously clever means can Nature devise
so as to avoid deviations from a multiply critical evolution of the
universe?  One ingenious master plan that Nature may have
implemented is the creation of life with the express ``purpose'' of evolving 
some (super intelligent?)
physicists that could ignite a ``vacuum bomb'' by first creating in some very
expensive accelerator the required
``seed'' of the ``correct'' vacuum having 
$\langle \phi(x^0,\vec{x}) \rangle_{other}$ that
subsequently would engulf the universe in a (for us) cataclysmic transition
to the ``other'' phase thereby permitting the continued evolution of a
``paradox-free'' universe!
 
\section{Conclusion}

We attempt to justify the assertion that fine-tuning in Nature 
seems to imply a fundamental form of non-local interaction. 
This could be manifested in a phenomenologically acceptable form as everywhere 
in spacetime 
identical interactions between any pair of spacetime points. This would be 
implemented by requiring the non-local action to be diffeomorphism invariant. 

Next we put forth our multiple point principle which states that coupling
parameters in the Standard Model tend to assume values that correspond to
the values of action parameters lying at the junction of a maximum number of
regulator induced phases (e.g., so-called ``lattice artifact phases'') 
separated
from one another in action parameter space by first order transitions. 
The action
which of course is defined on a gauge group 
(e.g., the non-simple SM gauge group) governs fluctuation patterns
along the various subgroup combinations $(K,H)$ with $H \lhd  K \subseteq G$
that characterize the phases that come together at the multiple point.
We then consider extensive quantities that are functions 
of functionals $I_{f_j}[\phi(x)]$ that are essentially Feymann path histories
of the Universe for functions $f_j(\phi)$
of the fields $\phi (x)$ and derivatives of these fields. 
We then think of the generic situation in which
these extensive quantities can happen to be 
fixed 
at values  that require the universe to be realized as two or more
coexisting phases. We draw on the analogy to the forced coexistence of ice 
and liquid water that occurs for a whole range of possible total energies
because of the finite heat of melting (first order phase transition).
With our multiple point principle, the intensive quantities (couplings)
conjugate to extensive quantities fixed in this way  become fine-tuned in a 
manner analogous to the fine
tuning of temperature to $0^oC$ when the total energy of a system 
of $H_2O$ can only be realized as coexisting ice and liquid phases.
    
One generic way of having  coexisting phases in a quantum field
theory in 3+1 dimensions would be to have different phases in
different epochs of the lifetime of the Universe with phase transitions
occuring at various times in the course of the lifetime of the Universe.
If the transitions were first order, one would have fine-tuning of
(intensive) couplings conjugate to extensive quantitity values that can only be
realized by having  coexisting (i.e., more than one) phases. 
But such a fine-tuning would 
involve non-locality: the fine-tuned values of coupling constants would 
depend on future phase transitions into phases that do not even exist at
the time such couplings are fine-tuned.

Even non-locality of this sort (i.e., non-localy manifested as
a diffeomorphism invariant non-local action) can lead to paradoxes of the
``matricide paradox'' type.  We argue that such paradoxes are avoided when 
Nature chooses the multiple point principle solution to the problem of 
finetuning.

\newcommand{\Ltext}[1]{\ensuremath{\itindex{\mathcal{L}}{#1}}}
\newcommand{\diff}[1]{\mbox{d}#1}
\newcommand{\half}[1]{\ensuremath{\frac{#1}{2}}}
\newcommand{\intd}[1]{\int \!\! #1 \;}
\newcommand{\inv}[1]{\ensuremath{\frac{1}{#1}}}
\newcommand{\metrtilde}[1][]{\Tilde{g}_{\varphi \bar{\varphi} #1}}
\newcommand{\pathd}[1]{\mathcal{D} #1 \;}
\newcommand{\pathint}[1]{\int \mathcal{D} #1 \;}
\newcommand{\itindex}[2]{\ensuremath{#1_{\mbox{\scriptsize{\itshape #2}}}}} 
\newcommand{\varfrac}[2][]{\frac{\delta #1}{\delta #2}}
\title*{Dynamics of Glue-Balls in $N=1$ SYM Theory}
\author{Luzi Bergamin\footnote{bergamin@tph.tuwien.ac.at}}
\institute{%
Technische Universit\"at Wien\\
Institute of Theoretical Physics\\
Wiedner Hauptstrasse 8-10\\
1040 Wien, Austria
}

\titlerunning{Dynamics of Glue-Balls in $N=1$ SYM Theory}
\authorrunning{Luzi Bergamin}
\maketitle

\begin{abstract}
The extension of the Veneziano-Yankielowicz effective Lagrangian with terms
including covariant derivatives is discussed. This extension is important to
understand glue-ball dynamics of the theory. Though the superpotential remains
unchanged, the physical spectrum exhibits completely new properties.
\end{abstract}

\section{Introduction}
The low energy effective action of $N=1$ SYM theory is written in terms of a
chiral effective field $S = \varphi + \theta \psi + \theta^2 F$, which may be
defined from the local source extension of the SYM action
\cite{veneziano82,Shore:1983kh,burgess95,bergamin01}
\begin{align}
\label{eq:Sdef}
  S &\propto \varfrac{J} W[J, \bar{J}]\ , &
  e^{ i W[J, \bar{J}]} &= \intd{\pathd{V}} e^{i
  \intd{d^4x d^2 \theta} (J + \tau_0) \tr W^\alpha W_\alpha + \hc
  }\ .
\end{align}
With appropriate normalization $S$ is
equivalent to the anomaly multiplet $\bar{D}^{\dot{\alpha}} V_{\alpha \dot{\alpha}} =
D_\alpha S$. $J(x)$ is the chiral source
multiplet, with respect to which a Legendre transformation can be defined
\cite{burgess95,bergamin01}.
The resulting effective action is formulated in terms of the gluino condensate $\varphi
\propto \tr \lambda \lambda$, the glue-ball operators $F \propto \tr F_{\mu
  \nu} F^{\mu \nu} + i \tr F_{\mu \nu} \tilde{F}^{\mu \nu}$ and a spinor $\psi
\propto (\sigma^{\mu \nu}\lambda)_\alpha F_{\mu \nu}$. An effective Lagrangian
in terms of this effective field $S$ has the form \cite{veneziano82,Shore:1983kh}
\begin{equation}
  \label{eq:vylag}
  \Ltext{eff} = \intd{\diff{^4 \theta}} K(S, \bar{S}) - \Bigl( \intd{\diff{^2
  \theta}} S(\log \frac{S}{\Lambda^3} - 1) + \hc \Bigr)\ .
\end{equation}
The correct anomaly structure is realized by the superpotential and thus $K(S,\bar{S})$
is invariant under all symmetries.
In ref.\ \cite{veneziano82} the explicit ansatz $K = k
(\bar{S}S)^{1/3}$ had been made, which leads to chiral symmetry breaking due to $\langle S \rangle =
\Lambda^3$, but supersymmetry is not broken as  $\varphi$
and $\psi$ acquire the same mass $m = \Lambda / k$.
\section{Glue-balls and constraint K\"{a}hler geometry}
Though the spectrum found in ref.\ \cite{veneziano82} does not include any
glue-balls, such fields do appear in $F$. However, they drop out in the analysis of
\cite{veneziano82}, as $F$ is treated as an auxiliary field. Indeed, the
highest component of a chiral
superfield is auxiliary in standard SUSY non-linear\ $\sigma$-models, i.e.\ there appear no derivatives
acting onto this field and moreover its potential is not bounded from below,
but from above. In case of the Veneziano-Yankielowicz Lagrangian the part
depending on the auxiliary field reads
\begin{equation}
  \label{eq:auxlag}
  \Ltext{aux} = k (\bar{\varphi} \varphi)^{- \frac{2}{3}} \bar{F} F + \Bigl(
  \inv{3} \varphi^{-\frac{2}{3}} \bar{\varphi}^{-\frac{5}{3}} F \bar{\psi}
  \bar{\psi} - F \log\frac{\varphi}{\Lambda^3} + \hc \Bigr)\ ,
\end{equation}
and the supersymmetric spectrum is obtained, \emph{if and only if} $F$ is
eliminated by the algebraic equations of motion that follow from
\eqref{eq:auxlag}. This leads to the
unsatisfactory result that glue-balls cannot be introduced in a
straightforward way (cf.\ also \cite{farrar98}) which, in addition, contradicts
available lattice-data \cite{Peetz:2002sr}.

However, in the special case of $N=1$ SYM the elimination of $F$ is not consistent:
If $F$ is eliminated from  \eqref{eq:auxlag}, this implies that the theory
must be ultra-local in the field $F$ \emph{exactly}, i.e.\ even corrections to
the effective Lagrangian which are not included in \eqref{eq:vylag} are not
allowed to change the non-dynamical character of $F$. If this field would be
related to the fundamental auxiliary field, this restriction would be
obvious. But in $N=1$ SYM the situation is different: $S$ is the effective
field from a composite operator and $F$ is not at all related to the
fundamental auxiliary field $D$. As a consequence, the restriction of
ultra-locality on $F$ leads to an untenable constraint on the \emph{physical} glue-ball
operators (for details we refer to
\cite{bergamin01,bergamin02:1,Bergamin:2003ub}).

As shown in ref.\ \cite{Shore:1983kh}, the effective Lagrangian of
\cite{veneziano82} is not the most general expression compatible with all the
symmetries, but the constant $k$ may be generalized to a function $k(\frac{S^{1/3}}{\bar{D}^2 \bar{S}^{1/2}},
  \frac{\bar{S}^{1/3}}{D^2 S^{1/2}})$.
This non-holomorphic part automatically produces space-time derivatives onto
the field $F$, which is most easily seen when $K(S,\bar{S})$ is
rewritten in terms of two chiral fields \cite{Bergamin:2003ub}:
\begin{equation}
  \label{eq:constrkahler}
  K(S, \bar{S}) \rightarrow K(\Psi_0, \Psi_1; \bar{\Psi}_0, \bar{\Psi}_1)
\end{equation}
$\Psi_0$ and $\Psi_1$ are not independent, but they must obey the
constraints
\begin{align}
  \Psi_0 &= S^{\frac13} = \varphi^{\frac13} + \inv{3} \varphi^{-\frac23} \theta \psi +
  \inv{3} \theta^2 (\varphi^{-\frac23} F + \inv{3} \varphi^{-\frac53} \psi \psi)\ , \\
  \Psi_1 &= \bar{D}^2 \bar{\Psi}_0 = \inv{3} (\bar{\varphi}^{-\frac23} \bar{F} +
  \inv{3} \bar{\varphi}^{-\frac53} \bar{\psi} \bar{\psi}) - \frac i 3 \theta
  \sigma^\mu \partial_\mu (\bar{\varphi}^{-\frac23} \bar{\psi}) - \theta^2 \Box
  \bar{\varphi}^{\frac13}\ .
\end{align}
As $F$ appears as lowest component of $\bar{\Psi}_1$, the Lagrangian includes
a kinetic term for that field. In contrast to the situation in
\cite{veneziano82}, this is not inconsistent as the potential in $F$ may
include arbitrary powers in that field (instead of a quadratic term only) and
can be chosen to be bounded from below (instead of above). This way the field
$F$ is promoted to a usual physical field. It has been shown in
\cite{bergamin02:1} that there exist consistent models of this type. In \cite{Bergamin:2003ub} these
ideas have been applied to $N=1$ SYM, leading to an effective action of that
theory with dynamical glue-balls as part of the low-energy spectrum. Formally,
the effective potential looks the same as in the case of Veneziano and Yankielowicz:
\begin{equation}
  \label{eq:veff}
  \begin{split}
    V_{\mbox{\tiny eff}} &= - \metrtilde F \Bar{F} + \half{1}
    \metrtilde[, \bar{\varphi}] F (\Bar{\psi} \Bar{\psi}) + \half{1}
    \metrtilde[, \varphi] \Bar{F} (\psi \psi) -
    \inv{4} \metrtilde[, \varphi \Bar{\varphi}] (\psi
    \psi) (\Bar{\psi} \Bar{\psi}) \\[0.7ex]
    &\quad + c \bigl( F \log \frac{\varphi}{\Lambda^3} + \bar{F} \log \frac{\bar{\varphi}}{\bar{\Lambda}^3}
      - \inv{2 \varphi} (\psi \psi)  - \inv{2 \bar{\varphi}} (\bar{\psi}
    \bar{\psi})  \bigr)
  \end{split}
\end{equation}
However, in contrast to reference \cite{veneziano82} the K\"{a}hler
``metric''\footnote{This quantity is not equivalent to the true K\"{a}hler
  metric of the manifold spanned by $\Psi_0$ and $\Psi_1$, cf.\ \cite{Bergamin:2003ub}.}
is a function of $\varphi$ \emph{and} $F$, $\metrtilde(\varphi, F;
\bar{\varphi}, \bar{F})$. From eq.\ \eqref{eq:veff} the
consistent vacua can be derived, for explicit expressions we
refer to \cite{Bergamin:2003ub}. The most important properties of the
Lagrangian \eqref{eq:vylag} with \eqref{eq:constrkahler} are:

The effective potential is minimized with respect to \emph{all} fields
    $\varphi$, $\psi$ and $F$. Consequently, the dominant contributions that
    stabilize the potential must stem from the
    K\"{a}hler part, not from the superpotential: The superpotential is a
    holomorphic function in its fields and therefore its scalar part must have
    unstable directions. In the present context there exists no mechanism to transform
    these instabilities into stable but non-holomorphic terms.

Though the model has the same superpotential as the
Lagrangian of ref.\ \cite{veneziano82} its spectrum is completely different: Chiral symmetry breaks by a vacuum expectation value (vev) of $\varphi
        \propto \Lambda^3$, but this mechanism is more complicated than in
        \cite{veneziano82}. Any stable ground-state must have non-vanishing vev of
        $F$. But
        $\langle F\rangle$ is the order parameter of supersymmetry breaking and thus this
        symmetry is broken as well\footnote{The author of ref.\
    \cite{Shore:1983kh} concluded that this model cannot have a stable \emph{supersymmetric}
    ground-state. This is in agreement with our results, as the model breaks down as $F \rightarrow 0$.}. $\psi$ is a massless spinor, the Goldstino.

The supersymmetry breaking scenario is of essentially non-perturbative
    nature\footnote{The importance of such a breaking mechanism has been
      pointed out in \cite{bergamin01} already, but a concrete description was not yet
    found therein.}: it is not compatible with perturbative non-renormalization
    theorems, as the value of $V_{\mbox{\tiny eff}}$ in its minimum and the
    vev of ${T^\mu}_\mu$ are no longer
    equivalent. In particular, the former can be negative, while the latter is
    positively semi-definite due to the underlying current-algebra
    relations. To our knowledge this is the first model, where this type of
    supersymmetry breaking has found a concrete description (cf.\
    \cite{bergamin02:1,Bergamin:2003ub} for details).

Any ground state with $\langle \metrtilde \rangle \neq 0$ can be equipped with stable
    dynamics for $p^2 < |\Lambda|^2$. In the construction of concrete kinetic
    terms it is important to realize that \eqref{eq:constrkahler} may include
    expressions with explicit space-time derivatives. Again this is possible as
    $F$ is not interpreted as an auxiliary field. 

In summary, the Lagrangian of ref.\ \cite{Bergamin:2003ub} is the most
general one, which can be formulated in terms of the effective field
$S$. Consistent ground-states can be found together with broken supersymmetry
only. It would be interesting to compare these results with a \emph{different}
action, which has supersymmetric ground-states. But the "pi\`{e}ce de
r\'{e}sistance'' for such an action is the fact, that it cannot start from the
effective field $S$. 
\subsection*{Acknowledgements}
The author would like to thank U.~Ellwanger, J.-P.~Derendinger, E.~Kraus,
P.~Min\-kowski, Ch.~Rupp and E.~Scheidegger for interesting discussions. This
work has been supported by the Austrian Science Foundation (FWF) project P-16030-N08.


\title*{Quantization of Systems with Continuous Symmetries on 
the Classical Background: Bogoliubov Group Variables Approach} 
\author{Margarita V. Chichikina}
\institute{%
Moscow State University
Department of Quantum Field Theory and High Energy Physics
Leninskie Gory
119992 Moscow
Russian Federation\\
chich@goa.bog.msu.ru  
}

\titlerunning{Quantization of Systems ...}
\authorrunning{Margarita V. Chichikina}
\maketitle

\section{Introduction}
In present article the theory of quantization of nonlinear boson fields 
in a vicinity a nontrivial classical component by means of  the 
Bogoliubov group variables is reprezented. 
Among canonical transformations of a 
boson field always there is an separation $c $-numerical component, 
therefore operator of a boson field always potentially contains a 
classical boson field. 
 Therefore quantization of any nonlinear physical system should include 
quantization of a boson field in a vicinity of a classical background. If 
it is possible to suppose this field small, connected with 
it the effects, as a rule, are satisfactorily described in the 
framework of standart perturbation theory, when the classical 
field is as a first approximation considered equal to zero.  If it is 
impossible to consider a classical field from the very 
beginning small (gravitation and extended particles), there is a task of 
the description of properties of physical system, in which main 
effect is the separation of a classical boson field.

The theories, existing on at present moment,  of quantization on a 
classical background meet  two basic difficulties.
Firstly, --- and this main --- a problem of the conservation laws.
The second  difficulty of quantization on a classical background is the 
zero-mode problem. 
 To bypass these problems  we propose to use idea of the Bogoliubov,  which 
was proposed in work of N.N.Bogoliubov. He  has proposed to carry out 
quantization in the terms new variable.  The transformations of the 
 Bogoliubov are widely used in a quantum field theory. However, if the group 
of invariancy of system includes transformation of time, there are 
difficulties, connected with following:  for reception of expression for 
Hamiltonian  as generator of time translation  it is necessary to know 
equations of motion, and for record of equations of motion in an explisit 
form it is necessary to know expression for Hamiltonian. This difficulty 
seriously limited a  field  of application of the Bogoliubov group 
variables, since did not enable to consider non-stationary systems. 

In  this work we propose quantizations, new a way, which will allow to use 
the Bogoliubov group variables for systems having arbitrary symmetry group 
(provided that we we know representation of this group), including for 
non-stationary systems.

\section{The basic ideas of a new method of quantization}

In this section we represent schematically the basic ideas of 
quantization of boson fields by means of  method of the Bogoliubov 
group variables. 
Firstly  we pass to new variable - 
Bogoliubov group variables - and all considered 
quantities we express in the terms new variable.

Let variable $x ^ {'} $ are related to $x $  
by group transformation:

$$ x ^ {\prime} =F (x, a), \qquad x ^ {\prime \prime} =F\left ( 
F (x, a), b\right) =F (x, c), \qquad c =\varphi \left (a, b\right). $$ 

Variations of coordinates at a variation of parameters of group $a $ 
are:  $$\left ( \delta x ^ {\prime} \right) ^i =\xi _s^i (x ^ 
{\prime}) B_p^s (a) \delta a^p, $$ where $i=0,1,2,3 $- number of 
coordinate, $p=1..., r $, and $k $ - number of generators of group. 
The group properties of transformation are defined by tensor $B_p^s (a) $.

Let's define  Bogoliubov transformation as follows:
$$ f (x) =gv (x ^ {'}) + u (x ^ {'}), $$
dimensionless parameter $g $ is supposed to be large,
and $ (u (x ^ {'}), a) $ are independent new variable 
(Bogoliubov group variable). Explisitly  separated  large 
component  depends on $x ^ {'} $ as well as a quantum 
part.  Thus we restore invariancy with respect to transformations 
group, which was  broken by explisit separation of classical 
component, as it was pointed out in introduction.  However 
consideration of variable $ \tau $ as independent 
leads to the fact  that  right-hand side of our 
equation contains now on $k $ variable more than  left one.  In 
order to equalize the number of variables we impose 
on $u (x ^ {'}) $ some invariant conditions. 

Secondly  we develop the perturbation theory.   All 
integrals of motions  are expressed in new variables. Analyzing  
received expressions, we obtain conditions, at which application of 
perturbation theory is correct, namely:  the 
classical part owes to satisfy to some equations. These 
equations turn out  coinside  with Eileur-Lagrange equations. 
However this result  is  not trivial: it was received not as a 
consequence of variational principle, but as a condition of 
development of a perturbative scheme for our system in the terms new 
variables,  while for reception of these equations we did not 
need to know  Hamiltonian structure  as generator of transaltions 
on time. 

So, at the second stage we obtain equations of motion for classical 
components. 

And then  we construct of  system state space .  
At this stage of construction of our scheme we achieve correct 
number of degrees of freedom  which will be equal to 
its real number.  During realization of a reduction we
obtain equations of motion for quantum correction for our field (the 
equations  appear  nonlinear).  Also at this stage the 
explisit expressions for creation-annihilation  operators of  
quantums of a field are obtained  as well as and spectrum of 
frequencies, that is actually, energy spectrum. Besides we 
obtain expressions for integrals of  motion as generators 
of corresponding symmetry transformations:

$$ O_0=i\Im ^ \alpha {\partial \over \partial \tau ^ \alpha}, $$

In particular, hamiltonian is generator of shift on time. 
Let's note, that equations of motion of a classical part and explisit form 
of hamiltonian are received independently, that is is overcome 
basic difficulty of the description of non-stationary systems, 
which was described in introduction.

Also in zero order we shall obtain Heizenberg equation in the terms 
new variables.

Totality of the received results:
\begin{itemize}
\item Equation of motion of a classical part;
\item Equation of motion of the quantum additive;
\item Energy spectrum;
\item Expressions for integrals of a motion;
\item Heisenberg equations;
\item Explicit  expression for the  field operator;
\end{itemize}

allows to assert, that our theory gives the complete description of 
system of boson fields, quantized  on  classical background.   While 
theory guarantees exact performance of the conservation laws in any 
order of perturbation theory and  also allows to avoid a zero-mode
problem.  


\title*{Singular Compactifications and Cosmology%
\thanks{Work supported by the `Schwerpunktprogramm Stringtheorie' of the DFG.}}
\author{Laur J\"arv, Thomas Mohaupt and Frank Saueressig} 
\institute{%
Institute of Theoretical Physics, Friedrich-Schiller-University Jena,\\
Max-Wien-Platz 1, D-07743 Jena, Germany\\ 
{\tt L.Jaerv, T.Mohaupt, F.Saueressig@tpi.uni-jena.de } }

\titlerunning{Singular Compactifications and Cosmology}
\authorrunning{Laur J\"arv, Thomas Mohaupt and Frank Saueressig}
\maketitle

\begin{abstract}
\noindent We summarize our recent results of studying
five-dimensional Kasner cosmologies in a time-dependent Calabi-Yau
compactification of M-theory undergoing a topological flop
transition. The dynamics of the additional
states, which become massless at the transition point and give
rise to a scalar potential, helps to stabilize the moduli and
triggers short periods of accelerated cosmological expansion.
\end{abstract}

During the last year a lot of effort has been made to explain the astronomical evidence for an inflationary epoch of the early universe and the current modest accelerated expansion by invoking a scalar potential derived from string or M-theory compactifications. So far two
mechanisms leading to potentials viable to describe accelerated cosmological expansion have been explored
 \cite{Emp&Tow}: (i)
compactifications on hyperbolic spaces
\cite{hyper_comp}
and (ii) compactifications with fluxes \cite{flux_comp}. 
Our recent work \cite{model,cosmo} gives the first example of (iii) compactification on a singular internal manifold.

In the case of smooth compactifications one usually has a moduli space of vacua
corresponding to the deformations of the internal manifold $X$ and
the background fields. For theories with eight or less supercharges this moduli space includes special points where $X$
degenerates, rendering the corresponding low energy effective action (LEEA)
discontinuous or singular. However, within the full string or
M-theory these singularities are believed to be artifacts, which
result from ignoring some relevant modes of the theory, namely the
winding states of strings or branes around the cycles of $X$.
Singularities of $X$ arise when such cycles are contracted to zero
volume, thereby introducing additional massless states. 
Incorporating these states leads to a smooth {\em gauged} supergravity action which entails a scalar potential.

In Calabi-Yau (CY) compactifications of M-theory undergoing a topological flop transition these additional states (`transition states') are given by $N$ charged hypermultiplets which become massless at the transition locus. There are two ways to include the effect of these extra states in the LEEA. The usual LEEA is obtained by dimensional reduction on the smooth CY and contains
only states which are generically massless. The flop manifests itself in a discontinuous change
of the vector multiplet couplings at the transition locus. We call this
description the `Out-picture' since the extra states are left
out. On the contrary, the `In-picture' is obtained by including the
transition states as dynamical fields in the Lagrangian.

In \cite{model} we constructed an In-picture LEEA for a generic M-theory flop by combining knowledge about the general ${\cal N} = 2, D = 5$ gauged supergravity action with information about the extra
massless states.\footnote{This strategy was first applied in \cite{SU2} in the case of $SU(2)$ enhancement.}  While the vector multiplet sector could be treated exactly we used a toy model based on the quaternion-K\"ahler manifolds $\frac{U(1+N,2)}{U(1+N) \times U(2)}$ to describe the
hypermultiplet sector. In order to find the gauging describing the flop we
worked out the metrics, the Killing vectors, and the moment maps
of these spaces. This data enabled us to construct a unique LEEA which has all the properties to model a flop: the extra hypermultiplets acquire a mass away
from the transition locus while the universal hypermultiplet remains massless.

In~\cite{cosmo} we considered an explicit model for a CY compactification undergoing a flop with $N=1$ and investigated the effect of the transition states on five-dimensional
Kasner cosmologies,\footnote{This setup was previously considered in \cite{BL}, but there the hypermultiplet manifold was taken to be hyper-K\"ahler which is not consistent with local supersymmetry.} 
\begin{equation}
ds^2 = - d \tau^2 + e^{2 \alpha(\tau)} d \vec{x}^2 + e^{2 \beta(\tau)}
d y^2 \;.
\end{equation}
Comparing the cosmological solutions of the Out- and the In-picture, we found that the inclusion of the dynamical transition states  has  drastic consequences for 
moduli stabilization and accelerated expansion.

As soon as we allow all light states to be excited
the scalar fields no longer show the usual run-away
behavior but are attracted to the flop region
where they oscillate around the transition locus.
Thus the ``almost singular'' manifolds close to the flop 
are dynamically preferred. 
This is somewhat surprising, because the potential has still
many unlifted flat directions meaning there is no energy barrier
which prevents the system from running away.
Hence this effect cannot be predicted by just analyzing the critical
points of the superpotential.
The following thermodynamic analogy helps to explain the situation.
Generically the available energy of the system is
distributed equally among all the light modes
(``thermalization''). Thus near the flop line the additional
degrees of freedom get their natural share of it. Once this has
happened, it becomes very unlikely that the system ``finds'' the
flat directions and ``escapes'' from the flop region. Our numerical 
solutions confirm this picture:
irrespective of the initial conditions the system finally settles
down in a state where all the fields either approach
finite values or oscillate around the transition region with
comparable and small amplitudes. From time to time one sees
``fluctuations from equilibrium'', i.e., some mode picks up a
bigger share of the energy for a while, but the system eventually
thermalizes again. 
In
an ideal scenario of moduli stabilization, however, one would like to have a damped system
so that the moduli converge to fixed point values.

The second important aspect is that the scalar potential
of the In-picture induces short periods of accelerated
expansion in the three-space. Yet 
the net effect of the accelerating
periods on cosmic expansion is not very significant.
Again, this feature can be understood
qualitatively in terms of the properties of the scalar potential. 
The point is that the potential 
 is only flat along the unlifted
directions 
 while along the non-flat ones it is too steep to support sustained accelerated expansion. 
Transient periods of acceleration occur when the scalar fields pass through their collective turning point, where running ``uphill'' the potential turns into running ``downhill'' and the potential energy momentarily dominates over the kinetic energy.\footnote{This behavior is also common to the models of hyperbolic and flux compactifications, where likewise the acceleration is not pronounced enough for primordial inflation.}
To get
a considerable amount of inflation via a slow-roll mechanism, one would need to lift some of the flat
directions gently without making them too steep.

In summary we see that the dynamics of the transition states is
interesting and relevant, and can be part of the solution
of the problems of moduli stabilization and inflation. 
One direction for further investigations is
to consider more general gaugings of our five-dimensional model. 
Once gaugings leading to interesting cosmological solutions
are found, one should  clarify whether these can
be derived from string or M-theory where they correspond to adding fluxes or branes. 
Another direction is to extend our construction to other topological transitions.  In particular it would be interesting to study the effect of transition states on four-dimensional cosmologies arising, e.g., from type II compactifications on singular CY manifolds. It is conceivable that a realistic cosmology derived from sting or M-theory will have to include both the effects of fluxes and branes, and the possibility of internal manifolds becoming singular.


\newcommand{\RLeq}[1]{Eq.~(\ref{#1})}

\title*{Fundamental Physics and Lorentz Violation}
\author{Ralf Lehnert}
\institute{%
CENTRA, \'Area Departamental de F\'{\i}sica,
Universidade do Algarve\\
8000-117 Faro, Portugal\\
rlehnert@ualg.pt}

\titlerunning{Fundamental Physics and Lorentz Violation}
\authorrunning{Ralf Lehnert}
\maketitle

\begin{abstract}
The violation of Lorentz symmetry
can arise in a variety of approaches to fundamental physics.
For the description of the associated low-energy effects,
a dynamical framework known as the Standard-Model Extension
has been developed.
This talk gives a brief review of the topic
focusing on Lorentz violation through varying couplings.
\end{abstract}

\section{Introduction}
\setcounter{equation}{0}

On the one hand,
the Standard Model (SM) of particle physics
is extremely successful phenomenologically.
On the other hand,
this conference is called
{\it What Comes Beyond the Standard Model}
because it is generally believed
that the SM is really the low-energy limit
of a more fundamental theory
incorporating quantum gravity.
Experimental research in this field
faces various challenges.
They include
the expected Planck suppression of quantum-gravity signatures
and the absence of a realistic underlying framework.

A promising approach for progress
in quantum-gravity phenomenology
is the identification of relations
that satisfy three principle criteria:
they must hold exactly in known physics,
they are expected to be violated in candidate fundamental theories,
and they must be testable with ultra-high precision.
Spacetime symmetries satisfy all of these requirements.
Lorentz and CPT invariance are key features
of currently accepted fundamental physics laws,
and they are amenable to Planck-sensitivity tests.
Moreover,
Lorentz and CPT breakdown
has been suggested in a variety of approaches to fundamental physics.
We mention
low-energy emergent Lorentz symmetry \cite{np83},
strings \cite{kps},
spacetime foam \cite{ell98},
nontrivial spacetime topology \cite{klink},
loop quantum gravity \cite{amu},
noncommutative geometry \cite{chklo},
and varying couplings \cite{klp03}.
The latter of these mechanism will be discussed
in more detail in this talk.

At presently attainable energies,
Lorentz and CPT violating effects
are described
by a general extension of the SM.
The idea is to include into the SM Lagrangian
Lorentz and CPT breaking operators
of unrestricted dimensionality
only constrained by coordinate independence \cite{sme}.
This Standard-Model Extension (SME)
has provided the basis
for many investigations
placing bounds on Lorentz and CPT violation.
For the best constraints in the matter and photon sectors,
see Ref.\ \cite{bear} and Refs. \cite{cfj,km},
respectively.
Note
that certain Planck-suppressed SME operators
for Lorentz and CPT breaking
provide alternative explanations
for the baryon asymmetry in our universe \cite{bckp}
and the observed neutrino oscillations \cite{neu}.

\section{Lorentz violation through varying couplings}
\setcounter{equation}{0}

Early speculations in the subject of varying couplings
go back to Dirac's numerology \cite{lnh}.
Subsequent theoretical investigations have shown
that time-dependent couplings arise naturally
in many candidate fundamental theories \cite{theo}.
Recent observational claims
of a varying fine-structure parameter $\alpha$ \cite{webb}
have led to a renewed interest in the subject \cite{jpuzan}.

Varying couplings are associated
with spacetime-symmetry violations.
For instance,
invariance under temporal and/or spatial translations
is in general lost.
Since translations are closely interwoven
with the other spacetime transformations
in the Poincar\'e group,
one anticipates
that Lorentz symmetry
might be affected as well.
This
is best illustrated by an example.
Consider the Lagrangian ${\cal L}$
of a complex scalar $\Phi$,
and suppose a spacetime-dependent parameter $\xi(x)$
is coupled to the kinetic term:
${\cal L}\supset\xi\partial_{\mu}\Phi\partial^{\mu}\Phi^*$.
An integration by parts
yields
${\cal L}\supset-\Phi(\partial_{\mu}\xi)\partial^{\mu}\Phi^*$.
If,
for instance,
$\xi$ varies smoothly on cosmological scales,
$(\partial_\mu\xi)=k_\mu$
is essentially constant locally.
The Lagrangian
then contains a nondynamical fixed 4-vector $k_\mu$
selecting a preferred direction
in the local inertial frame
violating Lorentz symmetry.

The above example can be generalized
to other situations.
For instance,
non-scalar fields can play a role,
and Lorentz violation can arise
through coefficients like $k_\mu$
in the equations of motion
or in subsidiary conditions.
Note
that the Lorentz breaking
is independent
of the chosen reference frame:
if $k_\mu\neq 0$
in a particular set of local inertial coordinates,
$k_\mu$ is nontrivial in any coordinate system.
In the next section,
we show
that varying couplings can arise
through scalar fields acquiring
expectations values in a cosmological context.
Note,
however,
that the above argument for Lorentz violation
is independent of the mechanism
driving the variation of the coupling.

\section{Four-dimensional supergravity cosmology}
\setcounter{equation}{0}

Consider a Lagrangian ${\cal L}$ with two real scalars $A$ and $B$
and a vector $F^{\mu\nu}$:
\begin{eqnarray}
\frac{4{\cal L}}{\sqrt{g}}
&=&
\frac{{\partial_\mu A\partial^\mu A
+ \partial_\mu B\partial^\mu B}}{B^2}
- 2 R
- M F_{\mu\nu} F^{\mu\nu}
- N F_{\mu\nu} \tilde{F}^{\mu\nu} \; ,
\nonumber\\
M &=& \frac{B (A^2 + B^2 + 1)}{(1+A^2 + B^2)^2 - 4 A^2}\; ,\quad
N = \frac{A (A^2 + B^2 - 1)}{(1+A^2 + B^2)^2 - 4 A^2}\; ,\qquad
\label{RLsugra}
\end{eqnarray}
where $g^{\mu\nu}$ represents the graviton
and $g=-\det (g_{\mu\nu})$, as usual.
We have denoted the Ricci scalar by $R$,
the dual tensor is
$\tilde{F}^{\mu\nu}=\varepsilon^{\mu\nu\rho\sigma}F_{\rho\sigma}/2$,
and the gravitational coupling has been set to one.
Then,
the Lagrangian (\ref{RLsugra})
fits into the framework of the pure $N=4$ supergravity
in four spacetime dimensions.

To investigate Lagrangian (\ref{RLsugra}) in a cosmological context,
we assume a flat Friedmann-Robertson-Walker universe
and model galaxies and other fermionic matter
by including the energy-momentum tensor $T^{\mu\nu}$ of dust,
as usual.\footnote{The dust can be accommodated
into the supergravity framework,
which also contains fermions uncoupled from the scalars.}
In such a situation,
the equations of motion can be integrated analytically \cite{klp03}
yielding a nontrivial dependence of $A$ and $B$
(and thus $M$ and $N$)
on the comoving time $t$.
Comparison with the usual electrodynamics Lagrangian
in the presence of a $\theta$ angle shows
$\alpha \equiv 1/4\pi M$ and $\theta \equiv 4\pi^2 N$,
so that the fine-structure parameter
and the $\theta$ angle
acquire related time dependences in our supergravity cosmology.

If mass-type terms
${\cal L}_m=-\sqrt{g}(m_A A^2 + m_B B^2)/2$
for the scalars
are included into Lagrangian \RLeq{RLsugra},
our simple model
can match
the observed
late-time
acceleration of the cosmological expansion \cite{klp03}.
Note also
that the scalars themselves
obey Lorentz-violating dispersion relation \cite{klp03,rl03}.

\section*{Acknowledgments}
I wish to thank C.\ Froggatt,
N.\ Manko\v{c}-Bor\v{s}tnik,
and H.B.\ Nielsen
for organizing a stimulating meeting.
This work was supported in part
by the the Centro Multidisciplinar de Astrof\'{\i}sica (CENTRA),
the Funda\c{c}\~ao para a Ci\^encia e a Tecnologia (FCT),
and an ESF conference grant.

\title*{Functional Approach to Squeezed States in Non-commutative
Theories}
\author{Lubo Musongela}
\institute{%
The Abdus Salam International Centre for Theoretical Physics\\
High Energy section\\
strada costiera , 11\\
P.O.Box 586\\
34100 Trieste\\
Italy}

\titlerunning{Functional Approach to Squeezed States in Non-commutative
Theories}
\authorrunning{Lubo Musongela}
\maketitle

\section{Summary}\label{s1}

We show how the difference of structure between the classical
Poisson brackets  and the quantum commutators for the non commutative
plane  generically leads to a
harmonic oscillator whose position mean values are not strictly periodic.
We also show that no state saturates
simultaneously all the non trivial Heisenberg uncertainties in this context.

This raises the question of the nature of quasi classical states in this
 model.
We propose an extension based on a variational
principle. 

\section{Periodicity of the harmonic oscillator.}

The non commutative plane attracted interest when it was realized it could
appear in the context of string theory \cite{witten}. It is defined by the commutation relations
\begin{eqnarray*}
[ \hat x_1 , \hat x_2] = i \, \theta  \quad , \quad [ \hat x_j , \hat p_k] = i
\, \hbar \, \delta_{jk} \quad ; \quad    \theta, \hbar \quad > 0 \quad .
\end{eqnarray*}
One can realize this algebra in a simple way: 
\begin{eqnarray*}
\hat x_1 &=& i \hbar \partial_{p_1} - \frac{1}{2} \frac{\theta}{\hbar} p_2
\quad , \quad \hat x_2 = i \hbar \partial_{p_2} + \frac{1}{2}
\frac{\theta}{\hbar} p_1 \quad  \quad , \nonumber\\
 \hat p_1 &=& p_1 \quad , \quad
\hat p_2 = p_2  \quad , \quad \quad \langle \phi \vert \psi \rangle = \int
d^2 p \, \phi^*(p) \, \psi(p) \quad .
\end{eqnarray*}

The time evolution of any operator in quantum mechanics is governed by
 the  equation $ \dot{\hat A} = i \hbar^{-1}
 [  \hat H, \hat A ]  $. Solving it for the position operator $x_1$ and computing its
 mean value in an arbitrary state, one finds \cite{03}
\begin{eqnarray*}
 \langle  \hat{x}_1(t) \rangle  =  \frac{1}{\hbar (\lambda_1 - \lambda_2)
(\lambda_1 + \lambda_2)} & & ( c_1 \cos{\lambda_1 t} + s_1 \sin{\lambda_1 t}
 +
 c_2 \cos{\lambda_2 t} + s_2 \sin{\lambda_2 t} )  
 \end{eqnarray*}
 where the frequencies are given by
\begin{eqnarray*}
 \lambda_{1,2} &=& \pm \sqrt{ \frac{k}{m} + \frac{k^2 \theta^2}{2 \hbar^2} \pm
            \frac{k^{3/2} \theta}{2 \hbar^2 \sqrt{m}}  \sqrt{4 \hbar^2 + k m \theta^2}}
	    \quad  .
\end{eqnarray*}
The coefficients $c_i,s_i$ depend on the state. 
The mean value of the position $x_1$ will not be periodic unless the ratio
of the two frequencies is rational. This is at odds with the usual quantum
mechanic result and originates from the fact that here the commutator 
 $[ \hat x_1, \hat x_2 ] = i \theta $ has not  
exactly the form of  the Poisson bracket $ \{ \hat x_1, \hat x_2 \}= 0. $

\section{Conflicting equalities}

Another difference of usual Q.M and non commutative theories lies in the fact
that in the later ones, all the non trivial uncertainties can not be satisfied 
simultaneously \cite{kosi}.
Habitually, the non trivial commutators imply only canonically conjugate pairs like
$x_1,p_1$. The squeezed or coherent states saturate the corresponding bounds:
$\Delta x_i \Delta p_i = \hbar/2$. On the non commutative plane, a state saturating 
all the non trivial uncertainties would obey the equations   
\begin{eqnarray*}
\Delta x_1 \Delta x_2 &=& \frac{\theta}{2}  \Longrightarrow
{\hat a_1} \vert \psi \rangle = ( \hat x_1 + i \lambda_1 \hat x_2 + \mu_1 ) \vert \psi \rangle = 0
\quad , \nonumber\\
\Delta x_1 \Delta p_1 &=& \frac{\hbar}{2}  \Longrightarrow  {\hat a}_2  \vert
\psi \rangle  = 0  \quad , \quad
\Delta x_2 \Delta p_2 = \frac{\hbar}{2}  \Longrightarrow  {\hat a}_3  \vert
\psi \rangle  = 0 \quad ,
\end{eqnarray*}
where ${\hat a}_2$ and $ {\hat a}_2$ are similar to ${\hat a}_1$.
It is easily obtained that $
[ \hat a_2 , \hat a_3 ] \vert \psi \rangle = i \theta 
\vert \psi \rangle = 0  $ so that no state saturates simultaneously all
the three bounds. This is a second difference with usual quantum mechanics.

\section{Generalization based on a functional approach.}

In usual quantum mechanics, the squeezed states saturate the uncertainty 
relations and so realize a minimum of the functional 
$ \sum_k( \Delta x_k \Delta p_k - h/2)^2 $. We have  verified
that our proposal works in the usual theory: we recover the known gaussian
functions and, besides them, other states which can be expressed as products
of gaussians with specific hypergeometrics \cite{03}.
The generalization  we consider here also rely on a functional involving only
the non trivial commutators: 
\begin{eqnarray*}
S &=&  \hbar^2 \left( \Delta x_1 \Delta x_2 - \frac{1}{2} \theta
\right)^2 + \theta^2 \left( \Delta x_1 \Delta p_1 - \frac{1}{2}  \hbar
\right)^2
+  \theta^2 \left( \Delta x_2 \Delta p_2 - \frac{1}{2}
\hbar \right)^2   \nonumber\\ 
&+& \lambda \left( \langle \psi \vert \psi
\rangle  - 1\right)  \quad .
\end{eqnarray*}
The normalization of the state is achieved thanks to a Lagrange multiplier.

Varying the action, one obtains that the desired wave function is an
 eigenfunction of the operator
\begin{eqnarray*}
\cal{O} &=& 
  \bar{a}_1 \partial^2_{p_1} + \bar{a}_2 \partial^2_{p_2}
 +  i  \left( \bar{a}_1
\frac{\theta}{\hbar^2}  p_2 + a_3 \right)
\partial_{p_1}
+ i  \left( -\frac{\theta}{\hbar^2} \bar{a}_2 p_1 
+ a_4 \right) \partial_{p_2} \nonumber\\
 &+&    ( \bar{a}_5 p_1^2 + \bar{a}_6 p_2^2
+ a_7 p_1 + a_8 p_2 + a_9 )   \quad .
\end{eqnarray*}
with zero eigenvalue.
The  $ \bar a_i$ are constants. They are related to the physical
characteristics of the state. Some exact solutions can be found. For example, introducing the variables
$y_1,y_2$ by the relations
\begin{eqnarray*}
p_1 = a_1 y_1 - \frac{a_4}{a_2^2} \quad , \quad p_2 = a_2 y_2 + \frac{a_3}{a_1^2}
\quad  ,
\end{eqnarray*}
the operator takes the simpler form
\begin{eqnarray*}
{\cal O} &=& - \partial^2_{y_1} -  \partial^2_{y_2} +
 i t_1^2 ( - y_2 \partial_{y_1} + y_1 \partial_{y_2})
 +
  ( t_2^2  y_1^2 + t_3^2   y_2^2
+ t_4  y_1 + t_5  y_2 + t_6 )   \quad . 
\end{eqnarray*}
One in particular find Gaussian solutions
\begin{eqnarray*}
\psi(p) = N \exp{( c_1 y_1^2 + c_2 y_2^2 +c_3 y_1 y_2 + c_4 y_1 + c_5 y_2)}
\end{eqnarray*}
provided that the  relation 
$ 2 c_1 + 2 c_2 + c_4^2 + c_5^2 - t_6 = 0 \quad $ holds.
One can similarly find other solution under other
assumptions \cite{03}.

It should be stressed that our use of a functional to define meaningful states 
is similar to the one of \cite{spindel}, although the physical requirements in the
two cases are different. 

\section{Conclusions}

We have suggested an approach towards  squeezed states which
relies on a functional method involving non trivial Heisenberg uncertainty
 relations. It is a  generalization of the squeezed states of usual quantum mechanics.
We have found special solutions to the second order differential equations 
obtained on the non
 commutative plane. 

One of the crucial points which remain to be addressed is the nature of the
critical points found here. To know if these states  are maxima or minima of the
 action, one has to ressort to a second order analysis. However, as the most general
solution of the second order partial differential equations involved are not known, such
a computation cannot tell us by itself if we are in front of an absolute 
minimum. 

Another crucial question about the  states found by the
 method presented here is the  other  properties of the usual coherent  states
they may possess, like  completeness \cite{pere,coh1,coh2,coh3,coh4}. If this was the case, they might be legitimate
candidates for the definition of a physically meaningful star product
\cite{voros}. The fact that some solutions obtained 
are special functions which are solutions of Sturm-Liouville systems is
promising.


\newcommand{\MPlesssim}{\stackrel{<}{_{\scriptstyle \sim}}}
\newcommand{\MPgtrsim}{\stackrel{>}{_{\scriptstyle \sim}}}

\title*{Constraining the Curvaton Scenario}
\author{Marieke Postma}
\institute{%
The Abdus Salam ICTP, 
Strada Costiera 11, 34100 Trieste, Italy\\
postma@ictp.trieste.it}

\titlerunning{Constraining the Curvaton Scenario}
\authorrunning{Marieke Postma}
\maketitle 

\begin{abstract}
We analyse the curvaton scenario in the context of supersymmetry.
Supersymmetric theories contain many scalars, and therefore many
curvaton candidates. To obtain a scale invariant perturbation
spectrum, the curvaton mass should be small during inflation $m \ll
H$. This can be achieved by invoking symmetries, which suppress the
soft masses and non-renormalizable terms in the potential. Other
model-independent constraints on the curvaton model come from
nucleosynthesis, gravitino overproduction, and thermal damping.  The
curvaton can work for masses $m \MPgtrsim 10^4 {\,\mbox{GeV}}$, and very small
couplings (e.g. $h \MPlesssim 10^{-6}$ for $m \MPlesssim 10^8 {\,\mbox{GeV}}$).
\end{abstract}

\section{The curvaton scenario}

It is now widely believed that the early universe went through a
period of rapid expansion, called inflation.  In addition to
explaining the homogeneity and isotro\-py of the observable universe,
inflation can provide the seeds for structure formation.  This makes
models of inflation predictive, but also restrictive. The observed,
nearly scale-invariant perturbation spectrum requires very small
coupling constants and/or masses, which renders many models unnatural.
For this reason it is worthwhile to explore alternative ways of
producing density perturbations.

In the curvaton scenario, the adiabatic perturbations are not
generated by the inflaton field, but instead result from isocurvature
perturbations of some other field --- the {\it curvaton} field.  After
inflation the isocurvature perturbations have to be converted into
adiabatic ones.  Such a conversion takes place with the growth of the
curvaton energy density compared to the total energy density in the
universe.  This alternative method of producing adiabatic did not
attract much attention until recently~\cite{curv}.

The usual implementation of the curvaton scenario is the following. If
the curvaton is light with respect to the Hubble constant during
inflation, it will fluctuate freely, leading to condensate formation.
In the post-inflationary epoch the field remains effectively frozen
until the Hubble constant becomes of the order of the curvaton mass,
$H \sim m_\phi$, at which point the curvaton starts oscillating in the
potential well.  During oscillations, the curvaton acts as
non-relativistic matter, and its energy density red shifts slower than
the radiation bath. Hence, the ratio of curvaton energy density to
radiation energy density grows $\rho_\phi/\rho_\gamma \propto a$, with
$a$ the scale factor of the universe, and isocurvature perturbations
are transformed into curvature perturbations. This conversion halts
when the curvaton comes to dominate the energy density, or if this
never happens, when it decays.

It seems natural to try to embed the curvaton scenario within
supersymmetric (SUSY) theories.  SUSY theories contain many flat
directions, and therefore many possible curvaton candidates.  The
problem, however, is that during inflation SUSY is broken dynamically
and soft mass terms are generated, which are typically of the order of
the Hubble constant.  But this is no good: $m_\phi \sim H$ during
inflation leads to a large scale dependence of the produced
perturbations, in conflict with observations. A way out of this is to
invoke symmetries.  Soft mass terms are for example suppressed in
$D$-term inflation, and in no-scale supergravities.  Another
possibility is to identify the curvaton with a pseudo-Goldstone
bosons.

\section{Constraints}

There are several model independent constraints on the curvaton
scenario.  We will discuss them briefly here; see the original paper
for more details~\cite{postma}.

First of all, the curvaton scenario should give rise to the observed
spectrum of density perturbations.  Curvature perturbations ${\mathcal
R}$ of the correct magnitude are obtained for $ {\mathcal R} \approx
({f}/{3 \pi}) ({H_*}/{\phi_*}) \approx 5 \times 10^{-5}$~\cite{curv}.
Here the subscript $*$ denotes the quantity at the time observable
scales leave the horizon. Further, $f = \rho_\phi / \rho_{\rm tot}$
evaluated at the time of curvaton decay.  If the curvaton contributes
less than 1\% to the total energy density, i.e., $f < 0.01$, then the
perturbations have an unacceptable large non-Gaussianity. If during
inflation $m_\phi \ll H$ --- which is required to get a nearly scale
invariant perturbation spectrum --- quantum fluctuations of the
curvaton grow until $m_\phi^2 \langle \phi^2 \rangle \sim H^4$, with
an exponentially large coherence length.  We will assume that this
sets the initial curvaton amplitude $\phi_* \sim \sqrt{\langle \phi^2
\rangle}$.  The non-detection of tensor perturbations puts an upper
bound on the Hubble scale during inflation $H_* \MPlesssim 10^{14}
{\,\mbox{GeV}}$.  Finally, in the curvaton scenario the adiabatic density
perturbations can be accompanied by isocurvature perturbations in the
densities of the various components of the cosmic fluid.  There are
particularly strong bounds on the isocurvature perturbations in cold
dark matter.

The initial curvaton amplitude should be $\phi_0 \MPlesssim {M_{\rm P}}$, to
avoid a period of curvaton driven inflation. Stronger constraints are
obtained if non-renormalizable operators are taken into account: $
V_{\rm NR} = {|\lambda|^2}{{M_{\rm P}}^{-n}} \phi^{4+n} $.
Non-renormalizable operators are unimportant for small enough masses,
$m_\phi \MPlesssim m_{\rm eff} = {V_{\rm eff}}''$.  For larger masses,
the curvaton slow-rolls in the non-renormalizable potential during and
after inflation.  In the post-inflationary epoch this leads to a huge
damping of the fluctuations, making it is impossible to obtain the
observed density contrast within the context of the curvaton
scenario~\cite{damping}.

The curvaton scenario should not alter the succesful predictions of
big bang nucleosynthesis (BBN).  This implies that the curvaton should
decay before the temperature drops below ${\,\mbox{MeV}}$.  To avoid gravitino
overproduction requires a reheat temperature $T_{\rm R} \MPlesssim 10^9
{\,\mbox{GeV}}$.  This also constrains the curvaton scenario, since isocurvature
perturbations are converted in adiabatic perturbations only after
inflaton decay.

Finally, one should take into account various thermal effects.  Large
thermal masses, $m_{\rm th} \MPgtrsim m_\phi$, induce early
oscillations, which are generically fatal for the curvaton scenario.
In addition, thermal evaporation of the curvaton scenario should be
avoided.

\begin{figure}
\centering
\includegraphics*[width=11cm,bbllx=10pt,bblly=10pt,bburx=290pt,bbury=230pt]{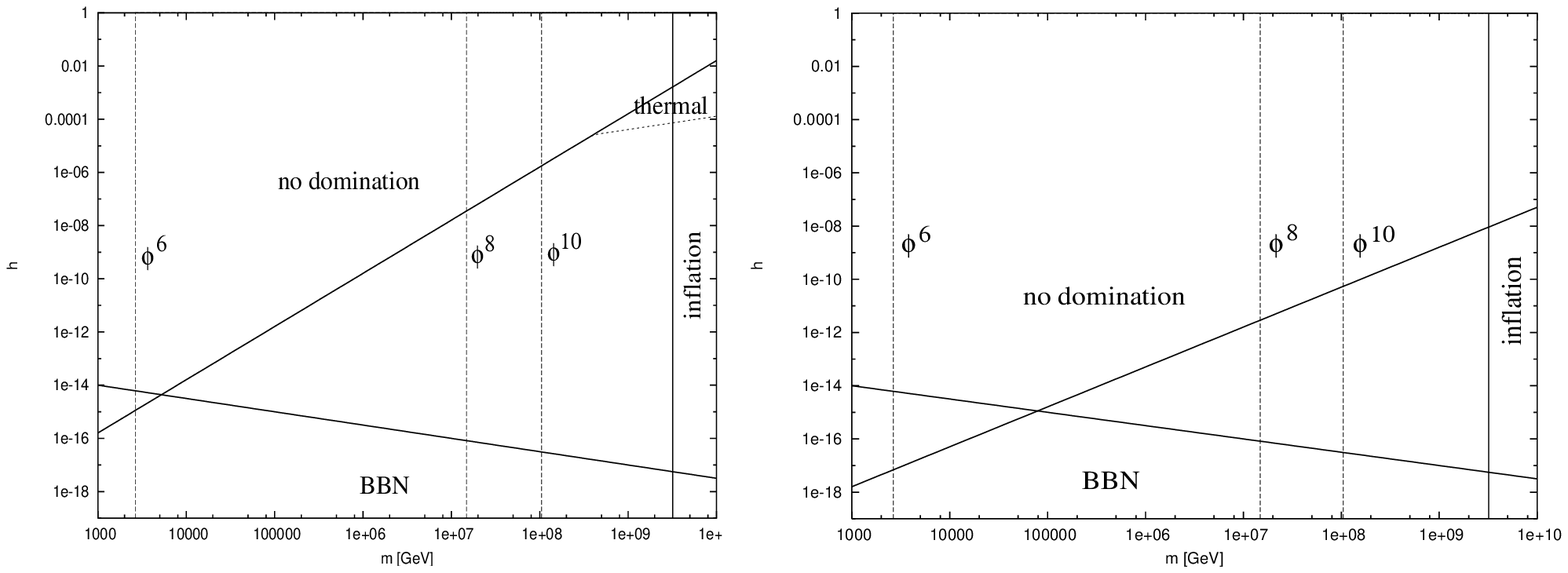}
\caption{Parameter space for curvaton domination ($f \MPgtrsim 0.5$).
In this plot $T_{\rm R}$ is unconstrained. The constraints from BBN,
domination, non-renormalizable terms, $\phi$-dominated inflation, and
thermal damping are shown.
\label{dom}}
\end{figure}

\begin{figure}
\centering
\includegraphics*[width=11cm,bbllx=290pt,bblly=10pt,bburx=600pt,bbury=230pt]{MPostma1.eps}
\caption{Parameter space for curvaton domination ($f \MPgtrsim 0.5$).
In this plot $T_{\rm R} \MPlesssim 10^9 {\,\mbox{GeV}}$. The constraints 
from BBN, domination, non-renormalizable terms, $\phi$-dominated inflation, and
thermal damping are shown.
\label{domR}}
\end{figure}

\section{Results}
Figs.~1 and~2 shows the parameter space for {\it curvaton domination}: $f
\MPgtrsim 0.5$.  In the figure~\ref{dom} the reheating temperature is
arbitrary high, whereas in the figure~\ref{domR} the gravitino
constraint is taken into account. In all parameter space $m_\phi \sim
10^{-4} H_*$.  Models with $V_{\rm NR} \sim \phi^{4+n}/{M_{\rm P}}^n$ and $n
\leq 2$ are ruled out.  For higher values of $n$, the curvaton
scenario can be succesfull for curvaton masses in the range $10^4 {\,\mbox{GeV}}
\MPlesssim m_\phi \MPlesssim 10^9 {\,\mbox{GeV}}$. Couplings have to be small $h
\MPlesssim 10^{-6}$, even $h \MPlesssim 10^{-10}$ if the gravitino
constraint is taken into account.  Thermal effects can be neglected
for such small couplings.

One can ask whether there are any natural canditates for the curvaton.
Moduli and other fields with only Planck suppressed couplings
generically decay after big bang nucleosynthesis, thereby spoiling its
succesfull predicitions.  MSSM flat directions typically have too
small masses and too large couplings to play the r\^ole of the
curvaton.  Better curvaton candidates are the right-handed sneutrino
and the Peccei-Quinn axion, which can have large masses and small
couplings.  In all cases though, considerable tuning of parameters is
needed.

\section*{Acknowledgments}
This work was supported by the European Union under the RTN contract
HPRN-CT-2000-00152 Supersymmetry in the Early Universe.

\title*{$D$-Branes and Unitarity of Noncommutative Field Theories}
\author{Alessandro Torrielli}
\institute{%
Dipartimento di Fisica ``Galileo Galilei''\\
Universit\`a di Padova\\
I.N.F.N. Sezione di Padova\\
torriell@pd.infn.it}

\titlerunning{$D$-Branes and Unitarity of Noncommutative Field Theories}
\authorrunning{Alessandro Torrielli}
\maketitle

\begin{abstract}
We review the original result we have obtained in the analysis of the breaking of perturbative unitarity in space-time noncommutative field theories in the light of their relations to D-branes in electric backgrounds
\end{abstract}

Since the seminal work of Seiberg and Witten\cite{sw} it was realized that open strings in the presence of an antisymmetric constant background are effectively described at low energy by certain noncommutative field theories, identified by a precise set of Feynman rules\cite{f}. The derivation by Seiberg and Witten was originally made for magnetic backgrounds only, corresponding to space-space noncommutativity. For electric backgrounds it is well known that problems are present when the electric field approaches a critical value $E_{cr}$, beyond which the string develops a classical instability\cite{b,bp,sst1}: tachyonic masses appear in the spectrum both for neutral (which is the case we will consider here) and charged open strings, connected with the vanishing of the effective tension and an uncontrolled growth of the oscillation amplitude of the modes in the direction parallel to the field. For purely charged strings this phenomenon (which has no analog in particle field theory) coexists with the quantum instability due to pair production which is the analog of the Schwinger phenomenon in particle electrodynamics\footnote{As a note, we recall that for neutral strings ($+q$-charge on one end, $-q$ on the other), one has $E_{cr}=1/(2\pi \alpha' |q|)$. For charges $q_1\neq q_2$ on the two boundaries, one finds that the pair production rate diverges at the same critical value of the classical instability $E_{cr} = 1/(2\pi \alpha' |max q_i |)$}. When one tries to perform the Seiberg-Witten zero-slope limit to noncommutative field theories one reaches a point in which the ratio between the electric field and its critical value becomes greater than one, and the string enters the region of instability. We will show that precisely at this point a flip-mechanism produces the appearance of the tachyonic branch cut due to the closed string sector in the non-planar diagrams, which is responsible for the lack of unitarity of the limit amplitude\footnote{Space-time noncommutative field theories present besides a series of exotic behaviours\cite{sst2,br}}. We relate the breaking of perturbative unitarity of these space-time noncommutative theories with the fact that they coincide with those obtained by conveying string theory in an unstable vacuum. This vacuum is likely to decay in the full string theory to a suitable configuration of branes. 

We take as action for a bosonic open string attached to a D-brane lying in the first $p+1$ dimensions with an antisymmetric constant background on its worldvolume the following
\begin{eqnarray}
\label{1}
S=\frac{1}{4\pi {\alpha}'}\int_{C_2}d^2 z \, (g_{ij} {\partial}_a X^i {\partial}^a X^j - 2 i \pi {\alpha}' B_{ij} {\epsilon}^{ab}{\partial}_a X^i {\partial}_b X^j).     
\end{eqnarray}
At one loop the string world-sheet is the cylinder $C_2 = \{0\leq \Re w\leq 1, w \equiv  w + 2 i \tau$\}. 

If one sets $w=x+iy$, the relevant propagator on the boundary of the cilinder ($x=0,1$) can be written as\cite{ad}
\begin{eqnarray}
\label{2}
G(y,y')={{1}\over{2}}{\alpha}' g^{-1} \log q - 2 {\alpha}' G^{-1} \log \Big[{{q^{{1}\over {4}}}\over {D(\tau)}} \, {\vartheta_4}({{|y-y'|}\over {2 \tau}}, {{i}\over {\tau}})\Big],\, \, \, x\neq x', 
\end{eqnarray}
\begin{eqnarray}
\label{3}
G(y,y')={{\pm i \theta }\over{2}} {\epsilon}_{\perp} (y-y') - 2{\alpha}' G^{-1} \log \Big[{{1}\over {D(\tau)}} {\vartheta_1}({{|y-y'|}\over {2 \tau}}, {{i}\over {\tau}})\Big],\, \, \, x=x',
\end{eqnarray}  
where $q=e^{-{{\pi}\over{\tau}}}$, $\pm$ correspond to $x=1$ and $x=0$ respectively, and ${\epsilon}_{\perp} (y) = sign(y) - {{y}\over{\tau}}$.
The open string parameters are as in \cite{sw}. With this propagator and the suitable modular measure, the non-planar two-point function can be written as follows: 
\begin{eqnarray}
\label{7}
A_{1.2}&=&{\cal{N}} {G_s}^2 \, 2^{{3 d} \over 2}{\pi}^{{{3 d}\over{2}}-2} {\alpha '}^{{{d}\over{2}}-3} \int_0^{\infty} dt \, t^{1-{{d}\over{2}}} \, {\Big[ \eta({{i t}\over{2\pi {\alpha '}}})\Big] }^{2-d}\times \nonumber \\
&&\, e^{- {{{\pi}^2}{\alpha '}^2 \over {t}} kg^{-1} k} 
\int_0^1 d\nu {\Bigg[{{e^{-{{{\pi}^2}{\alpha '}\over{2t}}} {\vartheta_4}(\nu, {{2 \pi i {\alpha '}}\over {t}})} \over { {{2 \pi{\alpha '}}\over{t}} {[\eta({2\pi i {\alpha '}\over {t}})]}^3 }}\Bigg]}^{- 2 {\alpha '} k G^{-1} k}.
\end{eqnarray}
Here ${\cal{N}}$ is the normalization constant, $d=p+1$, $G_s$ is the open string coupling constant. We have omitted the traces of the Chan-Paton matrices.

This is a form suitable for the field theory limit. We perform the zero-slope limit of Seiberg and Witten, which consists in sending ${\alpha}' \to 0$ keeping $\theta$ and $G$ fixed (in this case we will keep $t$ and $\nu$ fixed as well). This can be done setting ${\alpha}' \sim {\epsilon}^{1\over 2}$ and the closed string metric $g\sim \epsilon$, and then sending $\epsilon \to 0$ \cite{sw}. One obtains \cite{ad}
\begin{eqnarray}
\label{8}
A_{1.2}^{lim} ={\cal{N}} 2^{{3 d} \over 2}{\pi}^{{{3 d}\over{2}}-2} {g_f}^2 \int_0^{\infty} dt t^{1-{{d}\over{2}}} e^{- m^2 t \, + \, k\theta G\theta k/4 t}\int_0^1 d\nu e^{- t\, \nu (1 - \nu )\, k G^{-1} k}.
\end{eqnarray}
This reproduces the expression for the two-point function in the noncommutative $\phi^3$ theory\cite{dn,s,tesi} with coupling constant $g_f = G_s \, {{\alpha}'}^{{{d-6}\over {4}}}$. We choose now $d=2$. This case is peculiar, since the tachyon mass ${m^2} = {{2-d}\over {24 {\alpha}'}}$ goes to zero. In two dimensions the background is an electric one and noncommutativity is necessarily space-time. 

In order to interpret the field theory analysis, one continues the expression (\ref{7}) in the complex variable $k^2=k G^{-1} k$. It has a branch cut driven by the small-$t$ (closed-channel) behavior 
\begin{eqnarray}
\label{bigt}
\exp \Big( - k^2 \, {{{\pi}^2 {{\alpha '}^2 \, [1 - {(E / E_{cr})}^2 ]}}\over {t}}\Big).
\end{eqnarray}
The field-theory limit retains in turn another branch cut $Re [S] = Re [- k^2] > 0$ from the large-$t$ behaviour
\begin{eqnarray}
\label{smallt}
\exp \Big(- k^2 \, t \, \nu (1 - \nu)\Big).
\end{eqnarray}
$E_{cr} ={{g}\over{2 \pi {\alpha '}}}$ is the critical value of the electric field\footnote{We have taken unitary charges (see (\ref{1})) and a closed string metric $g_{\mu \nu } = g \, {\eta }_{\mu \nu }$}. The small-$t$ cut is on the physical side $Re [S] = Re [- k^2] > 0$. But we see that it is driven by a quantity which changes sign if the electric field overcomes its critical value, that is when the string enters the classical instability region. This is exactly what the Seiberg-Witten limit produces, since it scales ${(E / E_{cr})}^2 \sim {(\alpha'/g)}^2 \sim 1 / \epsilon$. In the field theory limit the branch cut flips to the unphysical side\cite{mio}, and in fact the amplitude (\ref{8}) for $m^2 = 0$ has two branch cuts, one of which is physical, the other one tachyonic, coalescing at the origin. By treating $m^2$ as a free parameter independent of $d$ and by taking it positive\footnote{This is the point of view of many works by Di Vecchia et al. in deriving field theory amplitudes from the bosonic string, and it is also the approach usually adopted in the noncommutative case}, one can see the shift of (\ref{bigt}) to a cut starting from $S > 4 m^2$ which gives the physical pair-production cut, and a closed branch cut  starting at $S > 4 m^2 / [1 - {({{E}/{E_{cr}}})}^2]$, which flips in the limit. 

\section*{Acknowledgments}
I wish to thank of heart Antonio Bassetto and Roberto Valandro. I wish to thank also L. Bonora and R. Russo for useful discussions, and C.-S. Chu, J. Gomis, S. Kar and M. Tonin for suggestions. Finally I wish to thank the organizers of the Euresco conference in Portoroz for such a beatiful and inspiring meeting.

\title*{Spinorial Cohomology and Supersymmetry}
\author{Dimitrios Tsimpis}
\institute{%
Department of Mathematics\\
King's College\\
The Strand\\
London WC2R 2LS\\
United Kingdom} 

\titlerunning{Spinorial Cohomology and Supersymmetry}
\authorrunning{Dimitrios Tsimpis}
\maketitle

\begin{abstract}
We review recently developed cohomological methods 
relating to the study of supersymmetric theories 
and their deformations.
\end{abstract}


String theory reduces at low energies 
to ordinary field theory supplemented by an infinite tower of 
higher-order curvature (derivative) corrections organised
in a double-series expansion in $\alpha'$ and $g_{string}$.
These corrections can in principle be derived 
by string-theoretical perturbative methods, but in certain cases 
supersymmetry alone is restrictive enough to detrmine them.

The question is more sharply posed
in the case of M-theory (see \cite{duff} for a review) whose 
low-energy effective
field-theory limit is captured 
by ordinary eleven-dimensional supergravity \cite{Cremmer:1978km}.
In order to gain insight into the nature of M-theory, one needs to
go beyond this limiting approximation; for example, by including
supersymmetric higher-order curvature 
corrections. In the absence of an
underlying perturbative analogue of string theory such corrections
cannot be found systematically, even in principle, but one might 
hope that supersymmetry would be sufficient to determine 
at least the first-order correction.

Higher-order corrections 
have far-reaching implications to a variety 
of physical problems at the heart of our 
understanding of M-theory: 
they can be used to derive modifications 
to macroscopic black hole
properties such as the entropy-area formula
and to test duality conjectures,
like the AdS/CFT correspondence,
beyond the leading-order approximation; they 
may give a mechanism 
for moduli stabilisation and 
they may 
provide a way to bypass no-go theorems concerning dS vacua.
However, even the first such deformation --corresponding 
to canonical mass dimension six operators ($R^4$)-- 
has proven notoriously difficult to determine.

The construction of the complete
$R^4$ terms in type II string theory or in M-theory 
remains an open problem, see 
\cite{Peeters:2000qj} for a review and
further references. A recent attempt was made in
\cite{deHaro:2002vk} to construct an $R^4$ action in type IIB
based on a chiral measure in on-shell IIB superspace, but it was
subsequently observed that such a measure does not exist
\cite{bh}. A completely different approach was taken by the
authors of \cite{Peeters:2000qj} who used partial results from
type II string theory and attempted to lift them to eleven
dimensions, but this proved to be  too difficult to carry out
completely.

Recently-developed techniques relating to the study of supersymmetric
theories, which go under the name of spinorial cohomology, 
offer perhaps 
the most promising and comprehensive approach 
to tackling these long-standing issues.
Spinorial cohomology, which is the subject 
of this brief review, was introduced in \cite{cnta}
and was further elaborated in \cite{cntc}. 
It provides a powerful and systematic way of
organising the field content and the possible deformations of
supersymmetric theories, as well as explicitly determining higher-order
curvature corrections. Pure-spinor cohomology \cite{paul}, 
which was recently used by Berkovits 
in a covariant approach to string theory \cite{berk,Berkovits:2002uc}, 
has been shown 
to be isomorphic to the concept of spinorial-cohomology 
with unrestricted coefficients \cite{h,ht}.

Spinorial-cohomology has already found applications 
in a variety of contexts; in
studies of ten-dimensional maximally-supersymmetric Yang-Mills 
theories \cite{cnta,cntb,cntd}, 
eleven-dimensional supergravity \cite{cntc,ht,cgnt}, 
and the world-volume effective
theory of the M2 brane \cite{hklt}. In \cite{cntb,hklt} 
the first-order curvature
correction to $D=10$, $N=1$ 
SYM and to the M2-brane action respectively, was
given explicitly. In \cite{ht} 
a method was outlined which makes it possible to
read off the first deformation of eleven-dimensional supergravity from the
five-brane anomaly-cancelling term. Moreover, it was argued that the
supersymmetric completion of this term is the unique anomaly-cancelling
invariant at this dimension which is at least quartic in the fields.

In order to give a geometrical definition of
spinorial cohomology (see \cite{ht} for
a more detailed discussion), we will suppose that we
have the usual machinery of supergravity, i.e. Lorentzian
structure group, connection, torsion and curvature. In particular
we suppose that the tangent bundle is a direct sum of the odd
and even bundles. 
The space of forms admits a natural bigrading according to the
degrees of the forms and their Grassmann character. The space of
forms with $p$ even and $q$ odd components is denoted by
$\Omega^{p,q}$.
The exterior derivative $d$ maps $\Omega^{p,q}$ to
$\Omega^{p+1,q}+\Omega^{p,q+1} +\Omega^{p-1,q+2}+\Omega^{p+2,q-1}$. Following
\cite{Bonora:mt} we split $d$ into its various components with
respect to the bigrading
 \begin{equation}
 d=d_0 + d_1 + \tau_0 + \tau_1~,
 \label{3.3}
 \end{equation}
where $d_0 (d_1)$ is the even (odd) derivative with bidegrees
$(1,0)$ and $(0,1)$ respectively, while $\tau_0$ and $\tau_1$ have
bidegrees $(-1,2)$ and $(2,-1)$. These two latter operators are
purely algebraic and involve the dimension-zero and
dimension-three-halves components of the torsion tensor
respectively.  The fact that $d^2=0$ implies in particular
that $\tau_0^2=0$. We can therefore consider the cohomology of
$\tau_0$ and set
 \begin{equation}
 H^{p,q}_{\tau}=\{\omega\in \Omega^{p,q}|\tau_0\omega=0\ {\rm mod}\ \omega=\tau_0 \lambda,
 \lambda\in \Omega^{p+1,q-2}\}~.
 \label{3.10}
 \end{equation}
We can now define a spinorial derivative $d_F$ which will act on
elements of $H^{p,q}_{\tau}$. If $\omega\in[\omega]\in H^{p,q}_{\tau}$ we set
 \begin{equation}
 d_F[\omega]:=[d_1 \omega]~.
 \label{3.11}
 \end{equation}
It is easy to check that this is well-defined, i.e. $d_1\omega$ is
$\tau_0$-closed, and $d_F [\omega]$ is independent of the choice of
representative. Furthermore it is simple to check that $d_F^2=0$.
This means that we can define the spinorial cohomology groups
\begin{equation}
H^{p,q}_F:=H^{p,q}(d_F|H_{\tau})~,
\end{equation}
in the obvious fashion.

To illustrate the meaning of the cohomology groups 
let us now turn to
$D=10$, $N=1$ SYM. The two lowest-dimensional components of the
Bianchi identity are
 \begin{eqnarray}
 d_1 F_{0,2} + \tau_0 F_{1,1}&=&0 \\
 d_1 F_{1,1} + \tau_0 F_{2,0}&=&0~.
 \end{eqnarray}
The ordinary (undeformed) theory is obtained
by setting $[F_{0,2}]=0$ and this is equivalent to specifying an
element of $H^{0,1}_F$. Hence we can set
$F_{0,2}=0$ in which case $F_{1,1}$ determines an element of
$H^{1,1}_F$. One can easily see that this cohomology
group is a spinor superfield $\lambda^\alpha$ satisfying the constraint
$D_\alpha \lambda^\beta = (\gamma^{ab})_\alpha{}^\beta F_{ab}$, i.e. it is the on-shell
field strength supermultiplet. 

If we relax the ordinary constrain $F_{\alpha\beta}=0$ 
by introducing a current $J_{\alpha\beta}$ on the right-hand side,
then the latter must be spinorially closed by virtue of the
Bianchi identity. On the other hand, if it is trivial, the
connection can be redefined to regain the original equations of
motion. This implies that $H^{0,2}_F$ describes the currents of
the theory (which are in one-to-one correspondence with the
anti-fields). If we are interested in looking at
deformations of the theory, we need to consider the
same cohomology group but with the coefficients restricted to
be tensorial functions of the field strength superfield $\lambda$ and
its derivatives. We denote this group by $H^{0,2}_F(phys)$. The
two groups are quite different; the former is dual to the physical
fields while the latter describes a composite multiplet in the
theory. The group $H^{0,2}_F(phys)$ has been explicitly computed 
in \cite{cnta} for the nonabelian SYM at order $\alpha'^2$ to
give the most general supersymmetric action at this order \cite{cntb}.
It has also been computed for the abelian SYM at 
order $\alpha'^3$ to prove rigorously the absence of any supersymmetric 
deformation \cite{cntd}.

Eleven-dimensional supergravity can also be understood in terms of
spinorial cohomolgy groups \cite{cntc,ht}. 
Apart from a purely geometrical
formulation in terms of the supertorsion \cite{hw}, 
the theory admits a formulation in terms of a closed four-form 
in superspace.
From this point of view, physical fields 
are elements of $H^{0,3}_F$, while deformations are parametrized
by the $H^{0,4}_F(phys)$ spinorial-cohomology group \cite{cntc,ht}.

One can also give a description in terms of a
`dual' super four-form obeying a deformed Bianchi identity
\cite{Duff:1995wd,cgnn}
 \begin{equation}
 dG_7=\half G_4^2 + \beta X_8~,
 \label{5.3}
 \end{equation}
where $\beta$ is a parameter of dimension $\ell^6$ and
$X_8=\tr(R^4)-{1\over4}(\tr(R^2))^2$. 
This equation uses imput from M-theory, 
namely the five-brane anomaly cancellation condition, and it 
can be used to read off the  
deformation consistent with $X_8$. One can
then make contact  with the four-form formulation
by solving the BI's up to dimension -1 \cite{ht}. 
 
The solution of the supertorsion Bianchi identities of eleven-dimensional
supergravity will be presented in \cite{cgnt}, completing the
work initiated in \cite{cgnn}. 
Moreover, provided a nontrivial element of the 
spinorial-cohomology group $H^{0,4}_F(phys)$
has been determined through the procedure described above, 
the information can be fed in the results of 
\cite{ht,cgnt} 
to give the explicit expression of the deformed equations of motion 
--and therefore the first-order ($R^4$) correction to the Lagrangian-- of
eleven-dimensional supergravity. This procedure, albeit tedious, is
straightforward and can be carried out in practice.

\section*{Acknowledgements}
I would like to acknowledge the 
support of EU contract HPRN-2000-00122 and
PPARC grants PPA/G/S/1998/00613 and PPA/G/O/2000/00451.


\backmatter
\thispagestyle{empty}
\parindent=0pt
\begin{flushleft}
\mbox{}
\vfill
\vrule height 1pt width \textwidth depth 0pt
{\parskip 6pt

{\sc Blejske Delavnice Iz Fizike, \ \ Letnik~4, \v{s}t. 2--3,} 
\ \ \ \ ISSN 1580--4992

{\sc Bled Workshops in Physics, \ \  Vol.~4, No.~2--3}

\bigskip

Zbornik konference `Euroconference on Symmetries Beyond the Standard Model', Portoro\v z, 12. -- 17. julij 2003

Proceedings to the 'Euroconference on Symmetries Beyond the Standard Model', 
Portoro\v z, July 12. -- 17., 2003

\bigskip

Uredili Norma Manko\v c Bor\v stnik, Holger Bech Nielsen, 
Colin D. Froggatt in Dragan Lukman 

Publikacijo sofinancira Ministrstvo za \v solstvo, znanost in \v sport 

Tehni\v{c}na urednica Andreja \v Cas

\bigskip

Zalo\v{z}ilo: DMFA -- zalo\v{z}ni\v{s}tvo, Jadranska 19,
1000 Ljubljana, Slovenija

Natisnila Tiskarna MIGRAF v nakladi 100 izvodov

\bigskip

Publikacija DMFA \v{s}tevilka 1551

\vrule height 1pt width \textwidth depth 0pt}
\end{flushleft}


\end{document}